\pdfoutput=1

\documentclass[11pt,twoside,a4paper,cmspaper,final,collab]{cms-tdr}
\def\svnVersion{194191}\def\cmsCernNo{2013-035}\def\cmsCernDate{\today}\def\cmsMessage{Published in the Journal of High Energy Physics as \href{http://dx.doi.org/10.1007/JHEP06(2013)081}{\doi{10.1007/JHEP06(2013)081.}}}

\begin{document}\cmsNoteHeader{HIG-12-036}

\hyphenation{had-ron-i-za-tion}
\hyphenation{cal-or-i-me-ter}
\hyphenation{de-vices}

\RCS$Revision: 186834 $
\RCS$HeadURL: svn+ssh://svn.cern.ch/reps/tdr2/papers/HIG-12-036/trunk/HIG-12-036.tex $
\RCS$Id: HIG-12-036.tex 186834 2013-05-21 08:20:19Z chiara $

\newlength\cmsFigWidth
\ifthenelse{\boolean{cms@external}}{\setlength\cmsFigWidth{0.95\columnwidth}}%
  {\setlength\cmsFigWidth{0.4\textwidth}}
\newlength\cmsFigWideWidth
\ifthenelse{\boolean{cms@external}}{\setlength\cmsFigWideWidth{0.95\columnwidth}}%
  {\setlength\cmsFigWideWidth{0.6\textwidth}}
\ifthenelse{\boolean{cms@external}}{\providecommand{\cmsLeft}{top}}{\providecommand{\cmsLeft}{left}}
\ifthenelse{\boolean{cms@external}}{\providecommand{\cmsRight}{bottom}}%
  {\providecommand{\cmsRight}{right}}
\ifthenelse{\boolean{cms@external}}{\providecommand{\color}[1]{\relax}}{}
\newcommand{\re}{\ensuremath{\cmsSymbolFace{e}}}

\providecommand{\Mjj}{\ensuremath{\mathit{m}(\mathrm{jj})}}
\providecommand{\ZllH}{\ensuremath{\cPZ(\ell\ell)\PH}}
\providecommand{\WlnH}{\ensuremath{\PW(\ell\cPgn)\PH}}
\providecommand\WtoLN {\ensuremath{\PW\to\ell\cPgn}}
\providecommand{\ZtoNN}{\ensuremath{\cPZ\to\cPgn\cPagn}}
\providecommand{\ZtoLL}{\ensuremath{\cPZ\to\ell\ell}}
\providecommand{\HBB}{\ensuremath{\PH\to \cPqb\cPqb}}
\providecommand{\ptjj}{\ensuremath{{\pt}(\mathrm{jj})}}
\providecommand{\dRJJ}{\ensuremath{\Delta R(\mathrm{j_1,j_2})}}
\providecommand{\dEtaJJ}{\ensuremath{|\Delta \eta(\mathrm{jj})|}}
\providecommand\dphiVH {\ensuremath{\Delta\phi(\mathrm{V,}\PH)}}
\providecommand{\dphiMJ}{\ensuremath{\Delta\phi(\mathrm{\MET,j})}}
\providecommand{\Naj}{\ensuremath{N_{\mathrm{aj}}}}
\providecommand{\Nal}{\ensuremath{N_{\mathrm{al}}}}
\cmsNoteHeader{HIG-12-036} 
\title{Observation of a new boson with mass near  125\GeV in pp collisions at $\sqrt{s} = 7$ and 8\TeV}

\date{\today}

\abstract{
A detailed description is reported of the analysis used by
the CMS Collaboration in the search for the standard model
Higgs boson in pp collisions at the LHC, which led to the
observation of a new boson.
The data sample corresponds to integrated luminosities
up to 5.1\fbinv at $\sqrt{s} = 7\TeV$, and up to 5.3\fbinv at
$\sqrt{s} = 8\TeV.$ The results for five Higgs boson decay
modes $\gamma \gamma$, $\cPZ\cPZ$, $\PW\PW$, $\tau \tau$, and $\cPqb\cPqb$, which show a
combined local significance of 5 standard deviations near 125\GeV, are reviewed.
A fit to the invariant mass of the two high resolution channels, $\gamma \gamma$ and $\cPZ\cPZ \to 4\ell$,
gives a mass estimate of $125.3\pm0.4\stat\pm0.5\syst\GeV$.
The measurements are interpreted in
the context of the standard model Lagrangian for the scalar Higgs
field interacting with fermions and vector bosons.
The measured values of the corresponding
couplings are compared to the standard model predictions.
The hypothesis of custodial symmetry is tested
through the measurement of the ratio of the couplings to
the $\PW$ and $\cPZ$ bosons.
All the results are consistent, within their uncertainties, with the
expectations for a standard model Higgs boson.
}

\hypersetup{%
pdfauthor={CMS Collaboration},%
pdftitle={Observation of a new boson with mass near  125 GeV in pp collisions at sqrt(s) = 7 and 8 TeV},%
pdfsubject={CMS},%
pdfkeywords={CMS, physics, Higgs}}

\maketitle                        

\newcommand{\MHmax}{{\color{black}145}}    

\newcommand{\ExpNFL}{{\color{black}110.0}}   
\newcommand{\ExpNFH}{{\color{black}145.0}}   
\newcommand{\ObsNFL}{{\color{black}121.5}} 
\newcommand{\ObsNFH}{{\color{black}128.0}}   

\newcommand{\ObsOneNNL}{{\color{black}110.0}}     
\newcommand{\ObsOneNNH}{{\color{black}111.5}}   
\newcommand{\ObsTwoNNL}{{\color{black}113.5}}   
\newcommand{\ObsTwoNNH}{{\color{black}121.0}}     
\newcommand{\ObsThreeNNL}{{\color{black}128.5}} 
\newcommand{\ObsThreeNNH}{{\color{black}145.0}}   

\newcommand{\MinLocalP}{{\color{black}$3 \times 10^{-7}$}}          
\newcommand{\MaxLocalZ}{{\color{black}5.0}}  
\newcommand{\Zgamgam}{{\color{black}4.1}}
\newcommand{\Zfourlepton}{{\color{black}3.2}}
\newcommand{\ZhighRes}{{\color{black}5.0}}
\newcommand{\Zww}{{\color{black}1.5}}
\newcommand{\Zbosonic}{{\color{black}5.1}}
\newcommand{\MaxLocalZseven}{3.2}  
\newcommand{\MaxLocalZeight}{{\color{black}3.8}} 

\newcommand{\MaxZmass}{{\color{black}125}}

\newcommand{\GlobalZsmall}{4.6} 
\newcommand{\GlobalZmedium}{4.5} 

\newcommand{\MASS}{\ensuremath{125.3 \pm 0.4\stat\pm 0.5\syst}}
\newcommand{\MASSH}{{\color{black}$125.1 \pm 0.5$}}

\newcommand{\MUHAT}{{\color{black}$0.87 \pm 0.23$}} 

\newcommand{\mH}{\ensuremath{m_{\PH}}}
\newcommand{\CLs}{\ensuremath{\mathrm{CL_s}\xspace}}

\newcommand{\ptgg}{\ensuremath{p_{\mathrm{T}}^{\gamma\gamma}}\xspace}
\newcommand{\mgg}{\ensuremath{m_{\gamma\gamma}}\xspace}

\newcommand{\Rwz}{{\color{black}$0.9~^{+1.1}_{-0.6}$}}     
\newcommand{\CV}{$1.0$}                                  
\newcommand{\CF}{$0.5$}                                  
\newcommand{\CVNF}{[0.7; 1.3]}   
\newcommand{\CFNF}{[0.2; 1.0]}   

\newcommand{\NiuMax}{{\color{red}XXX}}
\newcommand{\NiuMin}{{\color{red}XXX}}

\newcommand{\XXX}{{\color{red}XXX}}

\newcommand{\GAMMA}{\cPgg}
\newcommand{\ptV}{\ensuremath{\PT(\mathrm{V})}}
\newcommand{\ZmmH}{\ensuremath{\cPZ(\Pgm\Pgm)\PH}}
\newcommand{\Zudscg}{\ensuremath{\cPZ+\cPqu\cPqd\cPqs\cPqc\Pg}}
\newcommand{\ZeeH}{\ensuremath{\cPZ(\Pe\Pe)\PH}}
\newcommand{\ZnnH}{\ensuremath{\cPZ(\cPgn\cPgn)\PH}}
\newcommand{\WmnH}{\ensuremath{\PW(\Pgm\cPgn)\PH}}
\newcommand{\WenH}{\ensuremath{\PW(\Pe\cPgn)\PH}}
\newcommand{\Wudscg}{\ensuremath{\PW+\cPqu\cPqd\cPqs\cPqc\Pg}}
\newcommand{\Wbb}{\ensuremath{\PW\bbbar}}
\newcommand{\Zbb}{\ensuremath{\cPZ\bbbar}}
\newcommand{\dyll}{\ensuremath{\cPZ/\GAMMA^*{\to \ell^+\ell^-}}}
\newcommand{\wgamma}{\ensuremath{\W\GAMMA}}

\newcommand{\Wjets}{\ensuremath{\PW+\text{jets}}}
\newcommand{\Wt}{\PW\cPqt}
\newcommand{\Hww}{\Hi\to\WW}
\newcommand{\Hi}{\PH}
\newcommand{\W}{\PW}
\newcommand{\WW}{\PWpPWm}
\newcommand{\ZZ}{\cPZ\cPZ}
\newcommand{\WZ}{\ensuremath{\W\Z}}
\newcommand{\El}{\Pe}
\newcommand{\Elp}{\Pep}
\newcommand{\Elm}{\Pem}
\newcommand{\Elpm}{\ensuremath{\Pe^{\pm}}}
\newcommand{\Elmp}{\ensuremath{\Pe^{\mp}}}
\newcommand{\M}{\Pgm}
\newcommand{\Mp}{\Pgmp}
\newcommand{\Mm}{\Pgmm}
\newcommand{\Mpm}{\ensuremath{\Pgm^{\pm}}}
\newcommand{\Mmp}{\ensuremath{\Pgm^{\mp}}}
\newcommand{\Lep}{\ensuremath{\ell}}
\newcommand{\dyee}{\ensuremath{\mathrm{\Z}/\GAMMA^*\mathrm{\to \Pep\Pem}}}
\newcommand{\dymm}{\ensuremath{\mathrm{\Z/}\GAMMA^*\to\Pgmp\Pgmm}}
\newcommand{\dytt}{\ensuremath{\mathrm{\Z}/\GAMMA^* \to\Pgt^+\Pgt^-}}
\newcommand{\delphill}{\ensuremath{\Delta\phi_{\Lep\Lep}}}
\newcommand{\mll}{\ensuremath{m_{\Lep\Lep}}}
\newcommand{\ptlmax}{\ensuremath{p_{\mathrm{T}}^{\Lep,\mathrm{max}}}}
\newcommand{\ptlmin}{\ensuremath{p_{\mathrm{T}}^{\Lep,\mathrm{min}}}}
\newcommand{\met}{\ETm}
\newcommand{\delphimetll}{\ensuremath{\Delta\phi_{\met\Lep\Lep}}}
\newcommand{\Et}{\ET}

\newcommand{\usedLumi}{5.1\fbinv}
\newcommand{\usedLumiWithSyst}{\ensuremath{5.1 \pm 0.2 \fbinv}}

\newcommand{\xsecbr}{\sigma_\phi\cdot B_{\Pgt\Pgt}}
\newcommand{\xsecbrH}{\sigma_H\cdot B_{\Pgt\Pgt}}
\newcommand{\MT}{m_\mathrm{T}}
\newcommand{\Mvis}{m_\text{vis}}
\newcommand{\Mfit}{m_{\Pgt\Pgt}}
\newcommand{\emt}{\ensuremath{\Pe\Pgm\Pgt_h}}
\newcommand{\mmt}{\ensuremath{\Pgm\Pgm\Pgt_h}}
\newcommand{\WH}{\ensuremath{\W\Hi}}
\newcommand{\ZH}{\ensuremath{\Z\Hi}}
\newcommand{\hww}{\Hi\to\WW}
\newcommand{\htt}{\Hi\to\TT}
\newcommand{\Ztt}{\Z\to\TT}
\newcommand{\Zee}{\ensuremath{\cPZ\to\Pep\Pem}}
\newcommand{\tauh}{\ensuremath{\Pgt_\mathrm{h}}}
\newcommand{\taul}{\ensuremath{\Pgt_\ell}}
\newcommand{\Hmu}{\ensuremath{\PH\to\cPZ\cPZ^{(\ast)}\to 4\Pgm}}
\newcommand{\mmu}{\ensuremath{m_{4\Pgm}}}
\newcommand{\lnQ}{\ensuremath{\ln(Q)}}
\newcommand{\mlnQ}{\ensuremath{-2\ln(Q)}}
\newcommand{\LS}{\ensuremath{\mathcal{L}(S)}}
\newcommand{\LB}{\ensuremath{\mathcal{L}(B)}}
\newcommand{\LSB}{\ensuremath{\mathcal{L}(S+B)}}
\newcommand{\mZ}{\ensuremath{m_{\cPZ}}}
\newcommand{\Zgam}{\ensuremath{\cPZ/\Pgg^*}}
\newcommand{\mumu}{\ensuremath{\Pgmp\Pgmm}}
\newcommand{\Wo}{\ensuremath{\PW}}
\newcommand{\Wp}{\ensuremath{\PWp}}
\newcommand{\Wm}{\ensuremath{\PWm}}
\newcommand{\Zo}{\ensuremath{\cPZ}}
\newcommand{\Ho}{\ensuremath{\PH}}
\newcommand{\KD}{\mathrm{KD} }
\newcommand{\X}{\mathrm{X} }

\section{Introduction}\label{sec:introduction}
The standard model (SM)~\cite{Glashow:1961tr,Weinberg:1967tq,sm_salam} of particle physics accurately
describes many experimental results that probe elementary particles and their interactions up to
an energy scale of a few hundred \GeVns \cite{EWKlimits}.
In the SM, the building blocks of matter, the fermions, are comprised of quarks and leptons.
The interactions are mediated through the exchange of force carriers: the photon for electromagnetic
interactions, the $\PW$ and $\cPZ$ bosons for weak interactions, and the gluons for strong interactions.
All the elementary particles acquire mass through their interaction with the Higgs
field~\cite{Englert:1964et,Higgs:1964ia,Higgs:1964pj,Guralnik:1964eu,Higgs:1966ev,Kibble:1967sv,Nambu:1961tp,NambuNobel,GellMann:1960np}.
This mechanism, called the ``Higgs'' or ``BEH''
 mechanism~\cite{Englert:1964et,Higgs:1964ia,Higgs:1964pj,Guralnik:1964eu,Higgs:1966ev,Kibble:1967sv},
is the first coherent and the simplest solution for giving mass to \PW\ and \cPZ\ bosons, while still preserving the symmetry of the
Lagrangian. It is realized by introducing a new complex scalar field into the
model. By construction,  this field allows the $\PW$ and $\cPZ$ bosons to
acquire mass whilst the photon remains massless, and adds to the model one new scalar particle, the SM Higgs boson (\PH).
The Higgs scalar field and its conjugate can also give mass to the fermions,
through Yukawa interactions \cite{Nambu:1961tp,NambuNobel,GellMann:1960np}.
The SM does not directly predict the values of the masses of the elementary particles,
and in the same context there is no prediction for the Higgs boson mass.
The particle masses are considered parameters to be determined experimentally.
Nevertheless, a number of very general arguments~\cite{Cornwall:1973tb,Cornwall:1974km,LlewellynSmith:1973ey,Lee:1977eg}
have been used to narrow the range of possible values for the Higgs
boson mass to below approximately 1\TeV. The wealth of electroweak precision
data from the LEP and SLC colliders, the Tevatron, and other experiments
predicted the Higgs boson mass to be at approximately 90\GeV, with an upper limit of $152\GeV$ at
the 95\% confidence level (CL) \cite{EWKlimits}.  Direct searches at LEP excluded values lower
than $114.4\GeV$ at 95\% CL~\cite{LEPlimits}, and early Tevatron measurements
excluded the mass range 162--166\GeV at 95\%
CL~\cite{TEVHIGGS_2010}.

The discovery or exclusion of the SM Higgs boson is one of the primary scientific goals of the LHC.
Previous direct searches at the LHC were based on data from proton-proton collisions corresponding to
an integrated luminosity of 5.1\fbinv collected at a centre-of-mass
energy of 7\TeV.
The CMS experiment excluded
at 95\% CL masses from 127 to 600\GeV~\cite{Chatrchyan:2012tx}.
The ATLAS experiment excluded at 95\% CL the ranges 111.4--116.4,
119.4--122.1, and 129.2--541\unit{GeV}~\cite{ATLAScombJul2012_7TeV}.
Within the remaining allowed mass region, an excess of events between
2 and 3 standard deviations ($\sigma$) near
125\GeV was reported by both experiments.
In 2012, the proton-proton centre-of-mass energy was increased to 8\TeV, and by the end of June,
an additional integrated luminosity of more than 5.3\fbinv had been
recorded by each of the two experiments,
thereby enhancing significantly the sensitivity of the search for the Higgs boson.
The result was the observation by the ATLAS and CMS Collaborations of a new heavy boson with a mass of approximately $125\GeV$.
The two experiments simultaneously published the observation in concise
papers~\cite{ATLASobservation125,CMSobservation125}. The CMS publication~\cite{CMSobservation125} focused
on the observation in the five main decay channels in the low-mass
range from
$110$ to $145\GeV$: $\PH \to \Pgg\Pgg$, $\PH \to \cPZ \cPZ \to 4\ell$, $\PH \to \PW\PW \to
\ell\cPgn\ell\cPgn$, $\PH \to \Pgt\Pgt$, and  $\PH \to \cPqb\cPqb$, where  $\ell$ stands for electron or muon,
and for simplicity our notation does not distinguish
between particles and antiparticles.
In the summer 2012 the analysis of the full data set by the CDF and D0 Collaborations
resulted in an excess of events of about 3$\sigma$ in the mass
range $120 \le \mH \le 135\GeV$, while searching for a SM Higgs boson
decaying into \cPqb\ quarks~\cite{PhysRevLett.109.071804}.

The channels with the highest sensitivity for discovering the SM Higgs boson
with a mass near $125\GeV$ are $\PH \to \Pgg\Pgg$ and $\PH \to \cPZ
\cPZ \to 4\ell$.
The other three final states have poorer mass resolution and, therefore, necessitate more data
to achieve a similar sensitivity.
Among them, the  $\PH \to \PW\PW \to
\ell\cPgn\ell\cPgn$ channel has the largest signal-to-background ratio.
These five channels are complementary in the way they are measured in the detector, as is the
information they can provide about the SM Higgs boson.

A light Higgs boson has a natural width of a few \MeV~\cite{LHCHiggsCrossSectionWorkingGroup:2011ti},
and
therefore the precision of the mass measurement from fully
reconstructed decays would be limited
by the detector resolution.
The first two channels, $\PH \to \Pgg\Pgg$ and $\PH \to \cPZ\cPZ \to 4\ell$, produce a narrow mass
peak.
These two high-resolution channels were used to measure the mass of
the newly observed particle~\cite{CMSobservation125,ATLASobservation125}.

In the SM, the properties of the Higgs boson are fully determined once its mass is
known.  All cross sections and decay fractions are
predicted~\cite{LHCHiggsCrossSectionWorkingGroup:2011ti,Dittmaier:2012vm},
and thus the measured rates into each channel provide a test of the SM.
The individual measurements can be combined, and from them the coupling constants
of the Higgs boson with fermions and bosons can be extracted.
The measured values can shed light on  the nature of the newly
observed particle because the Higgs boson
couplings to fermions are qualitatively different from those to bosons.

The data described in this paper are
identical to those reported in
the observation publication~\cite{CMSobservation125}. The main focus of this paper
is an in-depth description of the five main analyses and a more detailed
comparison of the various channels with the SM predictions by
evaluating the couplings to fermions and vector bosons, as well as various coupling ratios.

The paper is organized into several sections. Sections 2 and 3 contain a short description of the CMS detector
and the event reconstruction of physics objects relevant for the Higgs boson search.
Section 4 describes the data sample, the Monte Carlo (MC) event generators used for the signal
and background simulation, and the evaluation of the signal sensitivity.
Then the analyses of the five decay channels
are described in detail in Sections 5 to 9.
In the last section, the statistical method used to combine the five channels and the statistical treatment of the systematic uncertainties
are explained. Finally, the results are combined and the first measurements of the couplings of the new particle to bosons and fermions
are presented.

\section{The CMS experiment}\label{sec:experiment}
The discovery capability for the SM Higgs boson is one of the
main benchmarks that went into optimizing the design of the
CMS experiment~\cite{Pimia:1990zy,DellaNegra:1992hp,Ellis:1994sq,Chatrchyan:2008aa}.

The central feature of the detector~\cite{Chatrchyan:2008aa} is a superconducting solenoid 13\unit{m} long,
with an internal diameter of 6\unit{m}.  The solenoid generates a uniform 3.8\unit{T} magnetic
field along the axis of the LHC beams.  Within the field volume are a silicon
pixel and strip tracker, a lead tungstate crystal electromagnetic calorimeter (ECAL),
and a brass/scintillator hadron calorimeter (HCAL).  Muons are identified and measured in gas-ionization
detectors embedded in the outer steel magnetic flux return yoke of the solenoid.
The detector is subdivided into a cylindrical barrel and endcap disks on each side of
the interaction point.  Forward calorimeters
complement the coverage provided by the barrel and endcap detectors.

The CMS experiment uses a right-handed coordinate system, with the origin at the nominal interaction point,
the $x$ axis pointing to the centre of the LHC, the $y$ axis pointing up (perpendicular to the LHC
plane), and the $z$ axis along the anticlockwise-beam direction.
The azimuthal angle $\phi$ is measured in the $x$-$y$ plane.
The pseudorapidity is defined as $\eta = -\ln[\tan{(\theta/2)}]$
where the polar angle $\theta$ is
measured from the positive $z$ axis.
The centre-of-mass momentum of the colliding partons in a proton-proton collision is subject to Lorentz boosts along the beam direction
relative to the laboratory frame.
Because of this effect, the pseudorapidity, rather than the polar angle, is a
more natural measure of the angular separation of particles in the rest
frame of the detector.

Charged particles are tracked within the pseudorapidity range
$|\eta|<2.5$.
The silicon pixel tracker is composed of 66~million pixels of area
$100\times150\mum^2$, arranged in three barrel
layers and two endcap disks at each end. The silicon strip tracker, organized in ten barrel
layers and twelve endcap disks at each end, is composed of 9.3 million strips with pitch between 80 and
205$\mum$, with a total silicon surface area of $198\unit{m}^2$. The performance of the tracker is essential to most
analyses in CMS and has reached the design performance in
transverse-momentum ($\pt$) resolution, efficiency,
and primary- and secondary-vertex resolutions.
The tracker has an efficiency larger than 99\% for muons with
$\pt >1\GeV$, a $\pt$ resolution
between 2 and 3\% for charged tracks of $\pt \approx 100$\GeV in the
central region ($|\eta| <$ 1.5),
and unprecedented capabilities for b-jet identification.
Measurements of the impact parameters of charged tracks and secondary vertices are used to
identify jets that are likely to contain the hadronization and decay
products of $\cPqb$ quarks (``$\cPqb$ jets'').
A b-jet tagging efficiency of more than 50\% is achieved with a
rejection factor for light-quark jets of ${\approx}200$, as measured with $\ttbar$ events in data~\cite{CMS-PAS-BTV-12-001}.
The dimuon mass resolution at the $\Upsilon$ mass, dominated by
instrumental effects, is measured to be 0.6\% in the barrel region~\cite{PhysRevD.83.112004}, consistent with the design
goal.
Due to the high spatial granularity of the pixel detector, the channel occupancy is less than $10^{-3}$,
allowing charged-particle trajectories to be measured
in the high-rate environment of the LHC without loss of performance.

The ECAL is a fine-grained, homogeneous calorimeter consisting of more than 75\,000
lead tungstate crystals, arranged in a quasi-projective geometry and distributed in a
barrel region ($|\eta| < 1.48$) and two endcaps that extend up to $|\eta| = 3.0$.
The front-face cross section of the crystals is approximately $22\times 22\unit{mm}^2$ in the barrel region
and $28.6\times 28.6\unit{mm}^2$ in the endcaps.
Preshower detectors consisting of two planes of silicon sensors interleaved with a total of
three radiation lengths of lead absorber are located in front of the endcaps.  Electromagnetic (EM)
showers
are narrowly distributed in the lead tungstate crystals (Moli\`ere radius of 21\unit{mm}), which have a transverse size
comparable to the shower radius.  The precise measurement of the transverse shower shape is the
primary method used for EM particle identification, and measurements in the surrounding crystals
are used for isolation criteria.
The energy resolution of the ECAL is the single most important performance
benchmark for the measurement of the Higgs boson decay into two photons and to a lesser extent
for the decay to \cPZ\cPZ\ that subsequently decay to electrons.
In the central barrel region, the energy resolution of electrons that
interact minimally with the tracker material indicates that the resolution of
unconverted photons is consistent with design goals.
The energy resolution for photons with transverse energy of ${\approx}60\GeV$ varies
between 1.1\% and 2.5\% over the solid angle of the ECAL barrel, and from 2.2\% to 5\% in the endcaps.
For ECAL barrel unconverted photons the diphoton mass resolution is estimated to be
1.1\GeV at a mass of 125\unit{GeV}.

The HCAL barrel and endcaps are sampling calorimeters composed of brass and
plastic scintillator tiles, covering $|\eta| < 3.0$. The hadron calorimeter thickness varies from 7 to 11
interaction lengths within the solenoid, depending on $|\eta|$; a scintillator ``tail catcher'' placed outside
the coil of the solenoid, just in front of the innermost muon detector, extends the instrumented thickness
to more
than 10 interaction lengths.  Iron forward calorimeters with quartz fibres, read
out by photomultipliers, extend the calorimeter coverage up to
$|\eta| = 5.0$.

Muons are measured in the range $|\eta| < 2.4$, with detection planes based on
three technologies: drift tubes ($|\eta| <$ 1.2), cathode strip chambers
($0.9 < |\eta| < 2.4$), and resistive-plate chambers ($|\eta| < 1.6$).
The first two technologies provide a precise position measurement and
trigger, whilst the third one provides precise timing
information, as well as a second independent trigger.
The muon system consists of four stations in the barrel and endcaps, designed to ensure robust
triggering and detection of muons
over a large angular range. In the barrel region, each muon station consists of twelve
drift-tube layers, except for the outermost station, which has eight layers. In the endcaps, each
muon station consists of six detection planes. The precision of the $r$-$\phi$
measurement is 100\mum in the drift tubes and varies from 60 to
140\mum in the cathode strip chambers, where $r$ is the radial distance from the beamline and $\phi$
is the azimuthal angle.

The CMS trigger and data acquisition systems ensure that data samples with potentially
interesting events are recorded with high efficiency. The first-level (L1)
trigger, composed of the calorimeter, muon, and global-trigger
processors, uses
coarse-granularity information to select the most interesting events in less than 4\mus. The
detector data are pipelined to ensure negligible deadtime up to a L1
rate of 100\unit{kHz}.
After L1 triggering, data are transferred from the readout electronics of all subdetectors
through the readout network to the high-level-trigger (HLT) processor farm, which
assembles the full event and executes global reconstruction algorithms.  The HLT filters the data, resulting
in an event rate of $\approx$500\unit{Hz} stored for offline processing.

All data recorded by the CMS experiment are accessible for offline analysis
through the world-wide LHC computing grid.
The CMS experiment employs a highly distributed computing infrastructure, with a
primary Tier-0 centre at CERN, supplemented by seven Tier-1, more than 50 Tier-2, and over 100
Tier-3 centres at national laboratories and universities throughout
the world.
The CMS software running on this high-perfor\-mance
computing system executes a multitude of crucial tasks, including the reconstruction
and analysis of the collected data, as well as the generation and
detector modelling of MC simulation.

\section{Event reconstruction}\label{sec:reconstruction}
\begin{figure}[htb]
\begin{center}
\includegraphics[width=0.47\textwidth]{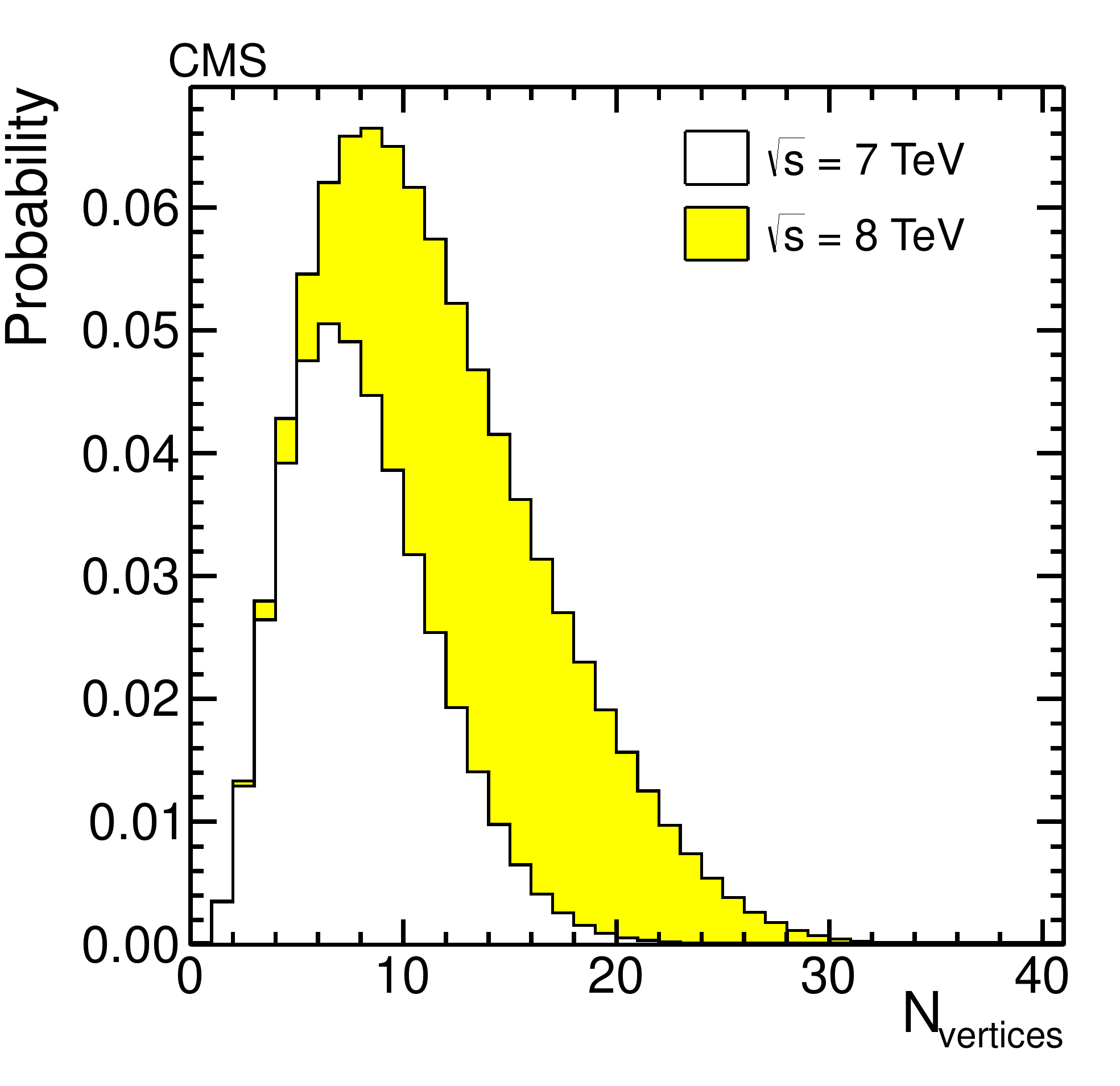}
\hspace{0.5cm}
\includegraphics[width=0.4\textwidth]{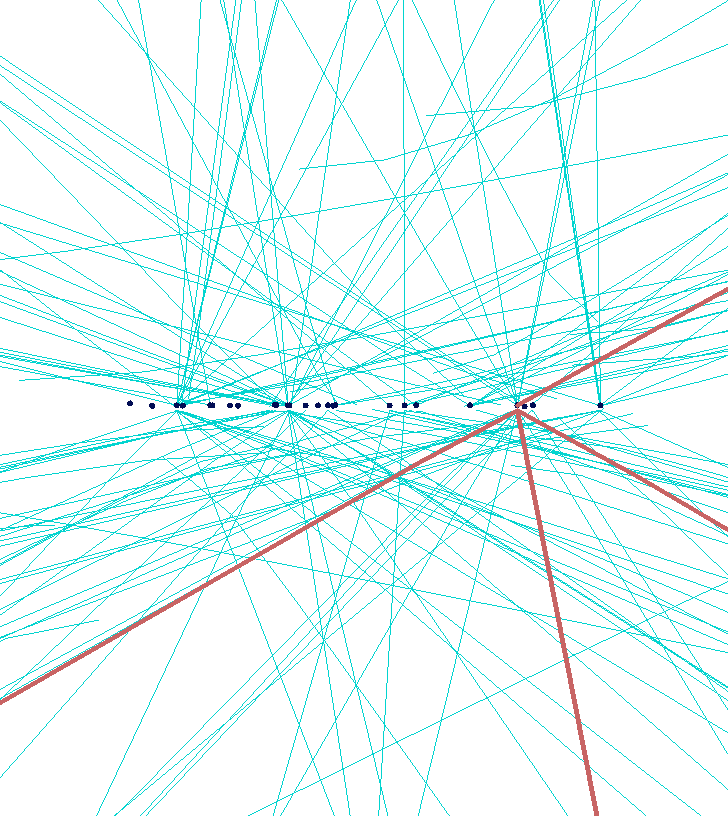}
\end{center}
\caption{Left: probability distribution for the number of vertices $N_\text{vertices}$ reconstructed per event in the 2011 and 2012 data.
The $\sqrt{s}=7$ and $8\TeV$ probability distributions are weighted by their equivalent integrated luminosity, and by the corresponding total cross section $\sigma(\Pp\Pp \to \PH+X)$ for a SM Higgs boson of mass 125\GeV.
Right: display of a four-lepton event recorded in 2012, with 24 reconstructed vertices.
The four leptons are shown as thick lines and originate from the vertex chosen for the hard-scattering process.}
\label{fig:reconstruction_vertices}
\end{figure}

Figure~\ref{fig:reconstruction_vertices} shows the distribution
of the number of vertices reconstructed per event in the 2011 and 2012 data, and the display of a four-lepton event recorded in 2012.
The large number of proton-proton interactions occurring per LHC bunch crossing  (``pileup''),
on average of 9 in 2011 and 19 in 2012,
makes the identification of the vertex corresponding to the hard-scattering process nontrivial,
and affects most of the physics objects: jets, lepton isolation, etc.
The tracking system is able to separate collision vertices as close as 0.5~\mm along the beam direction~\cite{IEEE_DetAnnealing}.
For each vertex, the sum of the $\pt^2$ of all tracks associated with the vertex
is computed. The vertex for which this quantity is the largest is assumed to correspond to the hard-scattering
process, and is referred to as the primary vertex in the event reconstruction.
In the $\PH\to\Pgg\Pgg$ final state, a large fraction of the transverse momentum produced in the collision is carried by the photons,
 and a dedicated algorithm,
described in Section~\ref{sec:hgg_vertex}, is therefore used to assign the photons to a vertex.

A particle-flow (PF) algorithm~\cite{CMS-PAS-PFT-09-001, CMS-PAS-PFT-10-002} combines the information
 from all CMS subdetectors to identify and reconstruct the individual particles emerging from all vertices:
charged hadrons, neutral hadrons, photons, muons, and electrons.
These particles are then used to reconstruct the missing transverse
energy,
jets, and hadronic $\tau$-lepton decays,
and to quantify the isolation of leptons and photons.

Electrons and photons can interact with the tracker material before reaching
the ECAL to create additional electrons and photons through pair
production and bremsstrahlung radiation.
A calorimeter superclustering algorithm is therefore used to combine
the ECAL energy deposits that could correspond to a photon or electron.
In the barrel region, superclusters are formed from five-crystal-wide areas
in $\eta$, centred on the locally most-energetic crystal and
having a variable extension in $\phi$.
In the endcaps, where the crystals are arranged according to an $x$-$y$ rather than $\eta$-$\phi$ geometry,
matrices of $5\times5$ crystals
around the most-energetic crystals are merged if they lie within
a narrow road in $\eta$.

The stability and uniformity of the ECAL response must be
calibrated at a fraction of a percent to maintain the excellent intrinsic
energy resolution of the ECAL \cite{ECAL-EnergyResol}. 
A dedicated monitoring system, based on the injection of laser light
into each crystal, is used to track and correct for channel response
changes caused by radiation damage and subsequent recovery of the crystals
\cite{ECAL-LaserMonit}. Response variations are a few percent in the
barrel region, and increase up to a few tens of percent in the most-forward
endcap regions.
The channel-to-channel response is equalized using several techniques
that exploit reference signatures from collision events (mainly
$\pi^0, \eta \to \gamma\gamma$) \cite{ECAL-Calibrations}. 
The residual miscalibration of the channel response
varies between 0.5\% in the central barrel to a few percent in
the endcaps \cite{ECAL-Role}.
At the reconstruction level, additional correction factors to the
photon energy are applied.
These corrections are sizeable for photons that convert before entering the ECAL, for
which the resolution is mainly limited by shower-loss
fluctuations. Given the distribution of the tracker material in front
of the ECAL, these effects are sizeable for $|\eta|>1$  \cite{ECAL-Role}.

Candidate photons for the $\PH\to\Pgg\Pgg$ search are reconstructed from the superclusters, and their identification is discussed in Section~\ref{sec:hgg_selection}.
The photon energy is computed starting from the raw supercluster energy.
In the region covered by the preshower detector ($|\eta| > 1.65$),
the energy recorded in that detector is added.
In order to obtain the best resolution, the raw energy is corrected
for the containment of the shower in the clustered crystals
and for the shower losses of photons that convert in
the tracker material before reaching the calorimeter.
These corrections are computed
using a multivariate regression technique based on the boosted
decision tree (BDT) implementation in \textsc{tmva}~\cite{Hocker:2007ht}.
The regression is trained on photons from a sample of simulated events using the ratio of
the true photon energy to the raw energy as the target variable.
The input variables are the $\eta$ and $\phi$ coordinates of the supercluster, a collection of
shower-shape variables, and a set of energy-deposit coordinates defined with respect to the supercluster.
A second BDT, using the same input variables, is trained on a separate sample of simulated photons to provide an
estimate of the uncertainty in the energy value provided by
the first BDT.

The width of the reconstructed $\cPZ$ resonance is used to quantify the
ECAL performance, using decays to two electrons whose energies are
measured using the ECAL alone, with their direction determined from the
tracks. In the 7\TeV data set, the dielectron mass resolution at the $\cPZ$
boson mass, fitting for the measurement contribution separately from the
natural width, is 1.56\GeV in the barrel and 2.57\GeV in the endcaps,
while in the 8\TeV data sample, reconstructed with preliminary calibration
constants, the corresponding values are 1.61\GeV and 3.75\GeV.

Electron reconstruction is based on two methods: the first where an ECAL
supercluster is used to seed the reconstruction of a
charged-particle trajectory in the tracker~\cite{Baffioni:2006cd,CMS-PAS-EGM-10-004},
and the second where a candidate track is used to reconstruct an ECAL supercluster~\cite{CMS-PAS-PFT-10-003}.
In the latter, the electron energy deposit is found by extrapolating the electron track to the ECAL,
and the deposits from possible bremsstrahlung photons are collected by extrapolating a straight line tangent to the electron track
from each tracker layer, around which most of the tracker material is concentrated.
In both cases, the trajectory is fitted with a Gaussian sum filter~\cite{Adam2005} using a dedicated modelling of the electron energy loss in the tracker material.
Merging the output of these two methods provides high electron reconstruction efficiency within $|\eta| < 2.5$ and $\PT>2$\GeV.
The electron identification relies on a \textsc{tmva} BDT that combines observables sensitive to the
amount of bremsstrahlung along the electron trajectory, the geometrical and momentum matching between the electron trajectory and the associated supercluster, as well as  the shower-shape observables.

Muons are reconstructed within $|\eta| < 2.4$ and down to a \PT of 3\GeV.
The reconstruction combines the information from both the silicon tracker and the
muon spectrometer.
The matching between the tracker and the muon system is initiated either ``outside-in'',
starting from a track in the muon system, or ``inside-out'', starting from a track
in the silicon tracker.
Loosely identified muons, characterized by minimal requirements
on the track components in the muon system and taking into account
small energy deposits in the calorimeters that match to the muon track,
are identified with an efficiency close to 100\% by the PF algorithm.
In some analyses, additional tight muon identification criteria are applied:
a good global muon-track fit based on the tracker and  muon chamber hits,
muon track-segment reconstruction in at least two muon stations,
and a transverse impact parameter with respect to the primary vertex smaller than 2\,mm.

Jets  are reconstructed from all the PF particles using the anti-\kt jet algorithm~\cite{Cacciari:2008gp}
implemented in \textsc{fastjet}~\cite{Cacciari:fastjet}, with a
distance parameter of 0.5.  
The jet energy is corrected for the contribution of particles created in pileup interactions and in the underlying event.
This contribution is calculated as the product of the jet area and an event-by-event \PT density $\rho$,
also obtained with \textsc{fastjet} using all particles in the event.
Charged hadrons, photons, electrons, and muons reconstructed by the PF algorithm have a calibrated
momentum or energy scale.  A residual calibration factor is applied to
the jet energy to account for imperfections in the
neutral-hadron calibration, the jet energy containment, and the estimation of the contributions from
pileup and underlying-event particles.
This factor, obtained from simulation, depends on the jet \PT and $\eta$, and is of the order of 5\% across the whole detector acceptance.
Finally, a percent-level correction factor is applied to match
the jet energy response in the simulation to the one observed in data.
This correction factor and the jet energy scale uncertainty are extracted from a comparison between the data and
simulation of $\gamma$+jets, \cPZ+jets, and dijet events~\cite{CMS-JME-10-011}.
Particles from different pileup vertices can be clustered into a
pileup jet, or significantly overlap a jet from the primary vertex below the \PT threshold applied in the analysis.
Such jets are identified and removed using a \textsc{tmva} BDT with the following input variables:
momentum and spatial distribution of the jet particles, charged- and neutral-particle multiplicities,
and consistency of charged hadrons within the jet with the primary vertex.

The missing transverse energy (\MET)  vector is calculated as the negative of the vectorial sum of the transverse momenta of all particles reconstructed by the PF algorithm.
The resolution $\sigma(E_{x,y}^{\text{miss}})$ on either the $x$ or $y$ component of the \MET vector is measured in $\cPZ\to\mu\mu$ events
and parametrized by $\sigma(E_{x,y}^{\text{miss}})=0.5\times\sqrt{{\Sigma {\ET}}}$,
where $\Sigma {\ET}$ is the scalar sum of the transverse momenta of all particles, with
$\sigma$ and $\Sigma {\ET}$ expressed in \GeVns{}.
In 2012, with an average number of 19 pileup interactions, $\Sigma {\ET}\approx 600\GeV$ for the analyses considered here.

Jets originating from b-quark hadronization are identified using different algorithms that exploit
 particular properties of such objects~\cite{CMS-PAS-BTV-12-001}.
These properties, which result from the relatively large mass and long lifetime of b quarks,
include the presence of tracks with large impact parameters, the presence of secondary decay vertices displaced from
the primary vertex, and the
 presence of low-\PT leptons from semileptonic b-hadron decays
 embedded in the jets~\cite{CMS-PAS-BTV-12-001}.
A combined secondary-vertex (CSV) b-tagging algorithm, used in the $\PH\to\cPqb\cPqb$ and $\PH\to\Pgt\Pgt$ searches,
makes use of the information about track impact parameters and secondary vertices within jets in a likelihood discriminant to provide separation
of b jets from jets originating from gluons, light quarks, and charm quarks.
The efficiency to tag b jets and the rate of misidentification of non-b jets depends on the algorithm used and the
operating point chosen. These are typically parameterized as a function of the transverse momentum and rapidity of the jets.
The performance measurements are obtained directly from data in samples that can be enriched in b jets,
such as $\ttbar$ and multijet events.

Hadronically decaying $\tau$ leptons ($\Pgt_h$) are reconstructed and identified using an algorithm~\cite{CMS-PAS-TAU-11-001}
which targets the main decay modes by selecting candidates with one charged hadron and up to two neutral pions,
or with three charged hadrons.
A photon from a neutral-pion decay can convert in the tracker material into an electron and a positron, which can then radiate
bremsstrahlung photons. These particles give rise to several ECAL energy deposits at the same $\eta$ value and separated in azimuthal angle,
and are reconstructed as several photons by the PF algorithm. To increase the acceptance for such converted photons,
the neutral pions are identified by clustering the reconstructed photons in narrow strips along the $\phi$ direction.
The $\Pgt_h$ from \PW, \cPZ, and Higgs boson decays are typically isolated from the other particles in the event, in contrast to misidentified $\Pgt_h$ from
jets that are surrounded by the jet particles not used in the $\Pgt_h$ reconstruction.
The $\Pgt_h$ isolation parameter $R_\text{Iso}^{\tau}$ is obtained from a multivariate discriminator,
 taking as input a set of transverse momentum sums $S_{j} = \sum_i p_{\mathrm{T}, i, j}$, where $p_{\mathrm{T}, i, j}$
 is the transverse momentum of a particle $i$ in a ring $j$ centred on the $\Pgt_h$ candidate direction
and defined in $(\eta, \phi)$ space.
Five equal-width rings are used up to a distance $\Delta R = \sqrt{(\Delta \eta)^2 + (\Delta \phi)^2}=0.5$ from the $\Pgt_h$ candidate, where
$\Delta \eta$ and $\Delta \phi$ are the pseudorapidity and azimuthal angle differences (in radians), respectively, between the particle
and the $\Pgt_h$ candidate direction.
The effect of pileup on the isolation parameter is mainly reduced by discarding from the $S_{j}$
calculation the charged hadrons with a track originating from a pileup vertex.
The contribution of pileup photons and neutral hadrons is handled by the discriminator,
which also takes as input the $\PT$ density $\rho$.

The isolation parameter of electrons and muons is defined relative to their transverse momentum $\PT^{\ell}$ as
\begin{equation}
R_\text{Iso}^{\ell} \equiv \left( \sum_\text{charged}  \PT + \mathrm{MAX}\left[ 0, \sum_\text{neut. had.}  \PT
                                        +  \sum_{\gamma}  {\PT} - \rho_\text{neutral} \times A_\text{eff}  \right] \right) /  \PT^{\ell},
\label{eq:reconstruction_isolation}
\end{equation}
where $\sum_\text{charged}  \PT$, $\sum_\text{neut. had.}  \PT$, and $\sum_{\gamma} \PT$ are, respectively, the scalar sums of the transverse momenta of charged hadrons, neutral hadrons, and photons located in a cone centred on the lepton direction in $(\eta, \phi)$ space.
The cone size $\Delta R$ is taken to be 0.3 or 0.4 depending on the analysis.
Charged hadrons associated with pileup vertices are not considered, and the contribution of pileup
photons and neutral hadrons is estimated as the product of the neutral-particle \PT density  $\rho_\text{neutral}$ and an effective cone area $A_\text{eff}$.
The neutral-particle \PT density is obtained with \textsc{fastjet} using all PF photons and neutral hadrons in the event, and the effective cone area is slightly different from the actual cone area,
being computed in such a way so as to absorb the residual dependence of the isolation efficiency on the number of pileup collisions.

\section{Data sample and analyses performance}\label{sec:searches}

The data have been collected by the CMS experiment at a centre-of-mass energy
of 7\TeV in the year 2011, corresponding to
an integrated  luminosity of about $\usedLumi$,  and a centre-of-mass
energy of 8\TeV in the year 2012, corresponding to
an integrated luminosity of about 5.3\fbinv.

A summary of all analyses described in this paper
is presented in Table~\ref{tab:channels}, where we list their main characteristics, namely:
exclusive final states, Higgs boson mass range of the search,
integrated luminosity used, and the approximate experimental mass resolution.
The presence of a signal in one of the channels
at a certain value of the Higgs boson mass, $\mH$, should manifest itself as an excess
in the corresponding invariant-mass distribution extending around that value for a range corresponding to the $\mH$ resolution.

\begin{table}
\begin{center}
\topcaption{Summary information on the analyses included in this paper.
The column ``\PH prod.'' indicates the production mechanism targeted by an analysis; it does not imply 100\% purity
(e.g.\ analyses targeting vector-boson fusion (VBF) production are
expected to have 30\%--50\% of their signal events coming from
gluon-gluon fusion production).
The main contribution in the untagged and inclusive categories is always gluon-gluon fusion. A final state can be further subdivided into multiple categories
based on additional jet multiplicity, reconstructed mass, transverse momentum,
or multivariate discriminators.
Notations used are:  $(\mathrm{jj})_{\mathrm{VBF}}$ stands for a dijet pair consistent with topology (VBF-tag);
$V$ = $\PW$ and $\cPZ$ bosons;
same flavour (SF) dileptons = $\Pe\Pe$ or $\mu\mu$ pairs;
different flavour (DF) dileptons = $\Pe\mu$ pairs; $\tau_h = \tau$
leptons decaying hadronically.
$V\PH$ stands for associated production with a vector boson.
}
\footnotesize
\label{tab:channels}
\begin{tabular}{l|c|l| ccc cc }
\hline
  \PH\ decay      & \PH\ prod.    & Exclusive final states    & No. of   & $m_{\PH}$ range   & $m_{\PH}$   & \multicolumn{2}{c}{ $\mathcal{L}$ (fb$^{-1}$)} \\
                 &             &    analysed               & channels  & (\GeVns)           & resolution  & 7\TeV       & 8\TeV             \\
\hline\hline
\multirow{3}{*}{$\gamma\gamma$}  & untagged   &     4 diphoton categories   & 4   & 110--150 & 1--2\%       & 5.1       & 5.3           \\
                                 & \multirow{2}{*}{VBF-tag}    &  $\gamma\gamma + (\rm{jj})_{\rm{VBF}}$  &  \multirow{2}{*}{1 or 2}    & \multirow{2}{*}{110--150}         & \multirow{2}{*}{1--2\%}       &  \multirow{2}{*}{5.1}                 & \multirow{2}{*}{5.3}  \\
                                 &             &  2 $m_{\rm{jj}}$ categories for 8\TeV &   &                      &      &                 &                 \\ \hline
$\cPZ\cPZ \to 4\ell$             & inclusive  &  $4\Pe, \, 4\mu, \, 2 \Pe 2\mu$                                                                          & 3         & 110--180         & 1--2\%       & 5.0                 & 5.3                \\  \hline

\multirow{3}{*}{$\PW\PW \to \ell\nu\ell\nu$}    & 0 or 1 jet  &  DF or SF dileptons             & 4         & 110--160         & 20\%        & 4.9                 & 5.1                  \\
  & \multirow{2}{*}{VBF-tag}    &  $ \ell\nu\ell\nu + (\mathrm{jj})_{\mathrm{VBF}}$ & \multirow{2}{*}{1 or 2}    & \multirow{2}{*}{110--160}         & \multirow{2}{*}{20\%}        & \multirow{2}{*}{4.9}                 & \multirow{2}{*}{5.1}                \\
                                                  &                             &  DF or SF dileptons for 8\TeV  &    &          &     &    &              \\
\hline
\multirow{4}{*}{$\tau\tau$} & \multirow{2}{*}{0 or 1 jet} & $(\Pe\tau_h, \, \mu\tau_h, \, \Pe\mu, \, \mu\mu) $                                             & \multirow{2}{*}{16}  & \multirow{2}{*}{110--145}  & \multirow{2}{*}{20\%}  & \multirow{2}{*}{4.9} & \multirow{2}{*}{5.1}   \\
       &                                  & 2 $p_T^{\tau\tau}$ categories and 0 or 1 jet                             &           &                  &             &                     &               \\
                                 & VBF-tag    &  $(\Pe\tau_h, \, \mu\tau_h, \, \Pe\mu, \, \mu\mu) + (\rm{jj})_{\rm{VBF}}$                                        & 4         & 110--145         & 20\%        & 4.9                 & 5.1                  \\
\hline
\multirow{2}{*}{$bb$}              & \multirow{2}{*}{$V\PH$-tag} &  $(\nu\nu, \, \Pe\Pe, \, \mu\mu, \, \Pe\nu, \, \mu\nu$ + 2 \cPqb\ jets) & \multirow{2}{*}{10}         & \multirow{2}{*}{110--135}         & \multirow{2}{*}{10\%}        & \multirow{2}{*}{5.0}                 & \multirow{2}{*}{5.1}                  \\
                             &                                   &  2 $\pt^{V}$ categories   &        &         &        &                &                 \\
\hline
\end{tabular}
\end{center}
\end{table}

\subsection{Simulated samples}

MC simulation samples for the SM Higgs boson signal and background processes are used
to optimize the event selection, evaluate the acceptance and systematic uncertainties, and
predict the expected yields.
They are processed through a detailed simulation of the CMS detector based on
\GEANTfour~\cite{Agostinelli:2002hh} and are reconstructed with the same algorithms used for the data.
The simulations include pileup interactions properly reweighted to match the
distribution of the number of such interactions observed in data.
For leading-order generators the default set of parton distribution functions
(PDF) used to produce these samples is CTEQ6L~\cite{CTEQ6L1}, while
CT10~\cite{Guzzi:2011sv} is employed for next-to-leading-order (NLO) generators.
For all generated samples the hadronization is handled by {\PYTHIA 6.4} ~\cite{Sjostrand:2006za}
or \HERWIG{++}~\cite{Gieseke:2006ga}, and the \TAUOLA \cite{TAUOLA} package is used for $\tau$ decays.
The {\PYTHIA} parameters for the underlying event and pileup interactions are set to the {Z2} tune \cite{1107.0330} for the
7\TeV data sample and to the {Z2*} tune \cite{1107.0330} for the 8\TeV data sample.

\subsection{Signal simulation}

The Higgs boson can be produced in pp collisions via four different processes:
gluon-gluon fusion, vector-boson fusion, associated production with a vector boson, and associated production with
a $\ttbar$ pair.
Simulated Higgs boson signals from gluon-gluon fusion ($\Pg\Pg \rightarrow \PH$),
and vector-boson fusion (VBF) ($\Pq\Pq \rightarrow \Pq\Pq \PH$), are generated with
{\POWHEG}~\cite{powheg,powheg1,powheg2} at NLO.
The simulation of associated-production samples uses {\PYTHIA},
with the exception of  the  $\PH\to\cPqb\cPqb$ analysis that uses {\POWHEG} interfaced to \HERWIG{++}.
Events at the generator level are reweighted according to the total cross section $\sigma(\Pp\Pp\rightarrow \PH)$,
which contains contributions from gluon-gluon fusion up to next-to-next-to-leading order (NNLO) and next-to-next-to-leading-log (NNLL) terms~\cite{LHCHiggsCrossSectionWorkingGroup:2011ti,deFlorian:2012yg,
Anastasiou:2012hx,Anastasiou:2008tj,deFlorian:2009hc,Baglio:2010ae,Djouadi:1991tka,Dawson:1990zj,Spira:1995rr,
Harlander:2002wh,Anastasiou:2002yz,Ravindran:2003um,Catani:2003zt,Actis:2008ug,Aglietti:2004nj,Degrassi:2004mx},
vector-boson fusion including NNLO quantum chromodynamic (QCD) and NLO
electroweak (EW)
terms~\cite{LHCHiggsCrossSectionWorkingGroup:2011ti,Ciccolini:2007jr,Ciccolini:2007ec,Figy:2003nv,Arnold:2008rz,Bolzoni:2010xr},
associated production V$\PH$ (where V $= \cPZ,\PW$) at NNLO QCD and
NLO EW ~\cite{Han:1991ia,Brein:2003wg,Ciccolini:2003jy,Hamberg:1990np,Denner:2011id,Ferrera:2011bk},
and the production in association with $\ttbar$ at NLO QCD~\cite{Beenakker:2001rj,Beenakker:2002nc,Dawson:2002tg,Dawson:2003zu}.

\begin{figure}[tbp]
  \begin{center}
    \includegraphics[width=0.48\textwidth]{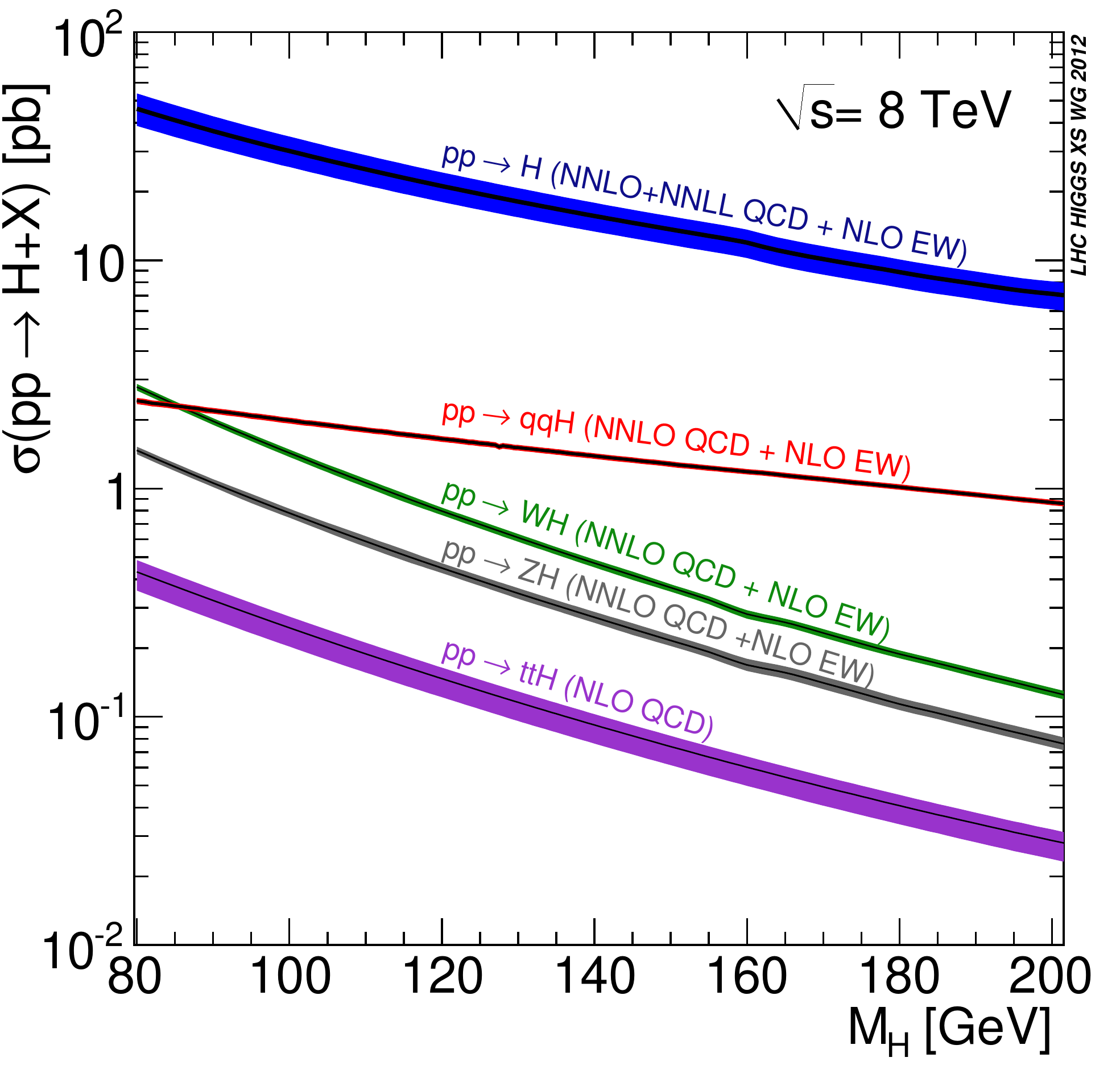}
    \includegraphics[width=0.48\textwidth]{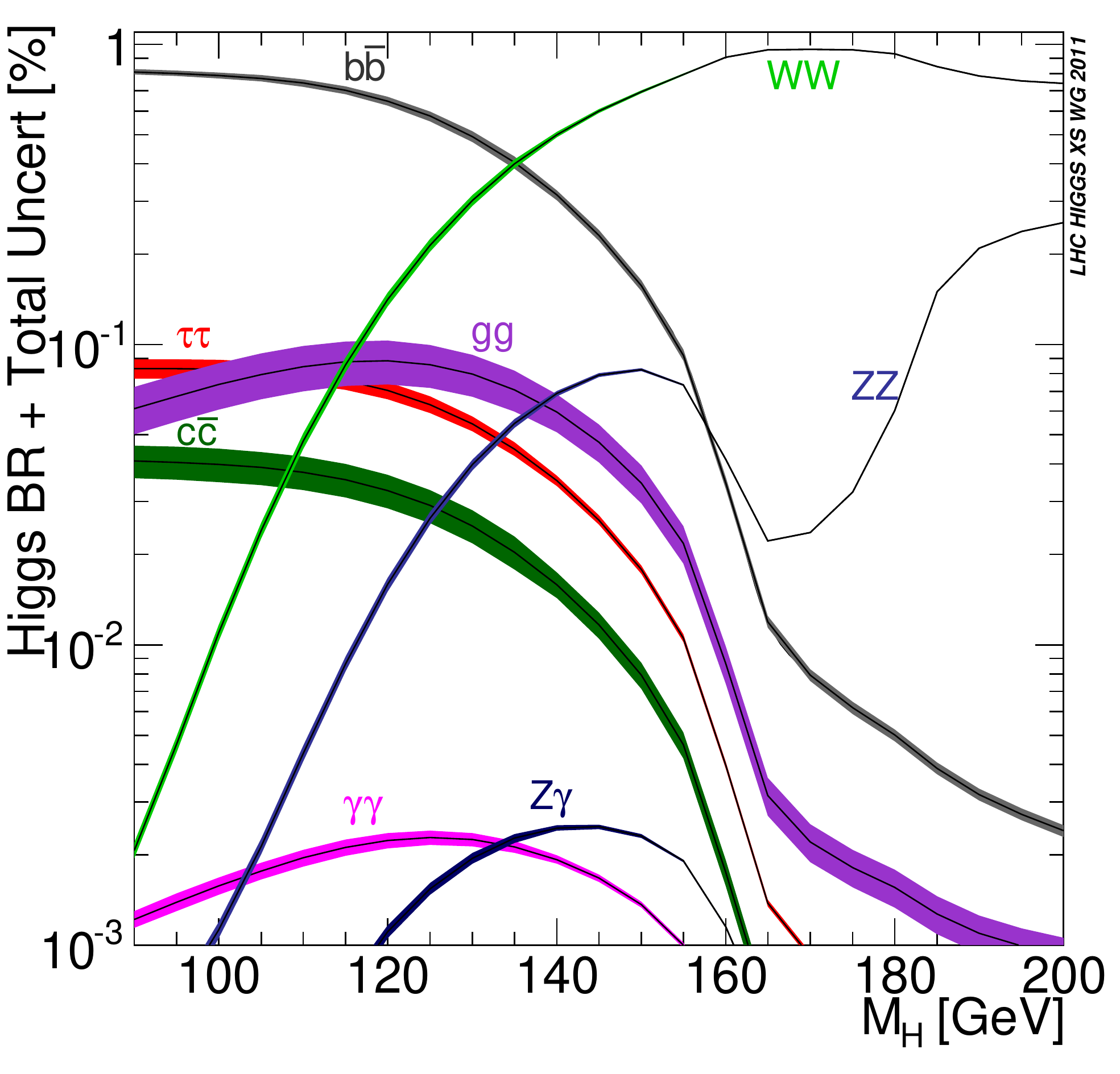}
    \caption{Higgs boson production cross sections  at $\sqrt{s}$ = 8\TeV (left) and branching fractions (right) as a function of the Higgs
boson mass from Refs.~\cite{LHCHiggsCrossSectionWorkingGroup:2011ti,Dittmaier:2012vm}. The width of the lines
represents the total theoretical uncertainty in the cross section and in the branching fractions.}
    \label{fig:xs_and_br_lm}
  \end{center}
\end{figure}

For the four-fermion final states the total cross section is scaled by the branching fraction $\mathcal{B}(\PH\rightarrow 4\ell)$ calculated
with the \textsc{prophecy4f} program \cite{Bredenstein:2006ha,Bredenstein:2006rh}.  The calculations include NLO QCD and EW corrections,
and all interference effects up
to NLO~\cite{LHCHiggsCrossSectionWorkingGroup:2011ti,Bredenstein:2006rh,Bredenstein:2006ha,Djouadi:1997yw,hdecay2,Actis:2008ts,Denner:2011mq,Dittmaier:2012vm}.
For all the other final states {\HDECAY} \cite{Djouadi:1997yw,hdecay2} is used, which includes NLO QCD and NLO EW corrections.
The predicted signal cross sections at 8\TeV and branching fraction for a low-mass Higgs boson are shown in the left and right plots
of Fig.~\ref{fig:xs_and_br_lm}, respectively~\cite{LHCHiggsCrossSectionWorkingGroup:2011ti,Dittmaier:2012vm}.

The uncertainty in the signal cross section related to
the choice of PDFs is determined with the PDF4LHC
prescription~\cite{Alekhin:2011sk,Botje:2011sn,Lai:2010vv,Martin:2009iq,Ball:2011mu}.
The uncertainty due to the higher-order terms
is calculated by varying the renormalization and factorization scales in each process,
as explained in Ref.~\cite{LHCHiggsCrossSectionWorkingGroup:2011ti}.

For the dominant gluon-gluon fusion process, the
transverse momentum spectrum of the Higgs boson in the 7\TeV MC simulation samples
is reweighted to match the
NNLL + NLO distribution computed
with \textsc{h}q\textsc{t}~\cite{Bozzi:2005wk,deFlorian:2011xf}
(and \textsc{fehipro}~\cite{FeHiPro1,FeHiPro2} for the high-$\pt$ range in the $\tau\tau$ analysis),
except in the $\PH\to\cPZ\cPZ$ analysis, where the reweighting is not necessary.
At 8\TeV,  \POWHEG was tuned to reach a good agreement of the $\pt$ spectrum with
the NNLL + NLO prediction in order to make reweighting unnecessary~\cite{Dittmaier:2012vm}.

\subsection{Background simulation}

The background contribution from \cPZ\cPZ\ production via $\Pq\Paq$ is generated at NLO with {\POWHEG}, while
other diboson processes (\PW\PW, \PW\cPZ) are generated with {\MADGRAPH}~\cite{Alwall:2011uj,Alwall:2007st}
with cross sections rescaled to NLO predictions. The {\PYTHIA} generator is also used to simulate all diboson processes.
The $\Pg\Pg \rightarrow$VV
contributions are generated with {\sc gg2vv}~\cite{Binoth:2008pr}.
The  V$+\text{jets}$  and V$\Pgg$ samples
are generated with {\MADGRAPH}, as are contributions to inclusive $\cPZ$ and  $\PW$ production,
with cross sections rescaled to NNLO predictions.
Single-top-quark and $\ttbar$ events are generated at NLO with {\POWHEG}.
 The {\PYTHIA} generator takes into account the  initial-state and final-state radiation effects that
can lead to the presence of additional hard photons in an event.
The {\MADGRAPH} generator is also used to generate samples of $\ttbar$ events.
QCD events are generated with {\PYTHIA}.
Table~\ref{tab:mc} summarizes the generators used for the different analyses.

\begin{table}
\begin{center}
\small
\topcaption{Summary of the generators used for the simulation of the main backgrounds for the analyses presented in this paper. }
\label{tab:mc}
\begin{tabular}{l|c|c}
\hline
   Analysis               &  Physics Process & Generator used  \\
\hline\hline
$ \PH \to \gamma\gamma$   & QCD &  \PYTHIA \\
                          & \cPZ+jet           & \MADGRAPH \\
\hline
$ \PH \to \cPZ\cPZ$       &   qq $\to 4\ell$   &    \POWHEG \\
                          & gg $\to 4\ell$    &    \textsc{gg2zz}  \\
                          &  \cPZ+jet            & \MADGRAPH \\
                          & $ \cPZ+\gamma$          &   \MADGRAPH \\
                          & $ \ttbar $         & \POWHEG \\
                          &  qq $\to \PW\PW,\PW\cPZ$     &    \MADGRAPH \\
\hline
$ \PH \to \PW\PW$         &   qq $\to \PW\PW$       &    \MADGRAPH \\
                          &  gg $\to \PW\PW$    &    \textsc{gg2ww}  \\
                          &   V+jet           &  \MADGRAPH        \\
                          & $ \ttbar$       &   \POWHEG \\
                          &  $ \cPqt\PW$        &  \POWHEG \\
                          & QCD               & \PYTHIA \\
\hline
$ \PH \to \tau\tau$       &  \cPZ+jet            & \MADGRAPH \\
                          & $ \ttbar$         & \MADGRAPH \\
                          & qq $\to \cPZ\cPZ,\cPZ\PW,\PW\PW$     &    \PYTHIA \\
                          & QCD               & \PYTHIA \\
\hline
$ \PH \to \cPqb\cPqb$             &   qq $\to \cPZ\cPZ,\cPZ\PW,\PW\PW$     &    \PYTHIA \\
                          &  \cPZ+jet            & \MADGRAPH \\
                          &  \PW+jet           & \MADGRAPH \\
                          & $ \ttbar$            & \MADGRAPH \\
                          &  $ \cPqt\PW$        &  \POWHEG \\
                          & QCD               & \PYTHIA \\
\hline
\end{tabular}
\end{center}
\end{table}

\subsection{Search sensitivities}

\begin{figure*}
\centering
\includegraphics[width=0.49\textwidth]{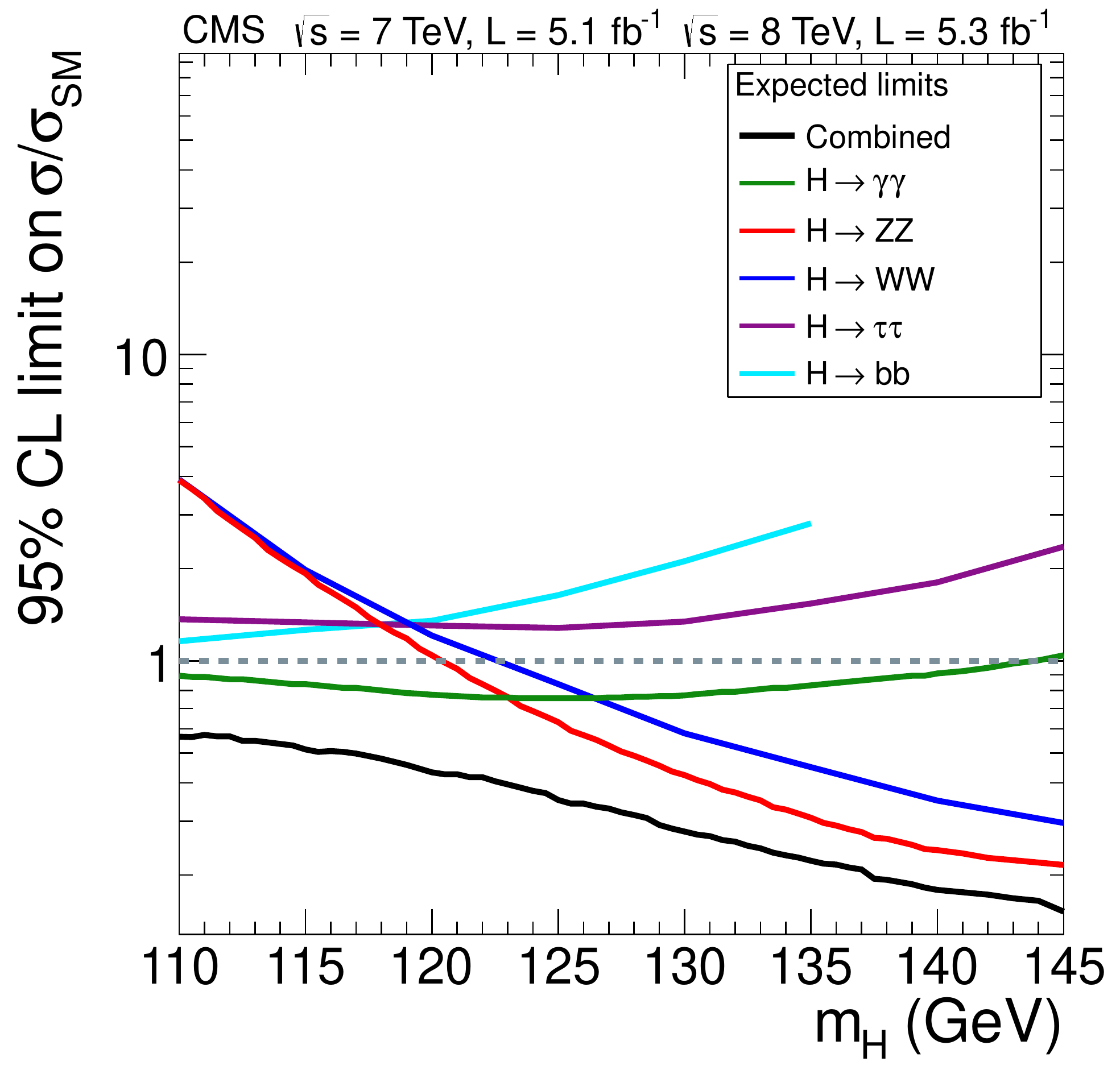} \hfill
\includegraphics[width=0.49\textwidth]{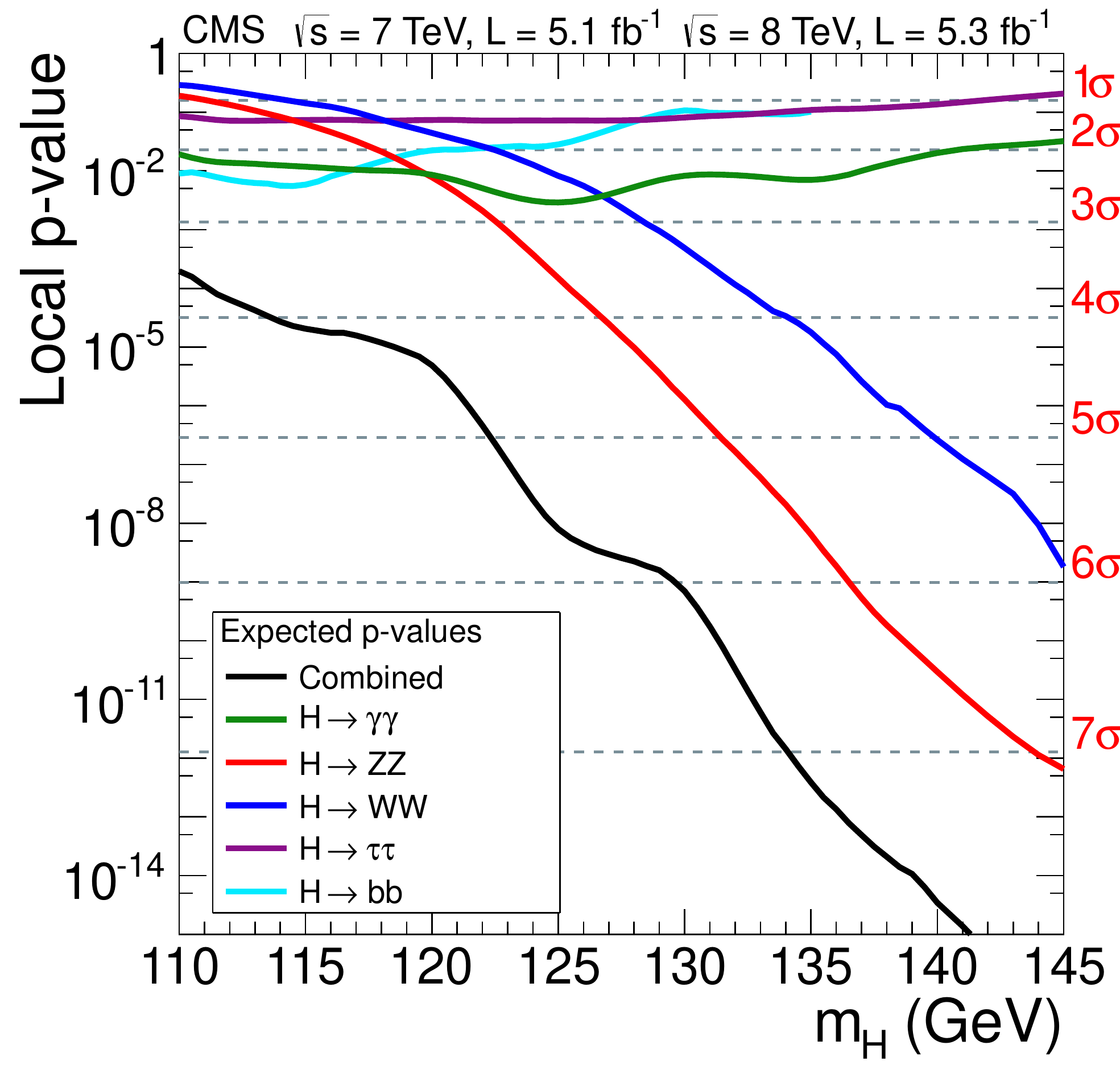}
\caption{ The median expected 95\% CL upper limits on the cross section ratio
$ \sigma / \sigma_{\mathrm{SM}}$ in the absence of a Higgs boson
(left) and the median expected local $p$-value for observing an excess,
assuming that a Higgs boson with that mass exists (right),
as a function of the Higgs boson mass for the five Higgs boson decay channels and their combination.
}
\label{fig:sensitivity}
\end{figure*}

The search sensitivities of the different channels, for the recorded
luminosity used in the analyses, expressed in terms
of the median expected 95\% CL upper limit on the ratio of the measured signal cross section, $\sigma$,
and the predicted SM Higgs boson cross section, $\sigma_{\mathrm{SM}}$,
are shown in Fig.~\ref{fig:sensitivity} (left)
as a function of the Higgs boson mass.
A channel showing values below unity (dashed horizontal line)
for a given mass hypothesis would be expected, in the absence of a Higgs boson
signal, to exclude the standard model Higgs boson at 95\% CL or more at that mass.
Figure~\ref{fig:sensitivity} (right) shows the expected sensitivities
for the observation of the Higgs boson in terms of local $p$-values
and significances as a function of the Higgs boson mass.
The local $p$-value is defined as the probability of a background fluctuation;
it measures the consistency of the data with the background-only hypothesis.

The overall statistical methodology used in this paper was developed by the ATLAS and CMS
Collaborations in the context of the LHC Higgs Combination Group~\cite{LHC-HCG-Report}.
A summary of our usage of this methodology in the
search for the Higgs boson is given in Section~\ref{sec:results}.

\section{\texorpdfstring{$\PH\to\Pgg\Pgg$}{H to gamma gamma}\label{sec:hgg}}

In the $\PH\to\Pgg\Pgg$ analysis, a search is made for a narrow peak,
of width determined by the experimental resolution of ${\sim}1\%$,
in the diphoton
invariant-mass distribution for the range 110--150\GeV,
on top of a large irreducible background
from the production of two photons originating directly from the hard-scattering process.
In addition, there is a sizable amount of reducible background in which one or both
of the reconstructed photons originate from the misidentification of
particles  in jets that deposit substantial energy in the ECAL, typically photons from
the decay of $\pi^0$ or $\eta$ mesons.
Early studies indicated this to be one of the most promising channels
in the search for a SM Higgs boson in the low-mass range~\cite{Seez1990a}.

To enhance the sensitivity of the analysis, candidate diphoton events are separated into mutually
exclusive classes with different expected signal-to-background ratios,
based on the properties of the reconstructed photons and
the presence or absence of two jets satisfying criteria aimed at selecting events
in which a Higgs boson is produced through the VBF process.
The analysis uses multivariate techniques
for the selection and classification of the events.
As independent cross-checks, two additional analyses are performed.
The first is almost identical to the CMS analysis described in Ref.~\cite{Chatrchyan:2012tw},
but uses simpler criteria based on the properties of the reconstructed photons to
select and classify events. The second analysis incorporates the same multivariate techniques
described here, however, it relies on a completely independent modelling of the background.
These two analyses are described in more detail in Section~\ref{sec:hgg_crosscheck}.

\subsection{Diphoton trigger}
\label{sec:hgg_dataReco}

All the data under consideration have passed at least one of a set of
diphoton triggers, each using transverse energy thresholds and
a set of additional photon selections, including criteria on the isolation and the shapes
of the reconstructed energy clusters. The transverse energy thresholds were chosen to be at least
10\% lower than the envisaged final-selection thresholds. This set of triggers
enabled events passing the later offline $\PH \to \gamma\gamma$ selection criteria
to be collected with a trigger  efficiency greater than  $99.5\%$.

\subsection{Interaction vertex location}
\label{sec:hgg_vertex}

In order to construct a photon four-momentum from the measured ECAL energies and the impact position
determined during the supercluster reconstruction, the photon production vertex, \ie the origin
of the photon trajectory, must be determined. Without incorporating any additional information, any of the
reconstructed pp event vertices is potentially the origin of the photon. If the distance
in the longitudinal direction
between the assigned and the true interaction point is
larger than 10\mm, the resulting contribution to the diphoton mass resolution becomes
comparable to the contribution from the ECAL energy resolution. It is, therefore, desirable
to use additional information to assign the correct interaction vertex for the photon with high probability.
This can be achieved by using the kinematic properties
of the tracks associated with the vertices and exploiting their correlation with
the diphoton kinematic properties, including the transverse momentum
of the diphoton
(\ptgg).
In addition, if either of the photons converts into an $\Pep\Pem$ pair
and the tracks from the conversion are reconstructed and identified, the direction of
the converted photon, determined by combining the conversion vertex position and the position of the
ECAL supercluster, can be extrapolated to identify the diphoton
interaction vertex.

For each reconstructed interaction vertex the following set of variables are calculated:
the sum of the squared transverse momenta of all tracks associated with the vertex
and two variables that  quantify the $\pt$ balance with respect to the diphoton system.
In the case of a reconstructed photon conversion, an additional
 ``pull'' variable is used,
defined as the distance between the vertex $z$ position and the beam-line extrapolated $z$ position
coming from the conversion reconstruction, divided by the uncertainty in this extrapolated $z$ position.
These variables are used as input to a BDT algorithm trained on simulated Higgs signal events
and the interaction point ranking highest in the constructed classifier
is chosen as the origin of the photons.

\begin{figure}[htbp]
  \begin{center}
    \includegraphics[width=0.49\linewidth]{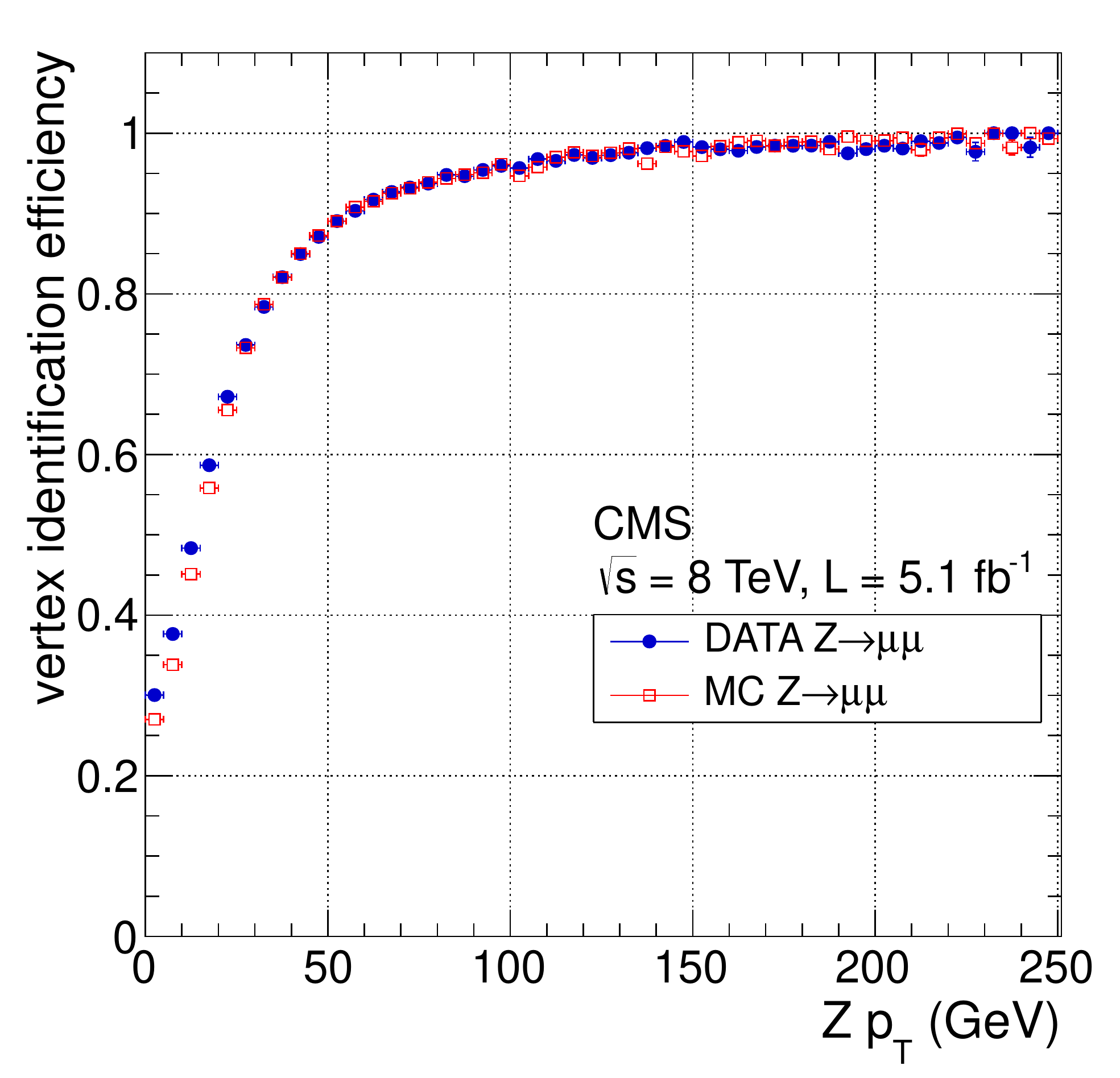}
    \caption{
      Comparison of the vertex-identification efficiency between data (circles)
      and MC simulated $\Z\to\mu\mu$ events (squares), as a function of the $Z$ boson \pt.
    }
    \label{fig:hgg_vtxeff}
  \end{center}
\end{figure}

The vertex-finding efficiency, defined as the efficiency to locate the vertex to
within 10\mm of its true position, is studied using $\Z\to\mu\mu$
events where the muon tracks were removed from the tracks considered,
and the muon momenta were replaced by the photon momenta.
The result
is shown in Fig.~\ref{fig:hgg_vtxeff}. The overall efficiency in signal events
with a Higgs boson mass of 120\GeV, integrated over its \pt spectrum, is
$(83.0\pm0.4)\%$ in the 7\TeV data set, and
$(79.0\pm0.2)\%$ in the 8\TeV data set. The statistical uncertainties in
these numbers are propagated to the uncertainties in the final result.

A second vertex related multivariate discriminant is employed to
estimate, event-by-event, the probability for the vertex
assignment to be within 10\mm of the diphoton interaction point.
This BDT is trained using simulated \HGG\ events. The input variables are
the classifier values of the vertex BDT described above for the three vertices with the highest score BDT values,
the number of vertices, the diphoton transverse momentum,
the distances between the chosen vertex and the second and third choices,
and the number of photons with an associated conversion track.
These variables allow for a reliable quantification of the probability
that the selected vertex is close to the diphoton interaction point.

The resulting  vertex-assignment probability from simulated events is used when constructing the Higgs boson signal models.
The signal modelling is described in Section~\ref{sec:hgg_smodeling}.

\subsection{Photon selection}
\label{sec:hgg_selection}

The event selection requires two photon candidates with transverse momenta satisfying
$\pt^{\gamma}(1) > \mgg/3$ and $\pt^{\gamma}(2) > \mgg/4$, where $\mgg$ is the diphoton invariant mass,
within the ECAL fiducial region $|\eta|$ $<~2.5$, and excluding the
barrel-endcap transition region $1.44 < |\eta| < 1.57$.
The fiducial region requirement  is applied to the supercluster position in the ECAL
and the $\pt$ threshold is applied after the vertex assignment.
The requirements on the mass-scaled transverse momenta are mainly motivated by the fact
that by dividing the transverse momenta by the diphoton mass,
turn-on effects on the background-shape in the low mass region are strongly reduced.
In the rare cases where the event contains more than two photons
passing all the selection requirements, the pair with the highest summed
(scalar) \pt is chosen.

The relevant backgrounds in the  \HGG\ channel consist of the irreducible background
from prompt diphoton production, \ie processes in which both photons originate directly
from the hard-scattering process, and the reducible
backgrounds from $\GAMJET$ and dijet events,
where the objects misidentified
as photons
correspond to particles in jets that deposit substantial energy in the ECAL, typically photons from
the decay of isolated $\pi^0$ or $\eta$ mesons.
These misidentified objects are referred to as \emph{fake} or \emph{nonprompt} photons.

In order to optimize the photon identification to exclude such nonprompt
photons, a BDT classifier is trained using simulated $\Pp\Pp\to\GAMJET$ event samples,
where prompt photons are used as the signal and nonprompt photons as the background.
The variables used in the training are divided into two groups.
The first contains information on the detailed electromagnetic shower topology,
the second has variables describing the photon isolation, \ie kinematic information on the
particles in the geometric neighbourhood of the photon. Examples of variables in the first group are the
energy-weighted shower width of the cluster of ECAL crystals assigned to the photon and the ratio of the
energy of the  most energetic $3\times3$ crystal cluster to the total cluster energy.
The isolation variables include  the magnitude of the sum of the transverse  momenta
of all other reconstructed particles inside a cone of size $\DR=0.3$ around the candidate photon direction.
In addition, the
geometric position of the ECAL crystal cluster, as well as the event energy density
$\rho$,
are used.
The photon ID classifier is based on the measured properties of a single photon and makes no use of the 
any properties that are specific to the production mechanism. 
Any small residual dependence on the production mechanism, e.g. through the isolation distribution, 
arises from the different event enviroments in Higgs decays and in photon plus jets events. 

\begin{figure}[htbp]
  \begin{center}
    \includegraphics[width=0.98\linewidth]{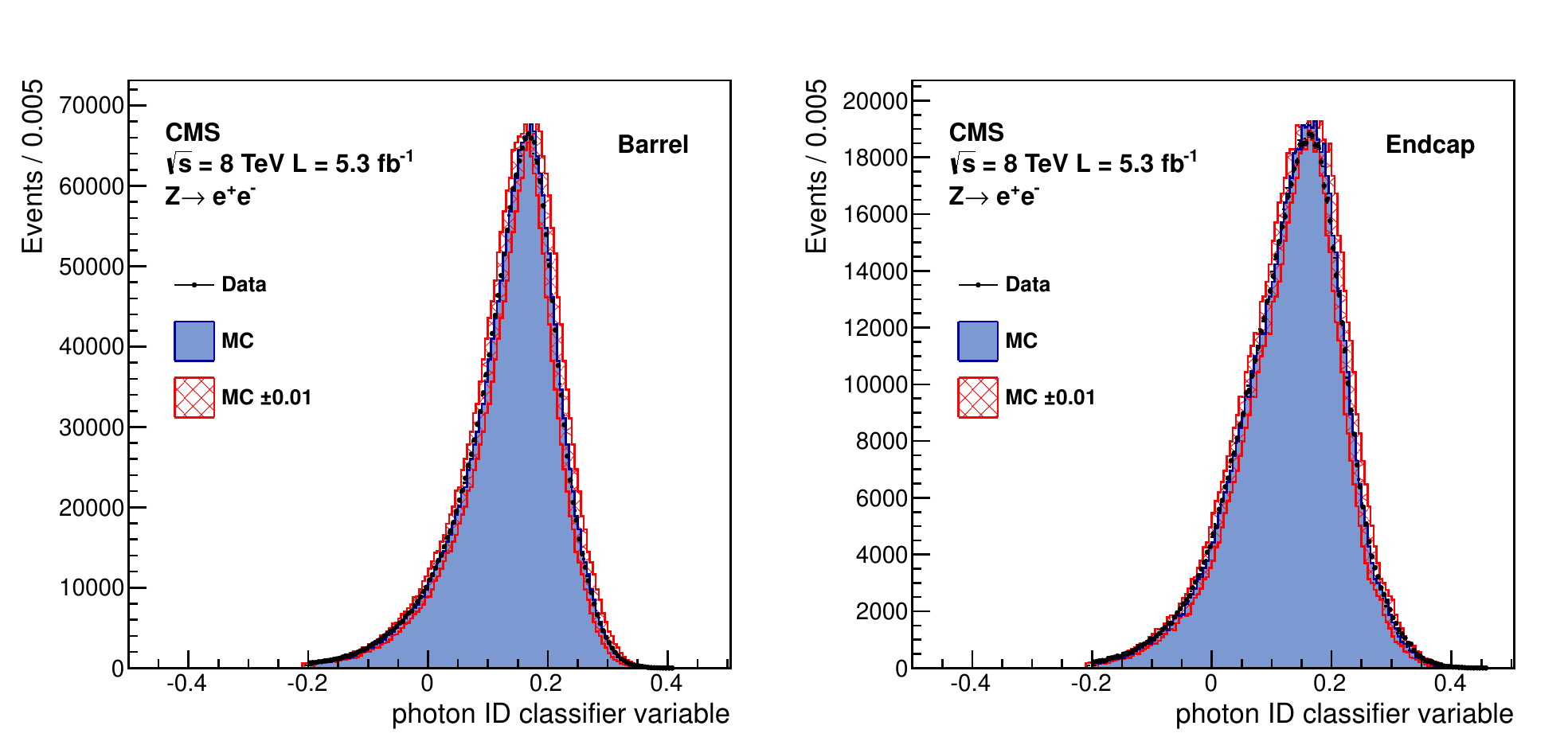}
    \caption{
      Comparison of the photon identification (ID) classifier variable distribution between 8\TeV
      data (points) and MC simulated events (histogram),
      separated into barrel (left) and endcap (right) electrons originating from $\cPZ\to \Pe\Pe$ events.
      The uncertainties in the distributions from simulation are shown by the cross-hatched histogram.
    }
    \label{fig:hgg_idmva}
  \end{center}
\end{figure}

Instead of having a requirement on the trained multivariate classifier value to select photons
with a high probability of being prompt photons, the classifier value itself is used
as input to subsequent steps of the analysis. To reduce the number of events, a loose requirement
is imposed on the  classifier value (${>}-0.2$) for candidate photons to be considered further.
This requirement retains more than $99\%$ of signal photons.
The efficiency of this requirement, as well as the differential shape of the classifier
variable for prompt photons, have been studied by comparing $\cPZ\to \Pe\Pe$
data to simulated events,  given the similar response of the detector to photon and electrons.
The comparisons between the differential shape in data and MC simulation for the 8\TeV analysis
are shown in Fig.~\ref{fig:hgg_idmva},
for electrons in the barrel (left) and endcap (right) regions.

\subsection{Event classification}
\label{sec:hgg_diphotonBDT}

The strategy of the analysis is to look for a narrow peak over the continuum in the diphoton invariant-mass
spectrum. To increase the sensitivity of the search, events are categorized according to their expected diphoton mass
resolution
and signal-to-background ratio. Categories with good
resolution and a large signal-to-background ratio dominate the sensitivity of the search.
To accomplish this, an event classifier variable is constructed based on multi-variate techniques, that assigns a high classifier value to events
with signal-like kinematic characteristics and good diphoton mass resolution, as well as
prompt-photon-like values for the photon identification classifier.
However, the classifier should not be sensitive to the value of the diphoton invariant mass, in order to avoid
biasing the mass distribution that is used to extract a possible signal.
To achieve this, the input variables to the classifier are made dimensionless.
Those that have units of energy (transverse momenta and resolutions) are divided by the diphoton invariant-mass value.
The variables used to train this diphoton event classifier are the scaled photon transverse momenta
($\pt^{\gamma}(1)/\mgg$ and $\pt^{\gamma}(2)/\mgg$), the photon pseudorapidities ($\eta(1)$ and $\eta(2)$),
the cosine of the angle between the two photons in the transverse plane ($\cos\left(\phi(1)-\phi(2)\right)$),
the expected relative diphoton invariant-mass resolutions under the hypotheses of selecting a correct/incorrect
interaction vertex ($\sigma_{m}^{\text{correct (incorrect)}}/\mgg$), the probability of selecting a correct vertex ($p_\text{vtx}$), and
the photon identification classifier values for both photons. The
$\sigma_{m}^{\text{correct (incorrect)}}/\mgg$ is computed using the
single photon resolution estimated by the dedicated BDT described in Section 3.
A vertex is being labeled as correct if the distance from the true
interaction point is smaller than 10\unit{mm}.

To ensure the classifier assigns a high  value to events with good mass resolution, the events
are weighted by a factor inversely proportional to the mass resolution,

\begin{equation}
w_\text{sig} = \frac{p_\text{vtx}}{\sigma_{m}^{\text{correct}}/\mgg} +
\frac{1-p_\text{vtx}}{\sigma_{m}^{\text{incorrect}}/\mgg}.
\end{equation}

This factor incorporates the resolutions under both correct- and incorrect-interaction-vertex
hypotheses, properly weighted by the probabilities of having assigned the vertex correctly. The training is
performed on simulated background and Higgs boson signal events. The training procedure makes full use 
of the signal kinematic properties that are assumed to be those of the SM Higgs boson.
The classifier, though still valid, would not be fully optimal for a particle produced with significantly 
different kinematic properties.

The uncertainties in the diphoton event classifier output come from potential mismodelling of the input
variables. The dominant sources are the uncertainties in the shapes of the photon identification (ID) classifier and
the individual photon energy resolutions, which are used to compute the relative diphoton invariant-mass resolutions.

\begin{figure}
  \begin{center}
    \includegraphics[width=0.49\textwidth]{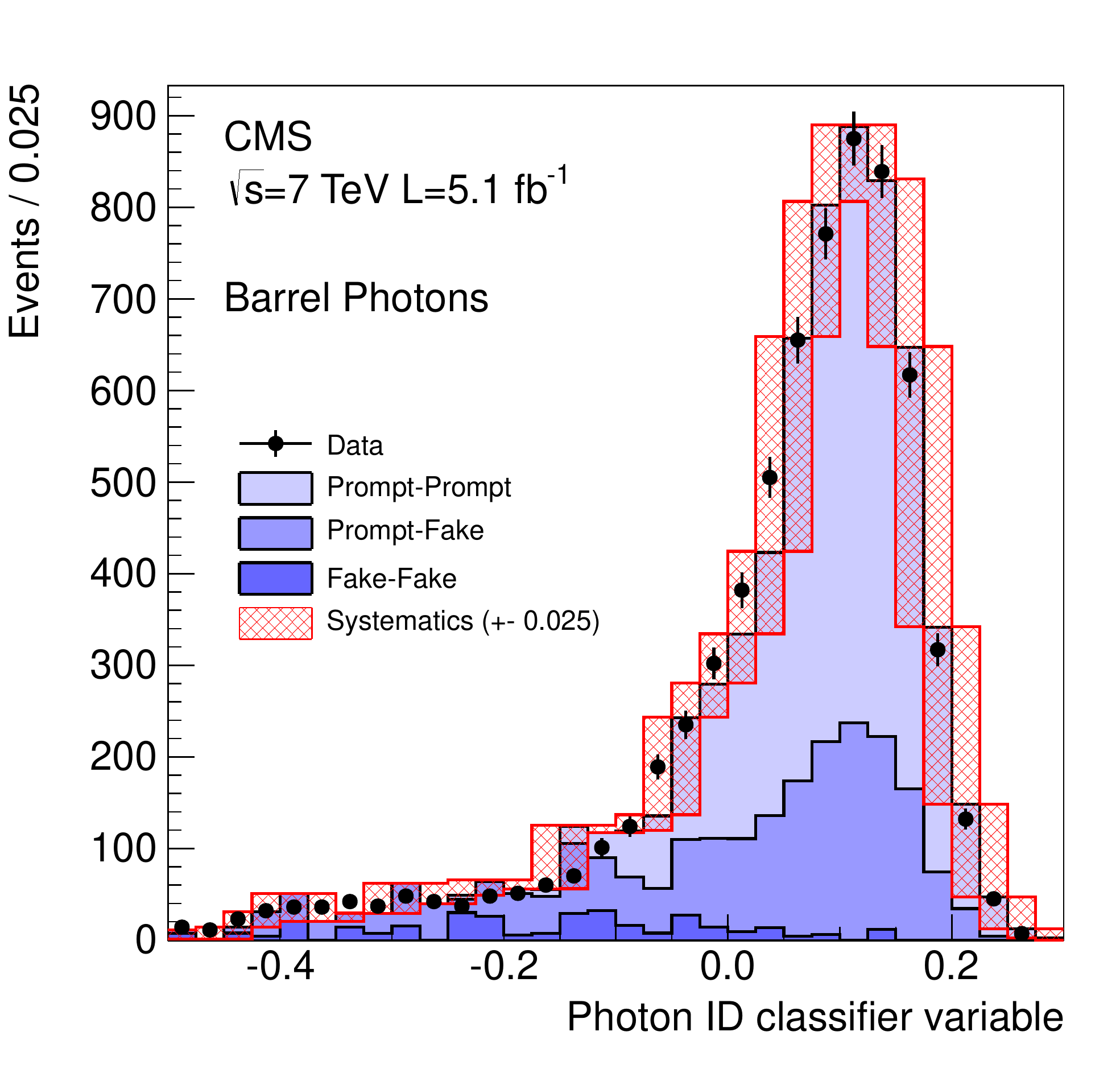}
    \includegraphics[width=0.49\textwidth]{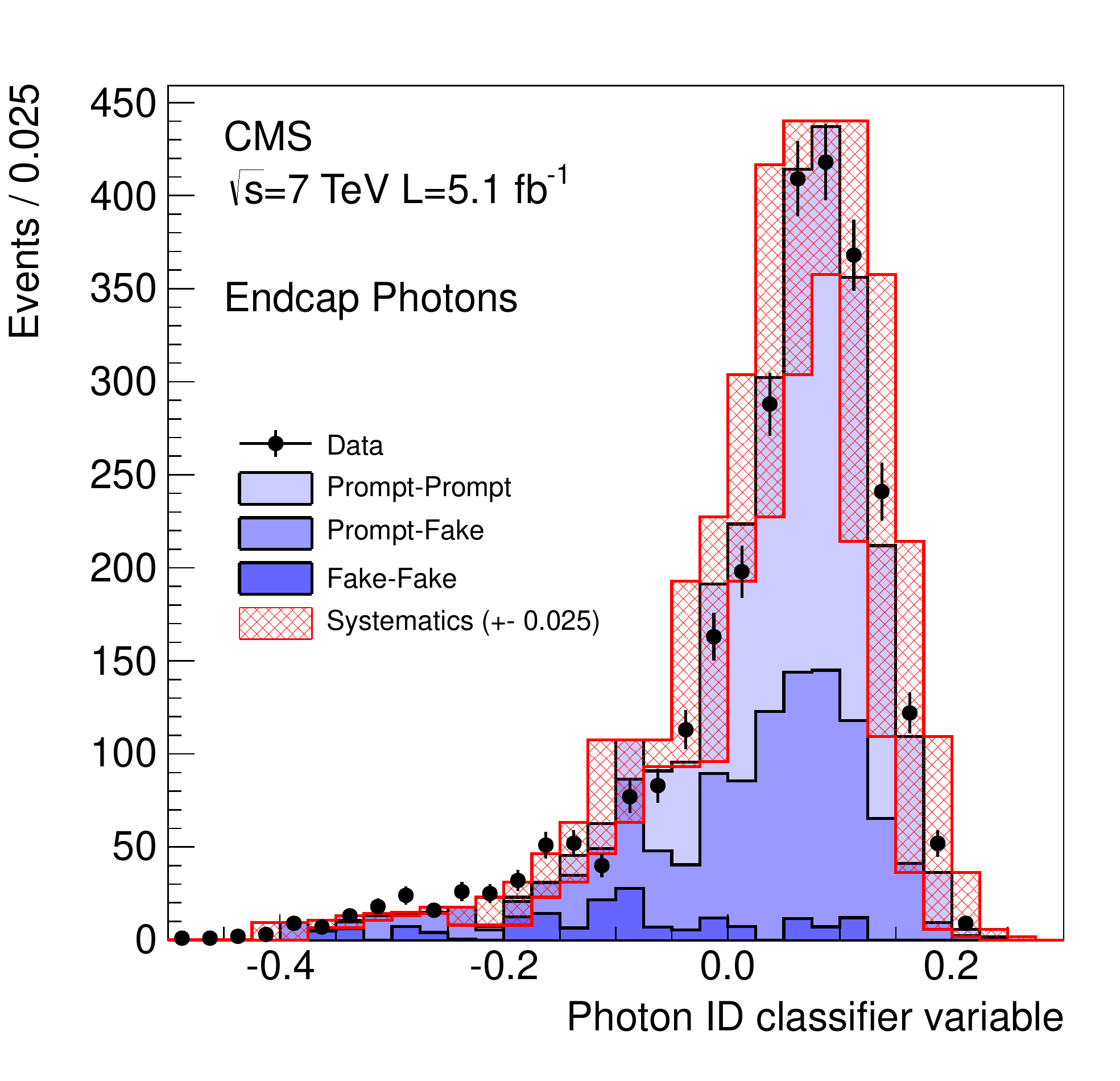}
    \caption{
Distribution of the photon ID classifier value for the larger transverse momentum photon in the
ECAL barrel (left) and endcaps (right) from candidate diphoton data
events (points) with $m_{\gamma\gamma}>160\GeV$.
The predicted distributions for the various diphoton backgrounds as determined from simulation are shown by the histograms.
The variations of the classifier value due to the systematic uncertainties are shown by the cross-hatched histogram.}
    \label{fig:hgg_phoidshift}
  \end{center}
\end{figure}

The first of these
amounts to a potential shift in the photon ID classifier value of at
most ${\pm}0.01$ in the 8\TeV and ${\pm}0.025$ in the 7\TeV analysis.
These values are
set looking to
the observed differences between the  photon ID classifier value distributions from data and simulation.
This comparison for the 7\TeV analysis
is shown in Fig.~\ref{fig:hgg_phoidshift}, where the distribution for the leading (highest $\pt$)
candidate photons in the ECAL barrel (left) and endcaps (right) are compared between data and MC simulation for
$\mgg>160\GeV$, where most photons are prompt ones.
In addition to the three background components described in Section~\ref{sec:hgg_selection} (prompt-prompt, prompt-nonprompt, and
nonprompt-nonprompt),
the additional component composed by Drell--Yan events, in which both final-state electrons are misidentified as photons, has been studied
and found to be negligible.
As discussed previously
a variation of the classifier value by ${\pm}0.025$, represented by the cross-hatched histogram, covers the
differences.

\begin{figure}
 \begin{center}
   \includegraphics[width=0.49\textwidth]{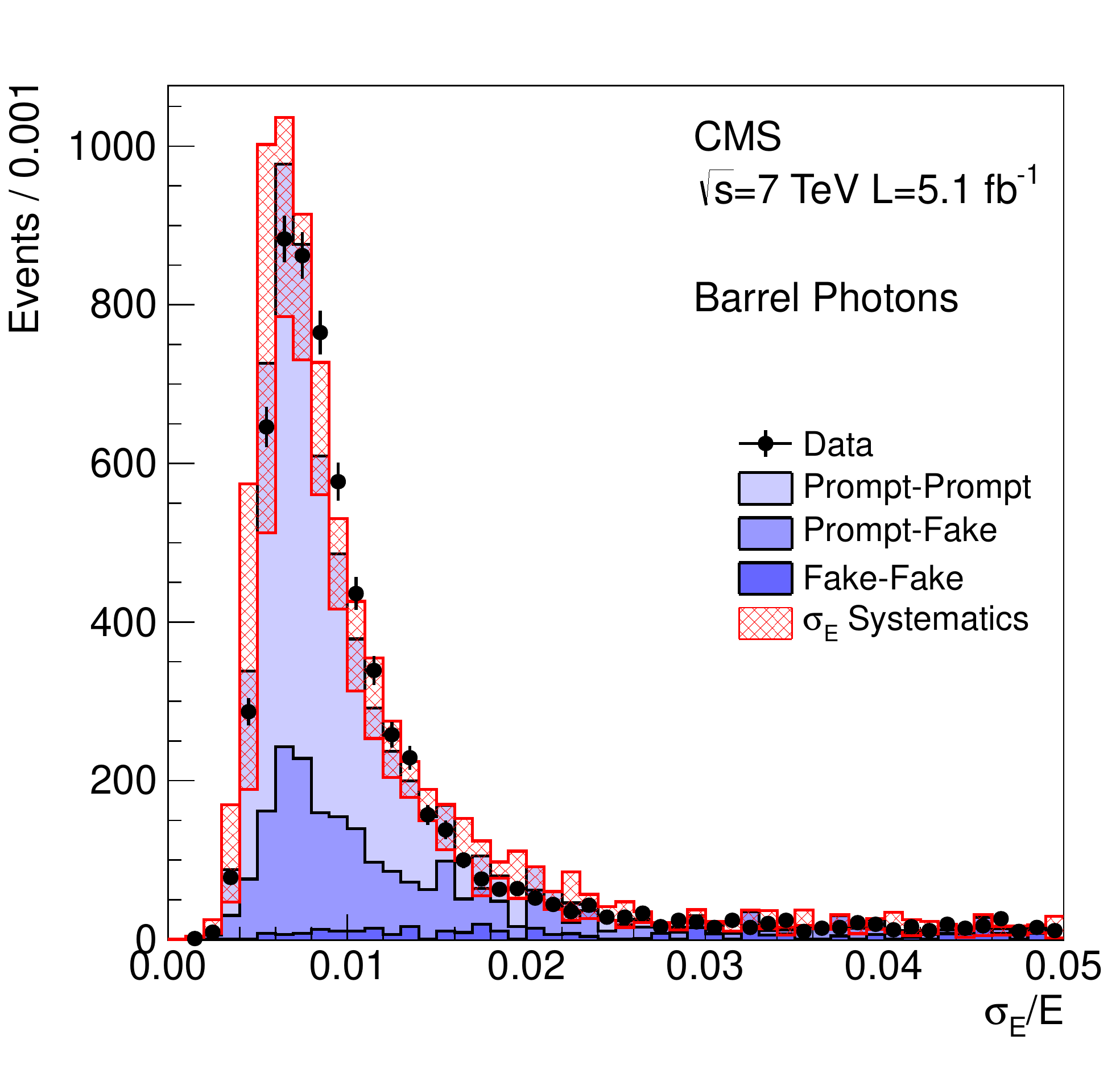}
   \includegraphics[width=0.49\textwidth]{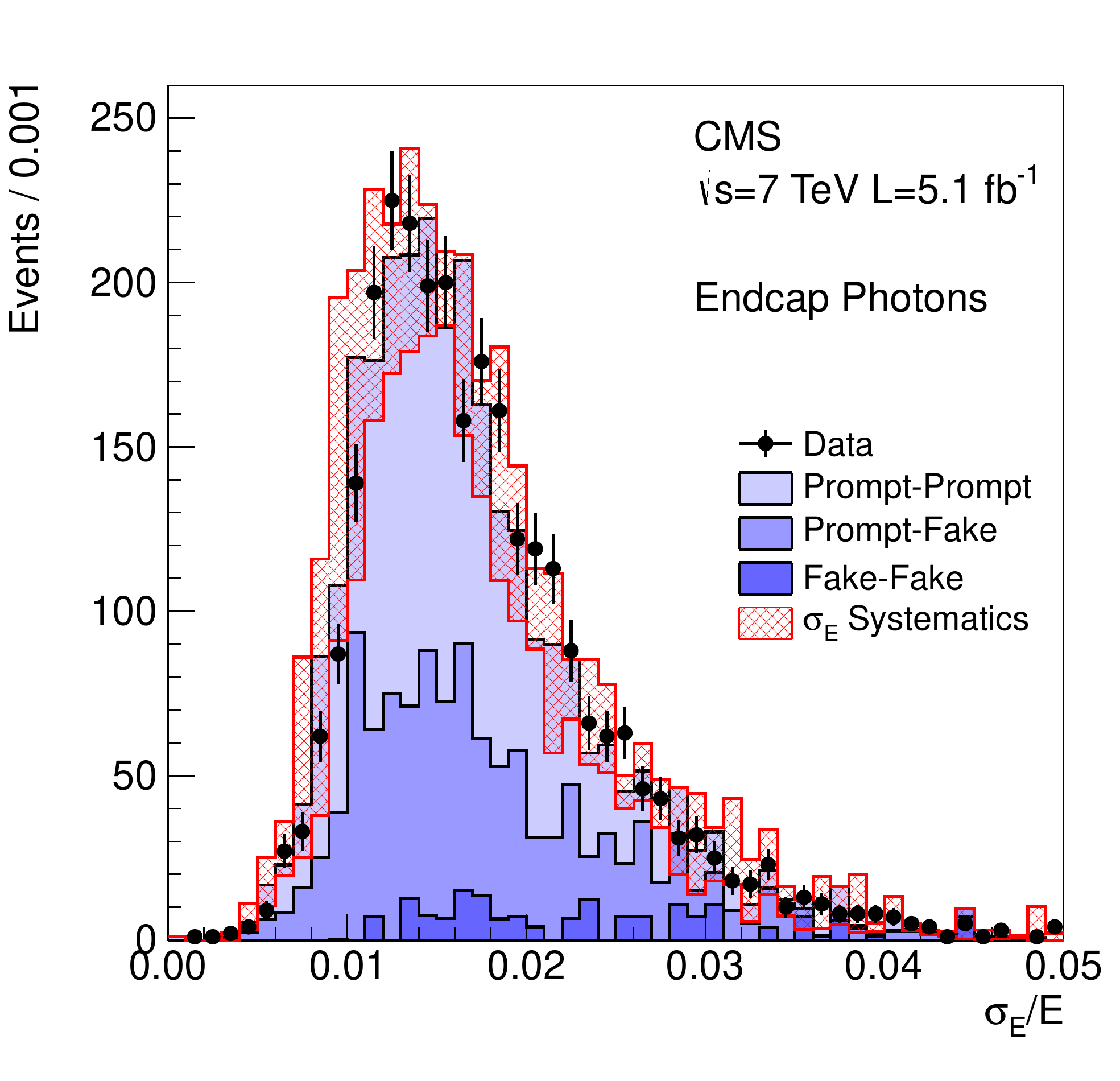}
   \caption{
Distribution of the photon resolution estimate $\sigma_E/E$ for the leading photon in the
ECAL barrel (left) and endcaps (right) from candidate diphoton data
events (points) with $m_{\gamma\gamma}>160\GeV$.
The predicted distributions for the various diphoton backgrounds, as determined from simulation, are shown by the histograms.
The variations of the resolution due to the systematic uncertainties are shown by the cross-hatched histogram.}
   \label{fig:hgg_sigescale}
 \end{center}
\end{figure}

For the second important variable, the photon energy resolution estimate (calculated by a BDT, as discussed in Section~3),
a similar comparison is shown in Fig.~\ref{fig:hgg_sigescale}. Again, the 7\TeV data distributions for candidate
photons in the ECAL barrel (left) and endcap (right) are compared to MC simulation for $\mgg>160\GeV$.
The systematic uncertainty of ${\pm}10\%$ is again shown as the
cross-hatched histogram.

\begin{figure}[htbp]
  \begin{center}
    \includegraphics[width=0.49\linewidth]{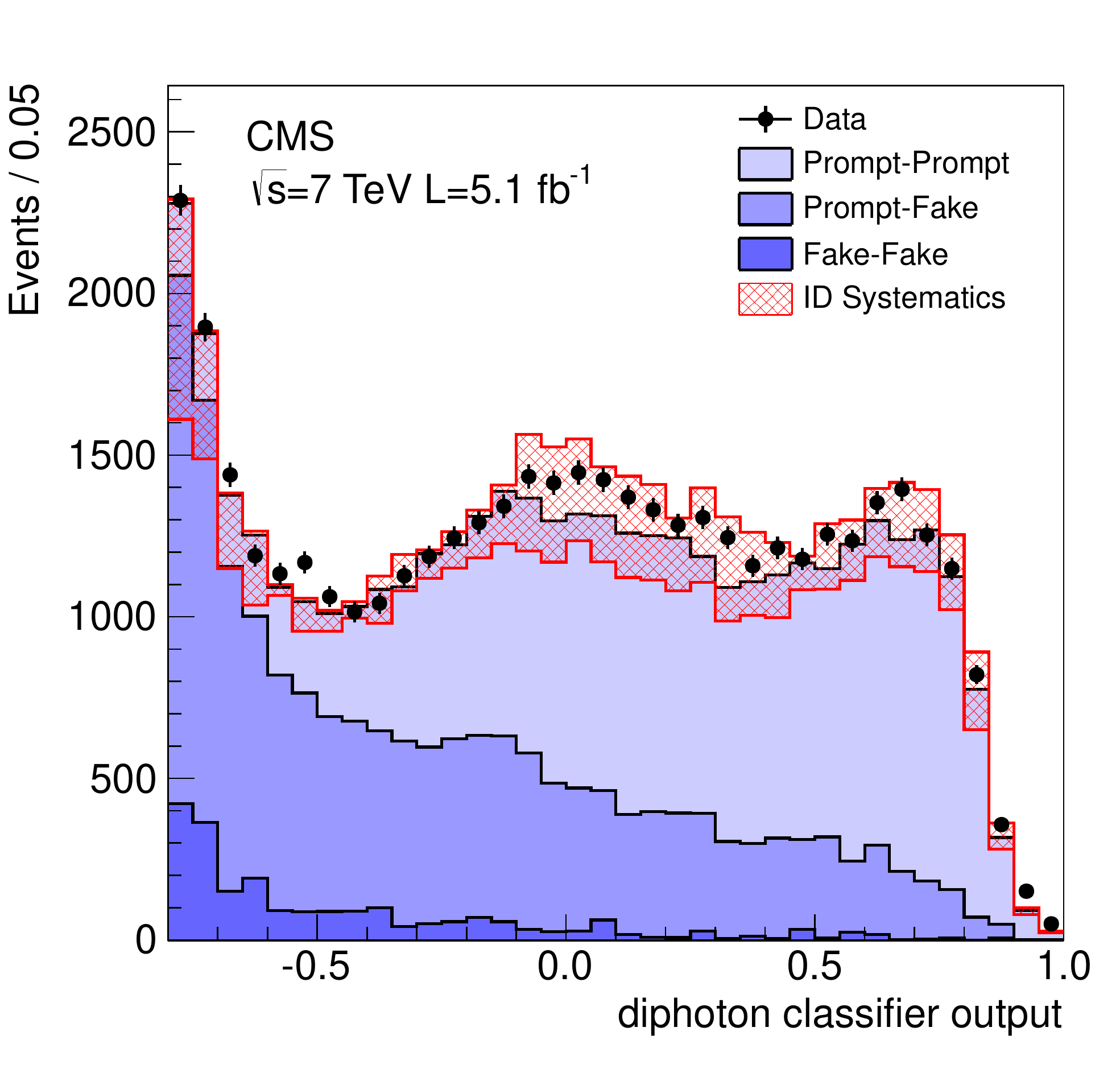}
    \includegraphics[width=0.49\linewidth]{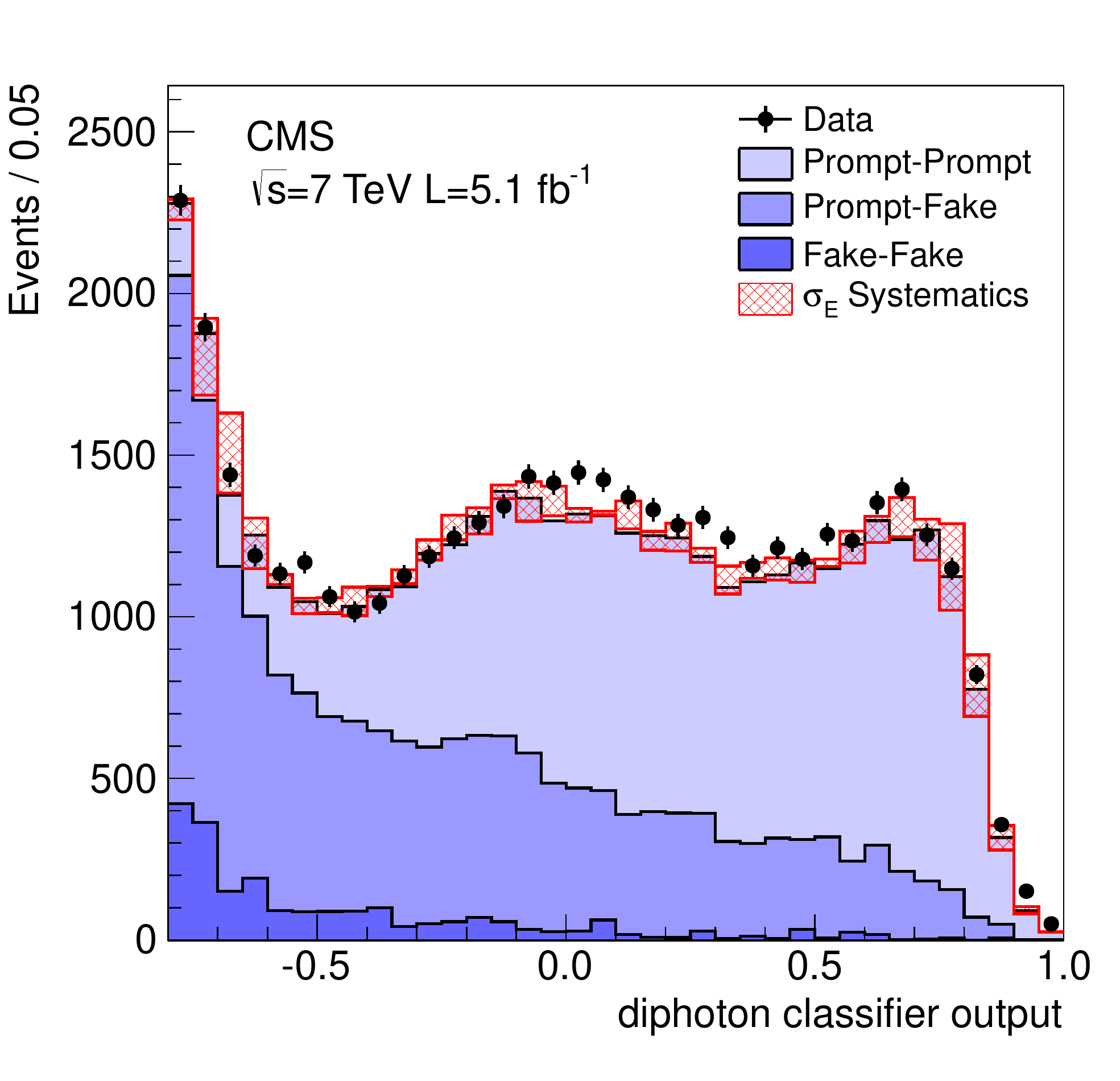}
    \caption{
      The effect of the systematic uncertainty assigned to the photon
      identification classifier output (left) and
      the photon resolution estimate (right)
      on the diphoton BDT output for background MC simulation ($100\GeV<\mgg<180\GeV$) and for data.
      The nominal BDT output is shown as a stacked histogram and the variation due to the
      uncertainty is shown as a cross-hatched band. These plots show only the systematic uncertainties
      that are common to both signal and background. There are additional significant uncertainties
      that are not shown here.
    }
    \label{fig:hgg_dpmvavalidation}
  \end{center}
\end{figure}

The effect of both these uncertainties propagated to the diphoton event classifier distribution
can be seen in Fig.~\ref{fig:hgg_dpmvavalidation}, where the 7\TeV data diphoton classifier variable is
compared to the MC simulation predictions. The data and MC simulation distributions in both the left and right plots of
Fig.~\ref{fig:hgg_dpmvavalidation} are the same. In the left plot, the uncertainty band arises from
propagating the photon ID classifier uncertainty, while in the right plot, it is
from propagating the energy resolution uncertainty.
From these plots one can see that the uncertainty in the photon ID classifier dominates the overall uncertainty,
and by itself almost covers the full difference between the data and MC simulation distributions.
Both uncertainties are propagated into the final result.

The diphoton event classifier output is then used to
divide events into different classes, prior to fitting the diphoton invariant-mass spectrum.
The procedure successively splits events into classes by
introducing a boundary value for the diphoton classifier output. The first boundary results in two classes,
and then these classes are further split.
Each split is introduced using the boundary value that gives rise to
the best expected exclusion limit. The procedure is terminated once additional
splitting results in a negligible (${<}1$\%) gain in sensitivity.
Additionally, the lowest score class is dropped since it does not
contribute significantly to the sensitivity. This procedure results
in four event classes for both the 7 and 8\TeV data sets. The systematic
uncertainties  in  the diphoton identification classifier and photon energy resolution
discussed above can cause
events to migrate between classes. In the 8\TeV analysis,
these class migrations are up to
$4.3\%$ and $8.1\%$, respectively. They are defined as the relative change of expected signal yield in
each category under the variation of the photon ID BDT classifier and the per-photon energy resolution estimate,
within their uncertainties as explained above.

The sensitivity of the analysis is enhanced by using
the special kinematics of Higgs bosons produced by the VBF process~\cite{Ballestrero:2008gf}.
Dedicated classes of events are selected using dijet-tagging criteria.
The 7\TeV data set has one class of dijet-tagged events, while
the 8\TeV data set has two.

In the 7\TeV analysis, dijet-tagged events are required to contain two jets with transverse
energies exceeding 20 and 30\GeV, respectively. The dijet invariant mass is required to
be greater than  350\GeV, and the absolute value of the difference of the pseudorapidities of the
two jets has to be larger than 3.5.
In the 8\TeV analysis, dijet-tagged events are required to
contain two jets and are categorized as ``Dijet tight" or ``Dijet loose".
The jets in Dijet tight events must have transverse energies above 30\GeV and  a
dijet invariant mass greater than  500\GeV.
For the jets in the Dijet loose events, the leading (subleading) jet transverse energy
must exceed 30 (20)\GeV and the dijet invariant mass be greater than
250\GeV, where leading and subleading refer to the jets with the highest and
next-to-highest transverse momentum, respectively.
The pseudorapidity separation between the two jets
is also required to be greater than 3.0. Additionally, in both analyses the difference between
the average pseudorapidity of the two jets and the pseudorapidity of the diphoton system must
be less than 2.5~\cite{Rainwater:1996ud}, and the difference in azimuthal
angle between the diphoton system and the dijet system is required to
be greater than 2.6\unit{radians}.
To further reduce the background in the dijet classes, the $\pt$ threshold on the leading
photon is increased to $\pt^{\gamma}(1) > \mgg/2$.

Systematic uncertainties in the efficiency of dijet tagging for
signal events arise from the uncertainty in the MC simulation modelling of the jet energy
corrections and resolution, and from uncertainties in simulating
the number of jets and their kinematic properties.
These uncertainties are estimated by using  different underlying-event tunes,
PDFs, and renormalization and factorization scales
as suggested in Refs.~\cite{LHCHiggsCrossSectionWorkingGroup:2011ti,Dittmaier:2012vm}.
A total systematic uncertainty of 10\% is assigned to the efficiency
for VBF signal events to pass the dijet-tag criteria, and an
uncertainty of 50\%, dominated by the uncertainty in the
underlying-event tune, to the efficiency for signal events
produced by gluon-gluon fusion.

Table~\ref{tab:ClassFracs} shows the predicted number of signal events for  a SM
Higgs boson with $\mH= 125\GeV$, as well as the estimated number of background events per \GeVns of invariant mass
at $\mgg = 125\GeV$, for each of the eleven event classes in the 7 and 8\TeV data sets.
The table also gives the fraction of each Higgs boson production process in each class (as
predicted by MC simulation) and the mass resolution, represented both as
$\sigma_\text{eff}$, half the width of the narrowest interval containing
68.3\% of the distribution, and as the full-width-at-half-maximum (FWHM) of the invariant-mass
distribution divided by 2.35.

\begin{table}[htbp]
\begin{center}
\topcaption{Expected number of SM Higgs boson events ($\mH=125\GeV$) and
estimated background (at $\mgg= 125\GeV$) for the event classes in the 7
 (5.1\fbinv) and 8\TeV (5.3\fbinv) data sets.
The composition of the SM Higgs boson signal in terms of the production
processes and its mass resolution are also given.}

\begin{tabular}{>{\small}c<{\small}|>{\small}r<{\small}||r|>{\small}r<{\small}>{\small}r<{\small}>{\small}r<{\small}>{\small}r<{\small}|>{\centering}b{1.3cm}<{\centering}|>{\centering}b{2.0cm}<{\centering}||r@{\,$\pm$\,}l}
\hline
\multicolumn{2}{c||}{\multirow{2}{*}{Event classes}} & \multicolumn{7}{c||}{SM Higgs boson expected signal ($\mH=125\GeV$)} &
\multicolumn{2}{c}{\multirow{2}{*}{\begin{minipage}[t]{2.5cm}\begin{center}Background \footnotesize{$\mgg=125\GeV$}\\\small{(events/\GeV)}\end{center}\end{minipage}}}\tabularnewline
\cline{3-9}
\multicolumn{2}{c||}{} & Events & ggH & VBF & VH & ttH & $\sigma_\text{eff}$ \small{(\GeVns{})} & \small{FWHM/2.35} \small{(\GeVns{})} & \multicolumn{2}{c}{} \tabularnewline
\hline\hline
\multirow{5}{*}{7\TeV}
& BDT 0       &  3.2 & 61\% & 17\% & 19\% & 3\% & 1.21 & 1.14 & \rule{6mm}{0mm} 3.3 & 0.4 \tabularnewline
& BDT 1       & 16.3 & 88\% &  6\% &  6\% &  -- & 1.26 & 1.08 &  37.5 & 1.3 \tabularnewline
& BDT 2       & 21.5 & 92\% &  4\% &  4\% &  -- & 1.59 & 1.32 &  74.8 & 1.9 \tabularnewline
& BDT 3       & 32.8 & 92\% &  4\% &  4\% &  -- & 2.47 & 2.07 & 193.6 & 3.0 \tabularnewline
& Dijet tag   &  2.9 & 27\% & 72\% &  1\% &  -- & 1.73 & 1.37 &   1.7 & 0.2 \tabularnewline
\hline
\multirow{6}{*}{8\TeV}
& BDT 0       &  6.1 & 68\% & 12\% & 16\% & 4\% & 1.38 & 1.23 &   7.4 & 0.6 \tabularnewline
& BDT 1       & 21.0 & 87\% &  6\% &  6\% & 1\% & 1.53 & 1.31 &  54.7 & 1.5 \tabularnewline
& BDT 2       & 30.2 & 92\% &  4\% &  4\% &  -- & 1.94 & 1.55 & 115.2 & 2.3 \tabularnewline
& BDT 3       & 40.0 & 92\% &  4\% &  4\% &  -- & 2.86 & 2.35 & 256.5 & 3.4 \tabularnewline
& Dijet tight &  2.6 & 23\% & 77\% &   -- &  -- & 2.06 & 1.57 &   1.3 & 0.2 \tabularnewline
& Dijet loose &  3.0 & 53\% & 45\% &  2\% &  -- & 1.95 & 1.48 &   3.7 & 0.4 \tabularnewline

\hline
\end{tabular}

\label{tab:ClassFracs}
\end{center}
\end{table}

\subsection{Signal and background modelling}
\label{sec:hgg_smodeling}

The modelling of the Higgs boson signal used in the estimation of the sensitivity has
two aspects. First, the normalization, \ie the
expected number of signal events for each of the considered Higgs boson production processes;
second, the diphoton invariant-mass shape. To model both aspects, including their respective
uncertainties, the MC simulation events and theoretical considerations described
in Section~\ref{sec:searches} are used. To account for the interference between the signal and
background diphoton final states~\cite{interference}, the expected
gluon-gluon fusion process cross section is reduced by 2.5\% for all
values of \mH.

Additional systematic uncertainties in the normalization of each event class
arise from potential class-to-class migration of signal events caused by
uncertainties in the diphoton event classifier value.
The instrumental uncertainties in the classifier value and their effect have been discussed previously.
The theoretical ones, arising from the uncertainty
in the theoretical predictions for the photon kinematics,
are estimated by measuring the amount of class migration under variation of the renormalization and factorization scales
within the range $\mH/2 < \mu < 2\mH$,
(class migrations up to 12.5\%) and the PDFs (class migrations up to 1.3\%).
These uncertainties are propagated to the final statistical analysis.

\begin{figure}[htbp]
  \begin{center}
    \includegraphics[width=0.55\linewidth]{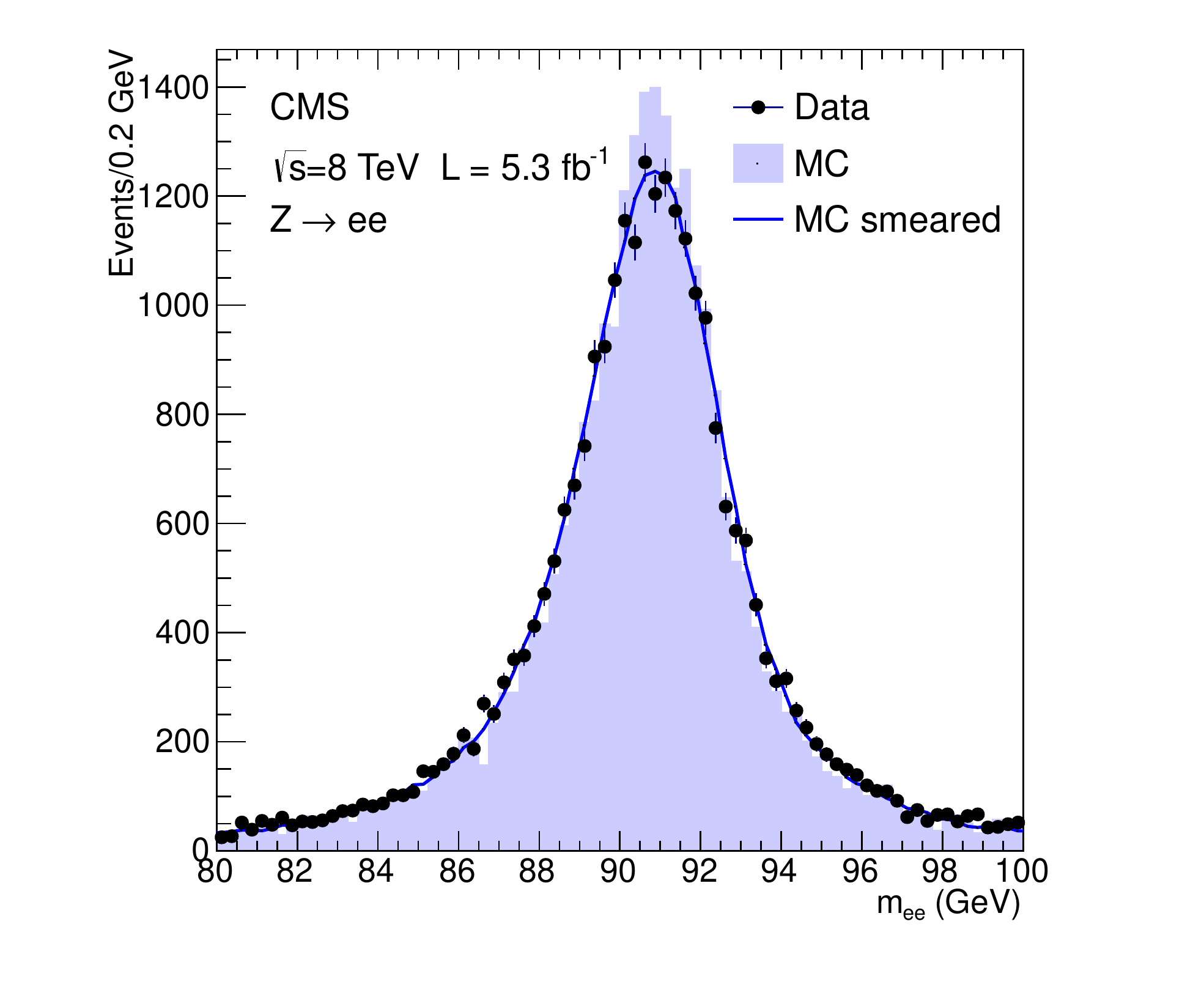}
    \caption{
      Comparison of the dielectron invariant-mass spectrum from \cPZ~$\to \Pe\Pe$ events between 8\TeV
      data (points) and the simulated events (histogram), where the selected electrons are reconstructed as photons.
      The simulated distribution after applying smearing and scaling corrections of the electron energies
      is shown by the solid line.
    }
    \label{fig:hgg_resolution}
  \end{center}
\end{figure}

To model the diphoton invariant-mass spectrum properly, it is essential that
the simulated diphoton mass and scale are accurately predicted.
This is done by comparing the dielectron invariant-mass distribution in
$\cPZ\to \Pe\Pe$ events between data and MC simulation, where the electrons have been
reconstructed as photons. This comparison is shown for the 8\TeV
data in Fig.~\ref{fig:hgg_resolution}, where the points represent
data, and the histogram MC simulation.
Before correction, the dielectron invariant-mass distribution from simulation is narrower than the one from data,
caused by an inadequate modelling of the photon energy resolution in the simulation.
To correct this effect, the photon energies in the Higgs boson signal MC simulation events
are smeared and the data events scaled, so that the dielectron invariant-mass scale and
resolution as measured in $\cPZ\to \Pe\Pe$ events agree between data and MC simulation.
These scaling and smearing factors are determined in a total of eight photon categories, \ie
separately for photons in four pseudorapidity regions
($|\eta|<1$, $1\leq|\eta|<1.5$, $1.5\leq|\eta|<2$, and $|\eta| \geq 2$),
and separately for high $R9$ (${>}0.94$)~and low $R9$ (${\leq}0.94$)~photons, where $R9$ is the ratio of the
energy of the  most energetic $3\times3$ crystal cluster and the total cluster energy.

Additionally, the factors are computed separately for different running periods
in order to account for changes in the running conditions, for example the change
in the average beam intensity.
These modifications reconcile the discrepancy between data and simulation,
as seen in the comparison of the dots and solid curve of
Fig.~\ref{fig:hgg_resolution}. The uncertainties
in the scaling and smearing factors, which
range from 0.2\% to 0.9\% depending on the photon properties, are taken as
systematic uncertainties in the signal evaluation and mass measurement.

The final signal model is then constructed separately for each event class
and each of the four production processes as the weighted sum of two submodels
that assume either the correct or incorrect primary vertex selection (as described in Section~\ref{sec:hgg_vertex}).
The two submodels are weighted by the corresponding probability
of picking the right ($p_\mathrm{vtx}$) or wrong ($1-p_\mathrm{vtx}$) vertex.
The uncertainty in the parameter $p_\mathrm{vtx}$ is taken as a systematic uncertainty.

To describe the signal invariant-mass shape in each submodel, two different approaches are
used. In the first, referred to as the parametric model, the MC simulated  diphoton invariant-mass
distribution is fitted to a sum of Gaussian distributions. The number
of Gaussian functions ranges from one to three depending on the event class, and whether the model is a
correct- or incorrect-vertex hypothesis.
The systematic uncertainties in the signal shape are estimated from the
variations in the parameters of the Gaussian functions.
In the second approach, referred to as the binned model,  the
signal mass shape for each event class is taken directly from the  binned histogram
of the corresponding simulated Higgs boson events.
The systematic uncertainties are included
by parametrizing the change in each bin of the histogram as a linear function under variation
of the corresponding nuisance parameter, \ie the variable that parametrizes this uncertainty
in the statistical interpretation of the data.
The two approaches yield consistent final results and serve as an additional verification
of the signal modelling. The presented results are derived using the parametric-model approach.

\begin{figure}[htbp]
  \begin{center}
    \includegraphics[width=0.49\linewidth]{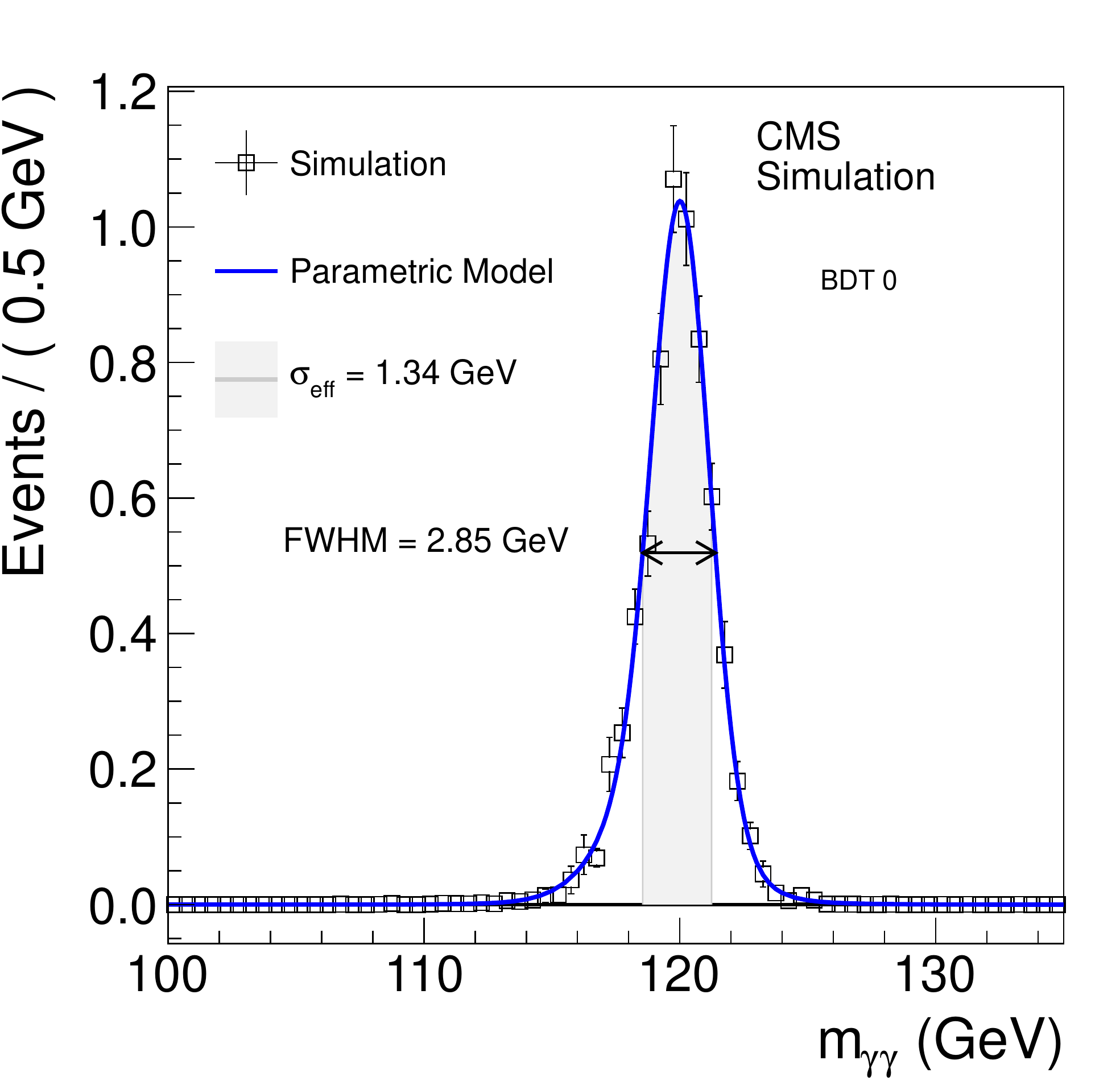}
    \includegraphics[width=0.49\linewidth]{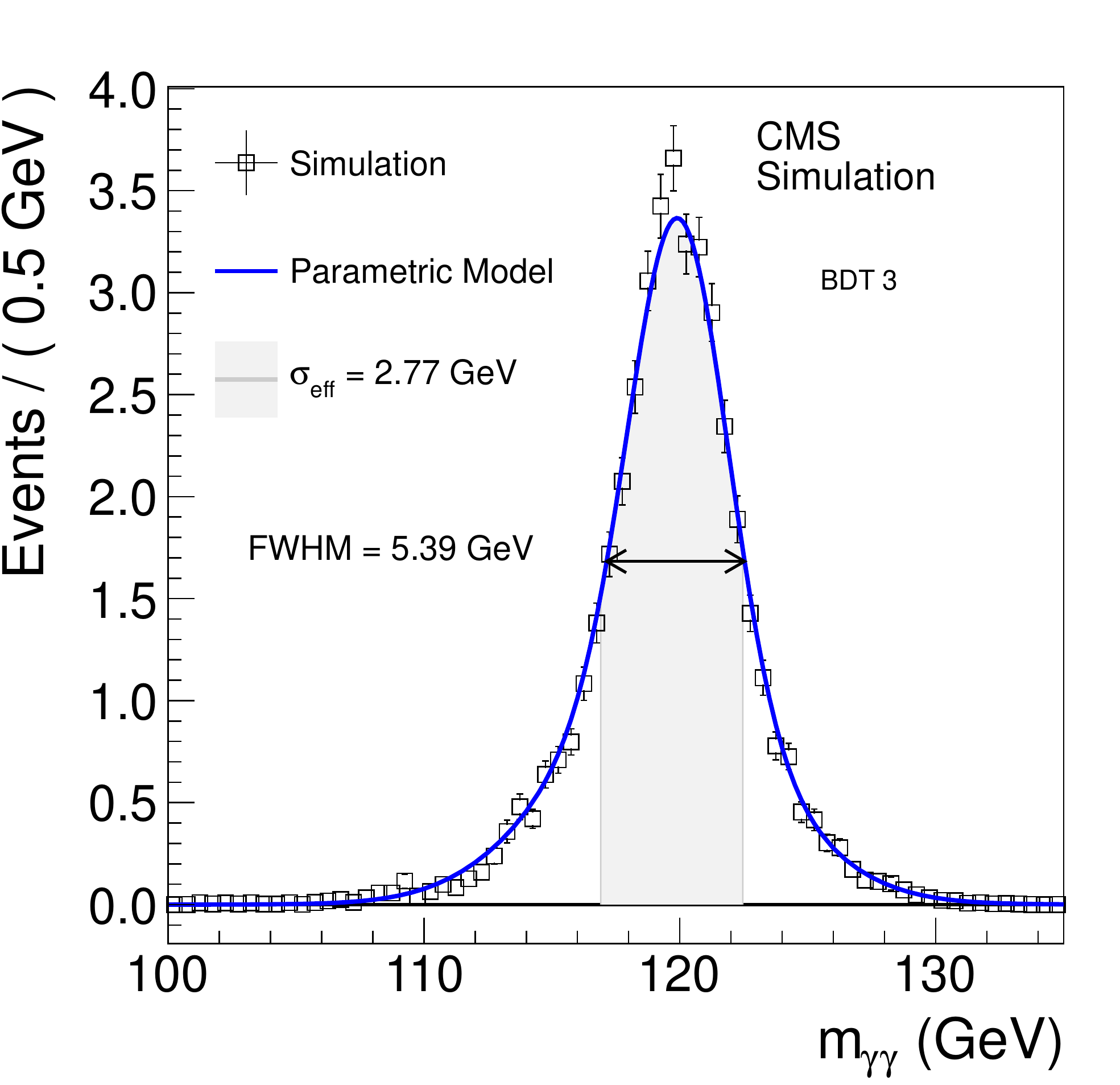}
    \caption{
      Comparison of the diphoton invariant-mass distribution from the parametric signal model (blue line) and simulated
      MC events (open squares) for a Higgs mass
      hypothesis of $\mH=120\GeV$ for two (BDT 0 on the left, BDT 3 on the right)
      of the four 8\TeV BDT event classes.
    }
    \label{fig:hgg_sigmodel}
  \end{center}
\end{figure}

The parametric signal models for a Higgs boson mass of
$120\GeV$ in two of the 8\TeV BDT event classes are shown in
Fig.~\ref{fig:hgg_sigmodel}. The signal models are
summed over the four production processes, each weighted by their respective
expected yield as computed from MC simulation.
The two plots in Fig.~\ref{fig:hgg_sigmodel} illustrate how the
diphoton invariant-mass resolution improves with increasing diphoton classifier value.
The left distribution is for classifier
values greater than 0.88 and has a mass
resolution $\sigma_{\text{eff}} = 1.34\GeV$, while the
right distribution is for classifier values
between $-0.05$ and 0.50 and has $\sigma_{\text{eff}} = 2.77\GeV$.
This is the intended behaviour of the event class implementation.

The uncertainties in the weighting factors for each of the production processes
arise from variations in the renormalization
and factorization scales, and uncertainties in the PDFs.
They range from several percent for associated production with W/Z to almost 20\%
for the gluon-gluon fusion process. The detailed values for the 8\TeV analysis, together with
all the other systematic uncertainties discussed above, are summarized in
Table~\ref{tab:hgg_systematics}. The corresponding uncertainties in the 7\TeV analysis are very similar,
with the exception of the already mentioned uncertainty on the photon ID classifier, which
was significantly larger in the 7\TeV analysis.
The reason for this is a worse agreement between data and MC simulation.

In addition to the per-photon energy scale uncertainties, that are derived in the eight $\eta-R9$
categories, additional fully correlated energy scale uncertainties are assigned in order to account
for possible non-linearity as a function of energy and for additional electron-photon
differences. The uncertainty associated with possible non-linearities in the energy measurement as a
function of the cluster energy are evaluated by measuring the energy scale of $\Z\to \Pe\Pe$ events as a
function of the scalar sum of  transverse momentum of the two electrons. The change in energy
scale due to possible non-linearities in the energy measurement is
estimated around $0.2\%$; since this correction is not applied, a
systematic uncertainty of $0.4\%$ is assigned.
An additional  fully correlated uncertainty related to difference of $0.25\%$
between  electron and photon is assigned,
amounting to half of the absolute energy scale difference between electrons and photons
for non-showering electrons/photons in the barrel. Adding these two numbers in quadrature results
in the additional energy scale uncertainty of $0.47\%$, that is treated as
fully correlated among all event classes.

\begin{table}[htbp]
\topcaption{Largest sources of systematic uncertainty in
the analysis of the 8\TeV data set. Eight photon categories are defined, depending on their $\eta$ and $R9$,
where $R9$ is the ratio of the energy of the  most energetic $3\times3$ crystal cluster and the total cluster energy.
The four pseudorapidity regions are: $|\eta|<1$ (low $\eta$), $1\leq|\eta|<1.5$ (high $\eta$) for the barrel,
and  $1.5\leq|\eta|<2$ (low $\eta$), $|\eta| \geq 2$ (high $\eta$) for the endcaps;
the two $R9$ regions are: high $R9$ ($>0.94$) and low $R9$ (${\leq}0.94$).}
\centering\small{
\begin{tabular}{ l r|c|c}
\hline
\multicolumn{2}{ l |}{\textbf{Sources of systematic uncertainty}} & \multicolumn{2}{ c }{\textbf{Uncertainty}}\\
\hline
\hline
\multicolumn{2}{ l |}{\textbf{Per photon}} & Barrel & Endcap \\
\hline
\multicolumn{2}{ l |}{Photon selection efficiency} & 0.8\% & 2.2\%\\
Energy resolution ($\Delta\sigma/E_{\mathrm{MC}}$) & $R9 > 0.94$ (low $\eta$, high
$\eta$) & 0.22\%, 0.60\% &0.90\%, 0.34\% \\
& $R9 \leq 0.94$ (low $\eta$, high $\eta$) & 0.24\%, 0.59\% &
0.30\%, 0.52\% \\
Energy scale ($(E_{\text{data}}-E_{\mathrm{MC}})/E_{\mathrm{MC}}$) & $R9 > 0.94$ (low $\eta$, high
$\eta$) & 0.19\%, 0.71\% & 0.88\%, 0.19\% \\
& $R9 \leq 0.94$ (low $\eta$, high $\eta$) & 0.13\%, 0.51\% &
0.18\%, 0.28\% \\
\hline
\multicolumn{2}{l |}{Energy scale (fully correlated)} &
\multicolumn{2}{ c }{$0.47\,\%$}\\
\multicolumn{2}{r|}{} &
\multicolumn{2}{ c }{} \\

\hline
\multicolumn{2}{l |}{Photon identification classifier} &
\multicolumn{2}{ c }{$0.01$}\\
\multicolumn{2}{r|}{} &
\multicolumn{2}{ c }{} \\

\hline
\multicolumn{2}{ l |}{Photon energy resolution BDT} &
\multicolumn{2}{ c }{$10\%$}\\
\multicolumn{2}{r|}{} &
\multicolumn{2}{ c }{} \\

\hline
\hline
\multicolumn{4}{ l }{\textbf{Per event}}\\
\hline
\multicolumn{2}{l|}{Integrated luminosity} & \multicolumn{2}{ c }{4.4\%} \\
\multicolumn{2}{l|}{Vertex finding efficiency} & \multicolumn{2}{ c }{0.2\%}\\
\multicolumn{2}{l|}{Trigger efficiency --- One or both photons
$R9 \leq 0.94$ in endcap} & \multicolumn{2}{ c }{0.4\%} \\
\multicolumn{2}{r|}{Other events} & \multicolumn{2}{ c }{0.1\%} \\

\hline
\hline
\multicolumn{4}{ l }{\textbf{Dijet selection}}\\
\hline
Dijet tagging efficiency & VBF  & \multicolumn{2}{ c }{10\%}\\
\multicolumn{2}{r|}{Gluon-gluon fusion } & \multicolumn{2}{ c }{50\%}\\
\multicolumn{2}{r|}{} &
\multicolumn{2}{ c }{} \\

\hline
\hline
\multicolumn{2}{ l |}{\textbf{Production cross sections}} & Scale & PDF \\
\hline
\multicolumn{2}{l|}{Gluon-gluon fusion} & +12.5\% -8.2\% & +7.9\% -7.7\% \\
\multicolumn{2}{l|}{VBF} & +0.5\% -0.3\% & +2.7\% -2.1\% \\
\multicolumn{2}{l|}{Associated production with W/Z} & 1.8\% & 4.2\% \\
\multicolumn{2}{l|}{Associated production with $\ttbar$} & +3.6\% -9.5\% & 8.5\% \\

\hline
\end{tabular}
}
\label{tab:hgg_systematics}
\end{table}

The modelling of the background relies entirely on the data.
The observed diphoton invariant-mass distributions for the eleven
event classes (five in the 7 and eight in the 8\TeV analysis)
are fitted separately over the range  $100 < \mgg < 180\GeV$.
This has the advantage that there are no systematic uncertainties due to potential
mismodelling of the background processes by the MC simulation.
The procedure is to fit the diphoton invariant-mass distribution to the sum of a
signal mass peak and a background distribution.
Since the exact functional form of the background
in each event class is not known, the parametric model has to be flexible
enough to describe an entire set of potential underlying functions.
Using a wrong background model can lead to biases in the measured
signal strength. Such a bias can, depending on the Higgs boson mass and the event class,
reach or even exceed the size of the expected signal, and therefore
dramatically reduce
the sensitivity of the analysis to any potential signal.
In what follows, a procedure for selecting the background function is described that
results in a potential bias small enough to be neglected.

If the true underlying background model could be used in the extraction of the signal strength,
and no signal is present in the fitted data, the median fitted signal strength would be zero in
the entire mass region of interest. The deviation of the median fitted signal strength from zero
in background-only pseudo-experiments can thus be used to quantify the potential bias.
These pseudodata sets
are generated from a set of hypothetical truth models, with each model using a different
analytical function that adequately describes the observed diphoton invariant-mass distribution.
The set of truth-models contains exponential and power-law functions,
as well as polynomials (Bernstein polynomials) and Laurent series of different orders.
None of these functions is required to describe the actual (unknown) underlying background distribution.
Instead, we argue that they span the phase-space of potential underlying
models in such a way that a fit model resulting in a negligible bias against all of them would also
result in a negligible bias against the (unknown) true underlying distribution.

The first step in generating such pseudodata sets consists of constructing a truth model, from which
the pseudodata set is drawn. This is done by fitting the data in each of the eleven event classes separately, and for each
of the four general types of background functions, resulting in four truth-models for each event class. The order
of the background function required to adequately describe the data for each of the models is determined
by increasing the order
until an additional increase does not result in a significant improvement of the fit to the
observed data. A $\chi^2$-goodness-of-fit is used to quantify the fit quality, and
an F-test to determine the termination criterion. ``Increasing the order'' here means adding additional
terms of higher order in the case of the polynomial and the Laurent series, and adding additional
exponential or power-law terms with different parameters in the case of the exponential and power-law
truth models.

Once the four truth models are determined for a given event class, ${\sim}40\,000$ pseudodata sets
are generated for each by randomly drawing diphoton mass values from them. The next step is then
to find a function (in what follows referred to as \emph{fit model}), that results in a negligible bias
against all four sets of toy data in the entire mass region of interest, \ie an analytical function that
when used to extract the signal strength in all the 40\,000  pseudodata sets, gives a mean value for the fitted strength
consistent with zero.

The criterion for the bias to be negligible is that it must be
five times smaller than the statistical uncertainty in the number of fitted events
in a mass window corresponding to the FWHM of the corresponding signal model.
With this procedure, any potential bias from the background fit function can be neglected in comparison with the statistical
uncertainty from the finite data sample. We find that only the polynomial background function produces a
sufficiently small bias for all four truth models. Therefore, we only use this background function to fit the data.
The required order of the polynomial function needed to reach the sufficiently small bias is determined separately for each
of the 11 event classes, and ranges from 3 to 5.

The entire procedure results in a background model for each of the event classes as a polynomial function
of a given, class-dependent order. The parameters of this polynomial, \ie the coefficients for each term,
are left free in the fit, and their variations are therefore the only source of uncertainty
from the modelling of the background.

\begin{figure}[htbp]
  \begin{center}
    \includegraphics[width=0.45\linewidth]{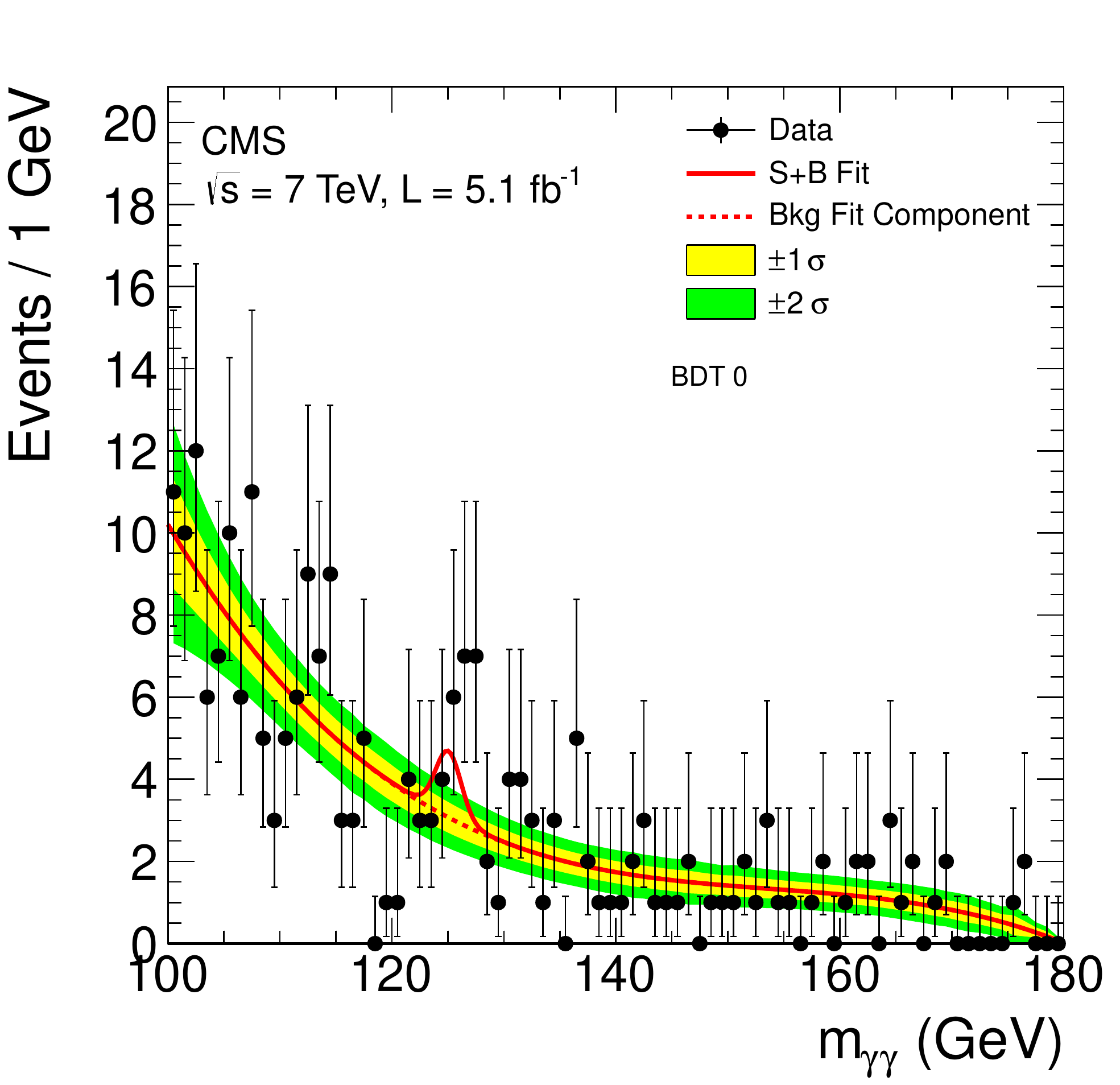}
    \includegraphics[width=0.45\linewidth]{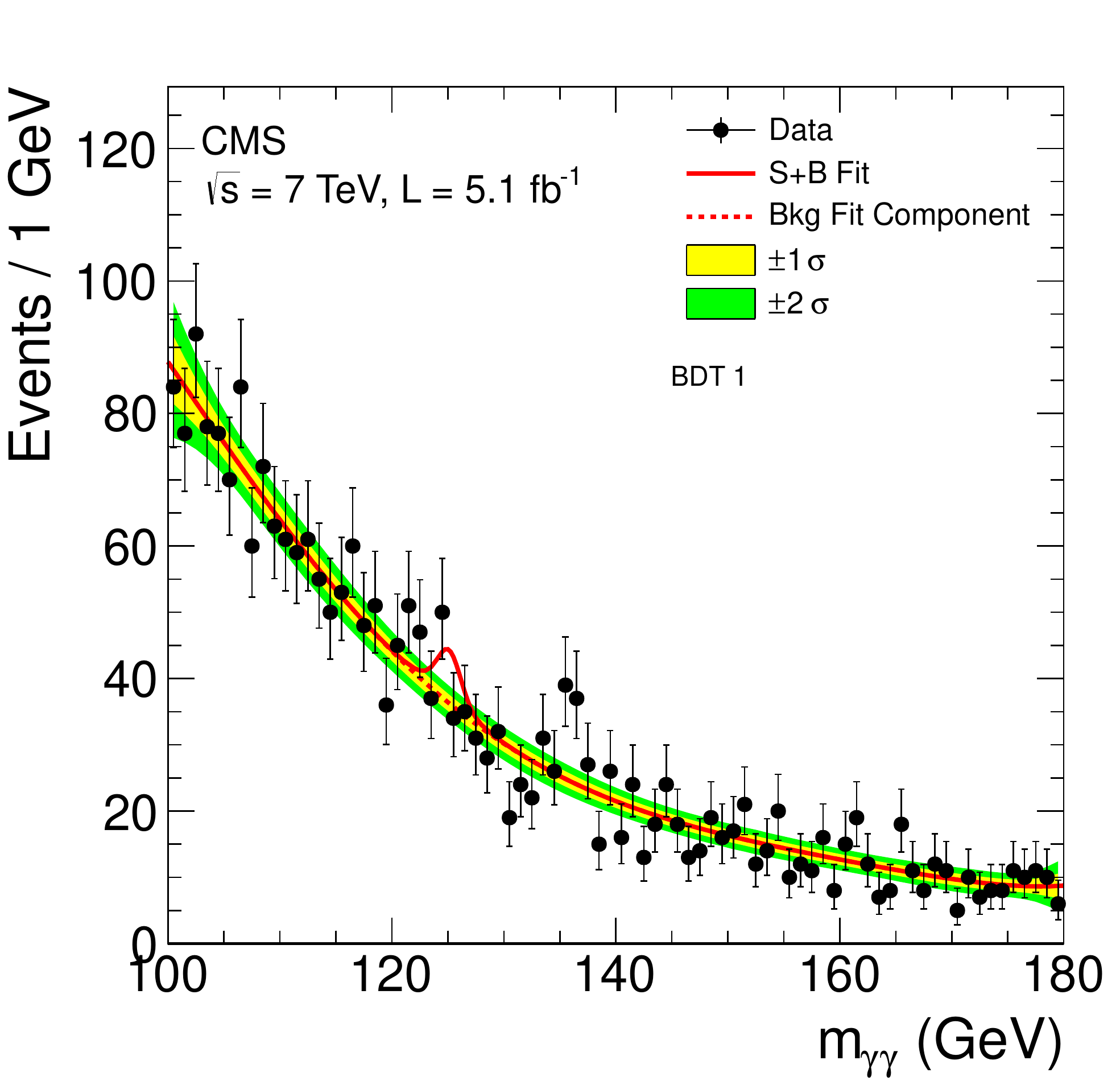}\\
    \includegraphics[width=0.45\linewidth]{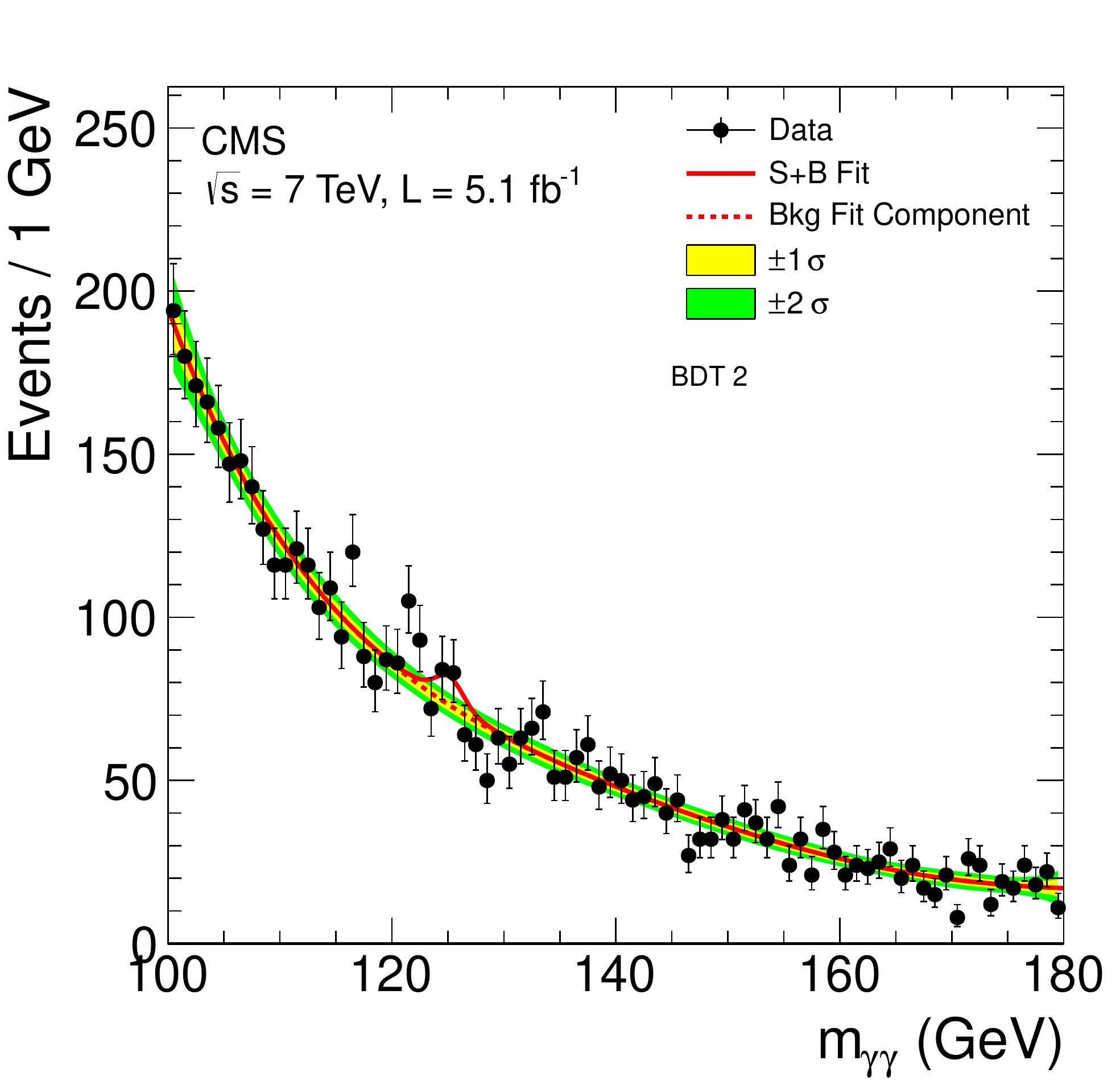}
    \includegraphics[width=0.45\linewidth]{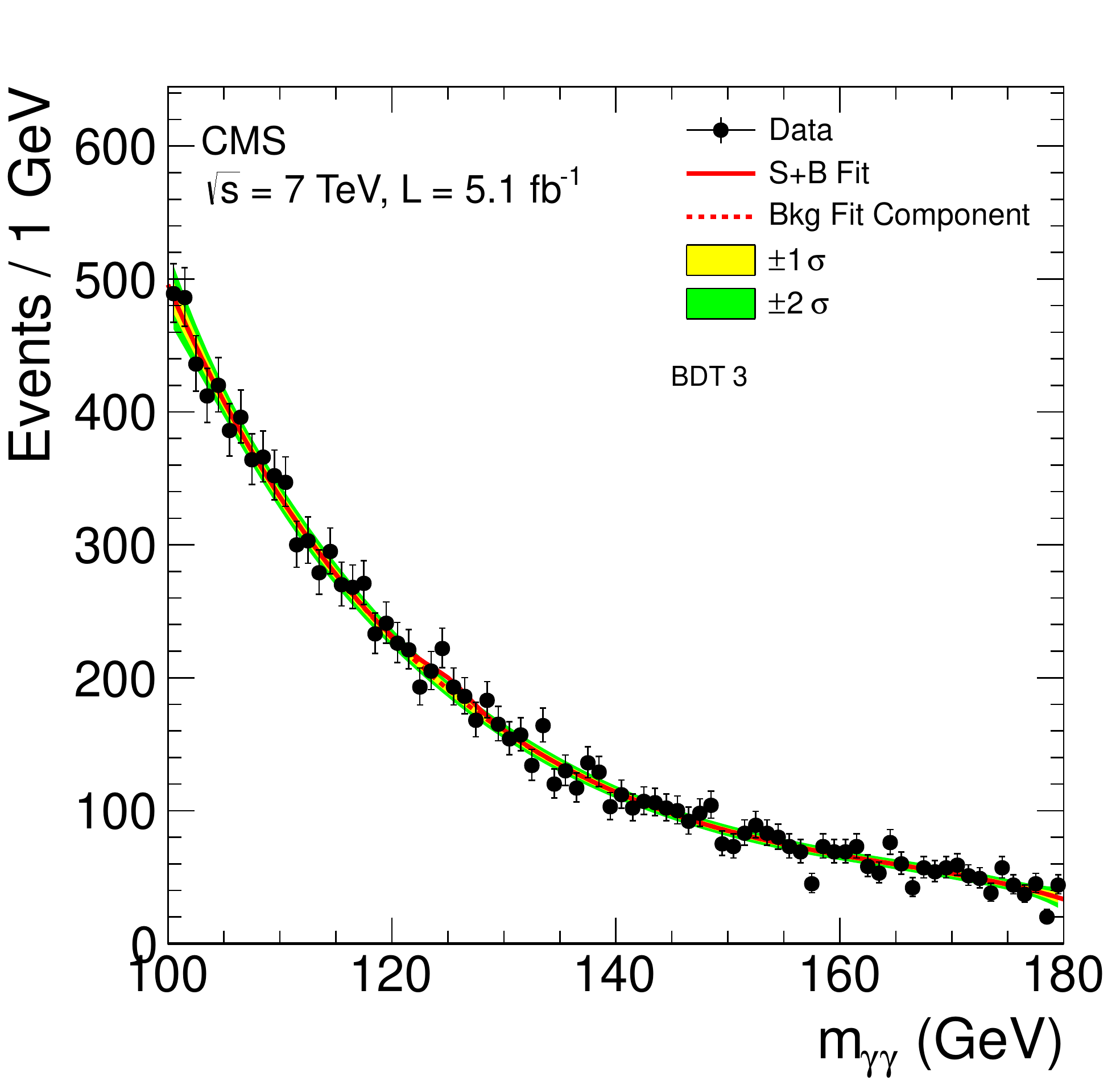}\\
    \includegraphics[width=0.45\linewidth]{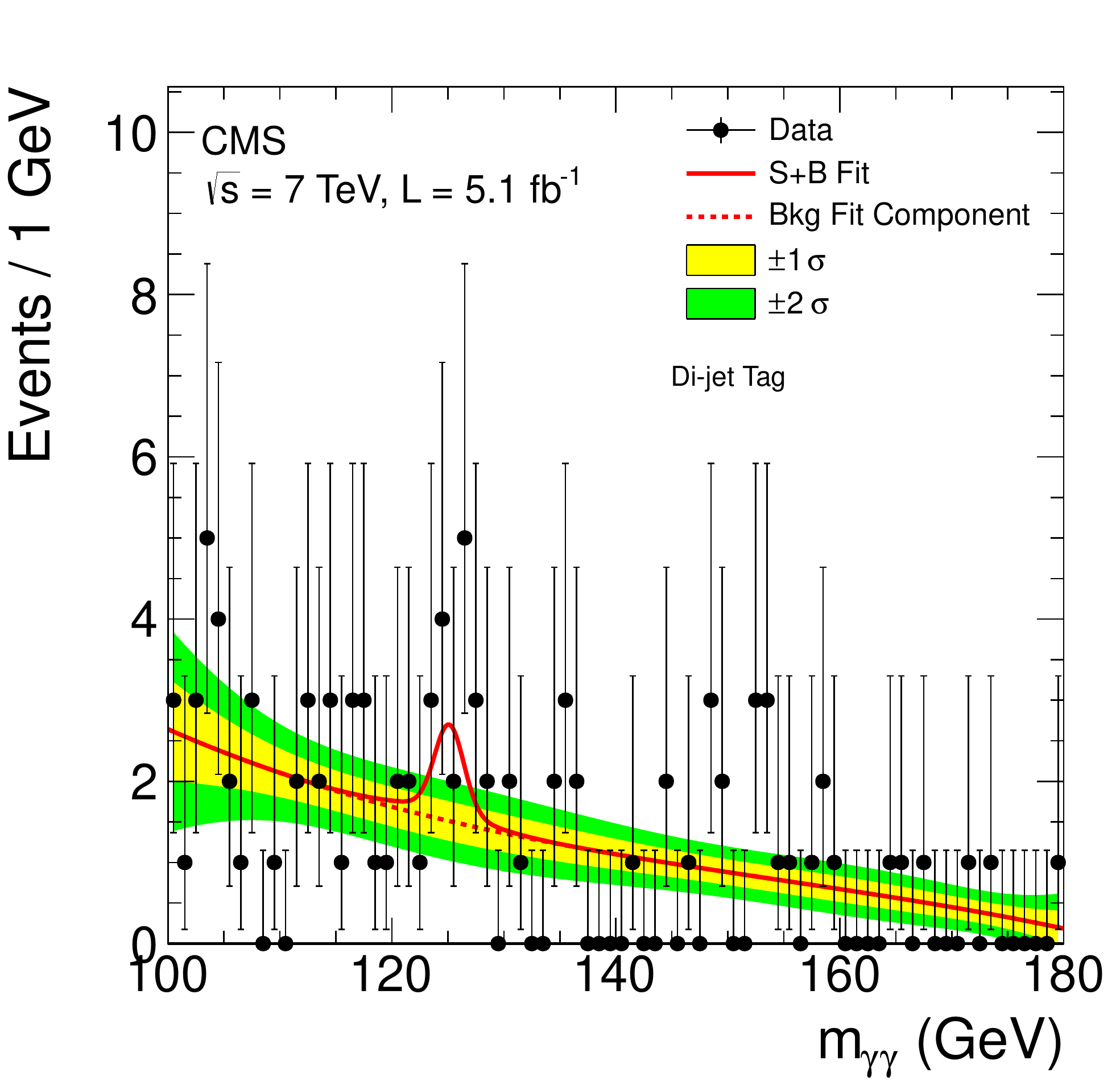}
    \caption{The diphoton invariant-mass distributions for the five classes of the 7\TeV data set (points) and the results of the
    signal-plus-background fits for $m_{\gamma\gamma}$ = 125\GeV (lines). The background fit components are shown by the dotted lines.
     The light and dark bands represent the ${\pm}$1 and ${\pm}$2 standard deviation
       uncertainties, respectively, on the background estimate.
    }
    \label{fig:hgg_BckSig7TeV}
  \end{center}
\end{figure}

\begin{figure}[htbp]
  \begin{center}
    \includegraphics[width=0.45\linewidth]{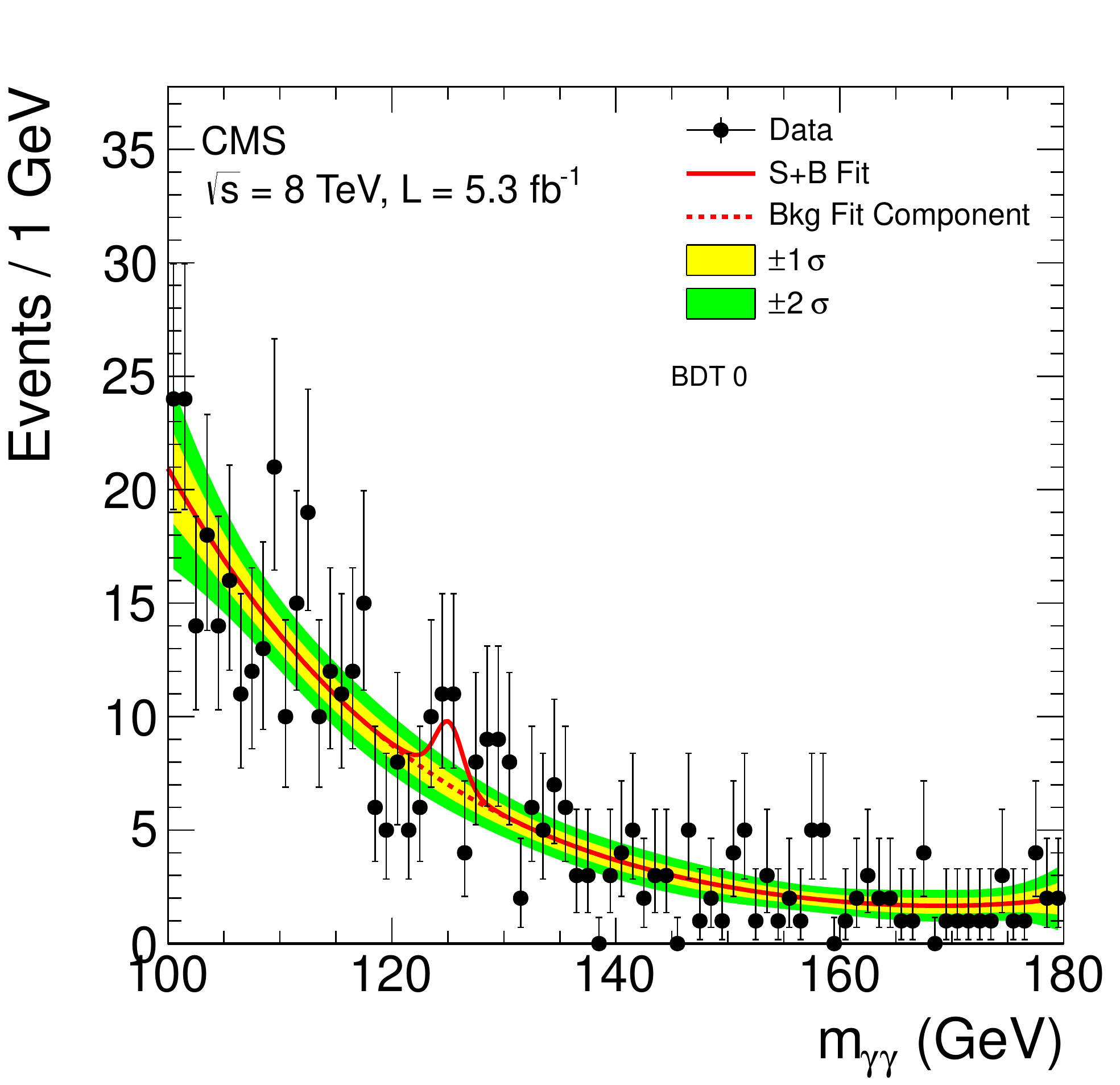}
    \includegraphics[width=0.45\linewidth]{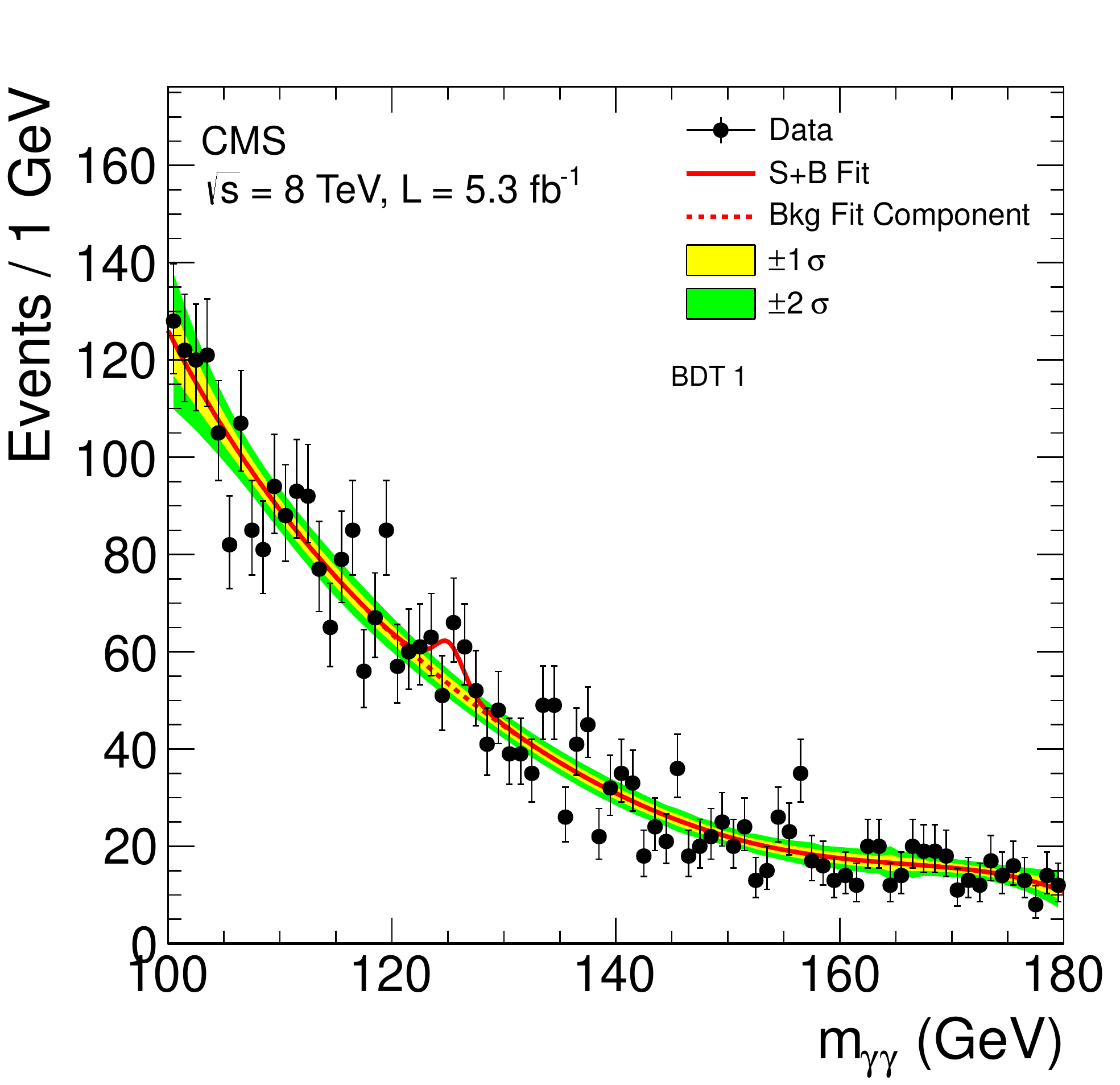}\\
    \includegraphics[width=0.45\linewidth]{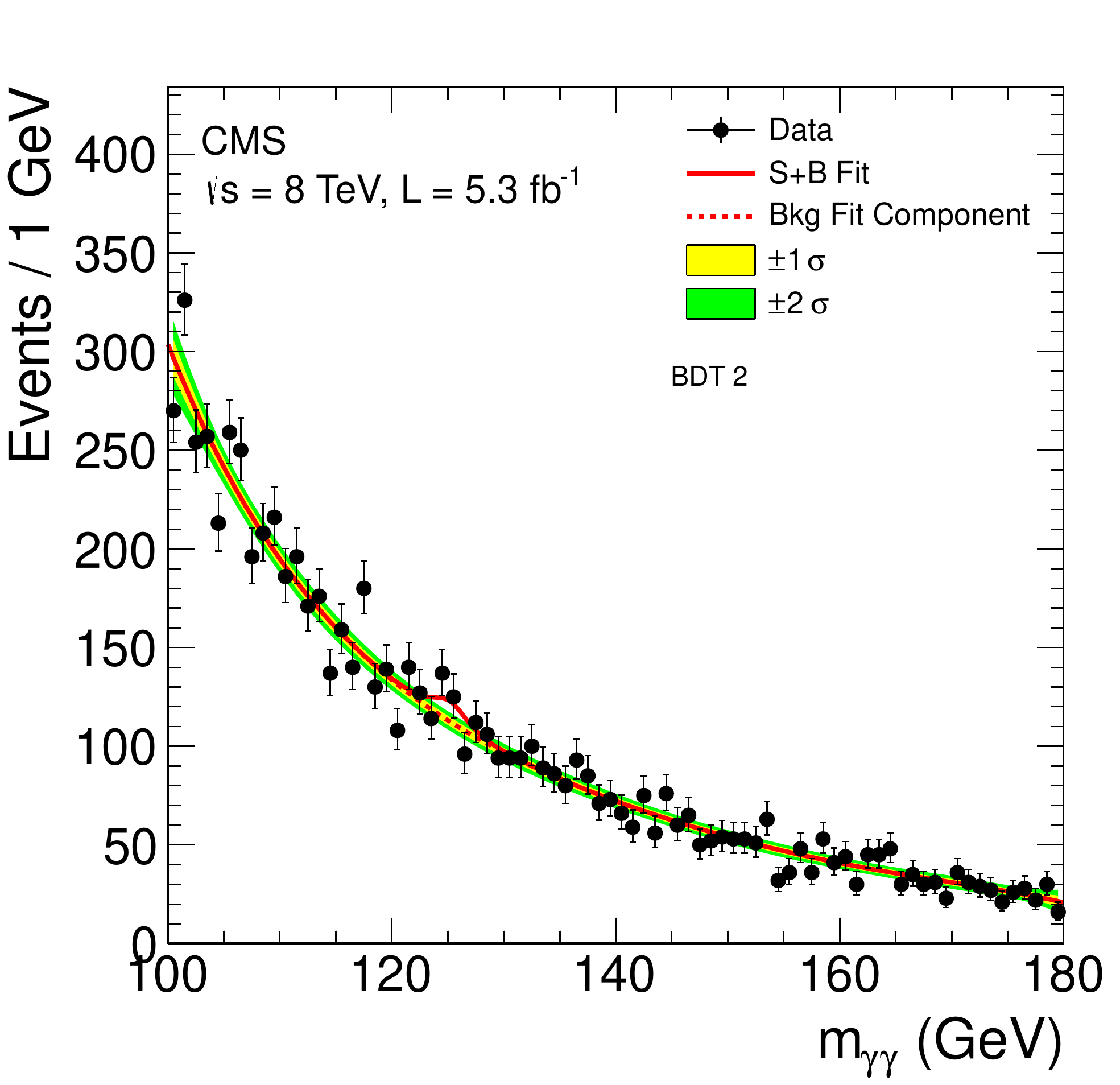}
    \includegraphics[width=0.45\linewidth]{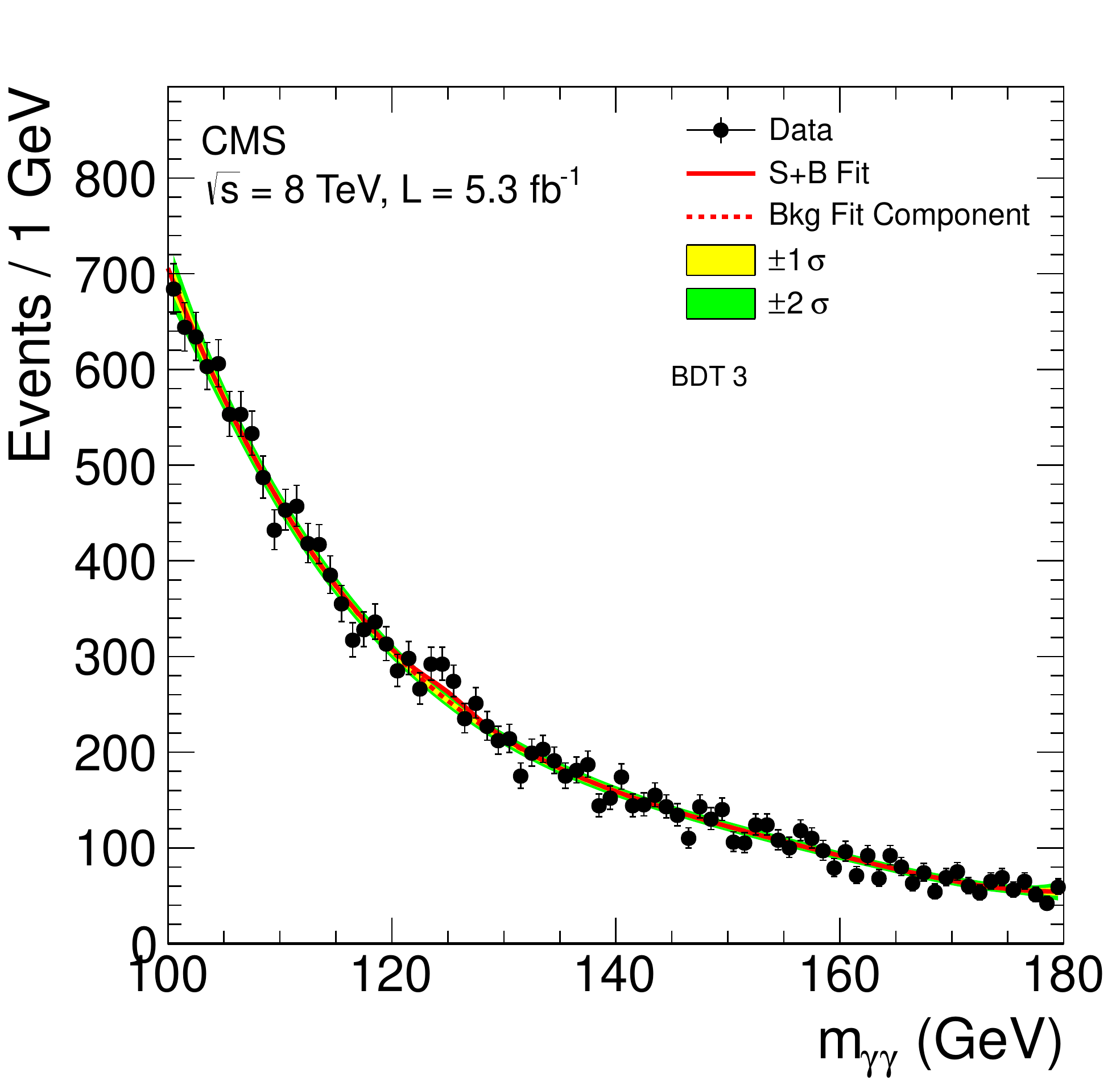}\\
    \includegraphics[width=0.45\linewidth]{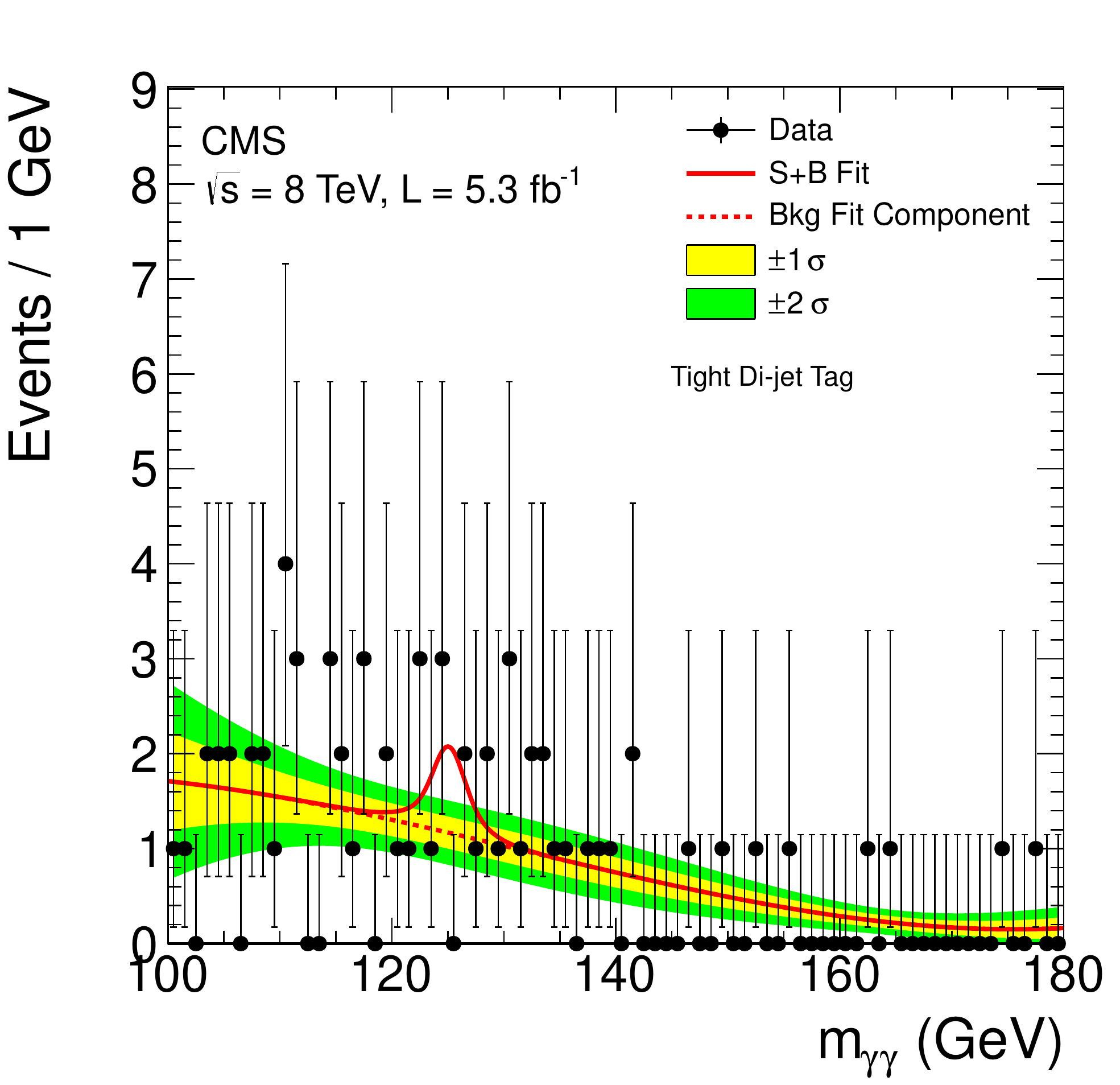}
    \includegraphics[width=0.45\linewidth]{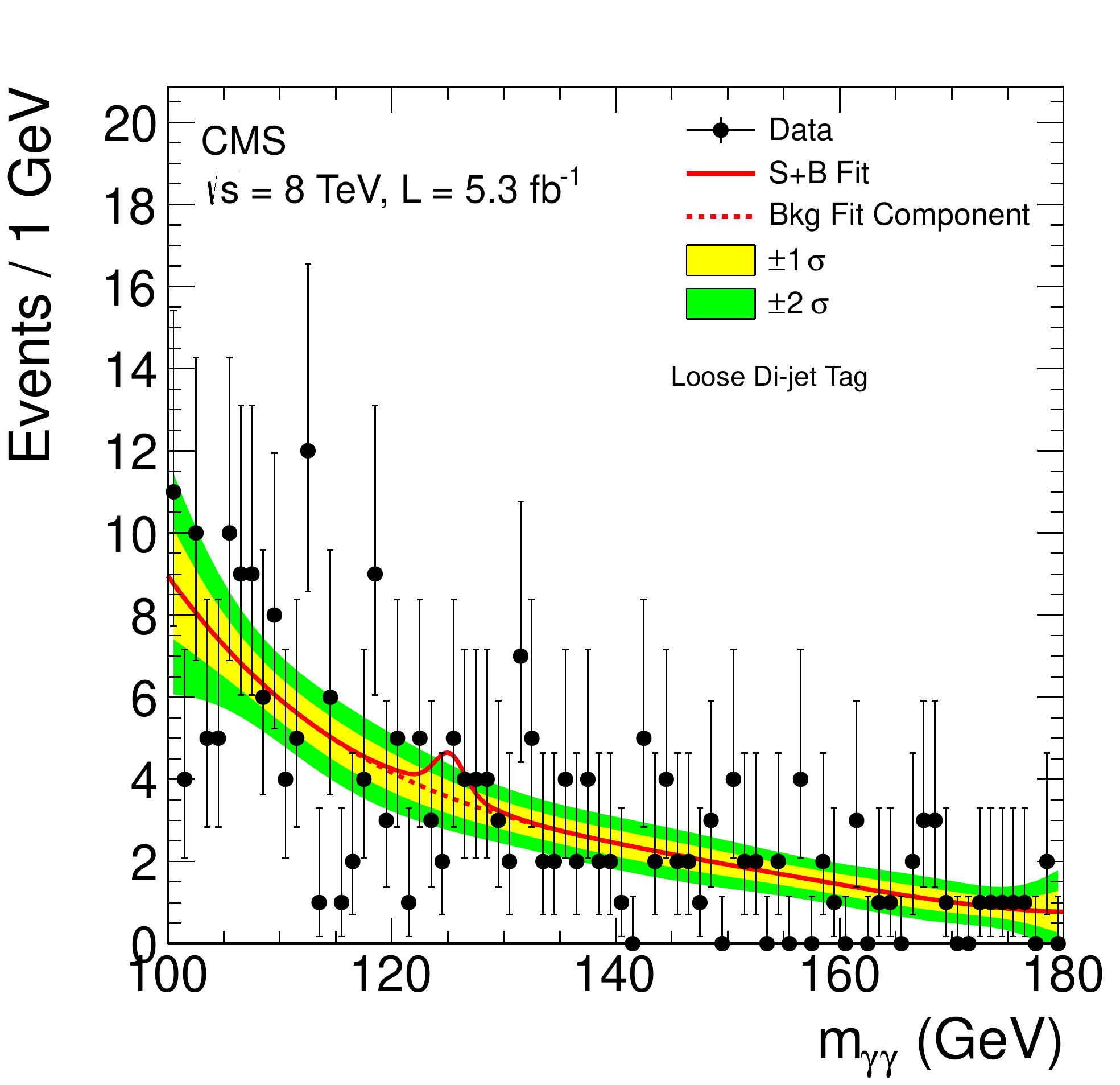}
    \caption{The diphoton invariant-mass distributions for the six classes of the 8\TeV data set (points) and the results of the
    signal-plus-background fits for $m_{\gamma\gamma}$ = 125\GeV (lines). The background fit components are shown by the dotted lines.
     The light and dark bands represent the ${\pm}$1 and ${\pm}$2 standard deviation
       uncertainties, respectively, on the background estimate.
    }
    \label{fig:hgg_BckSig8TeV}
  \end{center}
\end{figure}

The simultaneous fit to the signal-plus-background models, derived as explained above,
together with the \mgg distributions for the data, are shown
for the eleven event classes in Figs.~\ref{fig:hgg_BckSig7TeV} and \ref{fig:hgg_BckSig8TeV}
for the 7 and  8\TeV data samples, respectively.
The uncertainty bands shown in the background component of the fit arise from the
variation of the background fit parameters, and correspond to the uncertainties
in the expected background yield.
The fit is performed on the data from all event class distributions simultaneously, with an overall floating signal strength.
In these fits, the mass hypothesis is scanned in steps of 0.5\GeV between 110 and 150\GeV. At the
point with the highest significant excess over the background-only hypothesis ($\mH=125$\GeV),
the best fit value is $\sigma/\sigma_\mathrm{SM}=1.56\pm0.43$.

\begin{figure}
 \begin{center}
   \includegraphics[width=0.63\linewidth]{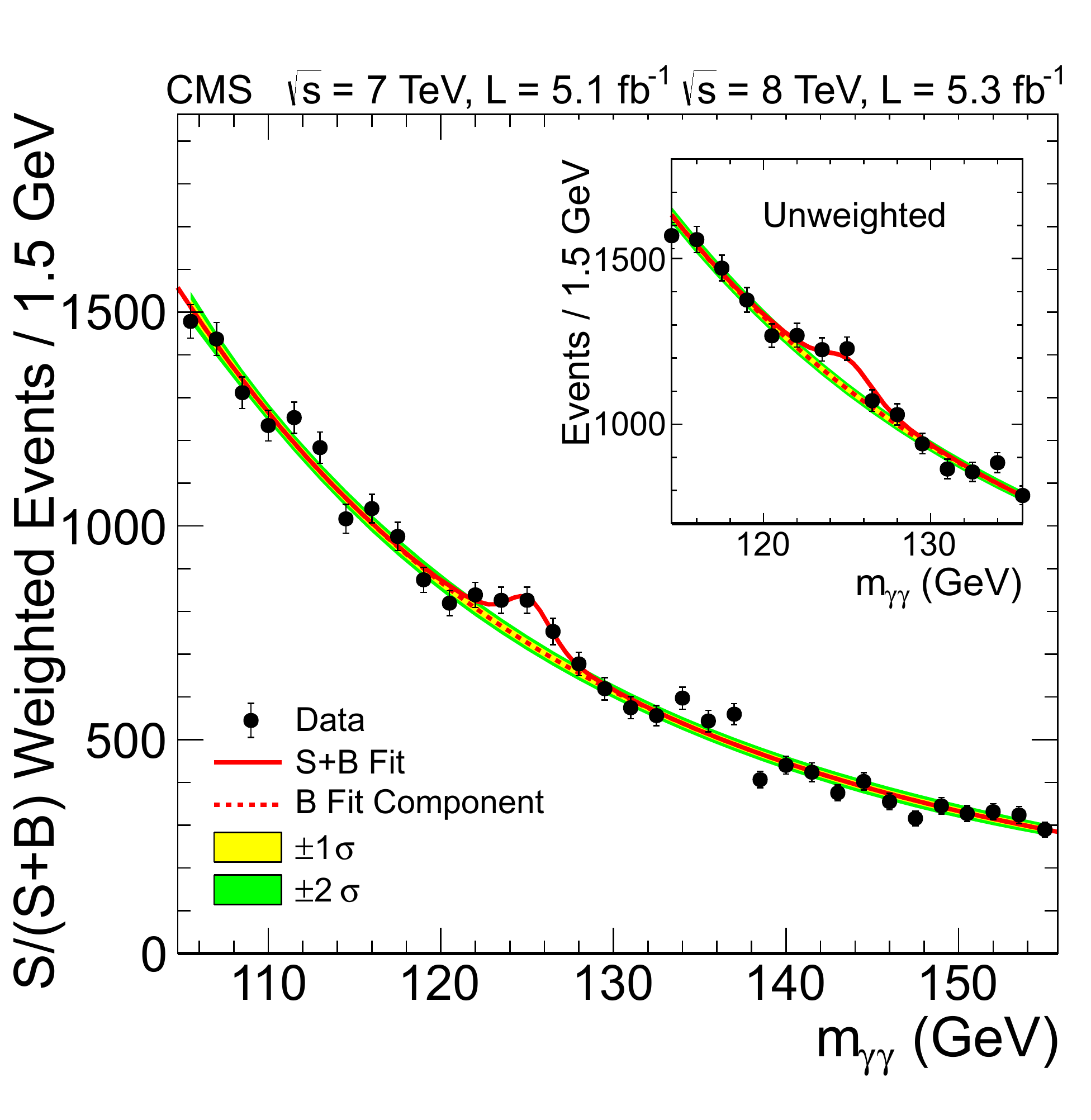}
   \caption{The diphoton invariant-mass
       distribution for the 7 and 8\TeV data sets (points), with each event weighted
       by the predicted $S/(S+B)$ ratio of its event class. The solid and dotted lines
       give the results of the signal-plus-background and background-only fit, respectively.
       The light and dark
       bands represent the $\pm$1 and $\pm$2 standard deviation
       uncertainties respectively on the background estimate.
       The inset shows the corresponding unweighted invariant-mass distribution around
      $m_{\gamma\gamma}$ = 125\GeV.
        }
   \label{fig:hgg_MassFactSoB}
 \end{center}
\end{figure}

In order to better visualize any overall excess/significance in the data,
each event is weighted by a class-dependent factor, and its corresponding diphoton invariant mass is plotted
with that weight in a single distribution. The weight depends on the event class and is proportional to $S/(S+B)$,
where $S$ and $B$ are the number of
expected signal and background events in a mass window corresponding
to $2\sigma_\text{eff}$, centered on $m_{\gamma\gamma}$ = 125\GeV and
calculated from the signal-plus-background fit to all data event classes simultaneously.
The particular choice of the weights is motivated in Ref.~\cite{Barlow:1986ek}.
The resulting distribution is shown in Fig.~\ref{fig:hgg_MassFactSoB}, where
for reference the distribution for the unweighted sum of events is shown as an inset.
The binning for the distributions is chosen to optimize the visual effect
of the excess at 125\GeV, which is evident in both the weighted and unweighted
distributions. It should be emphasized that this figure is for visualization
purposes only, and no results are extracted from it.

\subsection{Alternative analyses}
\label{sec:hgg_crosscheck}

\begin{figure}[htbp]
  \begin{center}
    \includegraphics[width=0.60\linewidth]{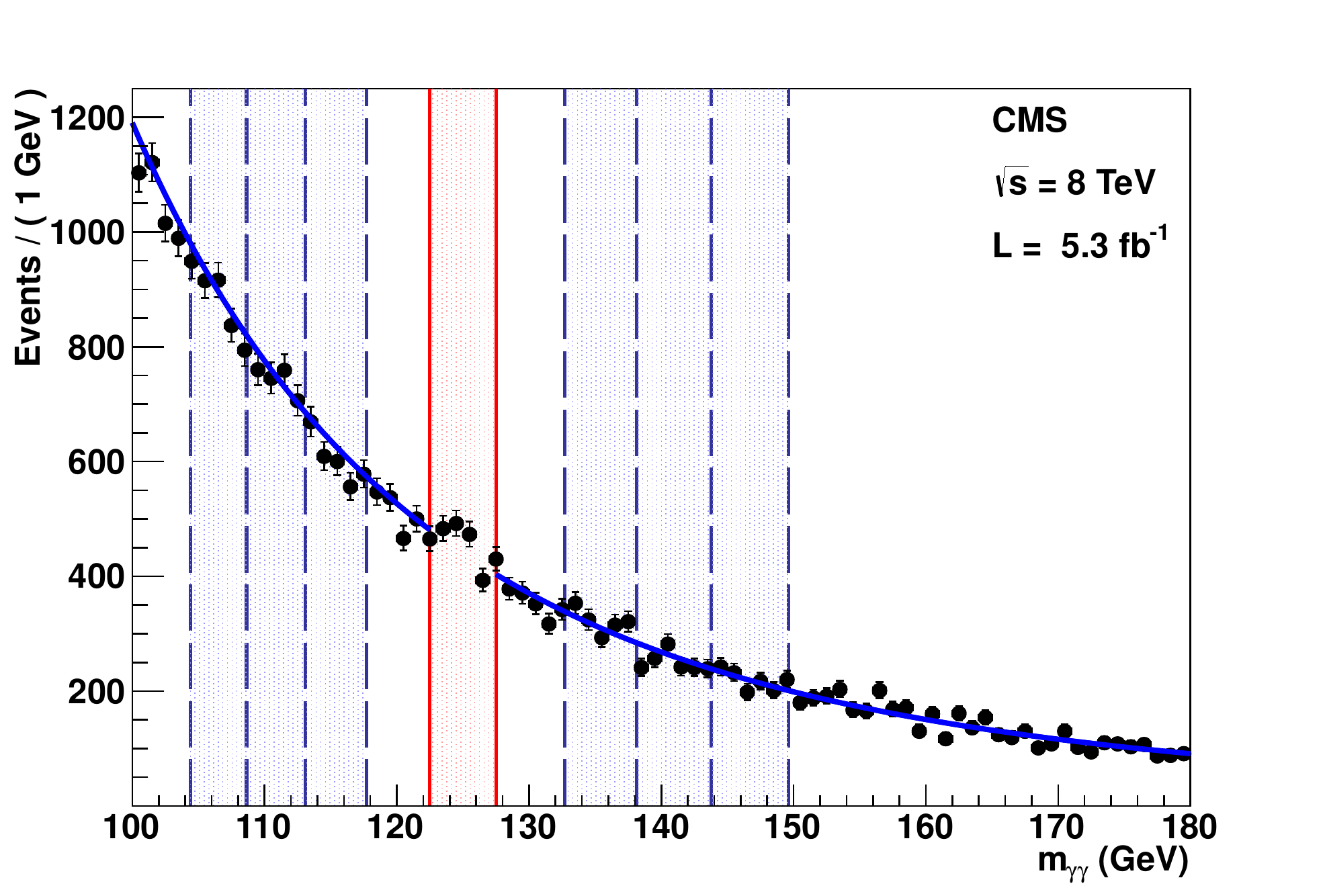}
    \caption{The six sidebands (dashed lines) around the signal region
      (solid line) in the sideband analysis. }
    \label{fig:hgg_sidebands}
  \end{center}
\end{figure}

In order to verify the results described above, two alternative analyses are performed.
The first (referred to as the {\it cut-based} analysis) refrains from relying on multivariate
techniques, except for the photon energy corrections described in Section~\ref{sec:reconstruction}.
Instead, the photon identification is performed by an optimized set of requirements on the
discriminating variables explained in Section~\ref{sec:hgg_selection}. Additionally, instead of using a BDT
event-classifier variable to separate events into classes, the event classes are built using requirements
on the photons directly. Four mutually exclusive classes are constructed by splitting the events according
to whether  both candidate photons are reconstructed in the ECAL barrel
or endcaps, and  whether the  $R9$ variable exceeds 0.94.
This categorization is motivated by the fact that photons in the barrel with high $R9$ values
are typically measured with better energy resolution than ones in the endcaps  with low $R9$.
Thus, the classification serves a similar purpose to the one
using the BDT event classifier: events with good diphoton mass resolution are grouped together into one class.
The four event classes used in this analysis are then:
\begin{itemize}
\item both photons are in the barrel, with $R9>0.94$,
\item both photons are in the barrel and at least one of them with $R9\leq0.94$,
\item at least one photon is in the endcap and both photons with $R9>0.94$,
\item at least one photon is in the endcap and at least one of them with $R9\leq0.94$.
\end{itemize}
The second alternative analysis (referred to as the {\it sideband} analysis)
uses the identical multivariate technique as the baseline analysis, as well as
an identical event sample, but relies on different procedures to model the signal and background contributions.
This approach uses data in the sidebands of the invariant mass distribution to model the background.
Consequently, this analysis is much less sensitive to the parametric form used to describe the diphoton
mass spectrum and allows the explicit inclusion of a systematic uncertainty for the possible bias in the
background mass fit.
For any given mass hypothesis \mH, a signal region is defined to be in
the range ${\pm}2\%$
on either side of \mH.
A contiguous set of sidebands is defined in the mass distribution on either side of the signal region, from
which the background is extracted.
Each sideband is defined to have the equivalent width of ${\pm}2\%$ relative to the mass hypothesis
that corresponds to the centre of the sideband.  A total of six sidebands are used in the analysis
(three on either side of the signal region), with the two sidebands adjacent to the
signal region omitted in order to avoid signal contamination, as
illustrated in Fig.~\ref{fig:hgg_sidebands}.

The result is extracted by counting events in the signal region, in classes
that are defined by the output distribution of a BDT.
This mass-window BDT takes two dimensionless inputs: the diphoton BDT output (as described in
Section \ref{sec:hgg_diphotonBDT}), and the mass, in the form $\Delta m/\mH$, where
$\Delta m =\mgg - \mH$ and $\mH$ is the Higgs boson mass hypothesis.
The output of the BDT is binned to define the event classes.
The bin boundaries are optimized to give the maximum expected significance in the presence of a Standard Model
Higgs boson signal, and the number of bins is chosen such that any additional increase in the number of bins
results in an improvement in the expected significance of less than 0.1\%.
The same bin boundaries are used for the signal region and for the six sidebands.
The dijet-tagged events constitute an additional bin (two bins for the 8\TeV data set)
appended to the bins of the mass-window BDT output value.

The background model (\ie the BDT output distribution for background events in the signal region) is constructed
from the BDT output distributions of the data in each of the six sidebands.
The only assumptions made concerning the background model shape, both verified within the assigned systematic errors,
are that the fraction of events in each BDT output bin varies linearly as a function of invariant mass
(and thus with sideband position), and that there is negligible signal contamination in the sidebands.
Only the overall normalization of the background model (the total number of background events in the signal region) is obtained from
a parametric fit to the mass spectrum. The signal region is excluded from this fit.  The bias incurred by the choice of the functional form used in the fit has been studied in a similar fashion to that described in Section \ref{sec:hgg_smodeling}, and is covered with a systematic uncertainty of 1\%.

The mass-window BDT is trained using simulated Higgs boson events with
$\mH=123\GeV$ and simulated background events, including prompt-prompt,
prompt-fake, and fake-fake processes.
The training samples are not used in any other part of the analysis,
except as input to the binning algorithm, thus
avoiding any biases from overtraining.

The signal region for mass hypothesis $\mH=125\GeV$
is estimated from simulation to contain 93\% of the signal.
The number of expected signal events in each bin is determined using MC simulation, as in the
baseline analysis. Systematic uncertainties in the signal modelling lead to
event migrations between the BDT  bins, that are accounted for as
additional nuisance parameters in the limit-setting procedure.

\begin{figure}[htbp]
  \begin{center}
    \includegraphics[width=0.49\linewidth]{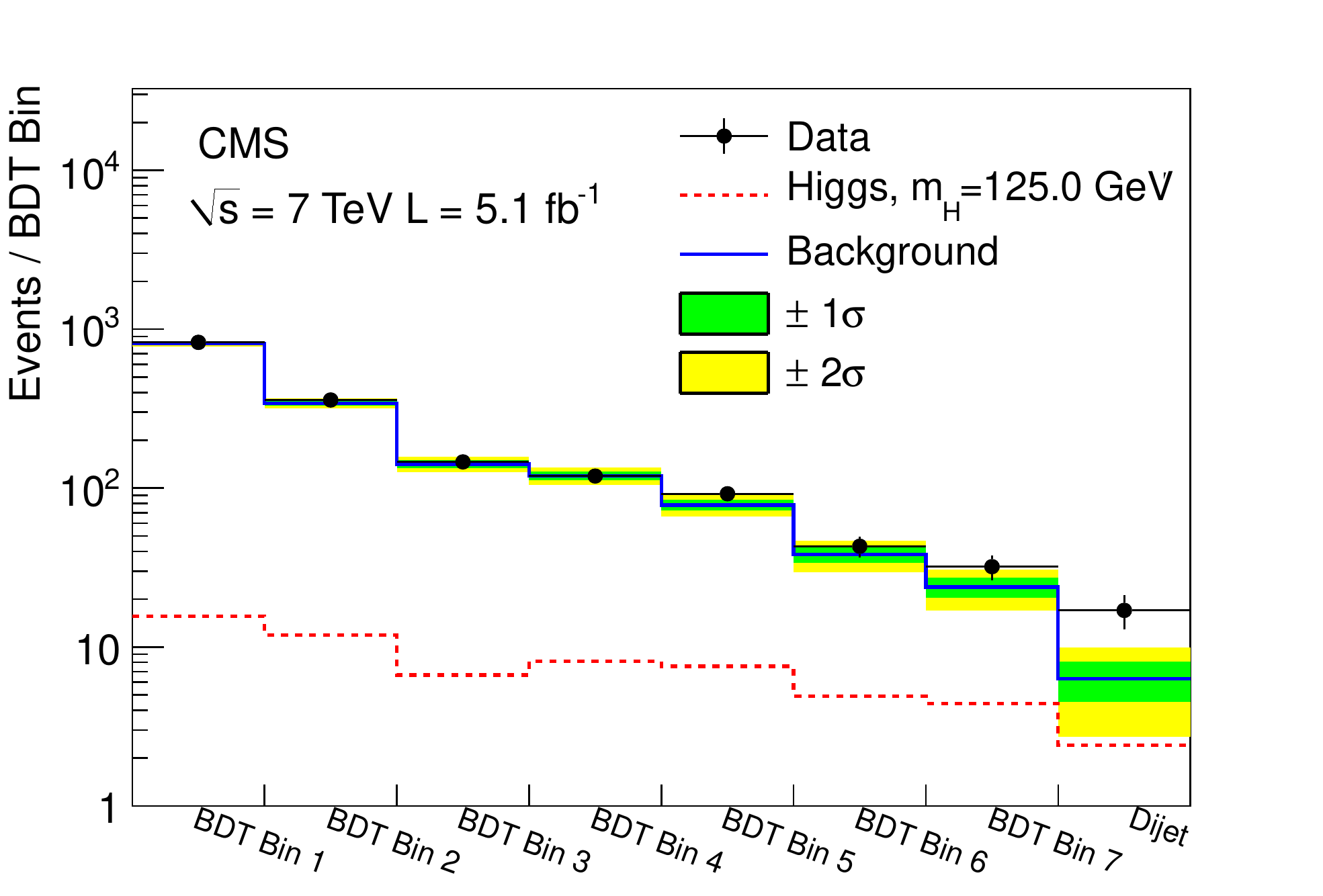}
    \includegraphics[width=0.49\linewidth]{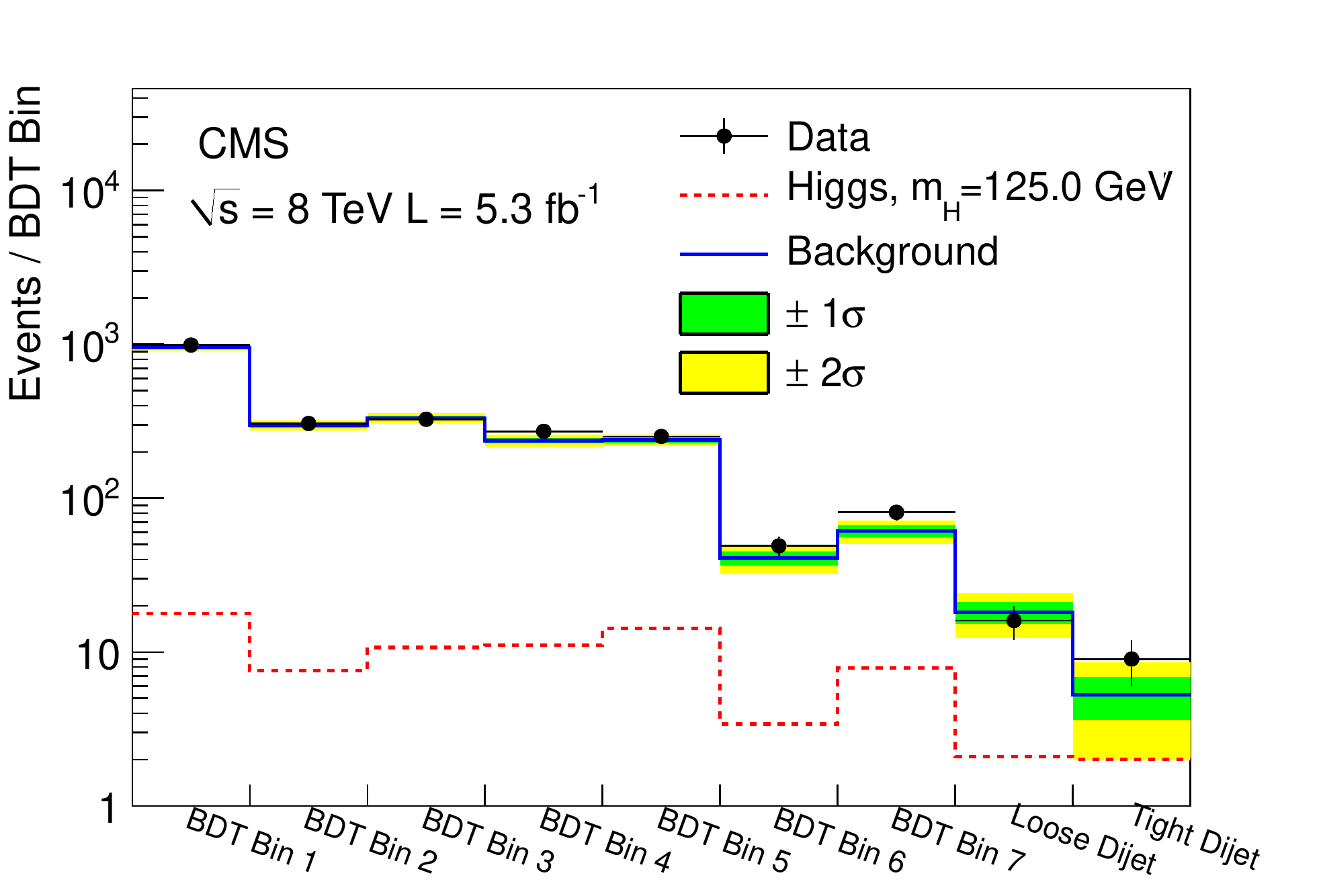}
    \caption{The number of observed events (points) for each of the mass-window BDT classes in the sideband analysis of
      $\PH \to \gamma\gamma$ for the 7 (left) and 8\TeV (right) data sets. The expected number of background events in
       each class, determined from the sidebands of the diphoton invariant-mass distribution, is shown by the solid line.
         The dark and light bands display the ${\pm}1$ and ${\pm}2$ standard deviation uncertainties in the background predictions,
         respectively. The expected number of signal events in each class for a 125\GeV Higgs boson, as determined from MC simulation,
         is shown by the dotted line.
    }
    \label{fig:hgg_MassWindowModel}
  \end{center}
\end{figure}

Examples of distributions in this analysis are shown in Fig.~\ref{fig:hgg_MassWindowModel},
for the 7 (left) and 8\TeV (right) data sets.
The different event classes are listed along the $x$ axis. The first seven classes
are the mass-window BDT classes.
They are ordered by increasing expected signal-to-background ratio.
The class labeled as ``Dijet'' contains the dijet-tagged events.
The number of data events, displayed as points, is compared to the expected background events determined from the sideband population,
shown by the histogram.
The expected signal yield for a Higgs boson mass of $\mH=125\GeV$ is shown
with the dotted line.

The statistical interpretation of the results is given in Section 10.

\section{\texorpdfstring{$\PH\to\cPZ\cPZ$}{H to ZZ}\label{sec:hzz4l}}
\subsection{Event selection and kinematics}

The search for the decay $\PH \rightarrow  \cPZ\cPZ   \rightarrow 4\ell$ with $\ell = \Pe, \mu$ is performed
by looking for a narrow  four-lepton invariant-mass peak in the presence of a small continuum background.
The background sources include an irreducible four-lepton contribution from direct \cPZ\cPZ\
($\cPZ\gamma^*$) production via the $\Pq\Paq$ annihilation and $\Pg\Pg$ fusion processes.
Reducible contributions arise from $\cPZ + \cPqb\cPaqb$ and $\ttbar$ production, where the final
state contains two isolated leptons and two $\cPqb$-quark jets that produce two nonprompt leptons.
Additional background arises from $\cPZ+$jets and $\PW\cPZ+$jets events, where jets are
misidentified as leptons.
Since there are differences in the reducible background rates and mass resolutions between the
subchannels $4\Pe$, $4\mu$, and $2\Pe2\mu$, they are analyzed separately and the results are then
combined statistically.

Compared to the first CMS $\cPZ\cPZ\ \to 4\ell$ analysis reported in Ref.~\cite{Chatrchyan:2012dg},
this analysis employs improved muon reconstruction, lepton identification
and isolation, recovery of final-state-radiation (FSR) photons, and the use of a kinematic discriminant
that exploits the expected decay kinematics of the signal events.
New mass and spin-parity results obtained from a $\PH \rightarrow  \cPZ\cPZ   \rightarrow 4\ell$
analysis using additional integrated luminosity at the centre-of-mass energy of 8\TeV are described
in a recent CMS publication~\cite{:2012br}, and not discussed further here.

Candidate events are first selected by triggers that require the presence of a pair of electrons or
muons. An additional trigger requiring an electron and a muon in the event is also used for the 8\TeV data.
The requirements on the minimum $\PT$ of the two leptons are 17 and 8\GeV. 
The trigger efficiency is determined by first adjusting the simulation to reproduce the efficiencies
obtained on single lepton legs in special tag-and-probe measurements, and then using the
simulation to combine lepton legs within the acceptance of the analysis.
The efficiency for a Higgs boson of mass  $> 120\GeV$, is greater than 99\% (98\%, 95\%) in
the $4\mu$ ($2\Pe 2\mu$, $4\Pe$) channel.
The candidate events are selected using identified and isolated leptons.
The electrons are required to have transverse momentum $\PT^{\Pe} > 7\GeV$ and pseudorapidity
within the tracker geometrical   acceptance of $|\eta^{\Pe}| < 2.5$.
The corresponding requirements for muons are $\PT^{\Pgm} > 5\GeV$ and $|\eta^{\Pgm}| < 2.4$.
No gain in expected significance for a Higgs boson signal
is obtained by lowering the $\PT$ thresholds for the leptons, since the
improvement in signal detection efficiency is accompanied by a large increase in the $\cPZ+$jets background.

The lepton-identification techniques have been described in Section~\ref{sec:reconstruction}.
The multivariate electron identification is trained using a Higgs boson MC simulation sample for the
$\PH\rightarrow\cPZ\cPZ$ signal and a sample of \PW+1-jet events from data for the background.
The working point is  optimized using a \cPZ+1-jet data sample.
For each lepton, $\ell = \Pe$, $\mu$, an isolation requirement of
$ R_\text{Iso}^{\ell} < 0.4$ is applied to suppress
the \cPZ+jet, \cPZ+$\cPqb\cPaqb$, and $\ttbar$ backgrounds.
In addition, the lepton impact parameter significance with respect
to the primary vertex, defined as
${\rm SIP_{3D}}= \frac{\rm IP}{\sigma_{\rm IP}} $,
with ${\rm IP}$ the impact parameter in three dimensions
and  $\sigma_{\rm IP}$ its uncertainty, is used to further reduce background.
The criteria of $| {\rm SIP_{3D}}  | < 4$ suppresses
the $\cPZ + \cPqb\cPaqb$ and $\ttbar$ backgrounds with negligible effect on the
signal efficiency.

\begin{figure}[htbp]
   \begin{center}
     \includegraphics[width=0.49\linewidth]{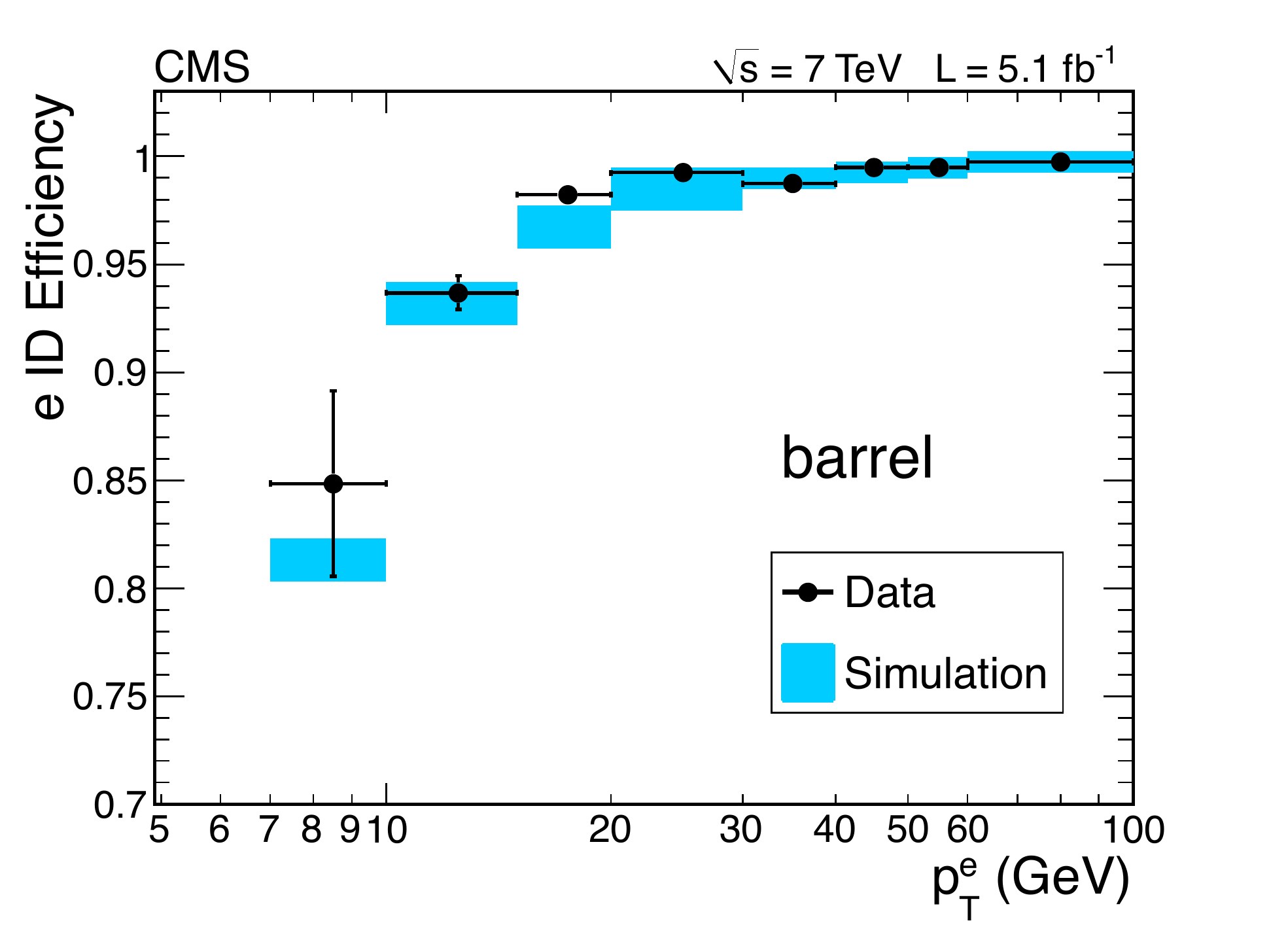}
     \includegraphics[width=0.49\linewidth]{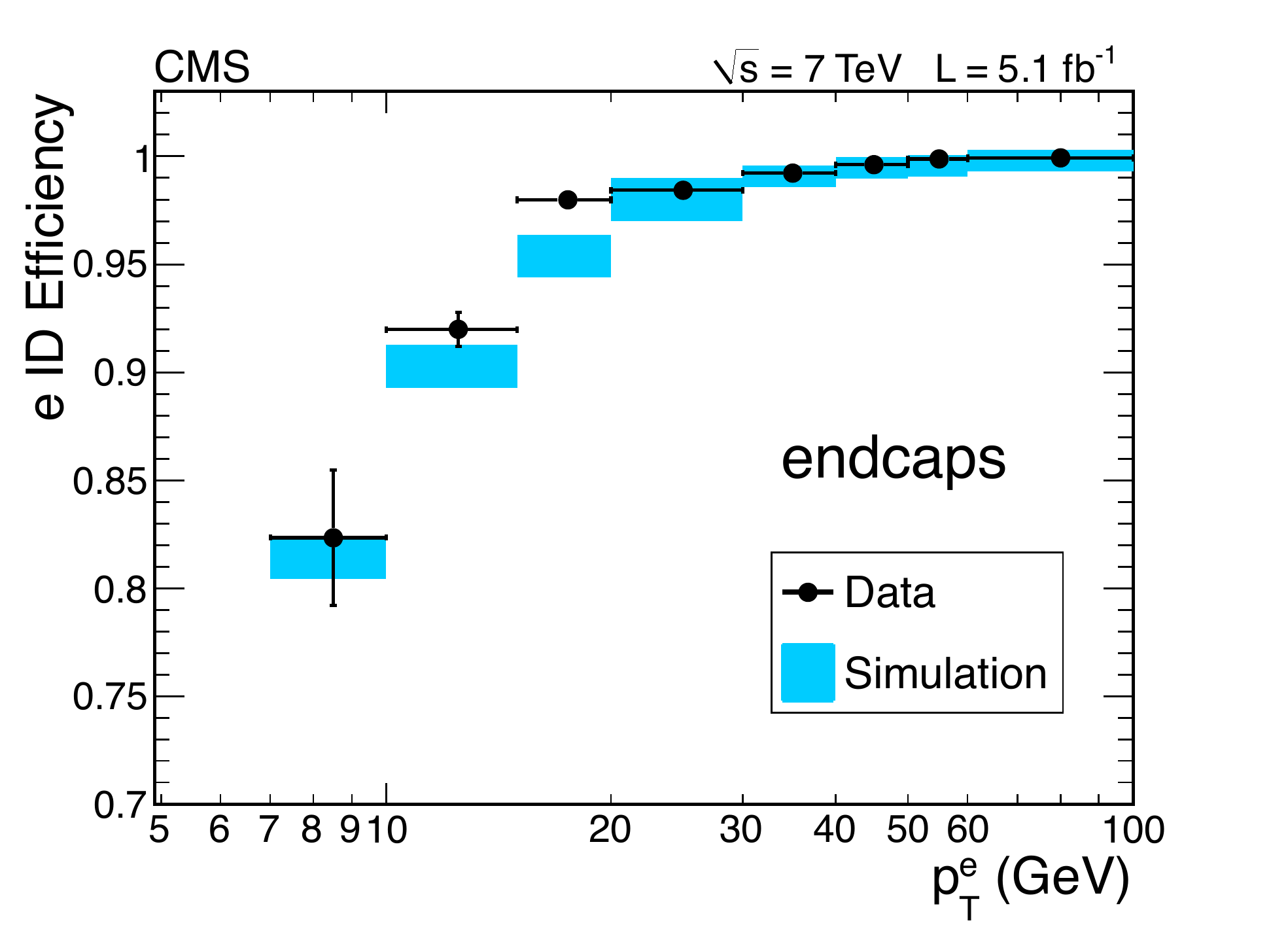}
     \includegraphics[width=0.49\linewidth]{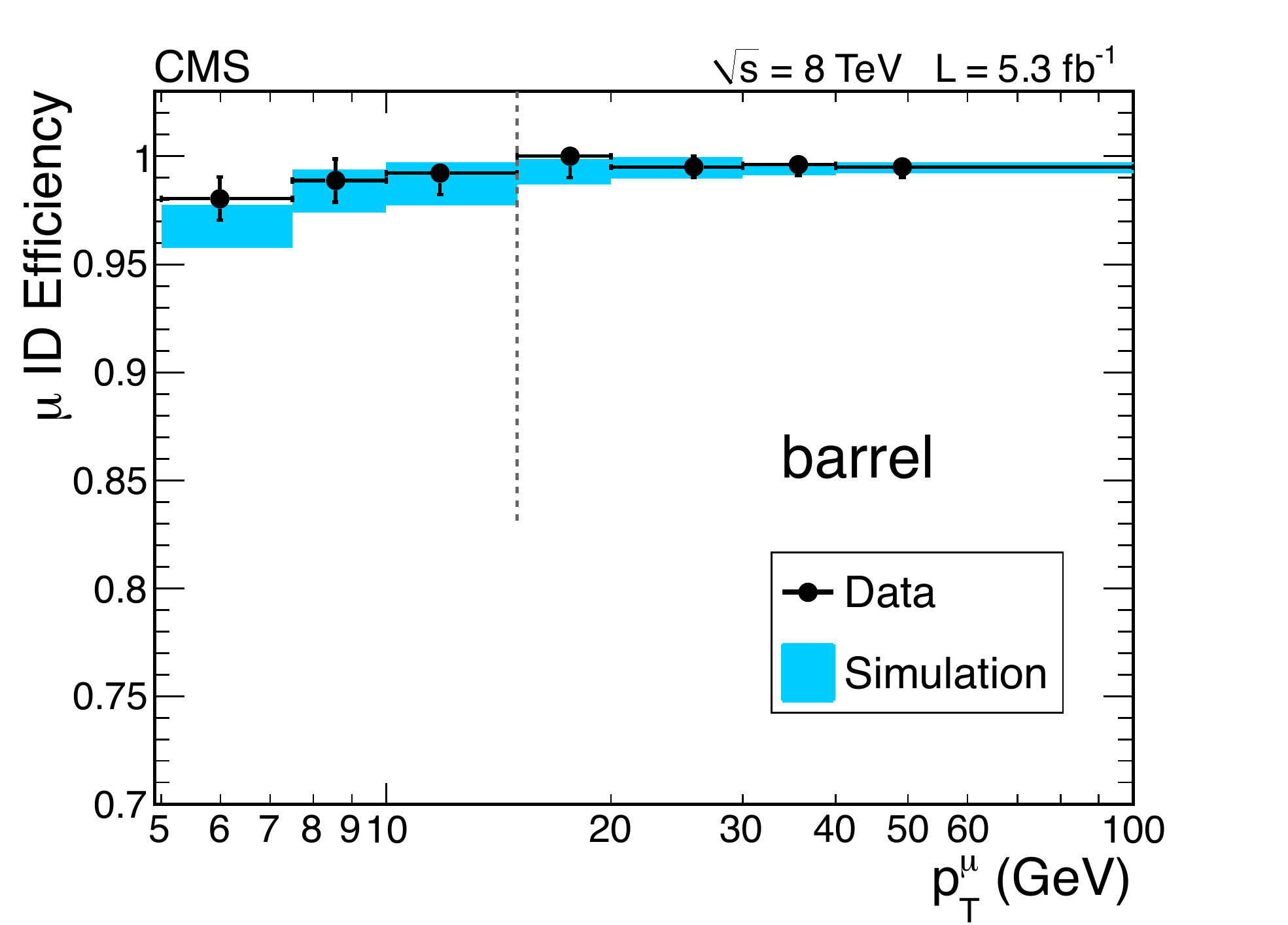}
     \includegraphics[width=0.49\linewidth]{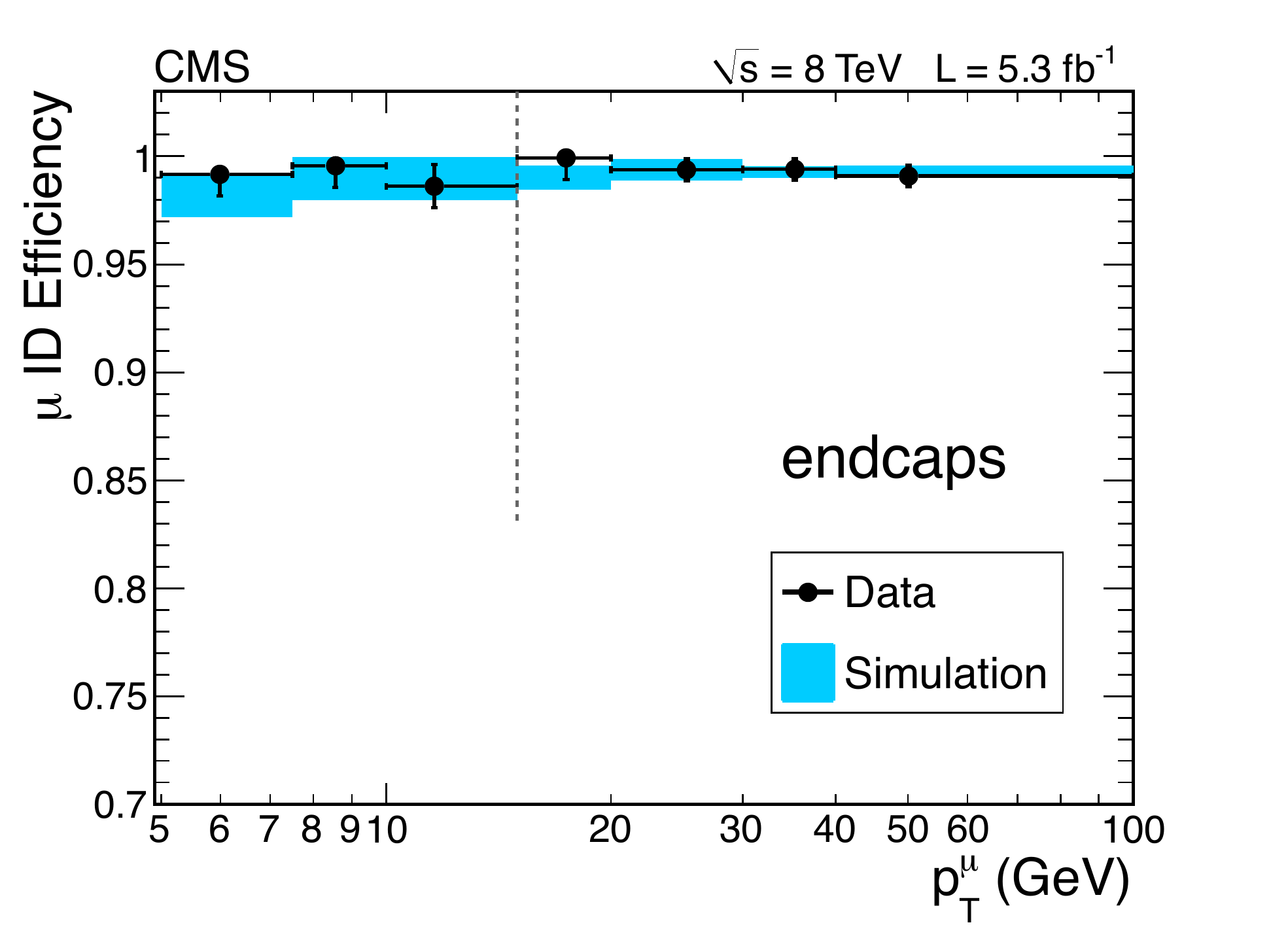}
     \caption{
     Measurements of the lepton identification efficiency using a tag-and-probe
     technique based on samples of \cPZ\  and $\cPJgy$ dilepton events. The measurements
     are shown for electrons (top) at 7 $\TeV$ and muons (bottom) at 8 $\TeV$ as a function of $\PT^{\ell} $ for
     the $ |\eta| $ regions of the barrel (left) and endcaps (right).
     For muons, the efficiencies at $\PT^{\mu} < 15\GeV$ (dashed line on bottom plots)
     is obtained using $\cPJgy$.
     The results obtained from data (points with error bars) are compared to results
     obtained from MC simulation (histograms), with the shaded region representing
     the combined statistical and systematic uncertainties.
           }
    \label{fig:leptonTP}
   \end{center}
\end{figure}

The efficiencies for reconstruction, identification, and isolation of electrons and muons
are measured in data, using a tag-and-probe technique~\cite{CMS:2011aa} based
on an inclusive sample of $\cPZ \to \ell\ell $ events.
The measurements are performed in bins of $\PT^{\ell} $ and $ |\eta| $.
Additional samples of dileptons with $\PT^{\ell} < 15\GeV$ from $\cPJgy$ decays are used
for the efficiency measurements (in the case of muons)
or for consistency checks (in the case of electrons).
Examples of tag-and-probe results for the lepton identification efficiencies
obtained with data and MC simulation
are shown for electrons (top) and muons (bottom) in Fig.~\ref{fig:leptonTP}.
The efficiencies measured with
data are in agreement with those obtained using MC simulation.
The  mean differences (at the percent level) are used to correct the MC simulation predictions, and the uncertainty in the difference is
propagated as a systematic uncertainty per lepton.
The overall lepton selection efficiencies are obtained as the product of the reconstruction, identification,
and isolation efficiencies.
The overall efficiency for selecting electrons in the ECAL barrel (endcaps) varies from
about  71\% (65\%) for $7 < \PT^{\Pe} < 10\GeV$ to  82\% (73\%) at $\PT^{\Pe} \simeq 10\GeV$,
and reaches 90\% (89\%) for $\PT^{\Pe} \simeq 20\GeV$.
The efficiency for electrons drops to about 85\% in the transition region, $1.44 < |\eta^e| < 1.57$, between
the ECAL barrel and endcaps.
The muons are selected with an efficiency above $98\%$ in the full $|\eta^{\Pgm}| < 2.4$ range
for $\PT^{\mu} > 5\GeV$.

Photons reconstructed with pseudorapidity  $\vert \eta^{\gamma} \vert < 2.4$ are possible FSR candidates.
The photon selection criteria are optimized as a function of the angular
distance between the photon and the closest lepton in $(\eta, \phi)$ space.
In an inner cone $\Delta R = 0.07$, photons are accepted if $\pt > 2\GeV$,
with no further requirements. In an outer annulus $0.07< \Delta R<0.5$, where
the rate of photons from the underlying event and pileup
is much larger, a tighter threshold of 4\GeV is used, and the photons are also required to be isolated:
the sum of the $\pt $ of all charged hadrons, neutral hadrons,
and photons in a cone of radius $\Delta R = 0.3$ centred on the photon should not exceed the $\pt $
of the photon itself.
In contrast to lepton isolation,  and in order to take into account the fact that the photon might come
from a pileup interaction, the photon isolation also uses the charged hadrons associated with other
primary vertices.
The selection criteria have been tuned to achieve approximately the same purity in the two angular regions.
When reconstructing the $\cPZ \to \ell\ell $ candidates, only FSR photons associated with the closest
lepton, and that make the dilepton-plus-photon invariant mass closer to the nominal
\cPZ\ mass than the dilepton invariant mass, are kept. The dilepton-plus-photon invariant mass must also be less
than 100\GeV.
The performance of the FSR selection algorithm is measured using
MC $\PH \to \cPZ\cPZ $  simulation samples, and the rate is verified
with inclusive $\cPZ$-boson  events in data.
Photons within the acceptance for the FSR selection are measured with an
efficiency of ${\simeq}50\%$ and a mean purity of $80\%$.
The FSR photons are selected in 5\% of  inclusive $\cPZ$-boson events
in the muon channel
and 0.5\% in the electron channels. In the case of electrons,
the FSR photons are often implicitly combined into the electron
superclusters, resulting in a lower FSR recovery efficiency.

\begin{figure}[htbp]
   \begin{center}
     \includegraphics[width=0.49\linewidth]{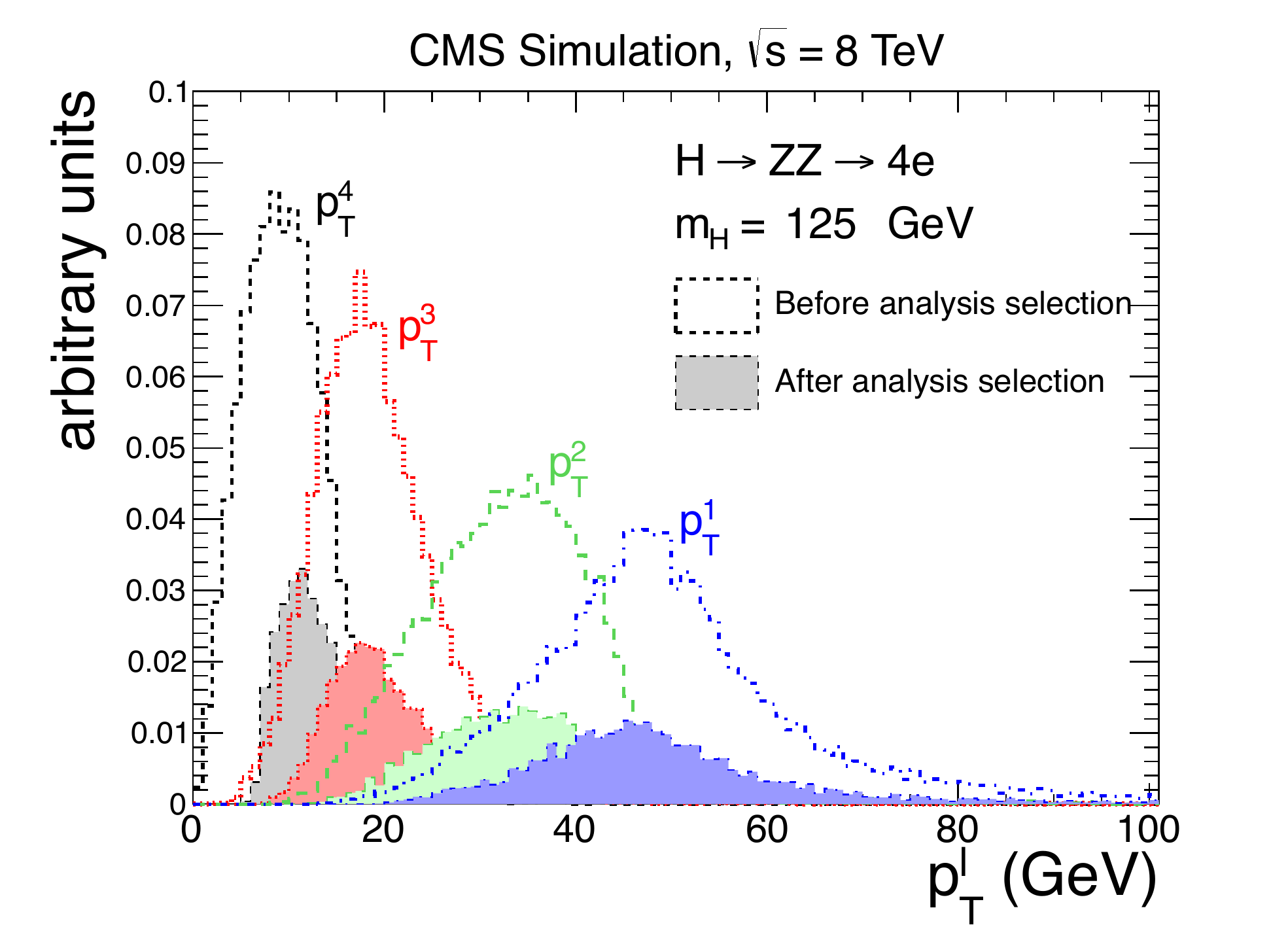}
     \includegraphics[width=0.49\linewidth]{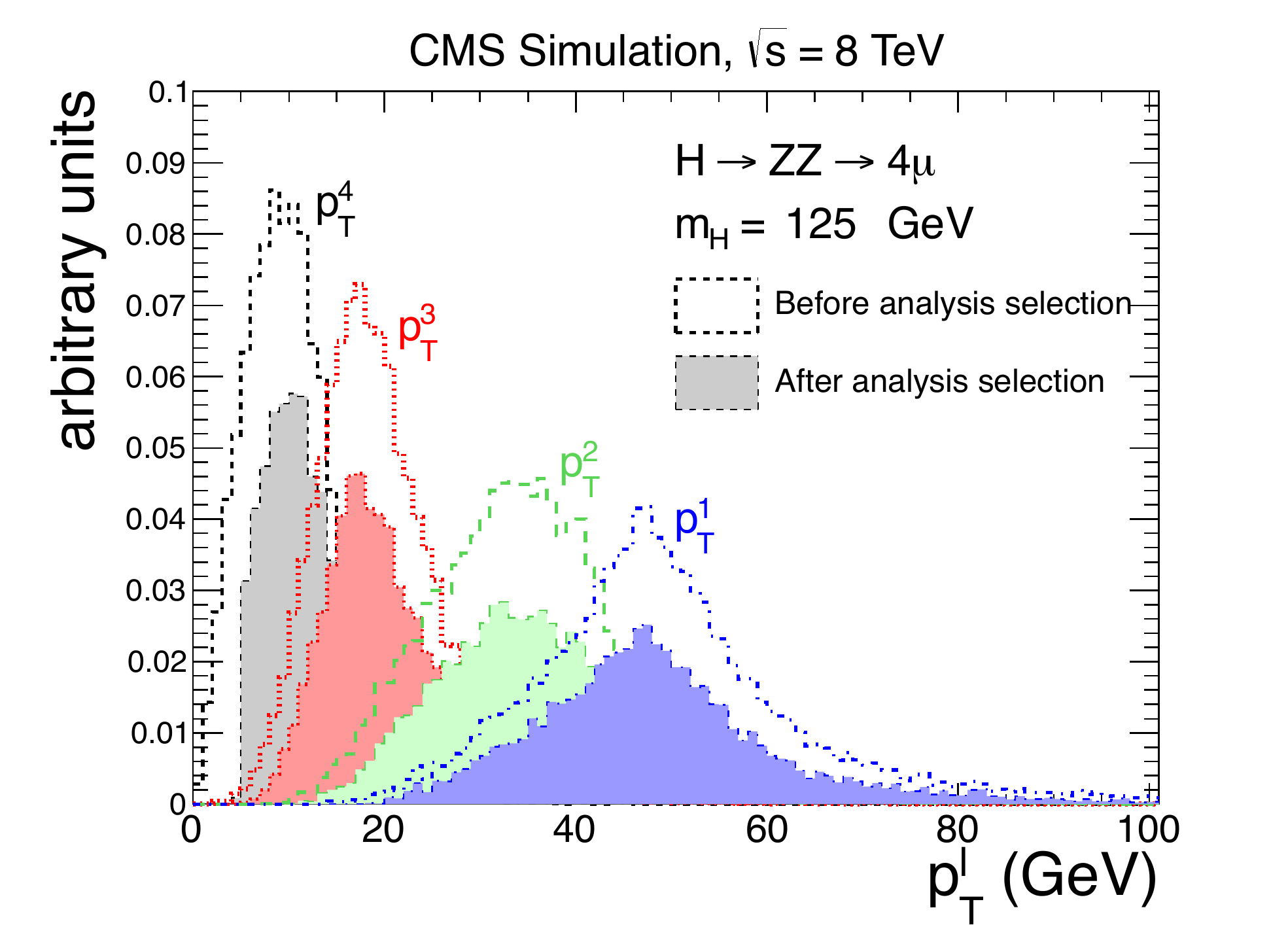}
     \includegraphics[width=0.49\linewidth]{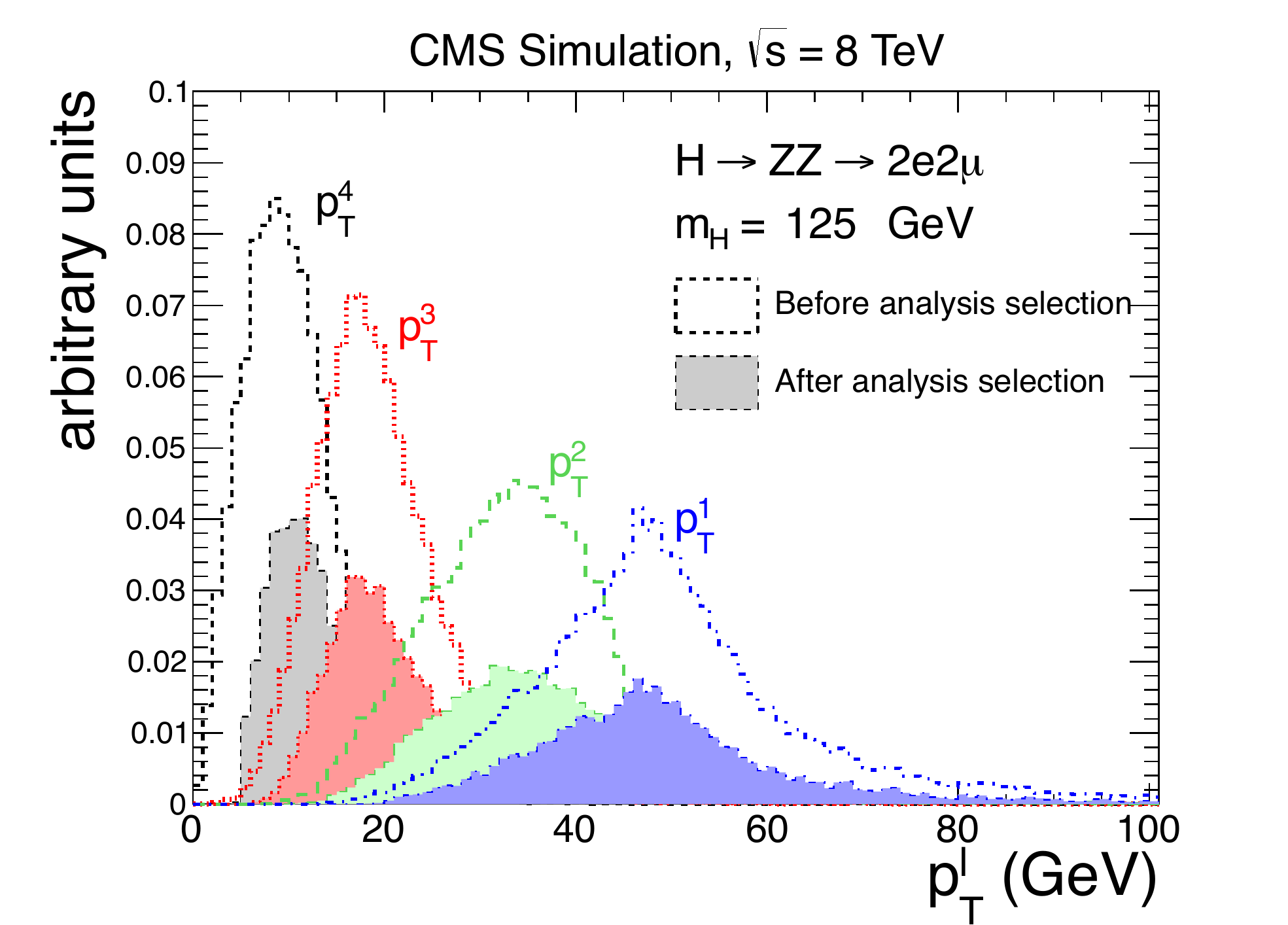}
     \includegraphics[width=0.49\linewidth]{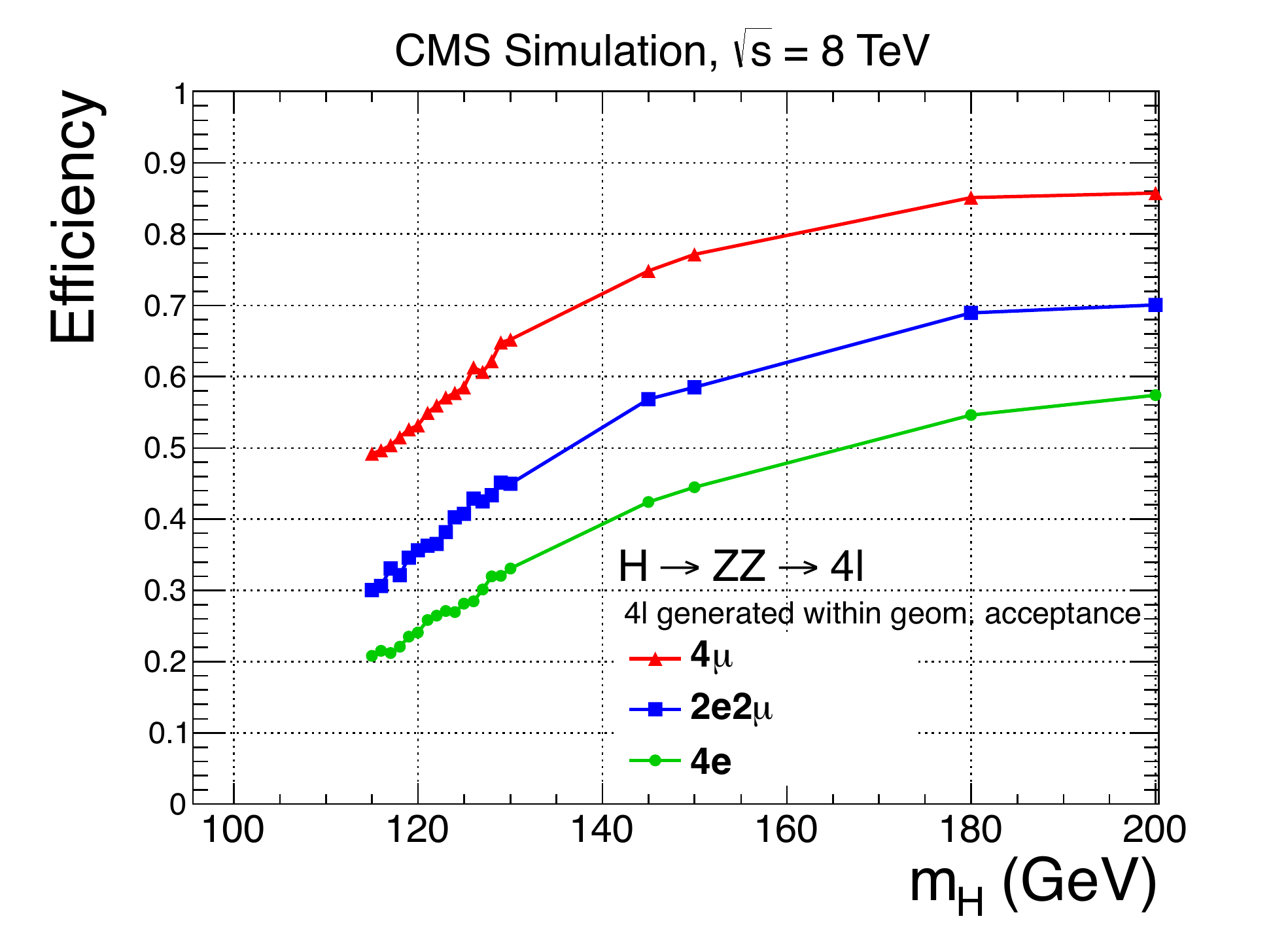}
     \caption{
The MC simulation distributions of the lepton transverse momentum $\PT^{\ell}$ for each of
the four leptons, ordered by $\PT^{\ell}$, from the process $\PH \to  \cPZ\cPZ \to 4\ell$
for a Higgs boson mass of 125\GeV in the $4\Pe$ (top left), $4\mu$ (top
right), and $2\Pe 2\mu$ (bottom left) channels.
The distributions are shown for events when all four leptons are within the geometrical
acceptance of the analysis (open histograms), and for events passing the final selection
criteria (solid histograms).The bottom-right plot displays the event selection efficiencies
for $\PH \to  \cPZ\cPZ \to 4\ell$  determined from MC simulation, as a function of the Higgs
boson mass, for the $4\Pe$, $4\mu$, and $2\Pe 2\mu$ channels. The efficiencies are
relative to events where all four leptons are within the geometrical acceptance.
Divergent contributions from $\cPZ\gamma^*$ with $\gamma^* \rightarrow \ell \ell$ at
generator level are avoided by requiring that all dilepton invariant masses are greater
than 1\GeV.}
\label{fig:leptonPT}
  \end{center}
\end{figure}

The \cPZ\ boson candidates are reconstructed from pairs of leptons of the same
flavour and opposite charge ($\ell^+\ell^-$).
The lepton pair with an invariant mass closest to the nominal \cPZ\  mass is denoted
as $\cPZ_1$ with mass $m_{\cPZ_1}$ and is retained if it satisfies $40 < m_{\cPZ_1} < 120\GeV$.
The invariant mass of the second \cPZ\ candidate, denoted $\cPZ_2$,
must satisfy
$12 < m_{\cPZ_2} < 120\GeV$.
The minimum value of $12\GeV$ is found from simulation to provide  the optimal sensitivity for a Higgs boson mass
in the range $ 110 < \mH < 160~\GeV$.
If more than one ${\cPZ_2}$ candidate satisfies all the criteria, we choose
the candidate reconstructed from the
two leptons with the highest scalar sum of their \PT.
Among the four selected leptons forming $\cPZ_1$ and $\cPZ_2$,
at least one is required to have $\PT > 20\GeV$ and another
$\PT > 10\GeV$.
These \PT thresholds ensure that the selected leptons
are on the high-efficiency plateau for the trigger.
To further reject leptons originating from weak semileptonic hadron decays
or decays of low-mass hadronic resonances, we require that all
opposite-charge pairs of leptons chosen from among the four selected leptons (irrespective of flavour)
have an invariant mass greater than  4\GeV.
The phase space for the Higgs boson search is defined by restricting the four-lepton
mass range to $m_{4\ell} > 100\GeV$.
The predicted lepton $\PT$ distributions from the  MC simulation for a Higgs boson with  $\mH = 125$\GeV
are shown in Fig.~\ref{fig:leptonPT} for the $4\Pe$, $4\mu$, and $2\Pe 2\mu$ channels.
Also given in Fig.~\ref{fig:leptonPT}  (bottom right) are the event selection efficiencies   for each
of the three lepton channels, as a function of the Higgs boson mass.
These distributions clearly emphasize the importance of low lepton-$\PT$ thresholds and high
lepton efficiencies.
The selection efficiencies shown in Fig.~\ref{fig:leptonPT} are relative to events where all
four leptons  are within the
geometrical acceptance and all dilepton invariant masses satisfy $m_{\ell\ell} > 1\GeV$.
The combined signal reconstruction and selection efficiency, for
a Higgs boson with $\mH = 125\GeV$,
is 18\% for the $4\Pe$ channel, 40\% for the $4\mu$ channel,
and 27\% for the $2\Pe2\mu$ channel.
The expected resolution on the per-event mass measurement is on average  2.2\%
for the $4\Pe$ channel, 1.1\% for the $4\mu$ channel, and 1.6\% for the $2\Pe2\mu$ channel.

The kinematics of the $\PH\rightarrow\cPZ\cPZ\rightarrow 4\ell$
process, as well as for any
boson decaying to  $\cPZ\cPZ$, has been extensively studied in the
literature~\cite{Soni:1993jc,Barger:1993wt,Choi:2002jk,Allanach:2002gn,Choi:2002jk,
Buszello:2002uu,Godbole:2007cn,Keung:2008ve,Antipin:2008hj,Hagiwara:2009wt,
Gao:2010qx,DeRujula:2010ys,Gainer:2011xz,Bolognesi:2012mm}.
Since the Higgs boson is spinless, the angular distribution of its decay products
is independent of the production mechanism.
In the Higgs boson rest frame, for a given invariant mass of the $4\ell$ system, the kinematics are fully
described by five angles, denoted $\vec{\Omega}$, and the invariant masses of the two lepton
pairs $\cPZ_1$ and $\cPZ_2$. 
These seven variables provide  significant discriminating power between signal and background.

A kinematic discriminant ($K_{D}$) is introduced using the full probability density in the dilepton
masses and angular variables, ${\cal P}(m_{\cPZ_1},m_{\cPZ_2},\vec{\Omega}|m_{4\ell})$.
The $K_{D}$ is constructed for each candidate event  based on the probability ratio of the signal and background
hypotheses, $K_{D}={\cal P_\mathrm{sig}}/({\cal P_\mathrm{sig}}+{\cal P_\mathrm{bkg}})$,
as described in Refs.~\cite{Chatrchyan:2012sn,CMSobservation125}.
For the signal, the phase-space and Z-propagator terms~\cite{Choi:2002jk} are
included in a fully analytic parametrization of the Higgs boson signal~\cite{Gao:2010qx}.
An analytic parametrization is also used for the background probability distribution for the mass
range above the $\cPZ\cPZ$ threshold, while it is  tabulated using a MC simulation of the
$\cPq\cPaq\to\cPZ\cPZ(\cPZ\gamma^*)$ process below this threshold.

\subsection{Background estimation and systematic uncertainties}

The small number of observed candidate events precludes a precise direct determination of the background
by extrapolating from the signal region mass sidebands.
Instead, we rely on MC simulation to evaluate the local density ($\Delta N /  \Delta m_{4\ell}$) of $\cPZ\cPZ$ background
events  expected as a function of  $m_{4\ell}$.
The cross section
for \cPZ\cPZ\ production at NLO is calculated with
\textsc{mcfm}~\cite{MCFM,Campbell:1999ah,Campbell:2011bn}.
This includes the dominant process from $\Pq\Paq$ annihilation,  as well as from gluon-gluon fusion.
The uncertainties in the predicted number of background events owing to the variation in the QCD renormalization and factorization
scales and PDF set
are on average 8\% for each final state~\cite{Dittmaier:2012vm}.
The number of predicted $\cPZ\cPZ\rightarrow 4\ell$  events and their systematic uncertainties after the signal
selection are given in Table~\ref{tab:SelectYieldsLowMass}.

The reducible $\cPZ+\bbbar$, $\ttbar$, $\cPZ+\text{jets}$,  $\cPZ+\gamma+\text{jets}$, and  $\PW\cPZ+\text{jets}$ backgrounds
contain at least one nonprompt lepton in the four-lepton final state.
The main sources of nonprompt leptons are electrons and muons coming from
decays of heavy-flavour quark, misidentified jets (usually originating from light-flavour quarks),
and electrons from photon conversions.
The lepton misidentification probabilities are measured in data samples of $\cPZ+\text{jet}$ events
with one additional reconstructed lepton, which are dominated by final states that
include a $\cPZ$ boson and a fake lepton.
The contamination from \PW\cPZ\ production in these events is suppressed by
requiring $\ETmiss <25$\GeV.
The lepton misidentification probabilities measured from these events are consistent with those
derived from MC simulation.
These misidentification probabilities are applied to dedicated
$\cPZ_1+X$ control samples, where $X$ contains two reconstructed leptons with relaxed isolation and identification
criteria.  Starting from these samples, two complementary approaches
are used  to extract  the corresponding reducible $\cPZ+X$ background
yield expected in the $4\ell$ signal region.
The first approach
avoids signal contamination in the background sample by reversing the
opposite-sign requirement on the $\cPZ_2$ lepton candidates, and then applies
the fake lepton efficiencies to the additional leptons to calculate the
expected number of background events in the signal sample.
The second approach  uses a control region defined by two opposite-sign leptons failing the isolation
and identification criteria, and using the misidentification probability to extrapolate to the signal region.
In addition, a control region with three passing leptons and one failing lepton is also used to
estimate the background with three prompt leptons and one misidentified lepton.
Comparable background predictions in the signal region are found  from
both methods within their uncertainties. The average of the two predictions is used
for the background estimate, with an uncertainty that includes the difference
between them (see Table~\ref{tab:SelectYieldsLowMass}).

Systematic uncertainties are evaluated from the data for the trigger (1.5\%), and the combined four-lepton reconstruction,
identification, and isolation efficiencies that vary from 1.2\% in the $4\mu$ channel at $\mH=150$\GeV
to about 11\% in the $4\Pe$ channel at  $\mH=120$\GeV.
The effects of the systematic uncertainties in the lepton energy-momentum calibration  (0.4\%) and energy resolution
on the four-lepton invariant-mass distribution are taken into account.
The accuracy of the absolute mass scale and resolution is validated using
$Z \to \ell\ell$, $Y \to \ell\ell$, and  $\cPJgy \to \ell\ell$ events.
The effect of the energy resolution uncertainty is taken into account by introducing a 20\%
variation on the simulated width of the signal mass peak.
An uncertainty of  50\% is assigned to the reducible background rate.
This arises from the finite statistical precision in the reducible background
control regions,  differences in the background composition between the
various control regions,
and  differences between the data samples used to measure  the lepton
misidentification probabilities.
Since all the reducible and instrumental background  are estimated using control regions in the data,
they are independent of the uncertainty in the integrated luminosity.
However, this uncertainty (2.2\% at 7\TeV~\cite{CMS-PAS-SMP-12-008} and 4.4\% at 8\TeV~\cite{CMS:2012jza})
does affect the prediction of the \cPZ\cPZ\ background and the
normalization of the signal in determining the Higgs boson cross section.
Finally, the systematic uncertainties in the theoretical Higgs boson cross section (17--20\%) and $4\ell$
branching fraction (2\%) are taken from Ref.~\cite{LHCHiggsCrossSectionWorkingGroup:2011ti}.

\subsection{Results}

The number of selected $\cPZ\cPZ\rightarrow 4\ell$ candidate events in the mass range $110 < m_{4\ell} < 160$\GeV
for each of the three final states is given in Table~\ref{tab:SelectYieldsLowMass}.
The number of predicted background events  in each of the three final states and their uncertainties are also given,
together with the number of signal events expected from a SM Higgs boson
of $\mH = 125$\GeV.

\begin{table}[htbp]
\begin{center}
\topcaption{The number of observed selected events, compared to the expected background yields and
the expected number of signal events ($\mH = 125$\GeV) for each lepton final state in the $\PH\to\cPZ\cPZ \to 4\ell$ analysis.
The estimates of the $\cPZ\cPZ$ background are from MC simulation and the $\cPZ+X$ background are based on data.
These results are given for the four-lepton invariant-mass range
from 110 to 160\GeV. The total expected background and the observed numbers of events are also given integrated
over the  three bins (``signal region'' defined as $121.5 < m_{4\ell} < 130.5$\GeV)
of Fig.~\ref{fig:ZZmass}, centred on the bin where the most significant excess is seen. The uncertainties shown include both
statistical and systematic components. }
\label{tab:SelectYieldsLowMass}
\begin{tabular}{l|c|c|c||c}
\hline
Channel & $4\Pe$ & $4\Pgm$ & $2\Pe2\Pgm$ & Total \\
\hline \hline
$\cPZ\cPZ$ background & 2.7 $\pm$ 0.3 & 5.7 $\pm$ 0.6 & 7.2 $\pm$ 0.8 & 15.6 $\pm$ 1.4 \\
$\cPZ+X$         &  $1.2 ^{ + 1.1}_{ - 0.8 }$ & $0.9 ^{ + 0.7 }_{ - 0.6 }$ & $2.3 ^{ + 1.8 }_{ - 1.4 }$ & $4.4 ^{ + 2.2 \phantom{^0}}_{ - 1.7\phantom{_0} }$ \\ %
\hline
All backgrounds \small{($110 < m_{4\ell} < 160$\GeV)} &  $3.9 ^{ + 1.1 }_{ - 0.8 }$ & $6.6^{ + 0.9 }_{ - 0.8 }$ & $9.5 ^{ + 2.0 }_{ - 1.6 }$ & $20.0 ^{ + 3.2}_{ - 2.6}$ \\ %
\hline
Observed \small{($110 < m_{4\ell} < 160$\GeV)} & 6 & 6 & 9 & 21\\
\hline \hline
Expected Signal \small{($\mH = 125$\GeV)} &  1.37  $\pm$ 0.44  &  2.75  $\pm$  0.56  &  3.44 $\pm$  0.81 & 7.6 $\pm$ 1.1 \\
\hline \hline
All backgrounds \small{(signal region)} &  $0.71^{+0.20}_{-0.15}$  &  $1.25^{+0.15}_{-0.13}$  &  $1.83^{+0.36}_{-0.28}$ & $3.79^{+0.47}_{-0.45}$\\
\hline
Observed \small{(signal region)} & 1 & 3 & 5 & 9\\
\hline
\end{tabular}
\end{center}
\end{table}

\begin{figure}[htbp]
   \begin{center}
     \includegraphics[width=0.7\linewidth]{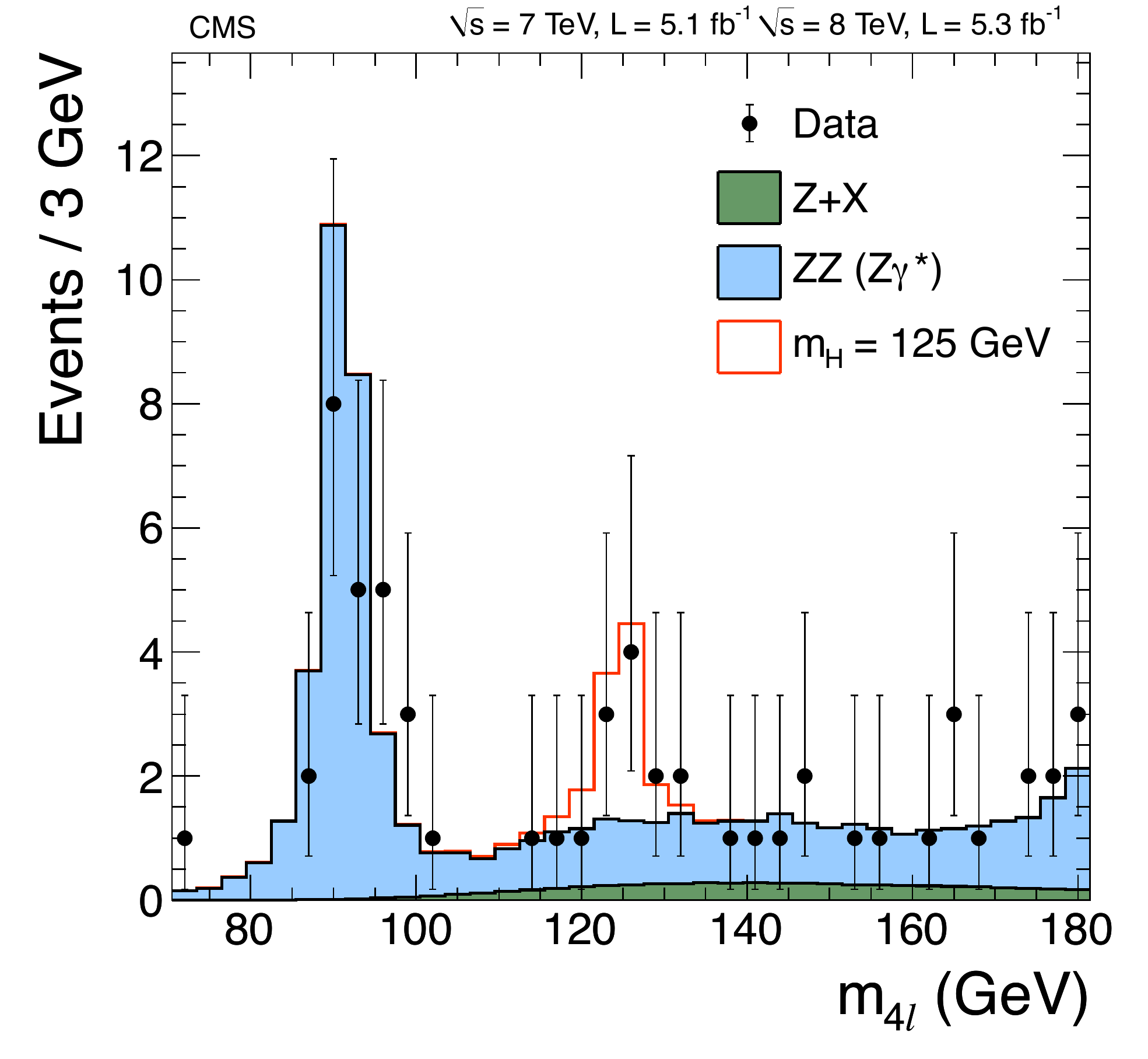}
     \caption{
     Distribution of the observed four-lepton invariant mass from the combined 7 and 8\TeV data
for the $\PH \to \cPZ\cPZ\to 4\ell$ analysis (points).
The prediction for the expected $\cPZ$+X and $\cPZ\cPZ(\cPZ\gamma^*)$ background are shown by the dark and
light histogram, respectively. The open histogram gives the expected distribution for a Higgs boson of mass 125\GeV.
     }
\label{fig:ZZmass}
   \end{center}
\end{figure}

The observed $m_{4\ell}$ distribution from data is shown in Fig.~\ref{fig:ZZmass}.
There is a clear peak 
at the \cPZ\ boson mass from the
decay $\cPZ \to 4\ell$~\cite{CMS:2012bw}. The size and shape
of the peak are consistent with those from the background
prediction. Over the full Higgs boson search region
from 110 to 160\GeV, the reducible background from
\cPZ+X events is much smaller than the
irreducible $\cPZ\cPZ (\cPZ\gamma^*)$ background. There is an excess
of events above the expected background near
125\GeV. The total number of observed events and
the expected number of background events in the three
bins centred on the excess ($121.5 < m_{4\ell} < 130.5$\GeV),
and referred to as the ``signal" region,
are given in Table 5. The expected four-lepton
invariant-mass distribution for a Higgs boson with
a mass of 125\GeV is shown by the open histogram
in Fig.~\ref{fig:ZZmass}.

\begin{figure}[htbp]
   \begin{center}
     \includegraphics[width=0.45\linewidth]{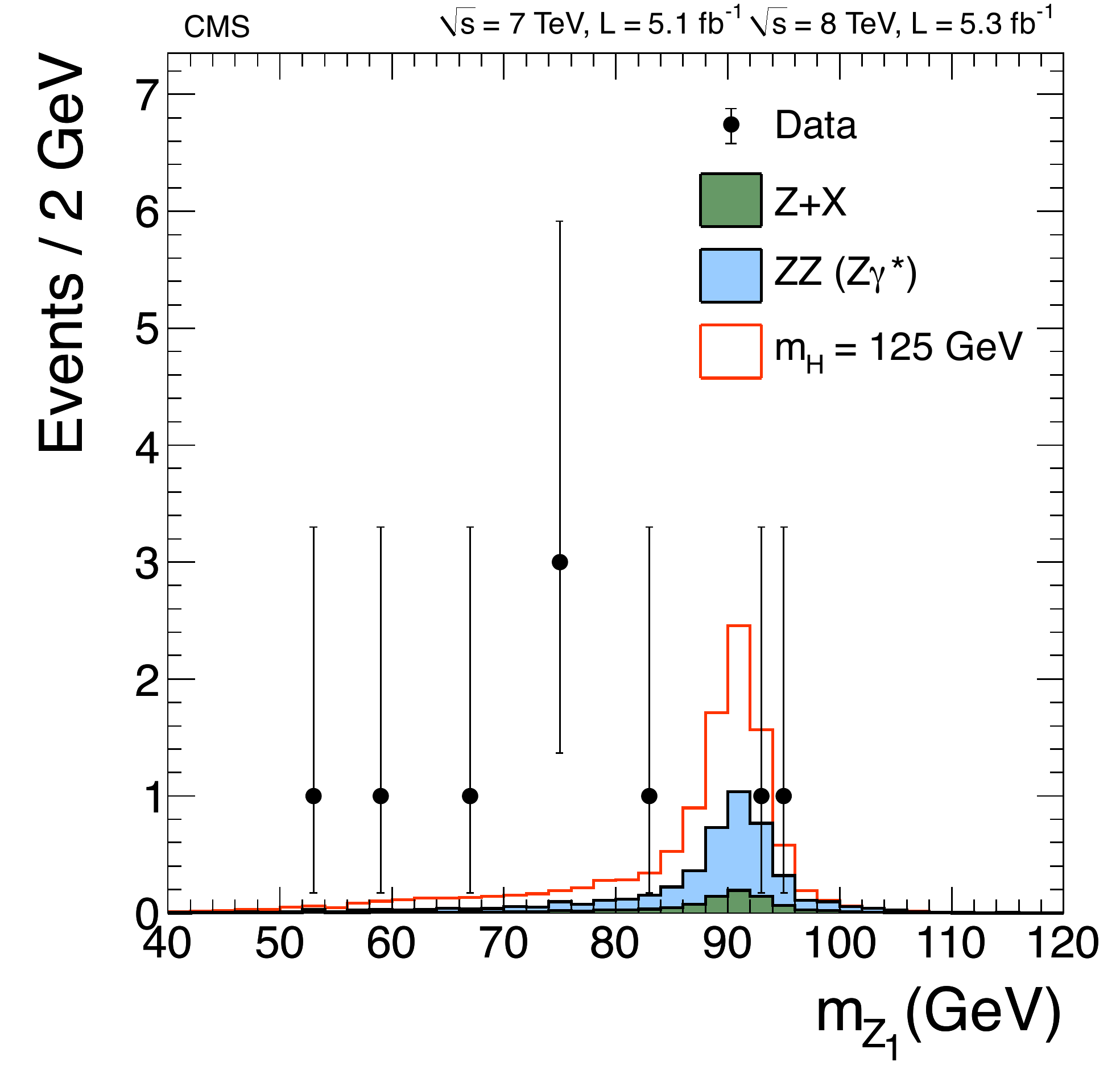}
     \includegraphics[width=0.45\linewidth]{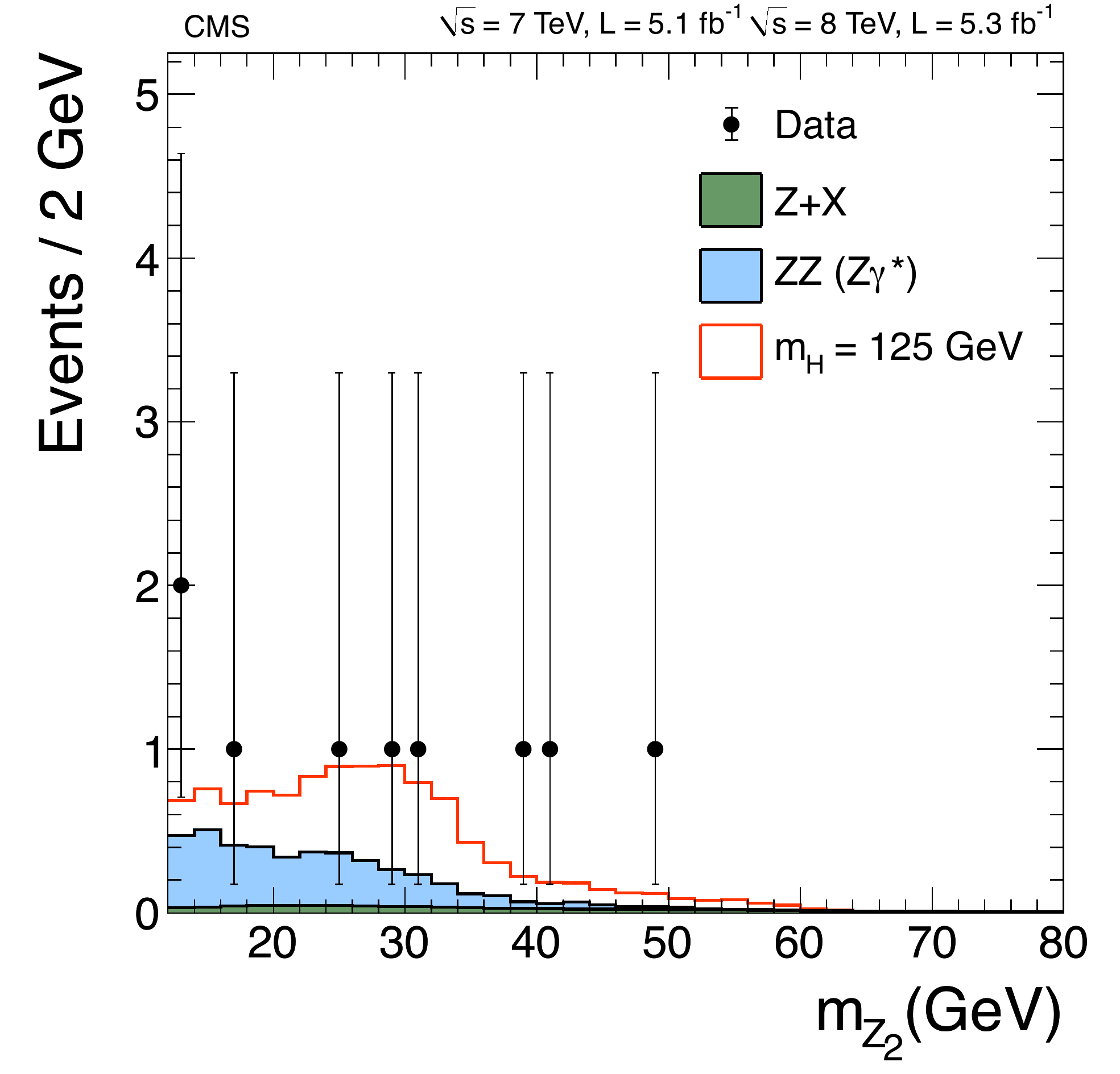}
     \caption{
     Distributions of the observed $\cPZ_1$ (left) and $\cPZ_2$ (right) dilepton invariant
     masses for four-lepton events in the mass range
     $121.5 < m_{4\ell} < 130.5$\GeV for the combined 7 and 8\TeV data (points) .
     The shaded histograms show the predictions for the background distributions, and the open histogram for
     a Higgs boson with a mass of 125\GeV.
     }
\label{fig:Z1Z2masses}
   \end{center}
\end{figure}

The distributions of the reconstructed $\cPZ_1$ and $\cPZ_2$ dilepton invariant masses for the events in the
signal region are shown in the left and right plots of Fig.~\ref{fig:Z1Z2masses}, respectively.
The $\cPZ_1$ distribution has a tail towards low invariant mass,
indicative that also the highest mass $\cPZ$ is often off-shell.

\begin{figure}[htbp]
   \begin{center}
     \includegraphics[width=0.7\linewidth]{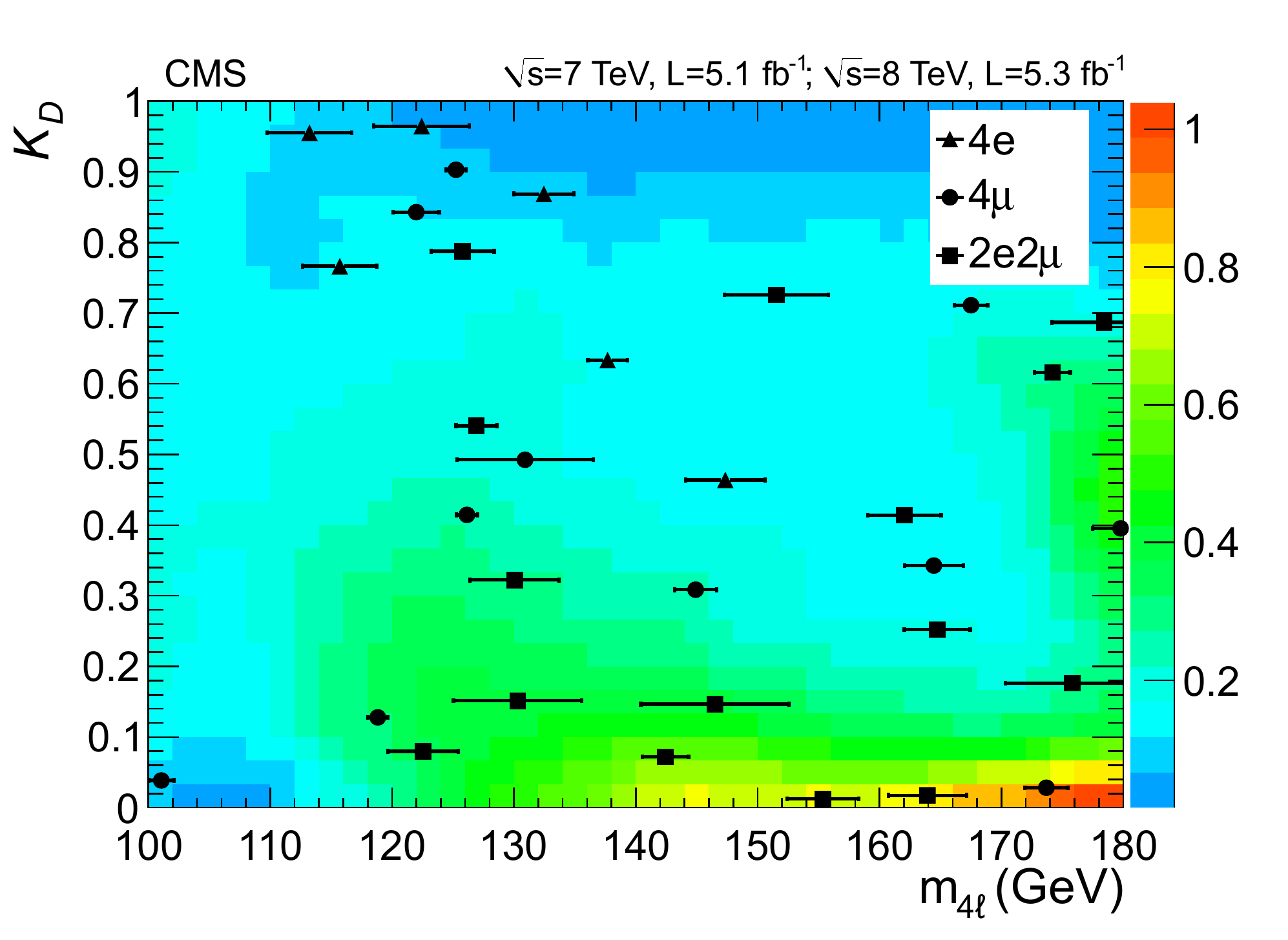}
     \includegraphics[width=0.7\linewidth]{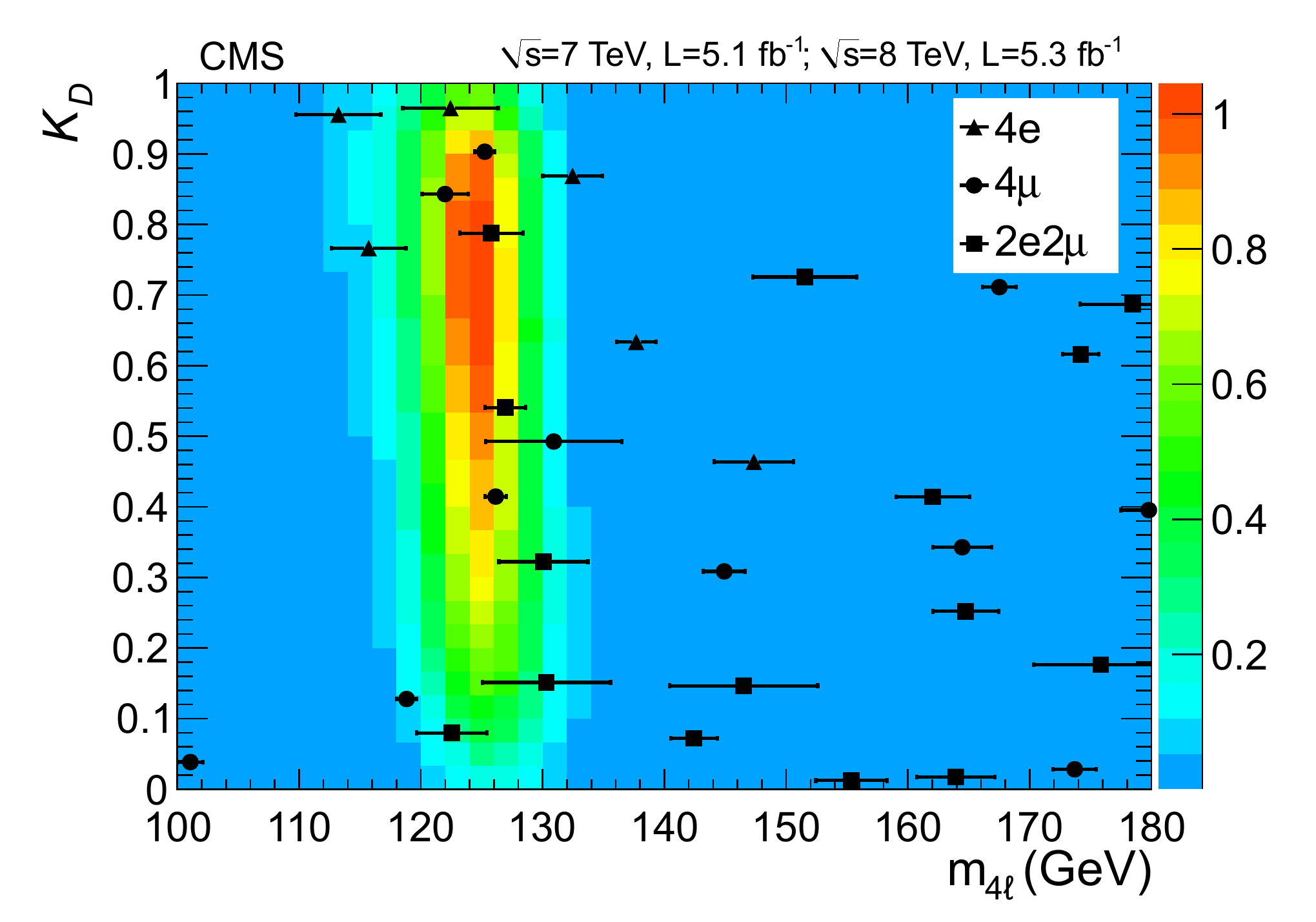}
    \caption{
    The two-dimensional distribution of  the kinematic
    discriminant $K_{D}$ versus $m_{4\ell}$ for selected $4\ell$ events in the combined 7 and 8\TeV data.
    Events in the three different final states are designated by symbols shown in the legend.
    The horizontal error bars indicate the estimated per-event mass resolution deduced from the
    combination of the per-lepton momentum uncertainties.
    The contours in the upper plot show the event density for the background expectation,
    and in the lower plot the contours for a
    SM Higgs boson with $\mH$ = 125\GeV (both in arbitrary units). }
\label{fig:Mass4lKD}
   \end{center}
\end{figure}

The two-dimensional distribution of the kinematic discriminant $K_{D}$ versus the four-lepton reconstructed mass
$m_{4\ell}$ is shown in Fig.~\ref{fig:Mass4lKD} for the individual selected events.
Superimposed on this figure are the contours of the
expected event density for the background (upper) and a
SM Higgs boson at $\mH$ = 125\GeV (lower).
A clustering of events is observed in the region around $m_{4\ell}$ = 125\GeV with $K_{D} \ge 0.7$.
The background expectation is low in this region and
the signal expectation is high, corresponding to the excess of events above background seen in the
one-dimensional $m_{4\ell}$ distribution.

\begin{figure}[!htb]
\begin{center}
\includegraphics[width=0.49\linewidth]{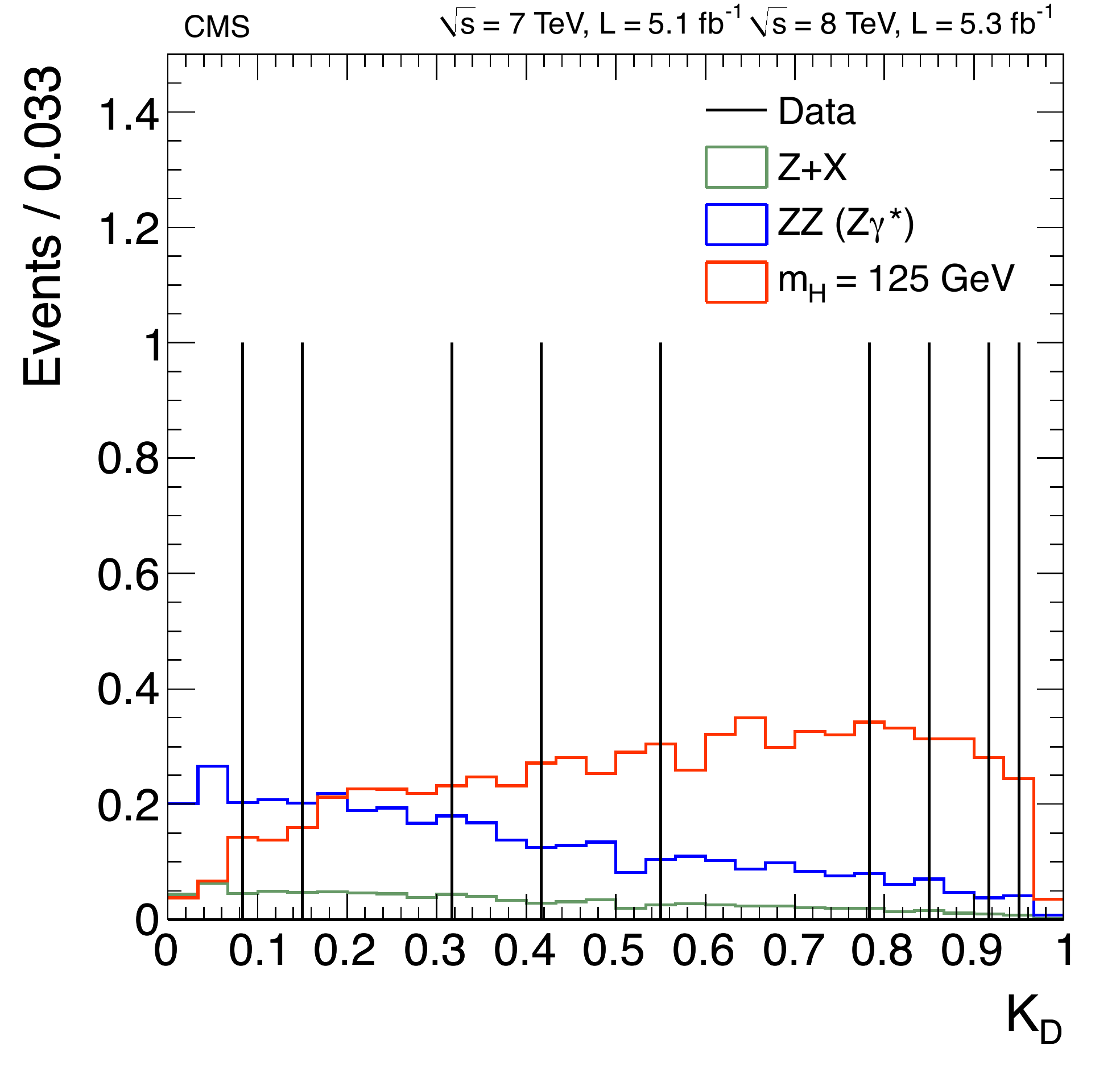}
\includegraphics[width=0.5\linewidth]{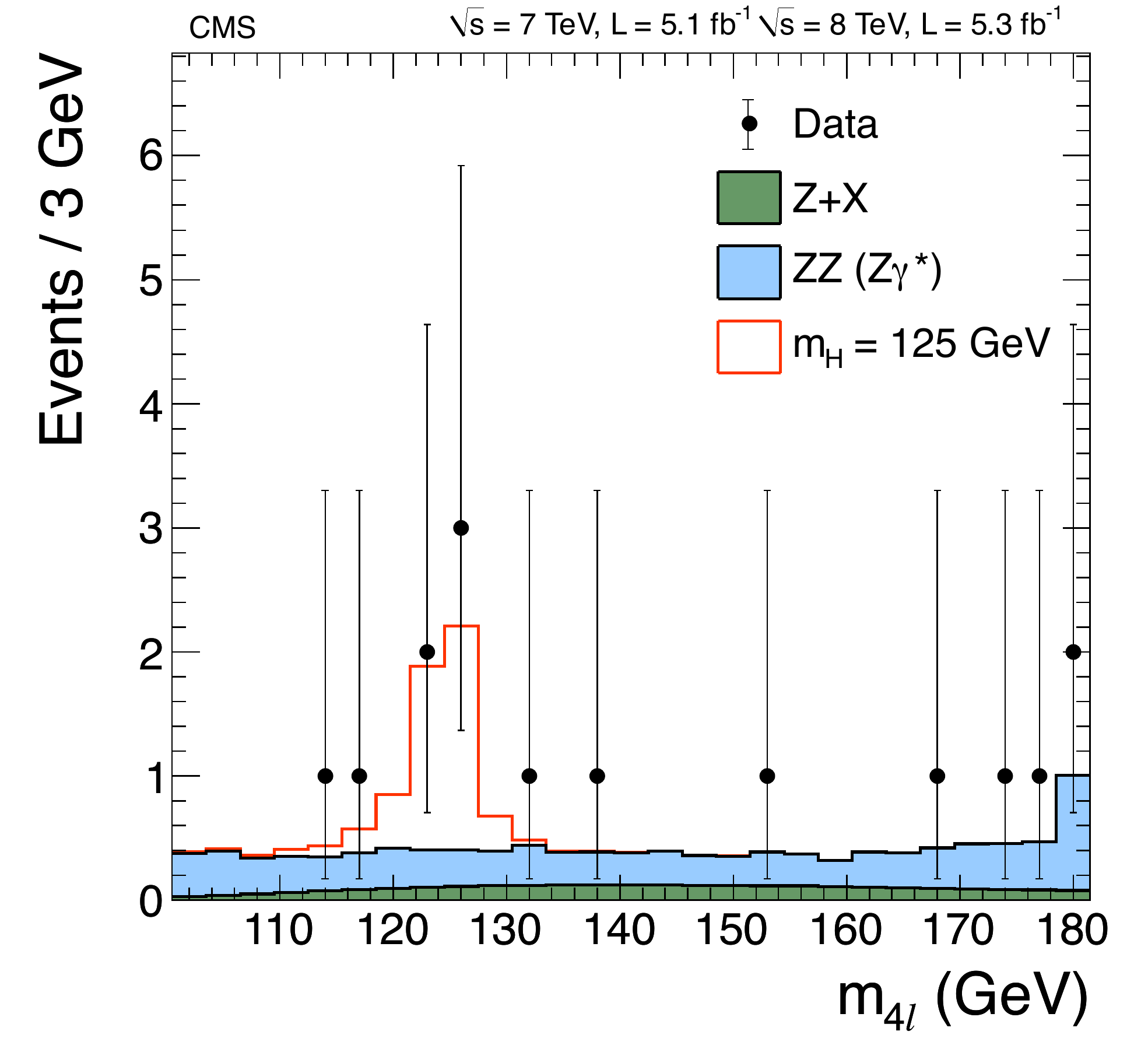}
\caption{
Left: Distribution of the kinematic discriminant $K_D$ for $\PH \to \cPZ\cPZ \to  4\ell$ candidate events
from the combined 7  and 8\TeV data (vertical lines) in the signal mass region $121.5 < m_{4\ell} < 130.5$\GeV.
The predicted distributions for the \cPZ+X and $\cPZ\cPZ(\cPZ\gamma^*)$ backgrounds and for a Higgs boson
with a mass of 125\GeV are shown by the histograms.
Right: The $m_{4\ell}$ distribution for data events with $K_D >$ 0.5 (points) and the predicted distributions
for the backgrounds and a Higgs boson with a mass of 125\GeV (histograms).
}
\label{fig:Mass4lKD05}
\end{center}
\end{figure}

The observed distribution of the $K_{D}$ discriminant values for invariant masses in
the signal range $121.5 < m_{4\ell} < 130.5\GeV$ is shown
in Fig.~\ref{fig:Mass4lKD05} (left).
The $m_{4\ell}$ distribution of events satisfying $K_{D} > 0.5$ is shown
in Fig.~\ref{fig:Mass4lKD05} (right). The clustering of events is clearly visible near $m_{4\ell}$$\approx$$125$\GeV.

\section{\texorpdfstring{$\PH\to\PW\PW$}{H to WW}\label{sec:hww2l2nu}}
The decay mode $\PH\to\PW\PW$ is highly sensitive to
a SM Higgs boson with a mass around the $\PW\PW$ threshold of
160~$\GeV$. With the lepton identification and
$\ETmiss$ reconstruction optimized for LHC pileup conditions, it is possible
to extend the sensitivity down to 120~$\GeV$. The search strategy for $\PH\to\PW\PW$
is based on the final state in which both $\W$ bosons decay leptonically,
resulting in a signature with two isolated, oppositely charged, high-$\pt$
leptons (electrons or muons) and large $\MET$ caused by the undetected neutrinos.
It is not possible to reconstruct the Higgs mass in this final state, nevertheless
there is some mass sensitivity via different kinematic distributions like the
dilepton mass or the invariant mass of leptons and $\met$.
The analysis of the 7\TeV data is described in Ref.~\cite{Chatrchyan:2012ty}
and remains unchanged, while the 8\TeV analysis is modified to cope with
the more difficult conditions induced by the higher pileup in the 2012 data
taking, and is explained below.

\subsection{\texorpdfstring{$\PW\PW$}{WW} event selection}
\label{sec:ww_evtsel}
To improve the signal sensitivity, events are separated by jet
multiplicity into three mutually exclusive categories, which are characterized
by different expected signal yields and signal-to-background ratios.
We call these the 0-jet, 1-jet, and 2-jet categories.
Jets are reconstructed using the selection described in
Section~\ref{sec:reconstruction}, and events are classified according to the
number of selected jets with $\Et>30\GeV$ and $|\eta|<$4.7.
To exclude electrons and muons from the jet sample, these
jets are required to be separated from the selected leptons in $\Delta R$
by at least $\Delta R^{\mathrm{jet-lepton}}>0.3$.
Events with more than 2 jets are only considered if there are no
additional jets above this threshold present in the pseudorapidity
region between the two highest-$\Et$ jets.
Furthermore, the search splits candidate signal events into
three final states, denoted by: $\Elp\Elm$, $\Mp\Mm$, and $\Elpm\Mmp$.

The bulk of the signal arises through direct $\PW\PW$ decays to dileptons
of opposite charge, where the small contribution
proceeding through an intermediate $\Pgt$ leptonic decays is implicitly included.
The events are selected by triggers that  require
the presence of one or two high-$\pt$ electrons or muons.
The trigger efficiency for signal events that pass the full event selection
is measured to be above 97\% in the $\Mp\Mm$ final state, and above 98\% in the
$\Elp\Elm$ and $\Elpm\Mmp$ final states for a Higgs boson mass of about $125\GeV$.
The trigger efficiencies increase along with Higgs boson mass. These
efficiencies are measured using $\dyll$ events~\cite{CMS:2011aa}, with associated
uncertainties of about 1\%.

Two oppositely charged lepton candidates are required, with $\pt > 20\GeV$ for
the higher-$\pt$ lepton ($\ptlmax$) and $\pt > 10\GeV$ for the lower-$\pt$ lepton ($\ptlmin$).
Only electrons (muons) with $|\eta| < 2.5 \, (2.4)$ are considered in the analysis.

A tight muon selection is applied,  as described in
Section~\ref{sec:reconstruction}. Muons are required to be isolated to
distinguish between muon candidates from \PW\ boson decays and those from QCD
background processes, which are usually in or near jets. For each muon
candidate, the scalar sum of the transverse energy of all particles
consistent with originating from the primary vertex is reconstructed in
cones of several widths around the muon direction, excluding the
contribution from the muon itself. This information is combined using a
multivariate algorithm that exploits the differences in the
energy deposition between prompt muons and muons from hadron decays inside
a jet.

Electron candidates are identified using the multivariate approach described in
Section~\ref{sec:reconstruction}. Electrons are required to be isolated by
applying a threshold on the sum of the transverse energy of the particles
that are reconstructed in a cone around them, excluding the contribution from
the electron itself.
For both electrons and muons, a correction is applied to account for the
contribution to the energy in the isolation cone from pileup, as
explained in Section~\ref{sec:reconstruction}.

In addition to high-momentum, isolated leptons and minimal jet activity, missing
transverse momentum is present in signal events, but generally not in the background.
In this analysis, a projected~$\MET$ variable is employed. It is
equal to the component of the $\MET$ vector transverse to the nearest lepton direction, if the
difference in azimuthal angle between this lepton and the $\MET$ vector is less
than $90^\circ$. If there is no lepton within  $90^\circ$ of the $\MET$ direction  in
azimuth, the value of $\MET$ is used.
Since the projected~$\MET$ resolution is degraded
by pileup, the minimum of two $\MET$ observables is used in the determination of the projected $\MET$ value:
the first is the standard
$\MET$, while the second uses only charged particles associated with the primary vertex to measure the missing transverse energy.
Events with projected~$\MET$ above 20$\GeV$ are selected for the analysis.

To suppress the top-quark background, a \textit{top-quark tagging} technique, based
on low-momentum muon identification and \cPqb-jet tagging~\cite{CMS-PAS-BTV-12-001}, is applied. The first
selection is designed to veto events containing muons from \cPqb\ hadrons coming from top-quark decays.
The second selection uses a \cPqb-jet tagging algorithm that looks for tracks with large impact parameter
within jets. The rejection when combining the two selections for the top-quark background is about 50\% in the 0-jet
category and above 80\% for events with at least one jet passing the selection criteria.

Various selection criteria are used to reduce the other background contributions. For the $\PW$+jets background,
a minimum dilepton transverse momentum ($\pt^{\ell\ell}$) of 45\GeV is required.
To reduce the background from $\WZ$ production, any event
that has a third lepton passing the identification and isolation requirements is rejected.
This requirement rejects less than 1\% of the $\PW\PW \to 2\ell2\nu$ events, while
rejecting around 35\% of the remaining $\WZ$ events.
The contribution from $\wgamma$ production,
where the photon converts into a electron pair, is reduced by about 90\%
in the dielectron final state by  requirements that reject $\gamma$ conversions.
Those requirements consist in finding tracks that associated with the
electron give good conversion candidates.
The background from low-mass resonances is rejected by requiring a dilepton mass ($\mll$) greater
than 12 $\GeV$.

The Drell--Yan process produces same-flavour lepton pairs ($\Elp\Elm$ and $\Mp\Mm$). In order
to suppress this background, a few additional requirements are applied in the same-flavour final states.
First, the resonant $\cPZ$ component of the Drell--Yan production is rejected by requiring
a dilepton mass outside a 30\GeV window centred on the $\Z$ mass.
Then, the remaining off-peak contribution is suppressed by exploiting different \MET-based approaches depending
on the number of jets and the Higgs boson mass hypothesis.
At large Higgs boson masses ($\mH > 140\GeV$), signal events are associated with large \MET\ and, thus,
to suppress the Drell--Yan background it is sufficient to require the minimum of the
two projected~$\MET$ variables to be greater than 45 $\GeV$.
On the contrary, in low-mass Higgs boson events ($\mH \leq 140\GeV$) it is more difficult to separate the signal from the Drell--Yan background;
therefore in this case, a dedicated multivariate selection, combining the missing transverse momentum with kinematic and topological variables, is used to
reject Drell--Yan events and maximize the signal yield.
A third approach is employed in events with two jets. Here, the dominant source of \MET\ is the mismeasurement of the hadronic jet energy,
and the optimal performance is obtained by requiring $\MET > 45\GeV$.
Finally, the momenta of the dilepton system and the most energetic jet must have an angle
in the transverse plane smaller than $165^\circ$. These selections reduce the Drell--Yan
background by three orders of magnitude, while rejecting less than 50\% of the signal, as determined from simulation.

\begin{figure}[h!t]
\begin{center}
   \includegraphics[width=0.49\textwidth]{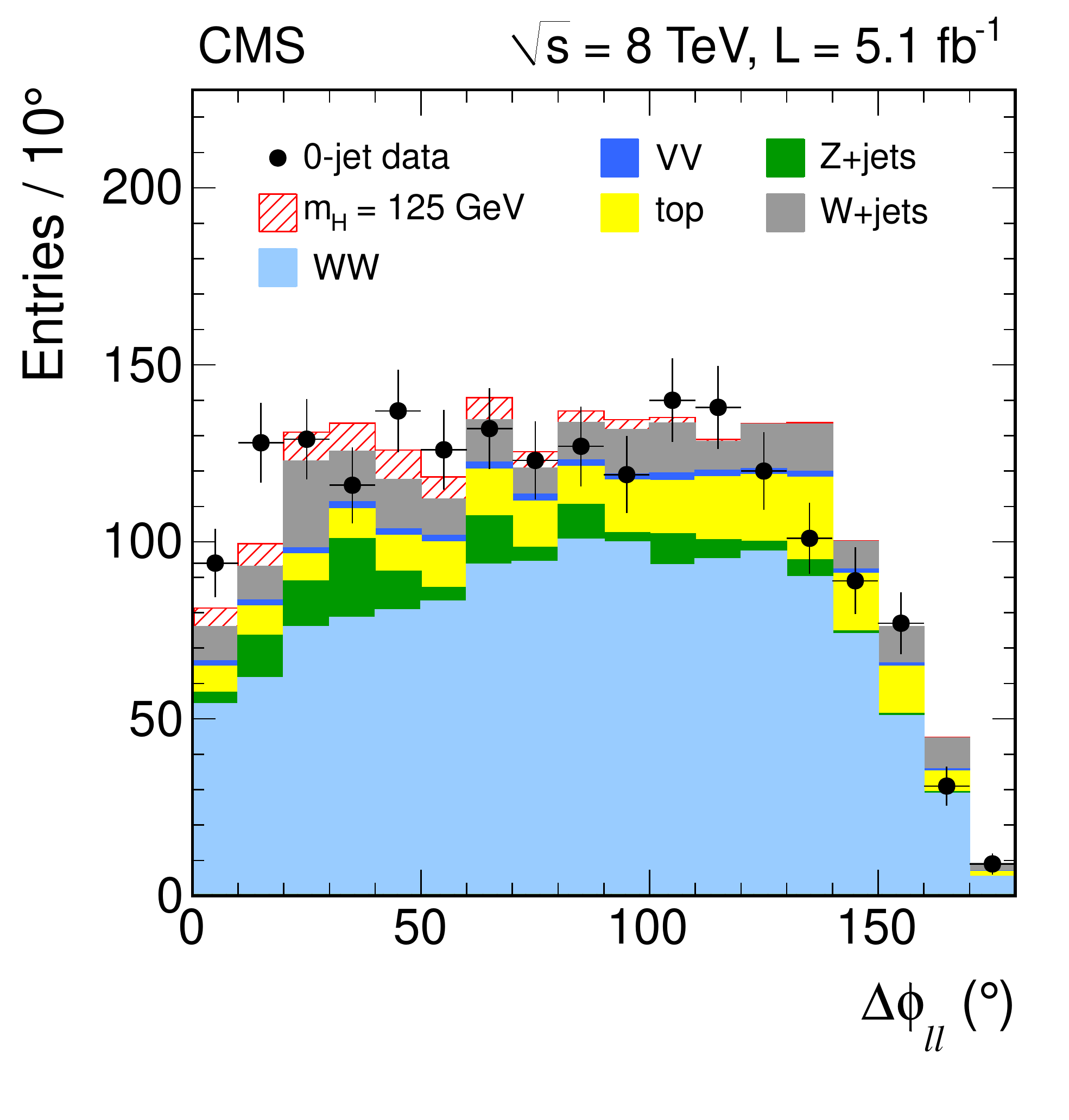}
   \includegraphics[width=0.49\textwidth]{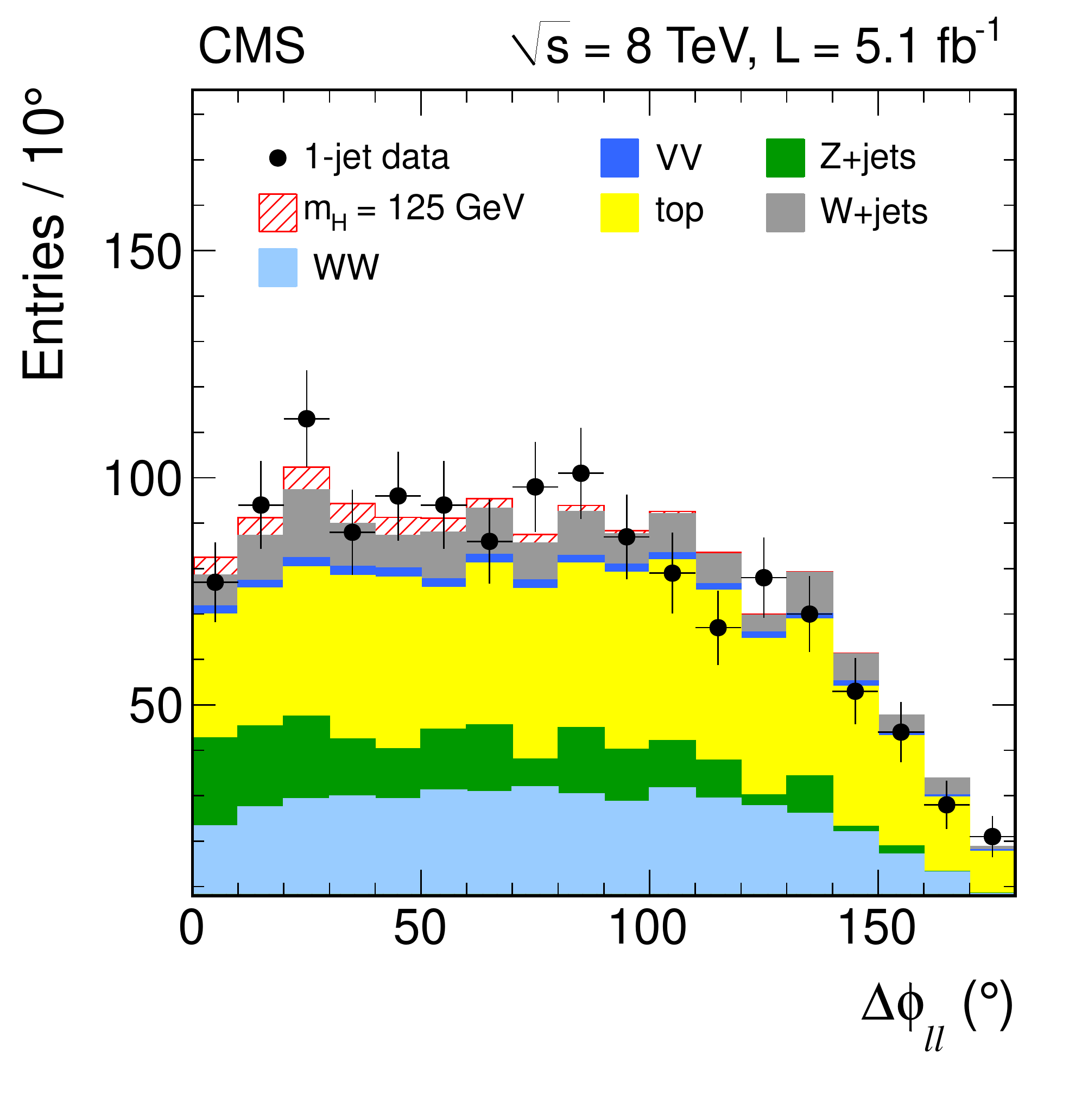}
	 \caption{Distributions of the azimuthal angle difference $\delphill$ between selected
leptons in the 0-jet (left) and 1-jet (right) categories, for data (points),
the main backgrounds (solid histograms), and a SM Higgs boson signal with $\mH= 125\GeV$
(hatched histogram) at 8\TeV. The standard $\PW\PW$
selection is applied.} \label{fig:wwpresel_nj_mh125_deltaphill}
\end{center}
\end{figure}

\begin{figure}[h!t]
\begin{center}
   \includegraphics[width=0.49\textwidth]{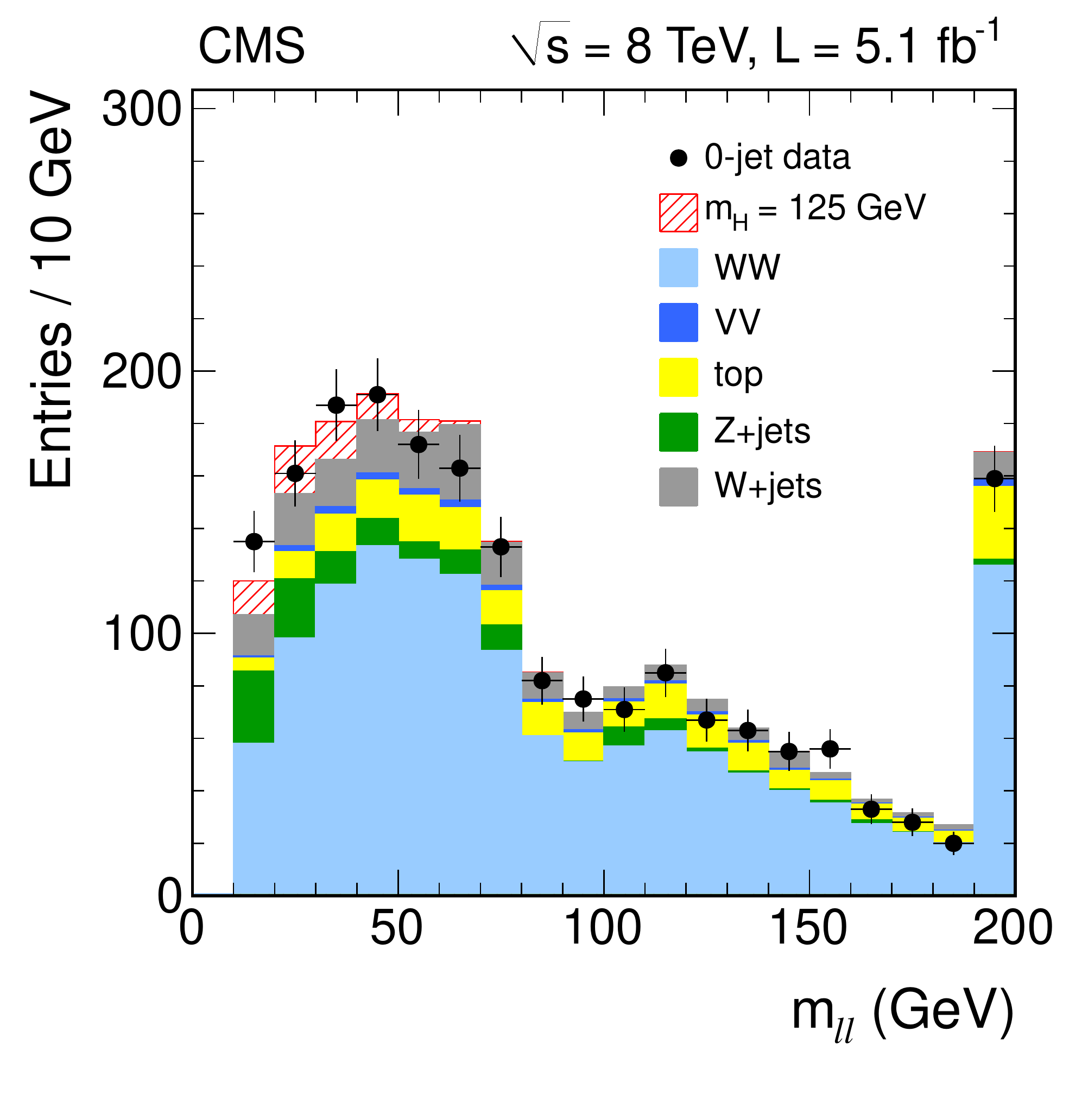}
   \includegraphics[width=0.49\textwidth]{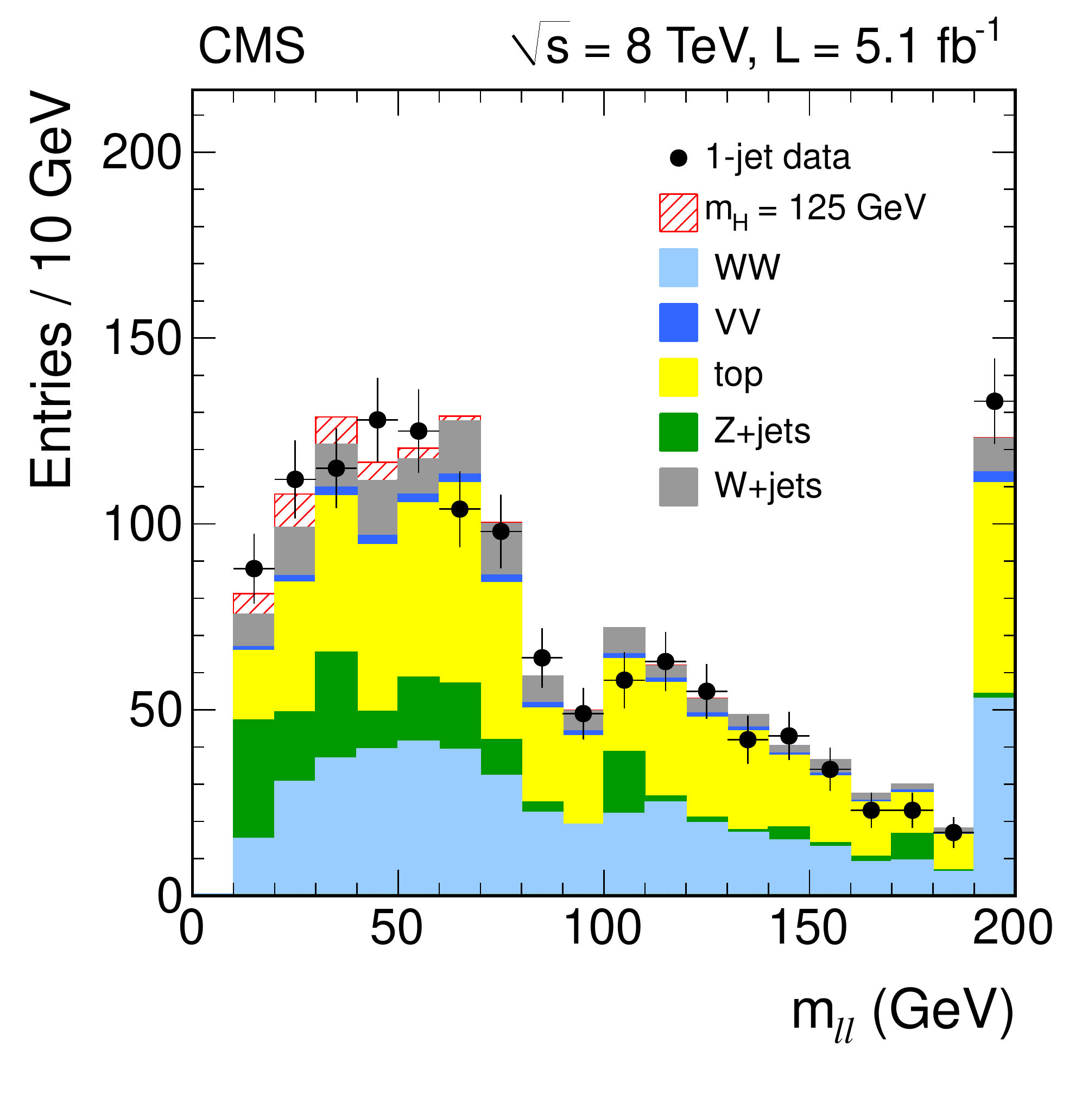}
	\caption{Distributions of the dilepton invariant mass $\mll$ of selected dileptons
	in the 0-jet (\cmsLeft) and 1-jet (\cmsRight) categories, for data (points),
the main backgrounds (solid histograms), and a SM Higgs boson with $\mH= 125\GeV$
(hatched histogram) at 8\TeV. The standard $\PW\PW$
selection is applied. The last bin contains overflows.}  \label{fig:wwpresel_nj_mh125_massll}
\end{center}
\end{figure}

After applying the full set of selection criteria,  referred to as the $\PW\PW$ selection,
the observed yields in the combined 7 and 8\TeV data set are 1594, 1186, and
1295 events in the 0-jet, 1-jet, and 2-jet categories, respectively. This sample is
dominated by nonresonant $\PW\PW$ events in the 0-jet category and by a
similar fraction of $\PW\PW$ and top events in the other two categories.
The main efficiency loss is due to the
lepton selection and the stringent $\MET$ requirements.
Figures~\ref{fig:wwpresel_nj_mh125_deltaphill} and~\ref{fig:wwpresel_nj_mh125_massll}
show the observed distributions of the azimuthal angle difference ($\delphill$) and
the dilepton mass ($m_{\ell\ell}$) after the $\PW\PW$ selection,
respectively, and the expected distributions for a SM Higgs boson with $\mH=125\GeV$
and for backgrounds in the 0- and 1-jet categories. The clear difference in
the shape between the $\Hi \to \PW\PW$ and the nonresonant $\PW\PW$
processes is because of the spin-0 nature of the Higgs boson.

\subsection{\texorpdfstring{$\Hi \to \PW\PW$}{Higgs to WW} search strategy}
\label{sec:hww}

\begin{figure}[h!t]
\begin{center}
   \includegraphics[width=0.49\textwidth]{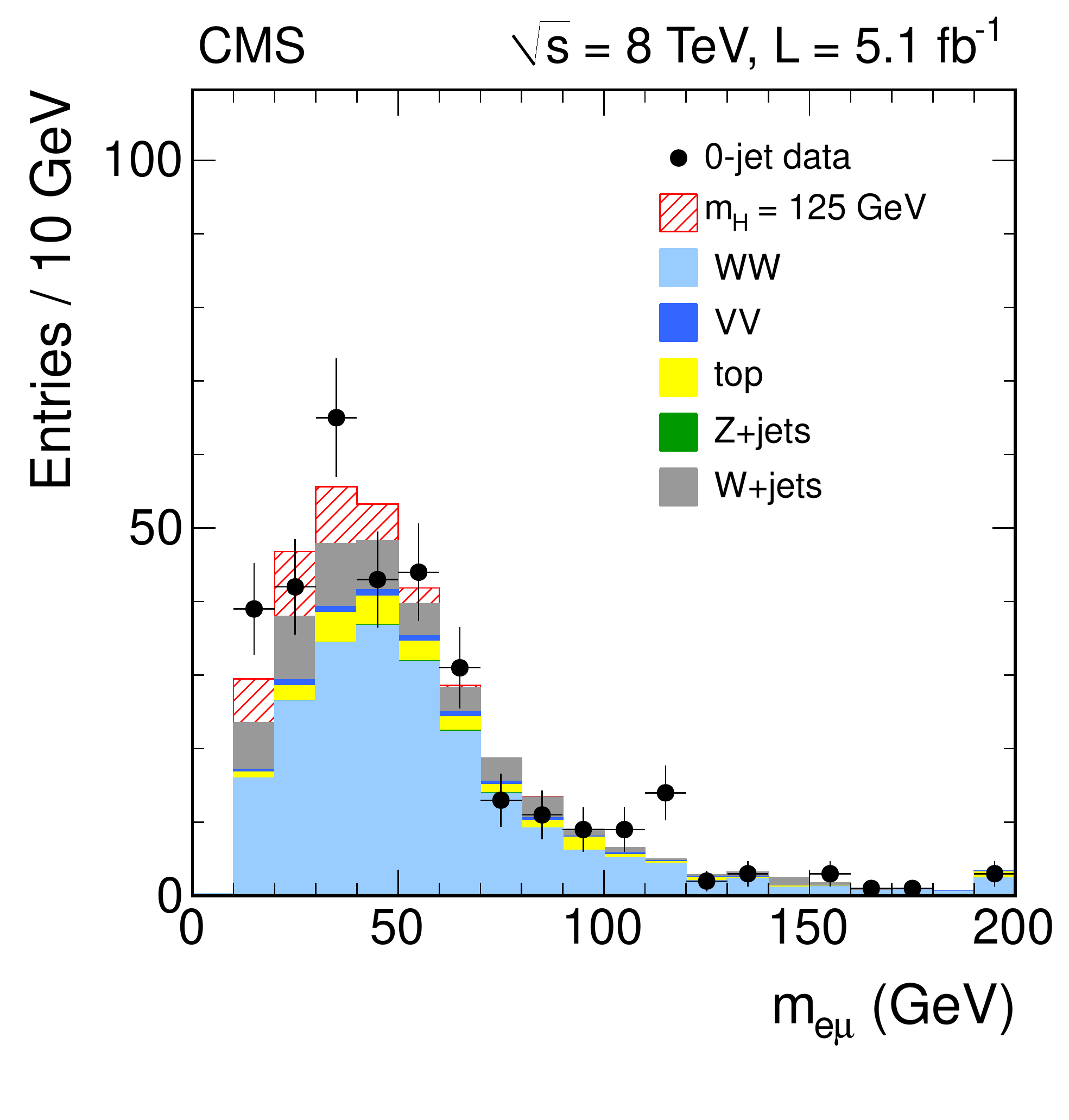}
   \includegraphics[width=0.49\textwidth]{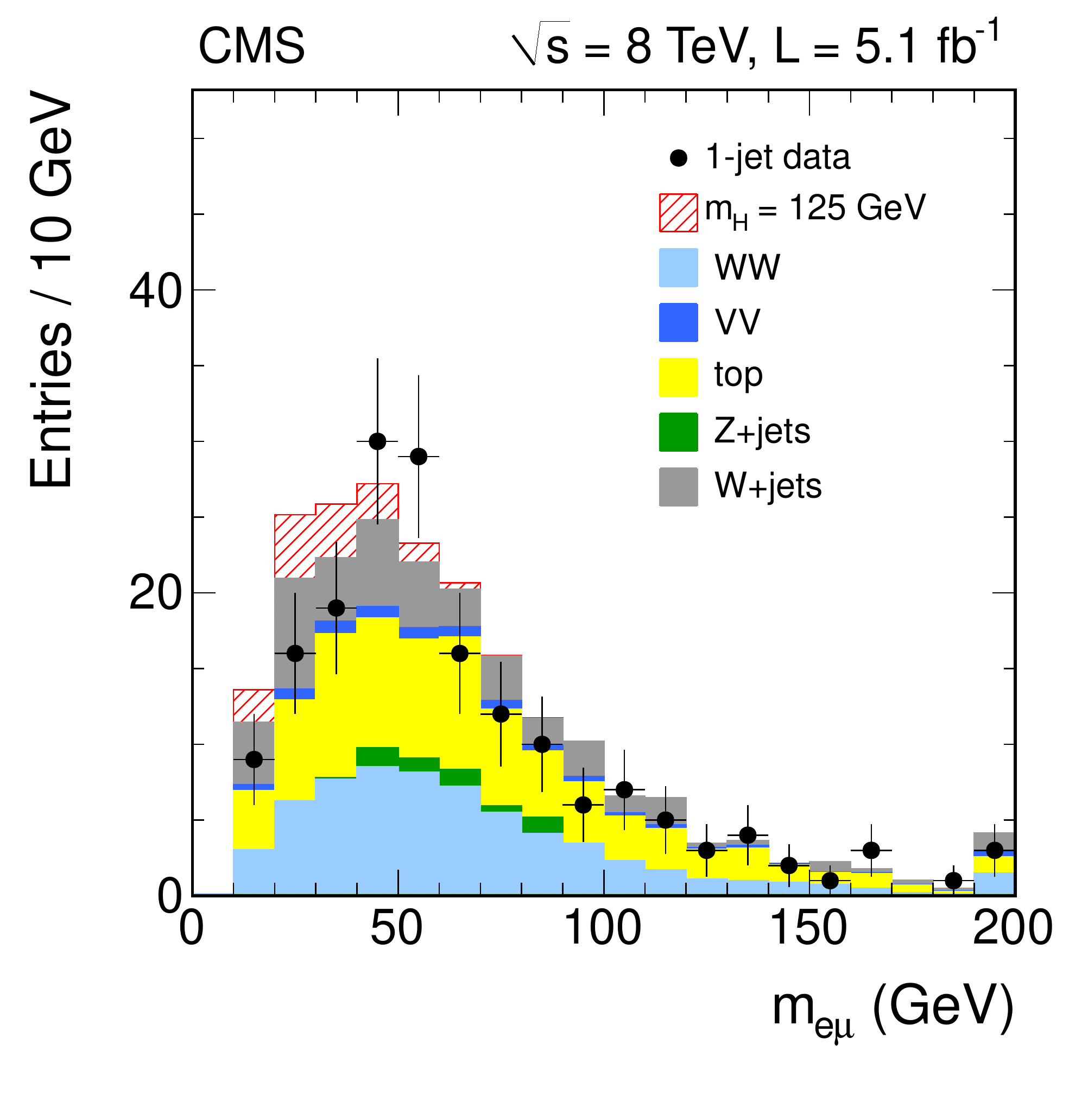}
	\caption{Dilepton invariant mass distribution from
	the 0-jet (\cmsLeft) and 1-jet (\cmsRight) $\Pe\mu$ events from the 8\TeV data (points with error bars),
        and the prediction for the various backgrounds (solid histograms), and for a SM Higgs boson
        with $\mH=125\GeV$ (hatched histogram) at 8\TeV.
       The cut-based $\PH \to \PW\PW$ selection, except for the requirement on the dilepton mass itself,
       is applied.}  \label{fig:hwwsel_nj_mh125_massem}
\end{center}
\end{figure}

To enhance the sensitivity for a Higgs boson signal, a cut-based approach is chosen
for the final event selection. Because the kinematics of signal events change
as a function of the Higgs boson mass, separate optimizations are performed for different
$\mH$ hypotheses.
The extra requirements, designed to optimize the
sensitivity for a SM Higgs boson, are placed
on $\ptlmax$, $\ptlmin$, $\mll$, $\delphill$ and
the transverse mass $m_\mathrm{T}$,
defined as $\sqrt{2 \pt^{\ell\ell} \MET (1-\cos\delphimetll)}$, where $\delphimetll$
is the difference in azimuthal angle between the $\MET$ direction, and the transverse momentum of the
dilepton system.
The requirements, which are the same for both the 0- and 1-jet categories,
are summarized in Table~\ref{tab:cuts_analysis}.
The $\mll$ distribution in the 0-jet (left) and 1-jet (right) categories for the $\Pe\mu$ candidate events
are shown in Fig.~\ref{fig:hwwsel_nj_mh125_massem}, along with the predictions for the background and
a SM Higgs boson with $\mH=125\GeV$.

\begin{table*}[h!t]
  \begin{center}
  \topcaption{Final event selection requirements for the cut-based analysis of the 0- and
  1-jet event samples. Values for other Higgs boson mass hypotheses follow a smooth behavior with respect to the
  reported values.}
 {\small
      \setlength{\extrarowheight}{1pt}
  \begin{tabular} {l|c|c|c|c|c}
  \hline
$\mH$ ($\GeVns{}$)      & $\ptlmax$ ($\GeVns{}$) & $\ptlmin$ ($\GeVns{}$) & $\mll$ ($\GeVns{}$)     & $\delphill$ (\de) & $m_\mathrm{T}$ ($\GeVns{}$) \\  \hline \hline
    125 & $>$23  & $>$10  & $<$43  & $<$100 & 80--123  \\
    130 & $>$25  & $>$10  & $<$45  & $<$90  & 80--125  \\
  \hline
  \end{tabular}
  }
   \label{tab:cuts_analysis}
  \end{center}
\end{table*}

The 2-jet category is mainly sensitive to VBF production
\cite{Ciccolini:2007jr, Ciccolini:2007ec, Arnold:2008rz,Cahn:1987}, whose
cross section is roughly ten times smaller than that from
gluon-gluon fusion. The VBF channel offers a different production mechanism
to test the consistency of a signal with the SM Higgs boson hypothesis.
The VBF signal can be extracted
using simple selection criteria, especially in the relatively low-background environment
of the fully leptonic $\PW\PW$ decay mode, providing additional search sensitivity.
The $\PH \to \PW\PW$ events from VBF production are characterized by two
energetic forward-backward jets and very little hadronic activity in
the rest of the event.
Events passing the $\PW\PW$ criteria are further required to satisfy $\pt>30\GeV$
for the two highest-$\Et$ jets, with no jets above this threshold
present in the pseudorapidity region between these two jets. Both leptons
are required to be within the pseudorapidity region between the two jets.
To reject the main background from top-quark decays,  the two jets
must have a pseudorapidity difference larger than 3.5 and a dijet
invariant mass greater than 450\GeV.
In addition, $m_\mathrm{T}$ is required to be between $30\GeV$ and
the Higgs boson mass hypothesis. Finally, a $\mH$-dependent upper limit on
the dilepton mass is applied. 

\subsection{Background predictions}
\label{sec:backgrounds}

A combination of techniques is used to determine the contributions from the background
processes that remain after the final selection.
The largest background contributions are estimated directly from data,
avoiding uncertainties related to
the simulation of these sources. The remaining contributions
estimated from simulation are small.

The $\PW$+jets and QCD multijet backgrounds arise from semileptonic
decays of heavy quarks, ha\-drons misidentified as leptons, and electrons
from photon conversions. Estimations of these contributions are derived
directly from data, using a control sample of events in which one lepton
passes the standard criteria and the other does not, but instead satisfies a
relaxed set of requirements (``loose" selection), resulting in a
``tight-loose" sample.
Then the efficiency, $\epsilon_\text{loose}$, for a lepton candidate that satisfies the loose
selection to also pass the tight selection is determined, using data from an
independent multijet event sample dominated by nonprompt leptons and
parametrized as a function of the $\pt$ and $\eta$ of the lepton. Finally, the background
contamination is estimated using the events in the ``tight-loose"
sample, weighted by \mbox{$\epsilon_\text{loose}$/$(1-\epsilon_\text{loose})$}. The
systematic uncertainty in the  determination of $\epsilon_\text{loose}$
dominates the overall uncertainty of this method, which is estimated
to be about 36\%. The uncertainty is obtained by varying the requirements to
obtain $\epsilon_\text{loose}$, and from a closure test, where the tight-loose rate derived
from QCD simulated events is applied to a $\Wjets$ simulated sample to predict the rate of
events with one real and one misidentified lepton.

The normalization of the top-quark background is estimated from data
by counting the number ($N_\text{tagged}$) of top-quark-tagged events and applying a
corresponding top-quark-tagging efficiency $\epsilon_\text{top}$. The top-quark-tagging efficiency
($\epsilon_\text{top}$) is measured with a control sample dominated by
$\ttbar$ and $\Wt$ events, which is selected by requiring a \cPqb-tagged jet in the event.
The number of top-quark background events in the signal region is then given
by: $N_\text{tagged} \times (1-\epsilon_\text{top})/\epsilon_\text{top}$. Background
sources from non-top events are subtracted by estimating the misidentification
probability from data control samples. The main uncertainty comes from the statistical
uncertainty in the \cPqb-tagged control sample and from the systematic uncertainties related to
the measurement of $\epsilon_\text{top}$. The uncertainty is about
20\% in the 0-jet category and about 5\% in the 1-jet category.

For the low-mass $\PH \to \PW\PW$ signal region, $\mH \leq 200\GeV$,
the nonresonant $\PW\PW$ background prediction is estimated from data. This contribution
is measured using events with a dilepton mass larger than 100\GeV,
where the Higgs boson signal contamination is negligible, and the MC simulation
is then used to extrapolate into the signal region. The total uncertainty is
about 10\%, where the statistical uncertainty of the data control region is the
largest component.
For larger Higgs boson masses there is a significant overlap between the
nonresonant $\PW\PW$ and Higgs boson signal, and the simulation is
used for the estimation of the background.

The $\dyll$ contribution to the $\Elp\Elm$
and $\Mp\Mm$ final states is estimated by extrapolating
the observed number of events with a dilepton mass
within $\pm7.5\GeV$ of the $\Z$ mass, with the residual
background in that region subtracted using $\Elpm\Mmp$ events.
The extrapolation to the signal region is then performed using the simulation.
The results are cross-checked with data, using the same algorithm and
subtracting the background in the $\Z$-mass region, estimated from the number of $\Elpm\Mmp$ events.
The largest uncertainty in the estimate is the statistical
uncertainty in the control sample, which is about 20\% to 50\%.
The $\dytt$ contamination is estimated using $\dyee$ and $\Mp\Mm$
events selected in data, where the leptons are replaced with simulated
$\Pgt$ decays, thus providing a better description of the process $\dytt$. The \TAUOLA~\cite{TAUOLA}
program is used in the simulation of the $\Pgt$ decays to account for $\tau$-polarization effects.

Finally, to estimate the $\wgamma^{*}$ background contribution
from asymmetric virtual photon decays to dileptons~\cite{wgammastart}, where one lepton escapes
detection, the \MADGRAPH generator~\cite{Alwall:2011uj} with dedicated cuts is
used. In particular, all the leptons are required to have a $\pt$ larger than 5 $\GeV$ and
the mass of each lepton is considered in the generation of the samples. To normalize the
simulated events, a control sample of high-purity $\wgamma^{*}$ events from data with three
reconstructed leptons is compared to the simulation prediction. A
normalization factor of $1.6\pm0.5$ with respect to the theoretical leading-order
$\wgamma^{*}$ cross section is found.

Other minor backgrounds from $\WZ$, $\ZZ$ (when the two selected leptons come from
different boson decays), and $\wgamma$ are estimated from simulation.
The $\wgamma$ background estimate is cross-checked in data using events passing
all the selection requirements, except the two leptons must have the
same charge; this sample is dominated by $\PW$+jets and $\wgamma$
events. The agreement between data and the background prediction in this test is at the 20\% level.

The number of observed events and the expected number of events from
all background processes after the $\PW\PW$ selection
are summarized in Table \ref{tab:wwselection_all}. The number of events
observed in data and the signal and
background predictions after the final selection are listed in Table~\ref{tab:hwwselection} for two
Higgs boson mass hypotheses.

\begin{table*}[h!t]
  \begin{center}
  \topcaption{Observed number of events and background estimates for
    the 8\TeV data sample, after applying the $\PW\PW$ selection requirements.
      The  uncertainties are  statistical only.}
   \label{tab:wwselection_all}
\footnotesize {
  \begin{tabular}{l|c|c|c|c|c|c|c|c}
\hline
& $\PW\PW$       & \ttbar+t$\PW$   & $\PW$+jets           & $\PW\cPZ+\cPZ\cPZ$            & $\cPZ/\gamma^*$       & $\wgamma^{(*)}$  &   tot. bkg. &  data            \\ \hline \hline
0-jet  & 1046.1 $\pm$ 7.2 & 164.2 $\pm$ 5.4    & 158.2 $\pm$ 7.1  & 32.6 $\pm$ 0.6   & 73 $\pm$ 17   & 27.1 $\pm$ 3.9 & 1501 $\pm$ 21  & 1594       \\
1-jet  & 381.0 $\pm$ 4.0  & 527.3 $\pm$ 8.4    & 122.6 $\pm$ 6.7  & 30.3 $\pm$ 0.6   & 77 $\pm$ 24   & 23.7 $\pm$ 5.2 & 1162 $\pm$ 27  & 1186         \\
2-jet  & 177.0 $\pm$ 2.8  & 886.5 $\pm$ 11.1   &  94.9 $\pm$ 6.4  & 20.8 $\pm$ 0.5   & 227 $\pm$ 20  & 5.6 $\pm$ 2.1  & 1412 $\pm$ 24  & 1295  \\ \hline
  \end{tabular}}
  \end{center}
\end{table*}

\begin{table*}[h!t]
  \begin{center}
  \topcaption{The signal predictions, background estimates, and  numbers
    of events in data for two different Higgs boson mass hypotheses
    with  the 8\TeV data set, after applying the final $\PH \to
    \PW\PW$ cut-based requirements, which depend on the Higgs boson
    mass hypothesis. The different jet categories and dilepton final
    states are shown separately.
  The combined statistical, experimental, and theoretical systematic uncertainties are given.
}
   \label{tab:hwwselection}
 {
\footnotesize
\setlength{\extrarowheight}{1pt}
\begin{tabular} {l|c|c|c|c|c|c|c|c}
  \hline
 $\mH$ & $\PH \to \PW \PW$  &  $\PW \PW$  & $\PW\cPZ+\cPZ\cPZ+\cPZ/\gamma^*$ & \ttbar+t$\PW$ & $\PW$+jets & $\wgamma^{(*)}$ & all bkg. & data\\ \hline \hline
\multicolumn{9}{c}{0-jet category $\Pe\mu$ final state } \\
\hline
 $125$ & $23.9\pm5.2$ & $87.6\pm9.5$ & $2.2\pm0.2$ & $9.3\pm2.7$ & $19.1\pm7.2$ & $6.0\pm2.3$ & $124.2\pm12.4$ & $158$ \\
 $130$ & $35.3\pm7.6$ & $96.8\pm10.5$ & $2.5\pm0.3$ & $10.1\pm2.8$ & $20.7\pm7.8$ & $6.3\pm2.4$ & $136.3\pm13.6$ & $169$ \\
\hline
\multicolumn{9}{c}{0-jet category $\Pe\Pe$/$\mu\mu$ final state} \\
\hline
 $125$ & $14.9\pm3.3$ & $60.4\pm6.7$ & $37.7\pm12.5$ & $1.9\pm0.5$ & $10.8\pm4.3$ & $4.6\pm2.5$ & $115.5\pm15.0$ & $123$ \\
 $130$ & $23.5\pm5.1$ & $67.4\pm7.5$ & $41.3\pm15.9$ & $2.3\pm0.6$ & $11.0\pm4.3$ & $4.8\pm2.5$ & $126.8\pm18.3$ & $134$ \\
\hline
\multicolumn{9}{c}{1-jet category  $\Pe\mu$ final state} \\
\hline
 $125$ & $10.3\pm3.0$ & $19.5\pm3.7$ & $2.4\pm0.3$ & $22.3\pm2.0$ & $11.7\pm4.6$ & $5.9\pm3.2$ & $61.7\pm7.0$ & $54$ \\
 $130$ & $15.7\pm4.7$ & $22.0\pm4.1$ & $2.6\pm0.3$ & $25.1\pm2.2$ & $12.8\pm5.1$ & $6.0\pm3.2$ & $68.5\pm7.6$ & $64$ \\
\hline
\multicolumn{9}{c}{1-jet category $\Pe\Pe$/$\mu\mu$ final state} \\
\hline
 $125$ & $4.4\pm1.3$ & $9.7\pm1.9$ & $8.7\pm4.9$ & $9.5\pm1.1$ & $3.9\pm1.7$ & $1.3\pm1.2$ & $33.1\pm5.7$ & $43$ \\
 $130$ & $7.1\pm2.2$ & $11.2\pm2.2$ & $9.1\pm5.4$ & $10.7\pm1.2$ & $3.7\pm1.7$ & $1.3\pm1.2$ & $36.0\pm6.3$ & $53$ \\
\hline
\multicolumn{9}{c}{2-jet category  $\Pe\mu$ final state} \\
\hline
 $125$ & $1.5\pm0.2$ & $0.4\pm0.1$ & $0.1\pm0.0$ & $3.4\pm1.9$ & $0.3\pm0.3$ & $0.0\pm0.0$ & $4.1\pm1.9$ & $6$ \\
 $130$ & $2.5\pm0.4$ & $0.5\pm0.2$ & $0.1\pm0.0$ & $3.0\pm1.8$ & $0.3\pm0.3$ & $0.0\pm0.0$ & $3.9\pm1.9$ & $6$ \\
\hline
\multicolumn{9}{c}{2-jet category $\Pe\Pe$/$\mu\mu$ final state} \\
\hline
 $125$ & $0.8\pm0.1$ & $0.3\pm0.1$ & $3.1\pm1.8$ & $2.0\pm1.2$ & $0.0\pm0.0$ & $0.0\pm0.0$ & $5.4\pm2.2$ & $7$ \\
 $130$ & $1.3\pm0.2$ & $0.4\pm0.2$ & $3.8\pm2.2$ & $2.0\pm1.2$ & $0.0\pm0.0$ & $0.0\pm0.0$ & $6.2\pm2.5$ & $7$ \\
\hline
  \end{tabular}
  }
  \end{center}
\end{table*}

\subsection{Efficiencies and systematic uncertainties}
\label{sec:systematics}

The signal efficiency is estimated using simulations.
All Higgs boson production mechanisms are considered:
gluon-gluon fusion, associated production  with
a $\PW$ or $\Z$ boson (VH), and VBF processes.

Residual discrepancies in the lepton reconstruction and identification
efficiencies between data and simulation are corrected for by
data-to-simulation scale factors measured using $\dyll$ events in the
$\Z$-peak region~\cite{CMS:2011aa}, recorded with dedicated unbiased triggers.
These factors depend on the lepton $\pt$ and $|\eta|$, and
are typically in the range 0.9--1.0. The uncertainties on the lepton and
trigger efficiencies are about 2\% per lepton leg.

Experimental effects, theoretical predictions, and the choice of MC event
generators are considered as sources of systematic uncertainty, and their impact on the signal
efficiency is assessed.
The experimental uncertainties in lepton efficiency, momentum scale and resolution, $\MET$
modelling, and jet energy scale are applied to the reconstructed objects in simulated events by smearing
and scaling the relevant observables, and propagating the effects to the kinematic variables used
in the analysis.
The 36\% normalization uncertainty in the $\Wjets$ background is included by varying
the efficiency for misidentified leptons to pass the tight lepton
selection and by comparing the results of a closure test using simulated samples.

The relative systematic uncertainty on the signal efficiency from pileup is evaluated to be $1\%$.
This corresponds to shifting the mean of the
expected distribution of the number of pp collision per beam-crossing that is used to reweight the simulation
up and down by one pp interaction.
The systematic uncertainty on the integrated luminosity measurement is $4.4\%$~\cite{CMS:2012jza}.

The systematic uncertainties from theoretical input  are separated
into two components, which are assumed to be
independent. The first component is the uncertainty in the fraction of
events classified into the different jet categories and the effect of migration between categories.
The second component is the uncertainty in
the lepton acceptance and the selection efficiency of the other
requirements. The effect of variations in the PDF, the
value of $\alpha_{s}$, and the higher-order corrections
are considered for both components, using the PDF4LHC
prescription~\cite{Botje:2011sn,Alekhin:2011sk,Lai:2010vv,Martin:2009iq,Ball:2011mu} and the recommendations
from~\cite{LHCHiggsCrossSectionWorkingGroup:2011ti}.
For the jet categorization, the effects of higher-order logarithmic terms via
the uncertainty in the parton shower model and the underlying event
are also considered by comparing different generators. These uncertainties range
between 10\% and 30\%, depending on the jet category.
The uncertainties related to the diboson
cross sections are calculated using the {\sc MCFM} program~\cite{MCFM}.

The systematic uncertainty in the overall signal efficiency is estimated to be about 20\%
and is dominated by the theoretical uncertainty in the missing
higher-order corrections and PDF uncertainties. The total uncertainty in the background estimations
in the $\PH\to\PW\PW$ signal region is about 15\%, dominated by the
statistical uncertainty in the observed number of events in the background-control regions.

The interpretation of the results in terms of upper limits on the
Higgs boson production cross section will be given in Section 10.
\section{\texorpdfstring{$\PH\to\Pgt\Pgt$}{H to tau tau}\label{sec:htt}}

The $\PH\to\tau\tau$ decay mode is sensitive to a SM Higgs boson with a mass below about $145\GeV$,
for which the branching fraction is large.
The search uses final states where the two $\tau$ leptons are identified either by their leptonic decay to an electron or muon,
or by their hadronic decay designated as $\Pgt_h$.
Four independent channels are studied: $\Pe\Pgt_h$, $\Pgm\Pgt_h$, $\Pe\Pgm$, and  $\Pgm\Pgm$.
In each channel, the signal is separated from the background, and in particular from the irreducible $\cPZ\to\tau\tau$ process,
using the $\tau$-lepton pair invariant mass $m_{\tau\tau}$, reconstructed from the four-momentum of the visible decay products of the two $\tau$ leptons and the \MET vector,  as explained in Section~\ref{sec:htt_mtautau}.
Events are classified by the number of additional jets in the final state, in order to enhance the contribution of different
Higgs boson production mechanisms.
The 0- and 1-jet categories select primarily signal events with a Higgs boson produced by gluon-gluon fusion,
or in association with a W or Z vector boson that decays hadronically.
These two categories are further classified according to the $\PT$ of the $\tau$-lepton decay products,
because high-$\PT$ events benefit from a higher signal-to-background ratio.
Events in the VBF category are required to have two jets separated by a large rapidity,
which preferentially selects signal events from the vector-boson fusion production mechanism and strongly enhances the signal purity.

\subsection{Trigger and inclusive event selection}

The high-level trigger requires a combination of electron, muon, and $\Pgt_h$ trigger objects~\cite{CMS-PAS-EGM-10-004,CMS-PAS-MUO-10-002,CMS-EWK-TAU}.
The electron and muon HLT reconstruction is seeded by electron and muon level-1 trigger objects, respectively,
while the $\Pgt_h$ trigger object reconstruction is entirely done at HLT stage.
A specific version of the particle-flow algorithm is used in the HLT to reconstruct these objects and quantify their isolation, as done in the offline reconstruction.
The identification and isolation criteria and the transverse momentum
thresholds for these objects were progressively
tightened as the LHC instantaneous luminosity increased over the data taking period.
In the $\Pe\Pgt_h$ and $\Pgm\Pgt_h$ channels, the trigger requires the presence of a lepton and a $\Pgt_h$,
both loosely isolated with respect to the offline isolation criteria described below.
In the $\Pe\mu$ and $\mu\mu$ channels, the lepton trigger objects are not required to be isolated.
For the $\Pe\Pgt_h$, $\Pgm\Pgt_h$, and $\mu\mu$ channels, the muon and electron trigger efficiencies
are measured with respect to the offline selection  in the data and the simulation using $\cPZ\to \ell\ell (\ell=\Pe,\Pgm)$ events passing a single-lepton trigger.
For the $\Pe\mu$ channel, they are determined using $\cPZ \to \tau\tau \to \Pe \mu $ events passing a single-lepton trigger.
The $\Pgt_h$ triggering efficiency is obtained using $\cPZ \to \tau\tau \to \mu \Pgt_h$ events passing a single-muon trigger.
In the analysis, simulated events are weighted by the ratio between the efficiency measured in the data and the simulation,
which are parametrized as a function of the lepton or $\Pgt_h$ transverse momentum and pseudorapidity.

To be considered in the offline event selection, electrons and muons must fulfill tight isolation criteria.
The electron and muon isolation parameter $R_\text{Iso}^{\ell}$ is calculated as in Eq.~(\ref{eq:reconstruction_isolation}) using a cone size $\Delta R=0.4$, but with the following differences.
The sum $\sum_\text{charged}  \PT$ is performed considering all charged particles associated with the primary vertex, including other electrons and muons.
The contribution of neutral  pileup particles is estimated as $0.5 \sum_{\rm charged, PU} \pt$,
where the sum is computed for all charged hadrons from pileup interactions in the isolation cone,
and where the factor 0.5 corresponds approximately to the ratio of neutral-to-charged
hadron energy in the hadronization process, as estimated from simulation.
Electrons and muons are required to have $R_\text{Iso}^{\ell}<0.1$.
This criterion is relaxed to 0.15 in the $\Pe\Pgm$ channel for leptons in the barrel,
and in the $\Pgm\Pgm$ channel for muons with $\PT<20\GeV$.
The $\tau$-isolation discriminator $R_\text{Iso}^{\tau}$ defined in Section~\ref{sec:reconstruction} is used to select loosely isolated $\tau_h$
so that the overall $\Pgt_h$ identification efficiency is 60--65\%,
for a jet misidentification probability of 2--3\%.
Finally, electrons and muons misidentified as $\Pgt_h$ are suppressed
using dedicated criteria based on the consistency between the tracker, calorimeter, and muon-chamber measurements.

In the $\Pe\Pgt_h$ and $\Pgm\Pgt_h$ channels, we select events containing either an electron
with  $\pt >$~20\GeV or a muon with  $\pt >17\GeV$, and $|\eta| < 2.1$,
accompanied by an oppositely charged $\Pgt_h$ with $\pt > 20\GeV$ and $|\eta| < 2.3$.
In the 8\TeV data set analysis, the electron and muon \pt thresholds are
increased to 24 and 20\GeV, respectively, to account for the higher trigger thresholds.
In these channels,
events with more than one loosely identified electron or muon with $\pt >15\GeV$ are rejected to reduce the Drell--Yan background.
In the $\Pe\Pgm$ channel, we demand an electron within $|\eta|<2.3$ and an oppositely charged muon within $|\eta|<2.1$.
The higher-\pt lepton must have $\pt >$ 20\GeV and the other lepton $\pt >10\GeV$.
In the $\Pgm\Pgm$ channel, the higher-\pt muon is required to have $\pt>20\GeV$ and the other muon $\pt>10\GeV$.
Both muons must be within $|\eta|<2.1$.

Neutrinos produced in the $\tau$-lepton decay are nearly collinear with the visible decay products
because the $\tau$-lepton energy is much larger than its mass after event selection.
Conversely, in
$\PW+$jets events where a jet is misidentified as $\Pgt_h$, one of the main backgrounds in the $\ell\Pgt_h$ channels,
the high mass of the
$\PW$ results in a neutrino direction approximately opposite to the lepton in the transverse plane.
In the $\Pe\Pgt_h$ and $\Pgm\Pgt_h$ channels, we therefore require the transverse mass
\begin{equation}
\MT = \sqrt{2 \pt \MET (1-\cos(\Delta\phi))}
\end{equation}
to be less than 40\GeV, where \pt is the lepton transverse momentum
and $\Delta\phi$ is the azimuthal angle difference between the lepton momentum and the \MET vector.
In the $\Pe\Pgm$ channel, instead of an $\MT$ requirement, we demand
$D_{\zeta} \equiv \not\!{p_\zeta}- 0.85 \cdot p_\zeta^{\text{vis}} > -25$\GeV, where
\begin{align}
\not\!{p_{\zeta}} &=\vec p_\mathrm{T,1} \cdot \hat \zeta + \vec p_\mathrm{T,2} \cdot \hat \zeta+ \VEtmiss \cdot \hat \zeta, \\
p_{\zeta}^{\text{vis}} &= \vec{p}_\mathrm{T,1} \cdot \hat \zeta + \vec{p}_\mathrm{T,2} \cdot \hat \zeta.
\end{align}

\begin{figure}[htbp]
\begin{center}
\includegraphics[width=0.35\textwidth]{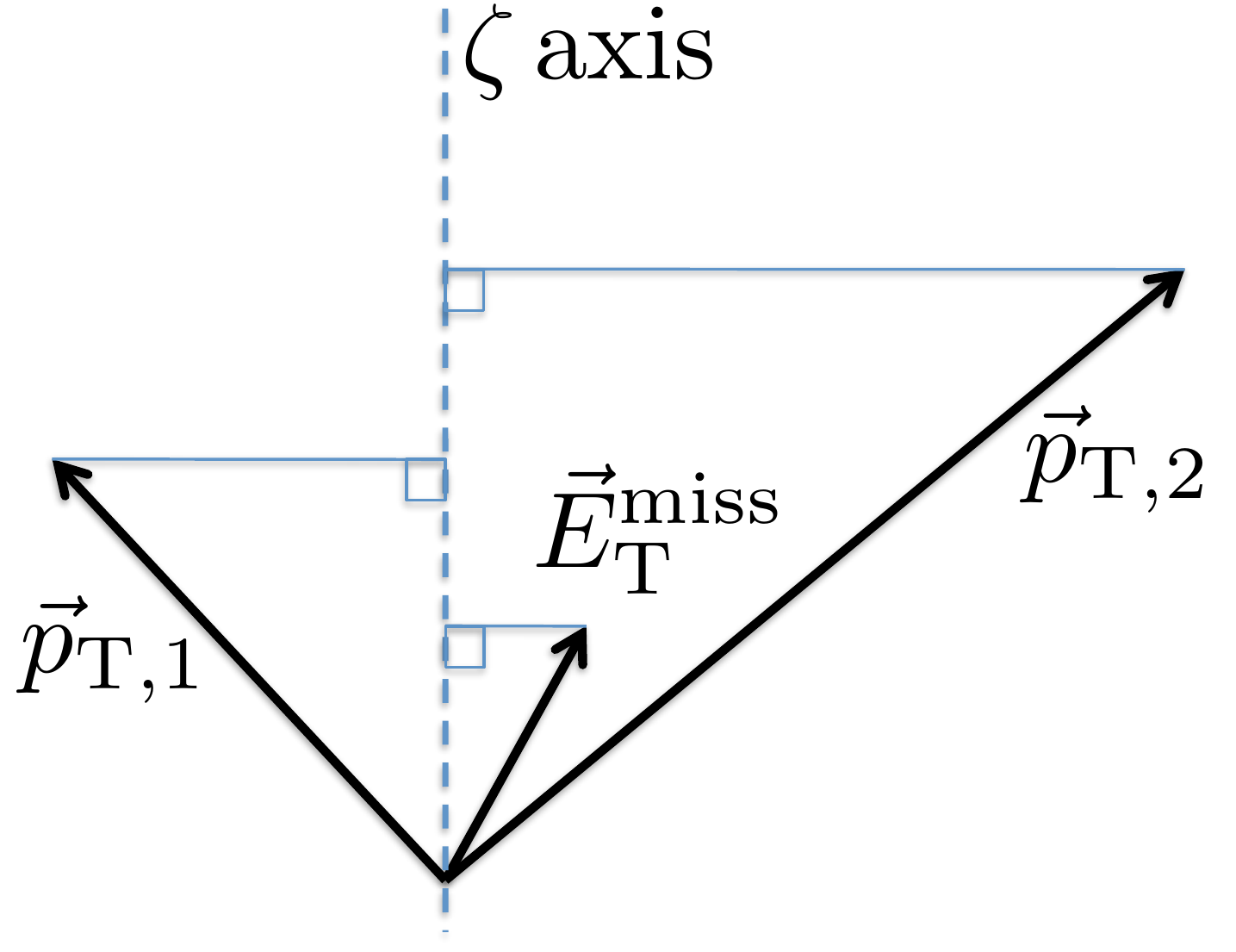}
\end{center}
\caption{The $\zeta$ axis and the projections onto this axis of $\VEtmiss$ and transverse momenta $\vec{p}_\mathrm{T,1}$ and $\vec{p}_\mathrm{T,2}$ of the two leptons.}
\label{fig:htt_zeta_drawing}
\end{figure}

Here, as illustrated in Fig.~\ref{fig:htt_zeta_drawing}, $\hat \zeta$ is a unit vector along the $\zeta$ axis, defined as the bisector of the lepton directions in the transverse plane~\cite{CRISTOBAL}, $\vec p_\mathrm{T,i}$ are the lepton transverse momenta, and $\VEtmiss$ is the missing transverse energy vector.

The $D_{\zeta}$ distribution is shown in Fig.~\ref{fig:htt_control}(b).
Requiring a large $D_{\zeta}$ rejects $\PW+$jets and $t\bar t$ events, for which the \MET vector is typically oriented in the opposite direction of the two-lepton system, resulting in a small $D_{\zeta}$.
Conversely, in $H\to \tau \tau$ or $Z \to \tau \tau$ events, the
neutrinos are emitted along the directions of the two $\tau$ leptons, resulting in a large $D_{\zeta}$. The 0.85 factor is introduced to bring the mean of the $D_{\zeta}$ distribution to 0 for $Z \to \tau \tau$.

In the $\Pgm\Pgm$ channel, the sample of dimuon events is largely dominated by the $Z\rightarrow\Pgm\Pgm$ background,
which is suppressed using a BDT discriminant combining a set of variables related to the kinematics of the dimuon system,
and the distance of closest approach between the two muons.

\subsection{The \texorpdfstring{$\tau\tau$}{tau tau} invariant-mass reconstruction}
\label{sec:htt_mtautau}

The invariant mass $m_\text{vis}$ of the visible decay products of the
two $\tau$ leptons can be used as an estimator of the mass of
a possible parent boson, in order
to separate the $\PH \to \tau \tau$ signal from the irreducible $\cPZ \to \tau \tau$ background.
However, the neutrinos from $\tau$-lepton decays can have substantial energy
limiting the separation power of this estimator.
An alternative approach is to reconstruct the neutrino energy using a collinear approximation~\cite{massRecoCollinearApprox},
which has the disadvantage of providing an unphysical solution for about 20\% of the events, in particular when the \MET and the parent
boson \PT are small.
The SVFit algorithm described below reconstructs the $\tau\tau$ invariant-mass $m_{\tau\tau}$ with improved resolution and gives a physical solution for every event.

Six parameters are needed to specify $\tau$-lepton decays to hadrons:
the polar and azimuthal angles of the visible decay product system in the $\tau$-lepton rest frame, the three boost parameters from the $\tau$-lepton rest frame to the laboratory frame, and the invariant mass $m_\text{vis}$ of the visible decay products.
In the case of a leptonic $\tau$-lepton decay, two neutrinos are produced,
and the invariant mass of the two-neutrino system constitutes a seventh parameter.
The unknown parameters are constrained by four observables that are the components of the four-momentum
of the system formed by the visible $\tau$-lepton decay products, measured in the laboratory frame.
For each hadronic (leptonic) $\tau$-lepton decay, 2 (3) parameters are thus left unconstrained.
We choose these parameters to be:
\begin{itemize}
\item $x$, the fraction of the $\tau$-lepton energy in the laboratory frame carried by the visible decay products.
\item $\phi$, the azimuthal angle of the $\tau$-lepton direction in the laboratory frame.
\item $m_{\nu\nu}$, the invariant mass of the two-neutrino system. For hadronic $\tau$-lepton decay, $m_{\nu\nu} \equiv 0$.
\end{itemize}
The two components $E_{x}^\text{miss}$ and $E_{y}^\text{miss}$ of the missing transverse energy vector provide two further constraints,
albeit with an experimental resolution of 10--15\GeV on each \cite{PFMEtSignAlgo}.

The fact that the reconstruction of the $\tau$-lepton pair decay kinematics is underconstrained by the measured observables
is addressed by a maximum-likelihood fit method.
The mass $m_{\tau\tau}$ is reconstructed by combining the measured observables $E_{x}^\text{miss}$ and $E_{y}^\text{miss}$ with a likelihood  model
that includes terms for the $\tau$-lepton decay kinematics and the \MET resolution.
The model gives the probability density $f(\vec{z} \vert \vec{y}, \vec{a_1}, \vec{a_2})$
to observe the values $\vec{z} = (E_{x}^\text{miss}, E_{y}^\text{miss})$ in an event,
given that the unknown parameters specifying the kinematics of the two $\tau$-lepton decays have values
$\vec{a_1} = (x_{1}, \phi_{1}, m_{\nu\nu,1})$ and $\vec{a_2} = (x_{2}, \phi_{2}, m_{\nu\nu,2})$,
and that the four-momenta of the visible decay products have the measured values $\vec{y} = (p^\text{vis}_{1}, p^\text{vis}_{2})$.
The likelihood model is used to compute the probability
\begin{equation}
P(m_{\tau\tau}^{i}) = \int \delta \left( m_{\tau\tau}^{i} - m_{\tau\tau}(\vec{y}, \vec{a_1}, \vec{a_2}) \right) f(\vec{z} \vert \vec{y}, \vec{a_1}, \vec{a_2})\, \rd\vec{a_1}\,\rd\vec{a_2},
\label{eq:mtautau}
\end{equation}
as a function of mass hypothesis $m_{\tau\tau}^{i}$.
The best estimate $\hat{m}_{\tau\tau}$ for $m_{\tau\tau}$ is taken to be the value of $m_{\tau\tau}^{i}$ that maximizes $P(m_{\tau\tau}^{i})$.

The probability density $f(\vec{z} \vert \vec{y}, \vec{a_1}, \vec{a_2})$ is the product of three likelihood functions.
The first two model the decay parameters $\vec{a_1}$ and $\vec{a_2}$
of the two $\tau$ leptons, and the last
one quantifies the consistency of a $\tau$-lepton decay hypothesis with the measured $\MET$.
The likelihood functions modelling the $\tau$-lepton decay kinematics are different for leptonic and hadronic $\tau$-lepton decays.
Matrix elements from Ref.~\cite{TauPol} are used to model the differential distributions in the leptonic
decays,
\begin{equation}
L_{\tau,l} = \frac{\rd\Gamma}{\rd x\, \rd m_{\nu\nu}\,\rd\phi} \propto \frac{m_{\nu\nu}}{4m_{\tau}^2} \big[(m_{\tau}^2 +2m_{\nu\nu}^2 )(m_{\tau}^2 - m_{\nu\nu}^2)\big],
\label{eq:likelihoodLepTauDecay}
\end{equation}
within the physically allowed region $0 \leq x \leq 1 \mbox{ and } 0 \leq m_{\nu\nu} \leq m_{\tau}\sqrt{1-x}$.
For hadronic $\tau$-lepton decays, a model based on two--body phase-space~\cite{PDG} is used,
treating all the $\tau$-lepton visible decay products as a single system,
\begin{equation}
L_{\tau,h} = \frac{\rd\Gamma}{\rd x\,\rd\phi} \propto \frac{1}{1- \frac{m^2_\text{vis}}{m^2_{\tau}}},
\label{eq:likelihoodHadTauDecay}
\end{equation}
within the physically allowed region $\frac{m_\text{vis}^{2}}{m_{\tau}^{2}} \leq x \leq 1$.
We have verified that the two-body phase space model is adequate for representing hadronic $\tau$-lepton decays
by comparing distributions generated by a parameterized MC simulation based on the two-body phase-space model
with the detailed simulation implemented in \TAUOLA.
The likelihood functions for leptonic (hadronic) $\tau$-lepton decays do not depend on the parameters $x$ and $\phi$ ($x$, $\phi$, and $m_{\nu\nu}$). The dependence on $x$ enters via the integration boundaries, and the dependence on $\phi$ comes from the \MET likelihood function.

The \MET likelihood function $L_{MET}$ quantifies the compatibility of a $\tau$-lepton decay hypothesis
with the reconstructed missing transverse momentum in an event,
assuming the neutrinos from the $\tau$-lepton decays are the only source of \MET, and is defined as
\begin{equation}
L_{\rm MET} (E_{x}^\text{miss}, E_{y}^\text{miss}) = \frac{1}{2 \pi \sqrt{\vert V \vert}}
\cdot \exp \left( -\frac{1}{2}
 \left( \begin{array}{c} E_{x}^\text{miss} - \sum p_{x}^{\nu} \\ E_{y}^\text{miss} - \sum p_{y}^{\nu} \end{array} \right)^{T}
\cdot V^{-1} \cdot
 \left( \begin{array}{c} E_{x}^\text{miss} - \sum p_{x}^{\nu} \\ E_{y}^\text{miss} - \sum p_{y}^{\nu} \end{array} \right)
\right).
\end{equation}
In this expression, the expected \MET resolution is represented by the covariance matrix $V$, estimated on an event-by-event basis using a
\MET-significance algorithm~\cite{PFMEtSignAlgo},
and $\vert V \vert$ is the determinant of this matrix.

\begin{figure}[htbp]
\begin{center}
\includegraphics[width=0.45\textwidth]{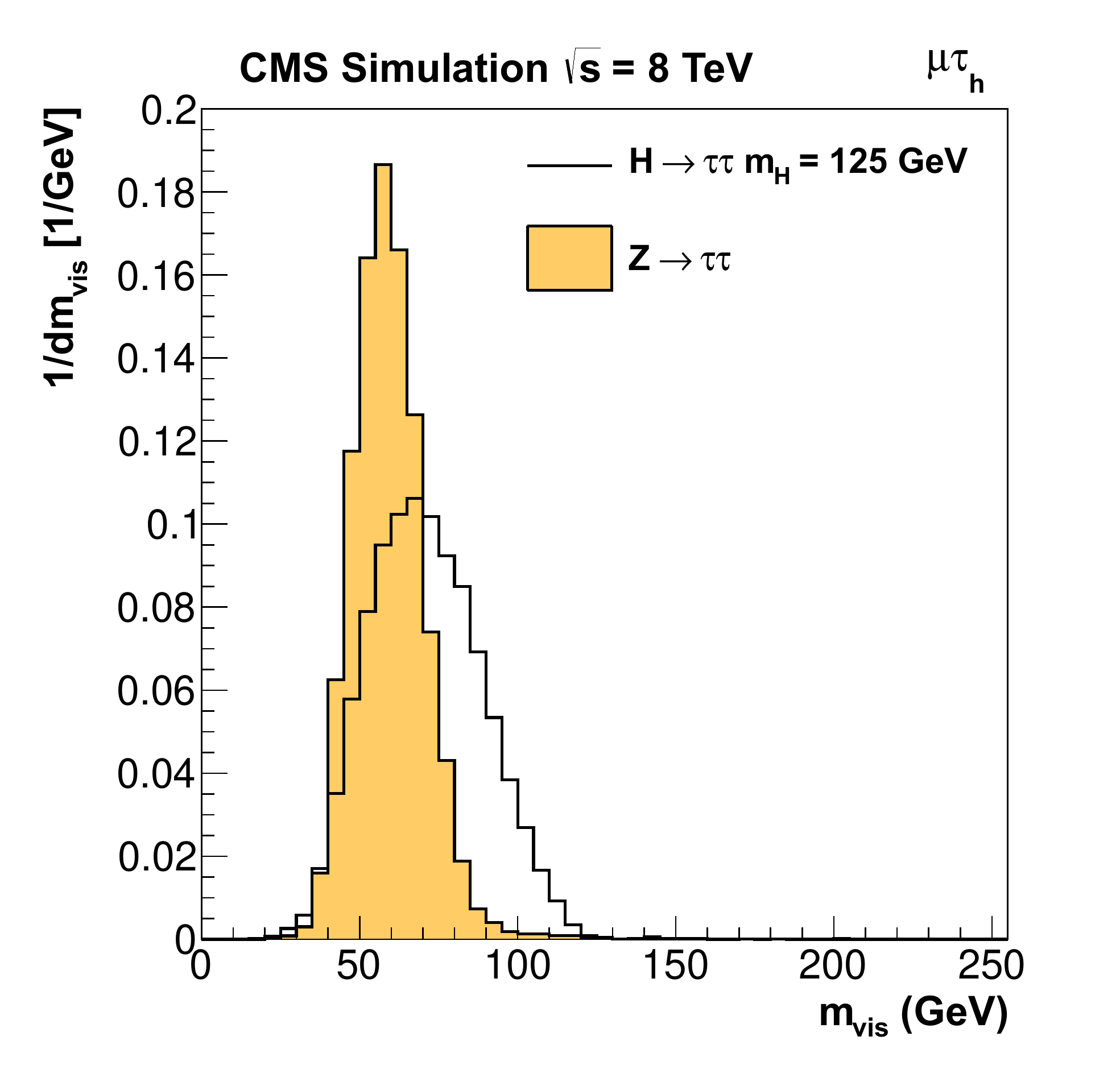}
\includegraphics[width=0.45\textwidth]{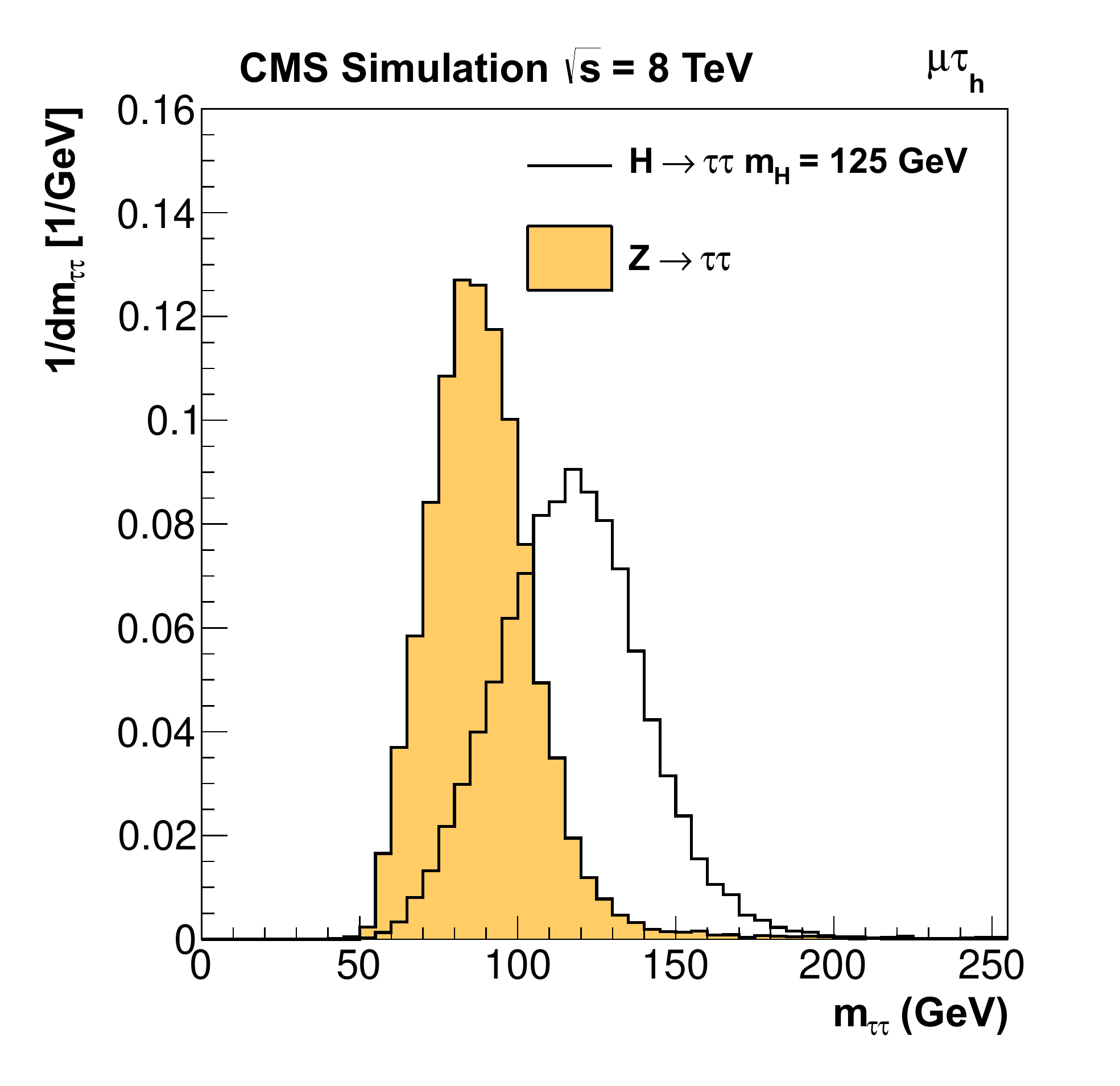} \\
\end{center}
\caption{Normalized distribution of the visible invariant mass $m_\text{vis}$ (left) and SVFit mass $m_{\tau\tau}$ (right) obtained from MC simulation in the
$\Pgm\Pgt_h$ channel for the $\cPZ\to \tau \tau$ background (solid histogram) and a SM Higgs boson signal of mass  $m_H=125\GeV$ (open histogram).}
\label{fig:htt_svfitperf}
\end{figure}

The $m_{\tau\tau}$  resolution achieved by the SVFit algorithm is estimated to be about
$20\%$ from simulation.
Figure~\ref{fig:htt_svfitperf} shows the normalized distributions of $m_\text{vis}$ and $m_{\tau\tau}$ in the $\Pgm\Pgt_h$ channel
from simulated  $\cPZ \to \tau \tau$ events and simulated SM Higgs boson events with $m_H=125\GeV$.
The SVFit mass reconstruction allows for a better separation between signal and background than $m_\text{vis}$.

\subsection{Event categories}

To further enhance the sensitivity of the search for the SM Higgs boson, the selected events are split into mutually exclusive categories based on the
jet multiplicity, and the transverse momentum of the visible $\tau$-lepton decay products.
The jet multiplicity categories are defined using jets within $|\eta|<5$.
In some cases, events are rejected if they contain a b-tagged jet, identified using the CSV algorithm described in Section~\ref{sec:reconstruction}.
From simulation, the efficiency for  b-jet tagging is $75$\%, with a misidentification rate of $1$\%. The event categories are:

\begin{itemize}
\item \textbf{VBF:}
In this category, two jets with $\pt>30$\GeV are required in the event.
A rapidity gap is demanded by requiring there be no third jet with $\pt>30$\GeV between these two jets.
A BDT discriminator is used to discriminate between VBF Higgs boson production and the background processes.
This discriminator takes as input the invariant mass of the two jets,
the differences in $\eta$ and $\phi$ between the directions of the two jets,
the \pt of the $\Pgt_h\Pgt_h$ system, the \pt of the $\Pgt_h\Pgt_h$-\MET system, the \pt of the dijet system,
and the difference in $\eta$ between the $\Pgt_h\Pgt_h$ system direction and the closest jet.
In the $\Pe\Pgm$ channel, the large \ttbar background is suppressed by rejecting events with a b-tagged jet with $\pt>20$\GeV.
\item \textbf{1-jet:} Events in this category are required to have $\ge$1 jet with $\pt>30$\GeV,
not fulfill the VBF criteria,
and not contain any b-tagged jet with $\pt>20$\GeV.
This category addresses the production of a high-\pt Higgs boson recoiling against a jet.
Events with high-\pt Higgs bosons typically have much larger \MET and thus benefit from a more precise measurement of $m_{\tau\tau}$,
owing to the improved \MET resolution.
In the $\Pe\Pgt_h$ channel, the large background from  $\cPZ \to \Pe\Pe$ + jets events with one electron misidentified as $\Pgt_h$ is reduced by requiring $\MET>30\GeV$.
\item \textbf{0-jet:} This category requires events to have no jet with $\pt>30$\GeV and no b-tagged jet with $\pt>20$\GeV.
In the $\Pe\Pgt_h$ channel, $\MET$ is required to be larger than 30\GeV as in the 1-jet category.
\end{itemize}

The 0- and 1-jet categories are each further divided into two subsets, using the \PT of the visible $\tau$-lepton decay products, either hadronic or leptonic. We label these subsets ``low-\pt" and ``high-\pt".
In the $\Pe\Pgt_h$ and $\Pgm\Pgt_h$ channels, the boundary between the two subsets is defined as $\PT(\Pgt_h)=40\GeV$.
In the $\Pe\Pgm$ and $\Pgm\Pgm$ channels, the threshold is at 35\GeV on the muon \PT and 30\GeV on the leading muon \PT, respectively.
Thus, five independent categories of events are used in the SM Higgs boson search: VBF, 1-jet/high-\PT, 1-jet/low-\PT, 0-jet/high-\PT, and 0-jet/low-\PT.

\subsection{Background estimation and systematic uncertainties}

For each channel and each category, Table~\ref{tab:htt_numevents} shows the overall number of events observed  in the 7 and 8\TeV data,
as well as the corresponding number of expected events from the various background contributions, in the full $m_{\tau\tau}$ range.
The expected number of events from a SM Higgs boson signal of mass $m_H=125\GeV$ is also shown.
The numbers in Table~\ref{tab:htt_numevents} cannot be used to estimate the global significance of a possible signal since
the expected significance varies considerably with $m_{\tau\tau}$, and the sensitive 1-jet/high-\PT category is merged with the 1-jet/low-\PT category.

\begin{table}[!hp]
\begin{center}
\topcaption{
Observed and expected numbers of events in the four $\PH \rightarrow \tau\tau$ decay channels and the 3 event categories,
for the combined 7 and 8\TeV data.
The uncertainties include the statistical and systematic uncertainties added in quadrature.
In the 0- and 1-jet categories, the low- and high-$\pt$ subcategories have been combined. The expected number of signal events
for a SM Higgs boson of mass $m_H=125\GeV$ is also given.
}
\begin{tabular}{c|c|c|c}
  \hline
  Process & 0-jet & 1-jet & VBF \\
  \hline
  \hline
  \multicolumn{4}{c}{$\Pe\tau_h$} \\
  \hline
  Z$\to\tau\tau$ & $2550 \pm 200$ & $2130 \pm 170$ & $53 \pm 6$ \\
QCD & $910 \pm 70$ & $410 \pm 30$ & $35 \pm 8$ \\
W+jets & $1210 \pm 70$ & $1111 \pm 75$ & $46 \pm 10$ \\
Z+jets & $560 \pm 99$ & $194 \pm 24$ & $13 \pm 2$ \\
$\ttbar$ & $162 \pm 21$ & $108 \pm 13$ & $7 \pm 2$ \\
Dibosons & $20 \pm 5$ & $60 \pm 14$ & $1.1 \pm 0.9$ \\

\hline
Total Background & $5410 \pm 270$ & $4020 \pm 220$ & $155 \pm 15$ \\
H$\to\tau\tau$ (125\GeVns{}) & $15 \pm 2$ & $26 \pm 4$ & $4.4 \pm 0.7$ \\
Data & 5273 & 3972 & 142 \\
  \hline
  \hline
  \multicolumn{4}{c}{$\mu\tau_h$} \\
  \hline
 $\cPZ\to\tau\tau$ & $50\,500 \pm 3800$ & $10\,570 \pm 830$ & $100 \pm 11$ \\
QCD & $14\,100 \pm 1600$ & $3980 \pm 510$ & $41 \pm 9$ \\
W+jets & $13\,300 \pm 1300$ & $5600 \pm 480$ & $72 \pm 15$ \\
Z+jets & $1620 \pm 230$ & $658 \pm 97$ & $2.5 \pm 0.6$ \\
$\ttbar$ & $651 \pm 82$ & $479 \pm 61$ & $15 \pm 3$ \\
Dibosons & $298 \pm 70$ & $256 \pm 58$ & $3 \pm 2$ \\
\hline
Total Background & $80\,400 \pm 4500$ & $21\,500 \pm 1200$ & $234 \pm 22$ \\
H$\to\tau\tau$ (125\GeV) & $141 \pm 21$ & $86 \pm 12$ & $8 \pm 1$ \\
Data & 80\,229 & 22\,009 & 263 \\

  \hline
  \hline
  \multicolumn{4}{c}{$\Pe\mu$} \\
  \hline
  $\cPZ\to\tau\tau$ & $22\,030 \pm 850$ & $5030 \pm 230$ & $56 \pm 5$ \\
QCD & $940 \pm 200$ & $550 \pm 120$ & $7 \pm 2$ \\
$\ttbar$ & $39 \pm 3$ & $831 \pm 86$ & $24 \pm 6$ \\
Dibosons & $796 \pm 96$ & $550 \pm 120$ & $11 \pm 2$ \\
\hline
Total Background & $23\,800 \pm 930$ & $6960 \pm 350$ & $99 \pm 9$ \\
H$\to\tau\tau$ (125\GeV)& $53 \pm 7$ & $35 \pm 4$ & $3.5 \pm 0.5$ \\
Data & 23\,274 & 6847 & 110 \\

  \hline
  \multicolumn{4}{c}{$\mu\mu$} \\
  \hline
$\cPZ\to\tau\tau$ & $9120 \pm 490$ & $1980 \pm 120$ & $5.3 \pm 0.4$ \\
QCD & $759 \pm 53$ & $341 \pm 27$ & ${<}1$   \\
W+jets & $145 \pm 10$ & $19 \pm 1$ & ${<}1$ \\
Z$\to\mu\mu$ & $(1263 \pm 73)\times10^3$ & $(380 \pm 24)\times10^3$ & $71 \pm 10$ \\
$\ttbar$ & $2440 \pm 200$ & $1330 \pm 130$ & $7 \pm 2$ \\
Dibosons & $1500 \pm 1100$ & $2210 \pm 790$ & $2.4 \pm 0.9$ \\
\hline
Total Background & $(1277 \pm 73)\times10^3$ & $(386 \pm 24)\times10^3$ & $85 \pm 11$ \\
$\PH\to\tau\tau$ (125\GeV)& $26 \pm 4$ & $16 \pm 2$ & $0.8 \pm 0.1$ \\
Data & 1\,291\,874 & 385\,494 & 83 \\

  \hline
\end{tabular}
\label{tab:htt_numevents}
\end{center}
\end{table}

The largest source of background is the Drell--Yan production of $\cPZ\to\Pgt\Pgt$.
This contribution is greatly reduced by the 1-jet and VBF selection criteria,
and is modelled using a data sample of $\cPZ\to\Pgm\Pgm$ events,
in which the reconstructed muons are replaced by the reconstructed particles from simulated $\tau$-lepton decays,
a technique called ``embedding''.
The background yield is rescaled to the $\cPZ\to\Pgm\Pgm$ yield in the data before any jet selection, thus,
for this dominant background, the systematic uncertainties in the efficiency of the jet-category selections and  the luminosity measurement are negligible.
In the $\Pe\Pgt_h$ and $\Pgm\Pgt_h$ channels, the largest remaining systematic uncertainty affecting this background yield is in the $\Pgt_h$ selection efficiency.
This uncertainty, which includes the uncertainty in the $\Pgt_h$ triggering efficiency, is estimated to be 7\% from an independent study based on a tag-and-probe technique~\cite{CMS:2011aa}.

The Drell--Yan production of $\cPZ\to \ell\ell$,
labelled as $\cPZ$+jets in Table~\ref{tab:htt_numevents},
is an important source of background in the $\Pe\Pgt_h$ channel,
owing to the 2--3\% probability for electrons to be misidentified as $\Pgt_h$~\cite{CMS-PAS-TAU-11-001},
and the fact that the reconstructed $\tau\tau$ invariant-mass distribution peaks in the Higgs boson mass search range.
The contribution of this background in the $\Pe\Pgt_h$ and $\mu\Pgt_h$ channels is estimated from simulation.
The simulated Drell--Yan yield is rescaled to the data using $\cPZ\to\Pgm\Pgm$ events,
and the efficiencies of the jet category selections are measured in a $\cPZ\to \mu\mu$ data sample.
The dominant systematic uncertainty in the background yield is from the $\ell \to \Pgt_h$ misidentification rate,
which is obtained by comparing tag-and-probe measurements from $\cPZ\to \ell\ell$ events in the data and the simulation,
and is 30\% for electrons and 100\% for muons.
The very small probability for a muon to be misidentified as $\Pgt_h$ makes
it difficult to estimate the systematic uncertainty in this probability, but also makes this background very small in the $\mu\Pgt_h$ channel.

\begin{figure}[t!]
\begin{center}
\includegraphics[width=0.4\textwidth]{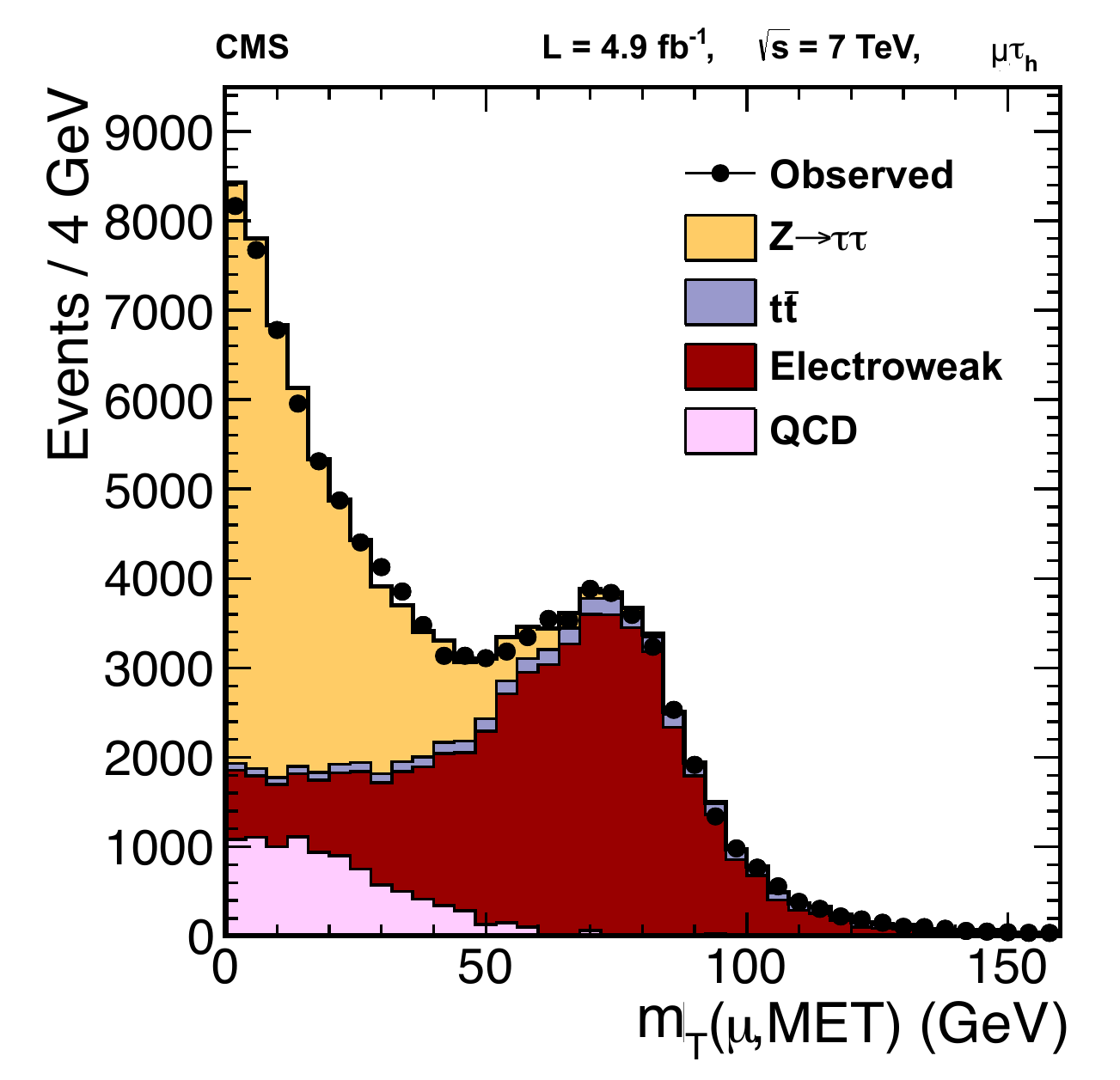}
\includegraphics[width=0.4\textwidth]{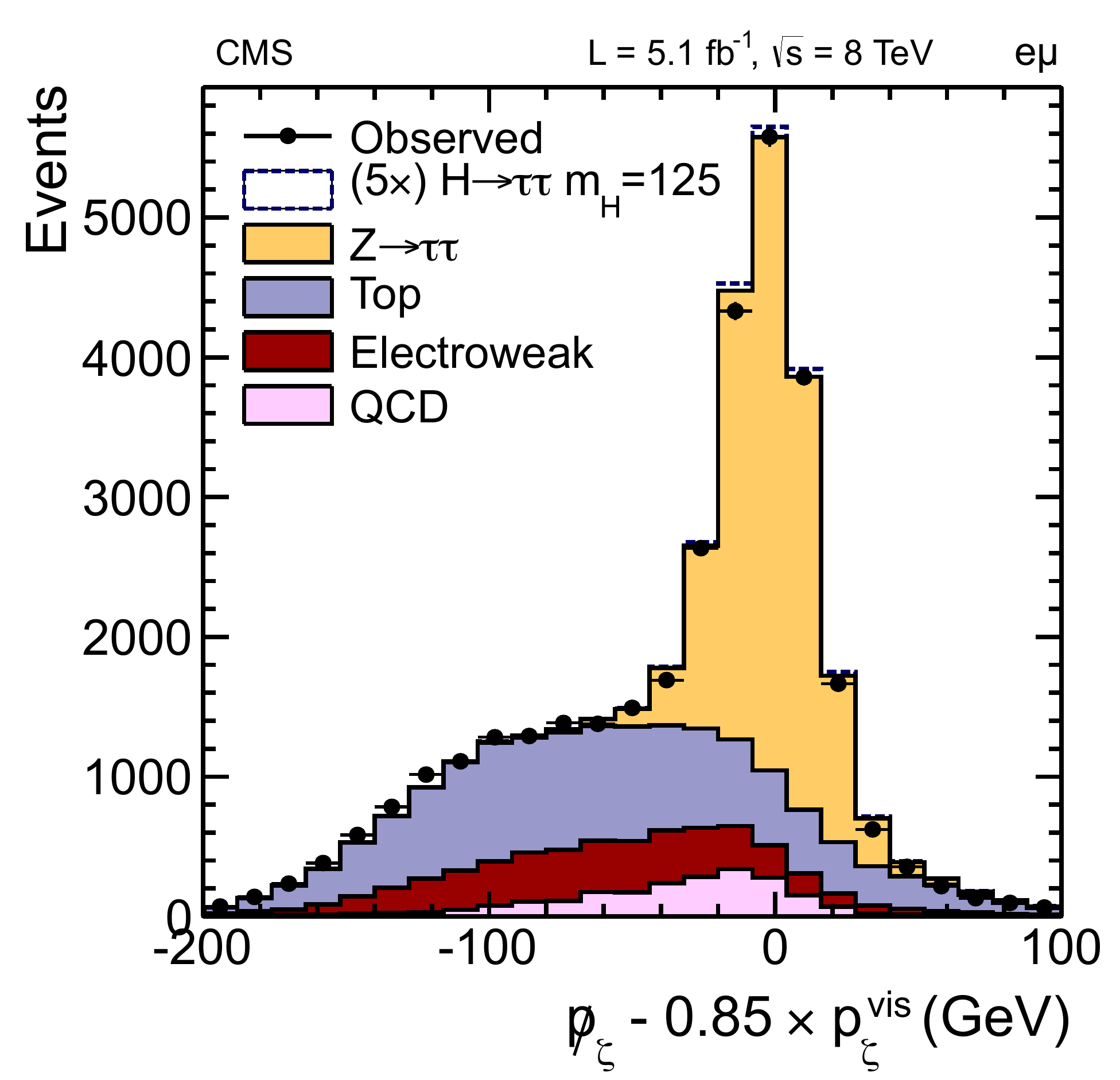} \\
\includegraphics[width=0.4\textwidth]{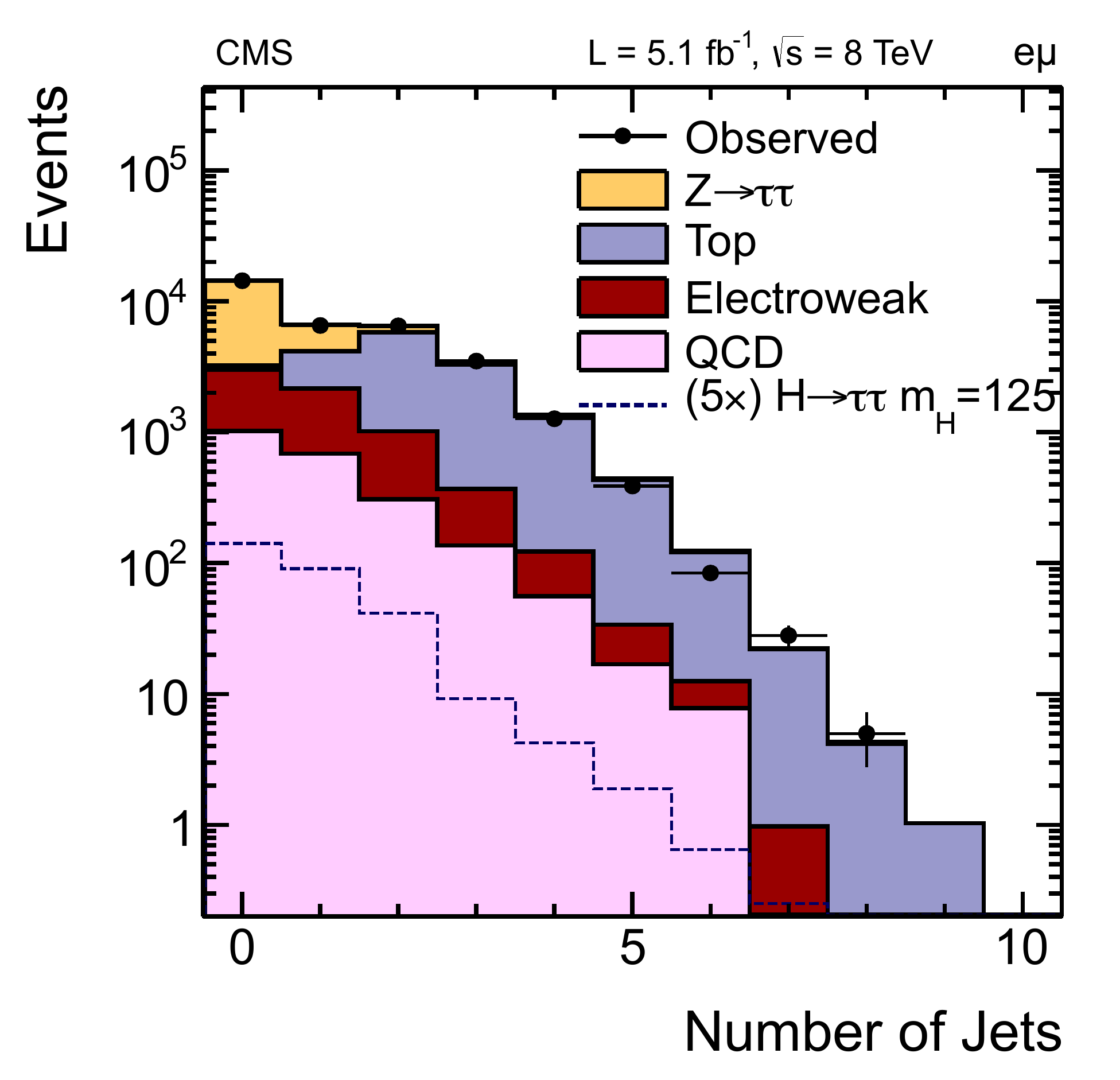}
\includegraphics[width=0.4\textwidth]{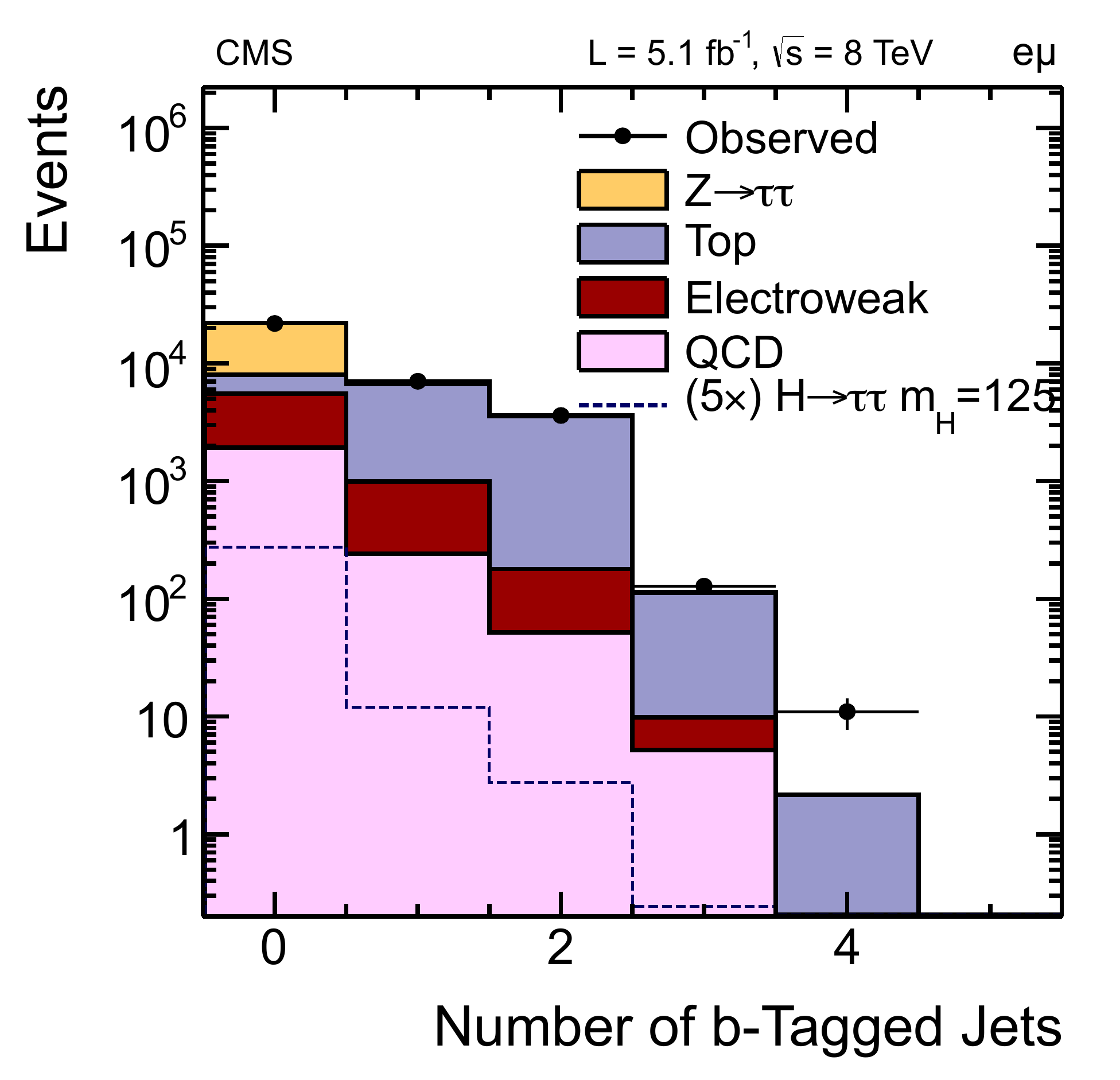} \\
\end{center}
\caption{The observed distributions (points with error bars) for the (upper left) transverse mass $\MT$ in the $\Pgm\Pgt_h$ channel at $\sqrt{s}=7\TeV$;
(upper right) $\not\! p_\zeta- 0.85 \cdot p_\zeta^{\mathrm{vis}}$, (lower left) number of jets, and (lower right) number of b-tagged jets in the $\Pe\Pgm$ channel at $\sqrt{s}=8\TeV$.
The expected distributions from the various background sources are shown by the shaded histograms.
In particular, the Electroweak background combines the expected contributions from $\PW$+jets, $\cPZ$+jets, and diboson processes.
The predictions for a SM Higgs boson with $\mH=125$\GeV are given by the dotted histograms, multiplied by a factor of 5 for clarity.}
\label{fig:htt_control}
\end{figure}

The background from $\PW$+jets production contributes significantly to the $\Pe\Pgt_h$ and $\Pgm\Pgt_h$ channels when the
$\PW$ boson decays leptonically and one jet is misidentified as a $\Pgt_h$.
The background is modelled for these channels using the simulation.
The $\PW$+jets background yield is normalized to the data in a high-$\MT$
control region dominated by the background in each of the five categories.
The factor for extrapolating to the low-$\MT$ signal region is obtained from the simulation,
and has a 30\% systematic uncertainty.
In the 1-jet/high-\pt and VBF categories, where the number of simulated events is marginal,
mass-shape templates are obtained by relaxing the $\Pgt_h$ isolation requirement,
ensuring that the bias introduced in the shape is negligible.
Figure~\ref{fig:htt_control}~(upper left) shows the $\MT$ distribution obtained in the $\Pgm\Pgt_h$ channel after the inclusive selection from data and simulation.
In the high-$\MT$ region, the agreement between the observed and expected yields comes from the normalization of the $\PW$+jets prediction to the data.
The agreement in shape indicates good modelling of \MET in the simulation.

The \ttbar production process is the main remaining background in the $\Pe\Pgm$ channel.
The predicted yield for all channels is obtained from simulation, with the yield rescaled to the one observed in the data from
a \ttbar-enriched control sample, extracted by requiring b-tagged jets.
The systematic uncertainty in the yield includes a 10\% systematic uncertainty in the b-tagging efficiency.
Figures~\ref{fig:htt_control} (upper right), (lower left), and (lower right) show the distributions of $D_\zeta$, the number of jets, and
the number of b-tagged jets in the $\Pe\Pgm$ channel.
There is good agreement between the data and the background predictions in the distributions at
low $D_\zeta$ values in Fig.~\ref{fig:htt_control} (upper right), and at high numbers of jets in
Fig.~\ref{fig:htt_control} (lower left) and (lower right), where the \ttbar process dominates.

QCD multijet events, in which one jet is misidentified as $\Pgt_h$ and another as a lepton, constitute another important source
of background in the $\Pe\Pgt_h$ and  $\Pgm\Pgt_h$ channels.
In the 0- and 1-jet categories, the QCD multijet background prediction is obtained using a control sample where the
lepton and the $\Pgt_h$ are required to have the same charge.
In this control sample, the QCD multijet distribution and yield are obtained by subtracting from the data the contribution of the Drell--Yan, \ttbar,
and $\PW$+jets processes, estimated as explained above.
The expected contribution of the QCD multijet background in the opposite-charge signal sample is then derived by rescaling the yield obtained in the same-charge control sample by a factor of 1.1,
which is measured in the data using a pure QCD multijet sample obtained by inverting the lepton isolation
and relaxing the $\Pgt_h$ isolation.
The 10\% systematic uncertainty in this factor covers its small dependence on $\PT(\Pgt_h)$ and the statistical uncertainty in its measurement, and dominates the uncertainty in this background contribution.
In the VBF category, the number of events in the same-charge control sample is too small to use this procedure.
Instead, the QCD multijet yield is obtained by multiplying the inclusive QCD yield
by the VBF selection efficiency measured in  data using a QCD-dominated sample in which the lepton and the $\Pgt_h$ are not isolated.
The mass shape template is obtained from data by relaxing the muon and $\Pgt_h$ isolation criteria.

The small background from $\PW+$jets and QCD multijet events in the $\Pe\Pgm$
channel is estimated from the number of events with one identified lepton and a second lepton
that passes relaxed selection criteria, but fails the nominal lepton selection.
This number is converted to the expected background yield using the efficiencies
for such loosely identified lepton candidates to pass the nominal lepton selection. These efficiencies
are measured in data using QCD multijet events.

Finally, the small background contribution in each channel from diboson and single top-quark production is estimated using the simulation.
The main experimental systematic uncertainties affecting the expected signal yield are from the $\Pgt_h$ identification efficiency (7\%),
the \MET scale (5\%), owing to the $\MT$ requirement and the \MET selection applied to the 0- and 1-jet categories of the $\Pe\Pgt_h$ channel,
the integrated luminosity (5\%), and the jet energy scale ($<$ 4\%).
The uncertainties in the muon and electron selection efficiencies, including trigger, identification, and isolation,
are both 2\%.
The theoretical uncertainty in the signal yield comes from the uncertainties in the PDFs,
the renormalization and factorization scales, and the modelling of the underlying event and parton showers.
The magnitude of the theoretical uncertainty depends on the production process (gluon-gluon fusion, VBF, or associated production)
and on the event category.
In particular, the scale uncertainty in the VBF production yield is 10\%.
The scale uncertainty in the gluon-gluon fusion production yield is 10\% in the 1-jet/high-\pt category
and 30\% in the VBF category.
The $\Pgt_h$ (3\%) and electron (1\%) energy scale uncertainties cause an uncertainty in the $m_{\tau\tau}$ spectrum shape, and are discussed in the next section.
The muon energy scale uncertainty is negligible.

\subsection{Results}

The statistical methodology described in Section~\ref{sec:method} is used to search for the presence of a SM Higgs boson signal,
combining the five categories of the four final states in the 7 and 8\TeV data sets as forty independent channels in a
binned likelihood based on the $m_{\tau\tau}$ distributions obtained for each channel.
Systematic uncertainties are represented by nuisance parameters in the likelihood.
A log-normal prior is assumed for the systematic uncertainties affecting the background normalization, discussed in the previous section.
The $\Pgt_h$ and electron energy scale uncertainties, which affect the shape of the $m_{\tau\tau}$
spectrum, are represented by nuisance parameters whose variation results in a
continuous change of this shape~\cite{Conway-PhyStat}.

\begin{figure}[htbp]
\begin{center}
\includegraphics[width=0.42\textwidth]{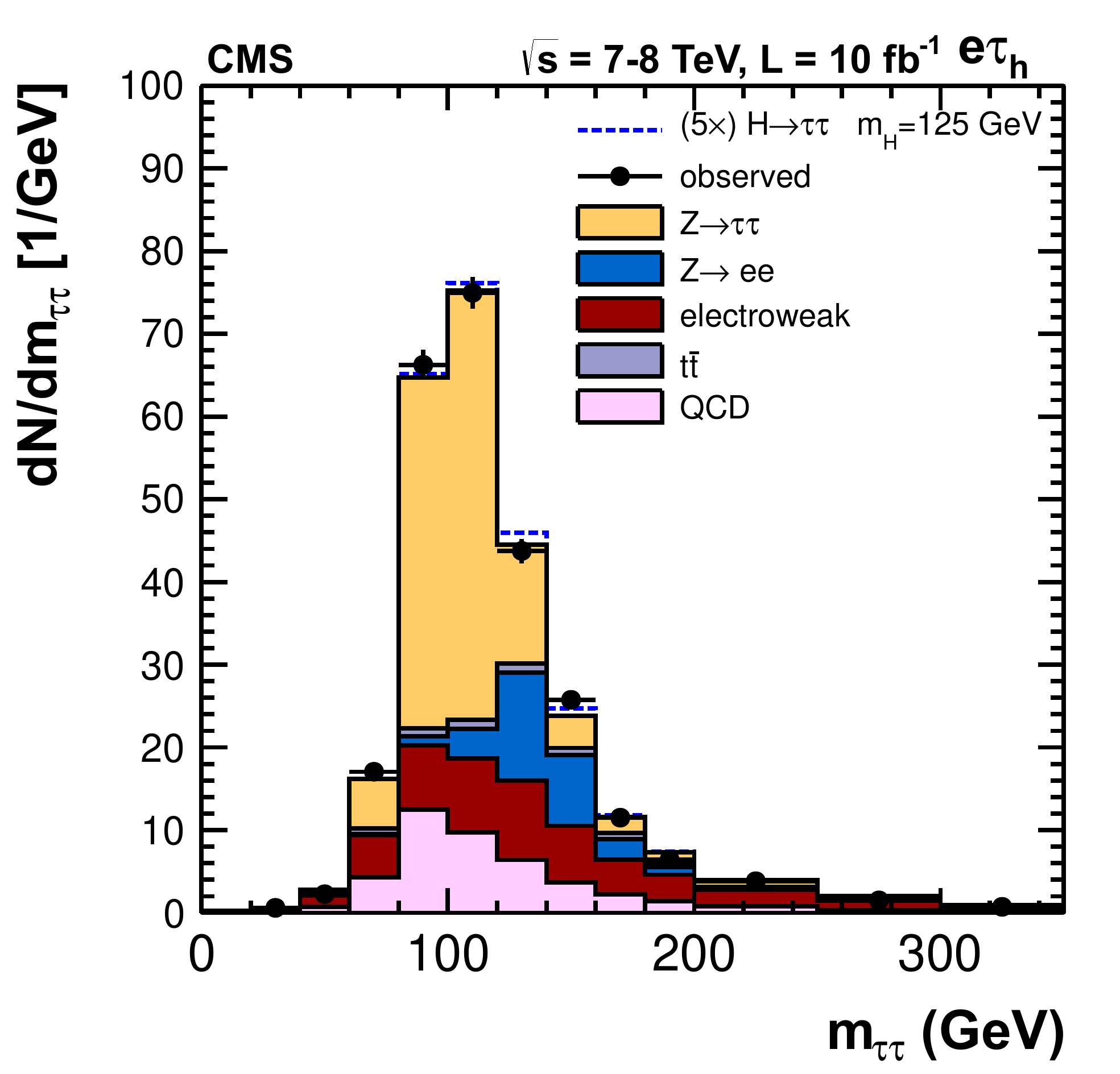}
\includegraphics[width=0.42\textwidth]{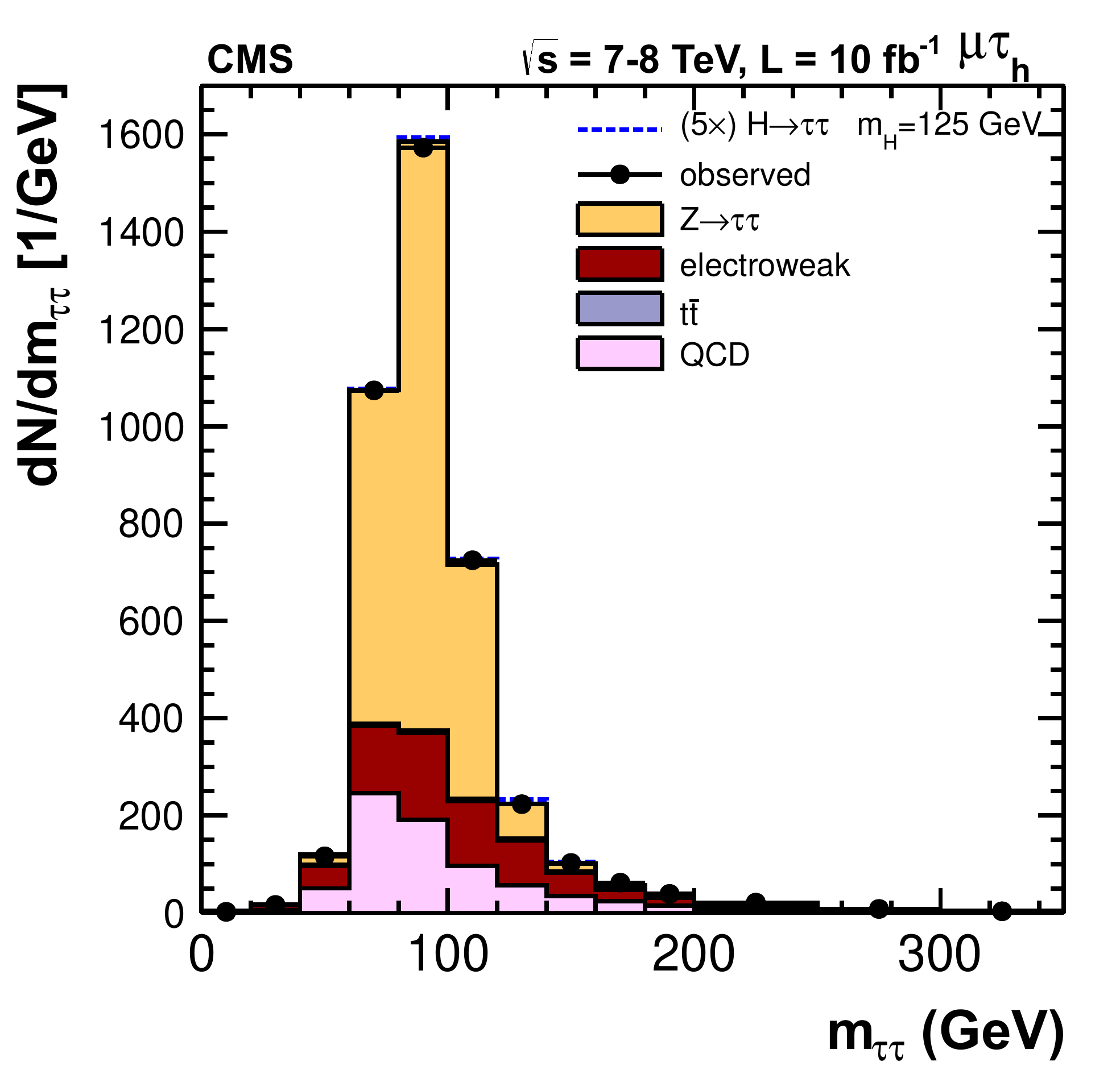} \\
\includegraphics[width=0.42\textwidth]{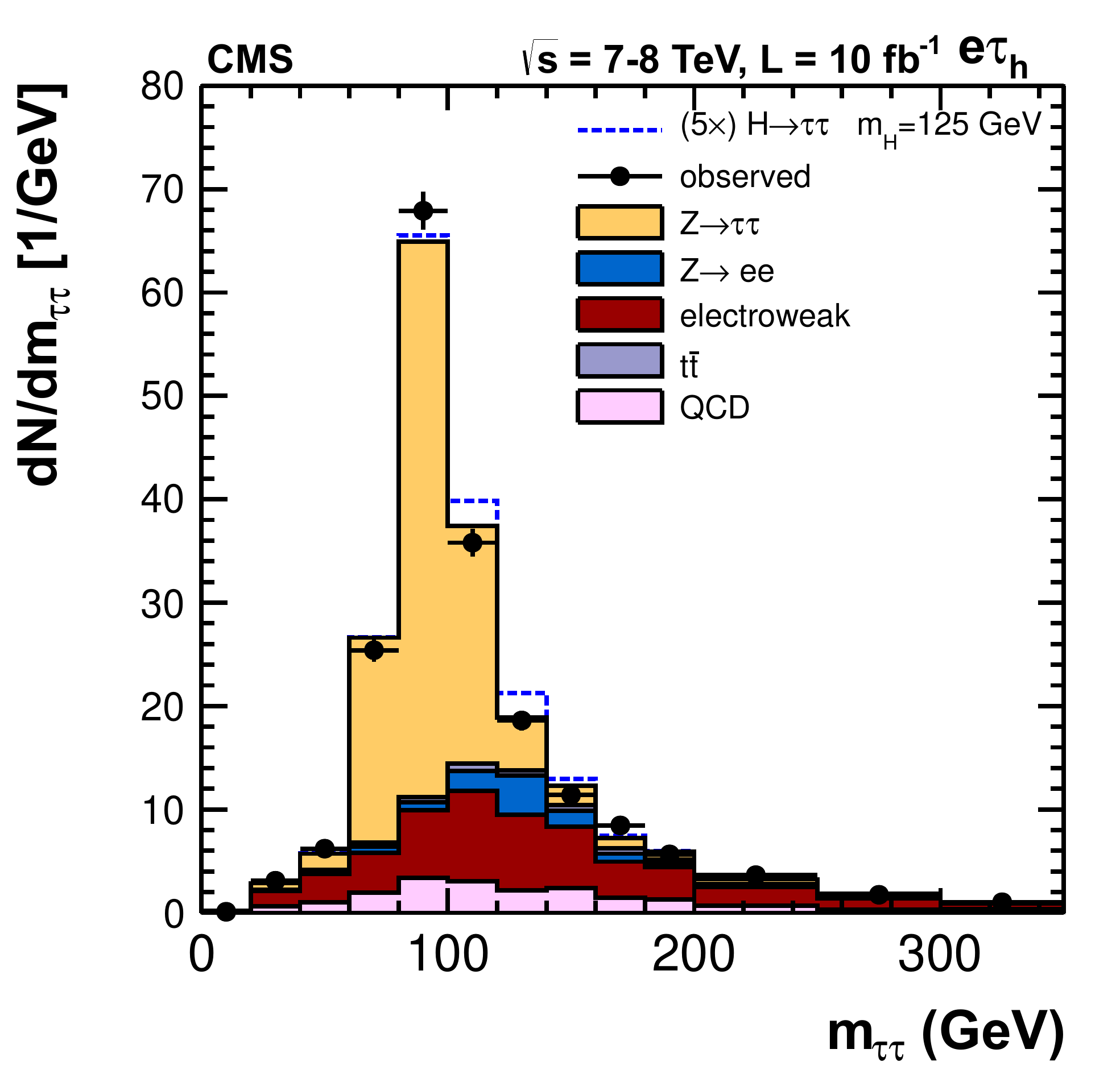}
\includegraphics[width=0.42\textwidth]{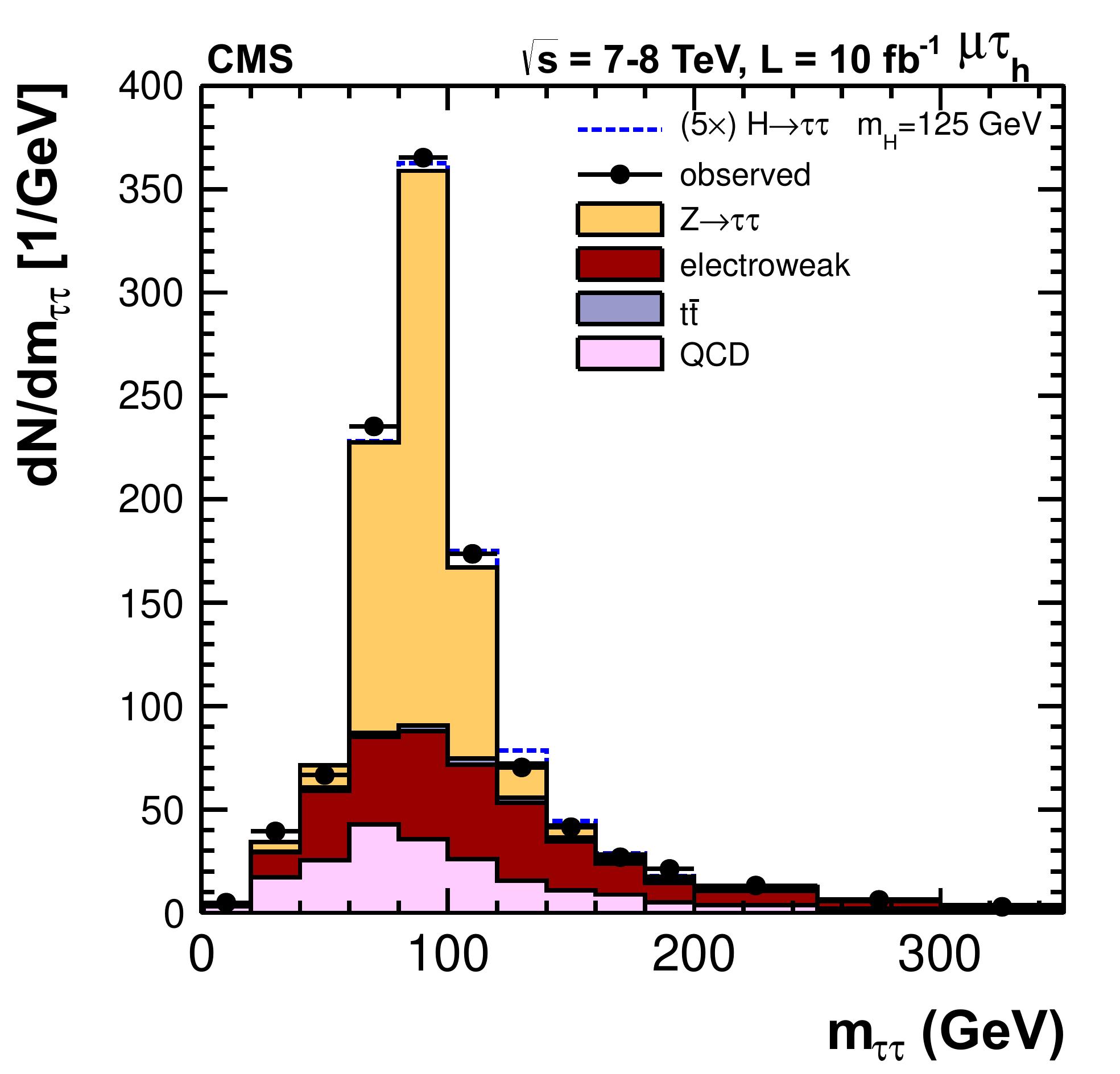} \\
\includegraphics[width=0.42\textwidth]{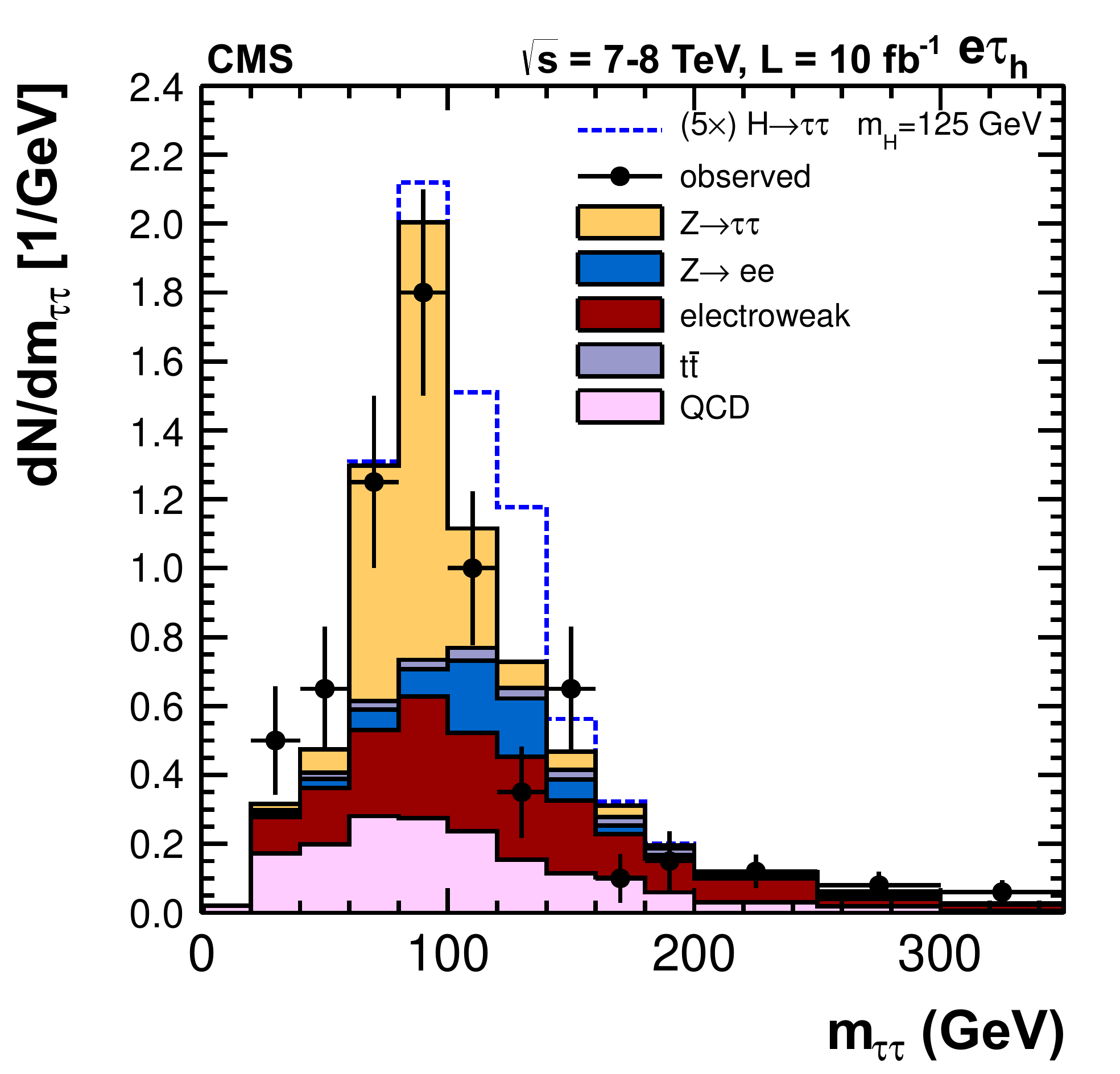}
\includegraphics[width=0.42\textwidth]{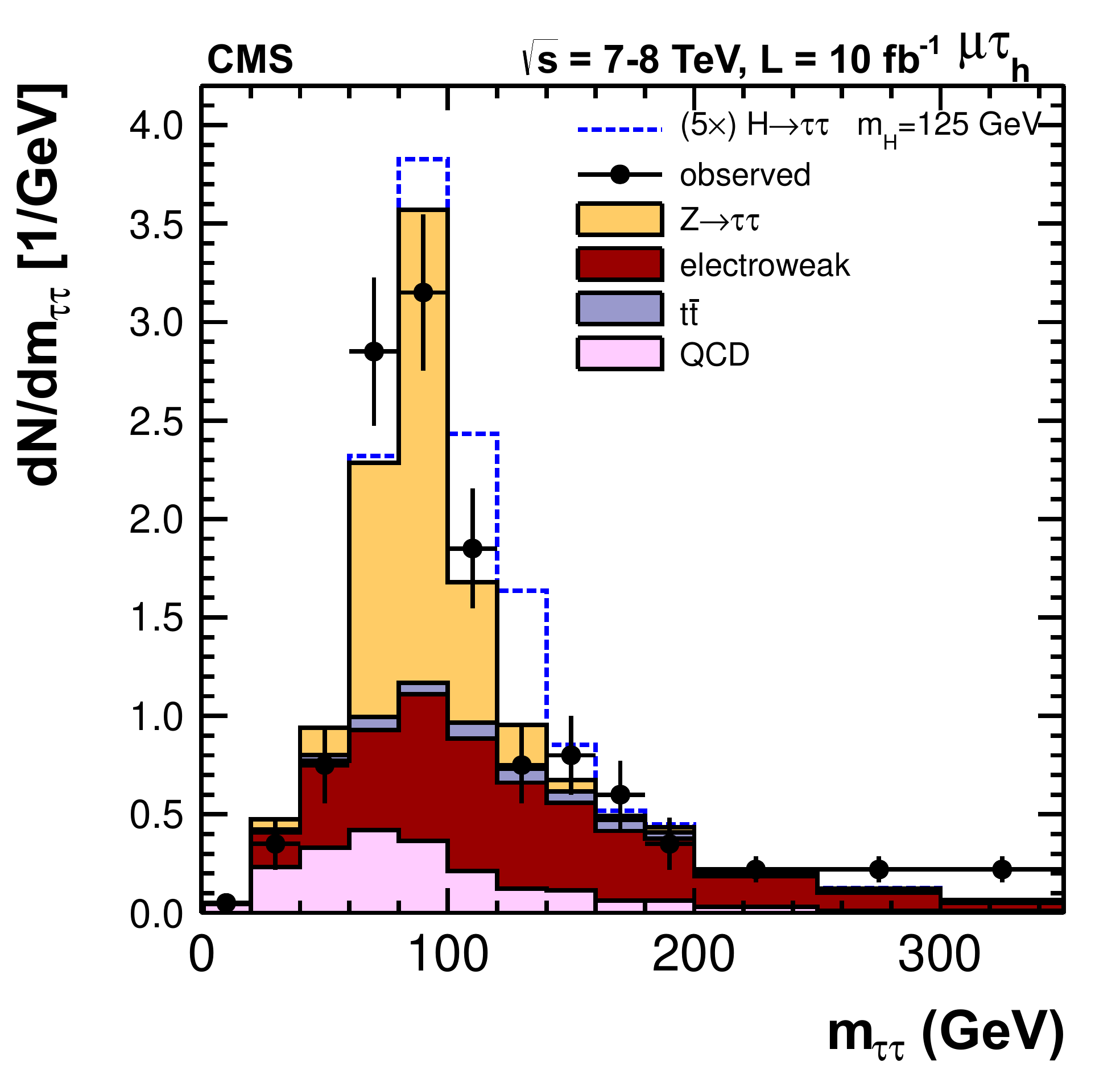}
\end{center}
\caption{Observed (points with error bars) and expected (histograms) $m_{\tau\tau}$ distributions for the $\Pe\Pgt_h$ (left) and $\mu\Pgt_h$ (right) channels, and,
from top to bottom, the 0-jet, 1-jet, and VBF categories for the combined 7 and 8\TeV data sets.
In the 0- and 1-jet categories, the low- and high-$\pt$ subcategories have been summed.
The electroweak background combines the expected contributions from $\PW$+jets, $\cPZ$+jets, and diboson processes.
In the case of $\Pe\Pgt_h$, the $\cPZ\to\Pe\Pe$ background is shown separately. The dotted histogram shows the expected distribution for a SM Higgs boson with $\mH=125$\GeV (multiplied by a factor of 5 for clarity).
}
\label{fig:htt_mtt_leptau}
\end{figure}

\begin{figure}[htbp]
\begin{center}
\includegraphics[width=0.42\textwidth]{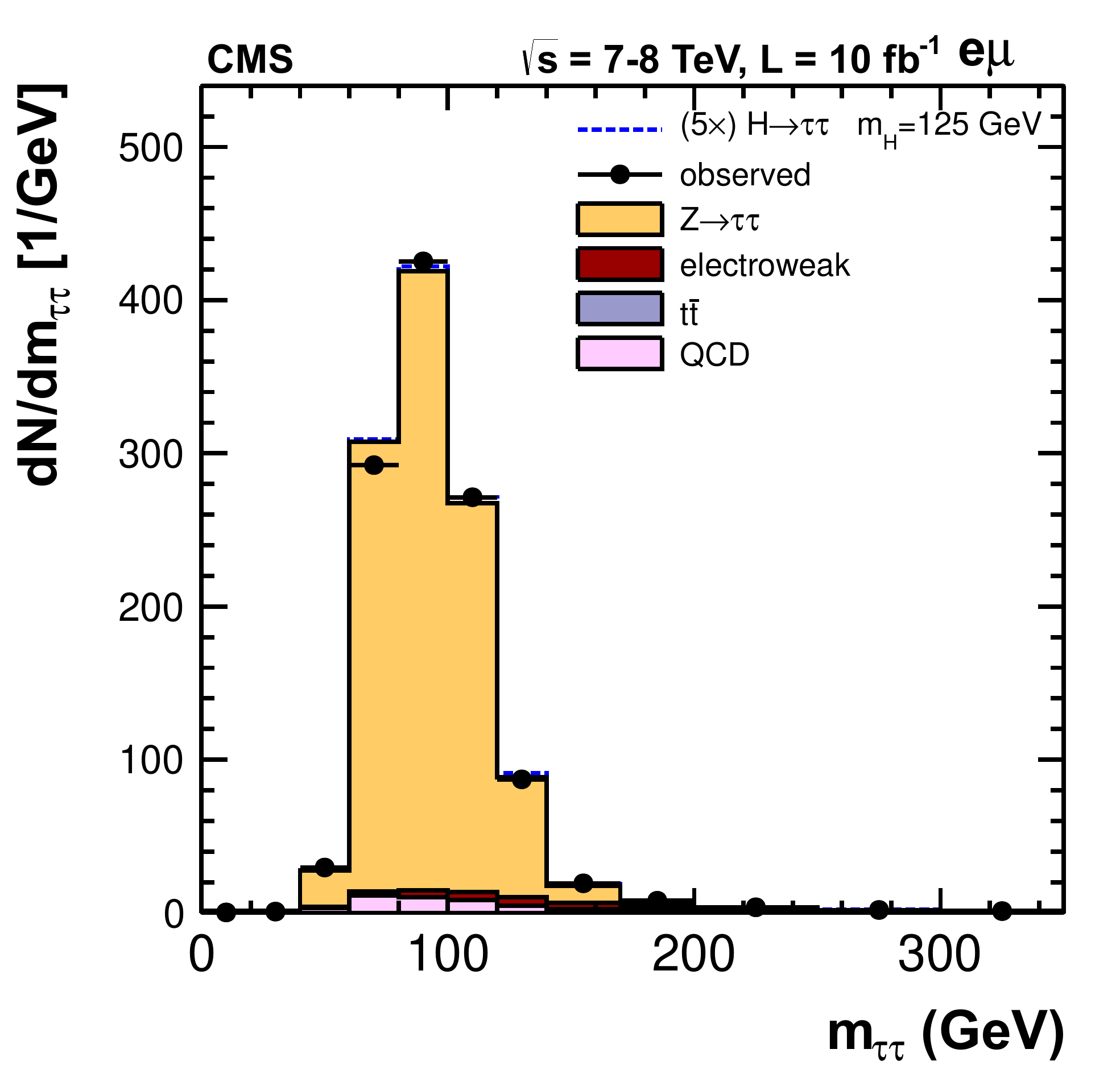}
\includegraphics[width=0.42\textwidth]{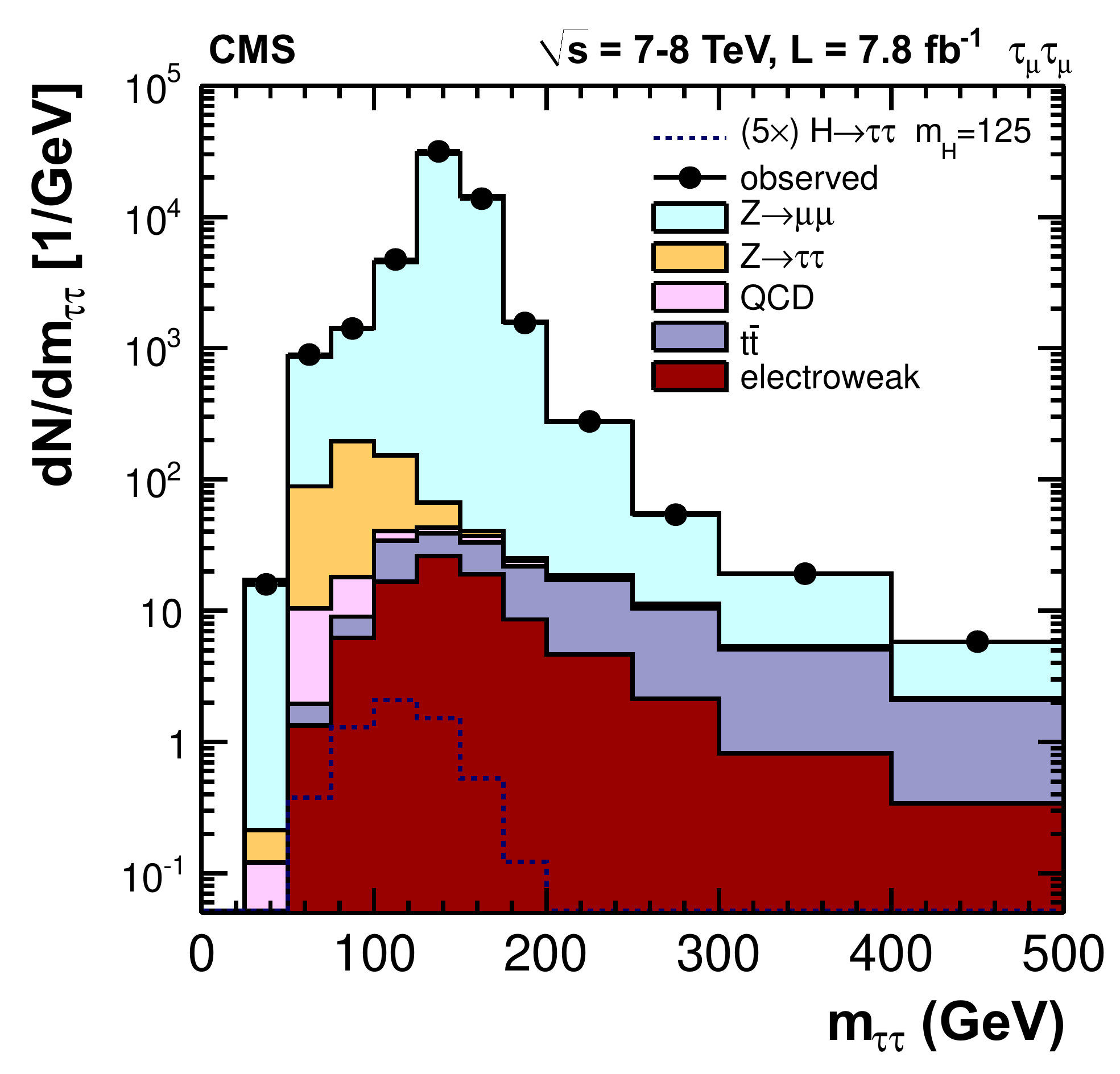} \\
\includegraphics[width=0.42\textwidth]{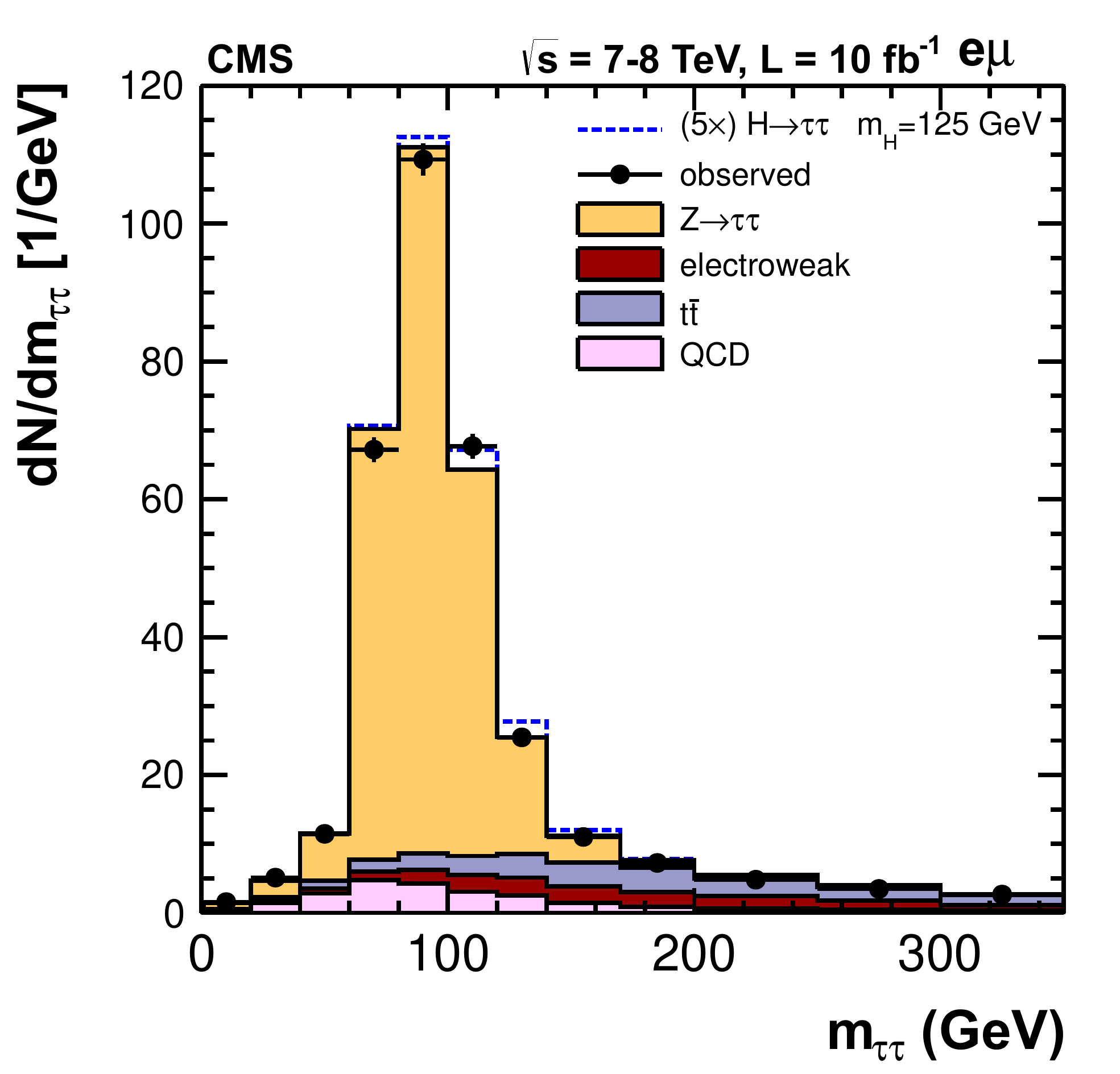}
\includegraphics[width=0.42\textwidth]{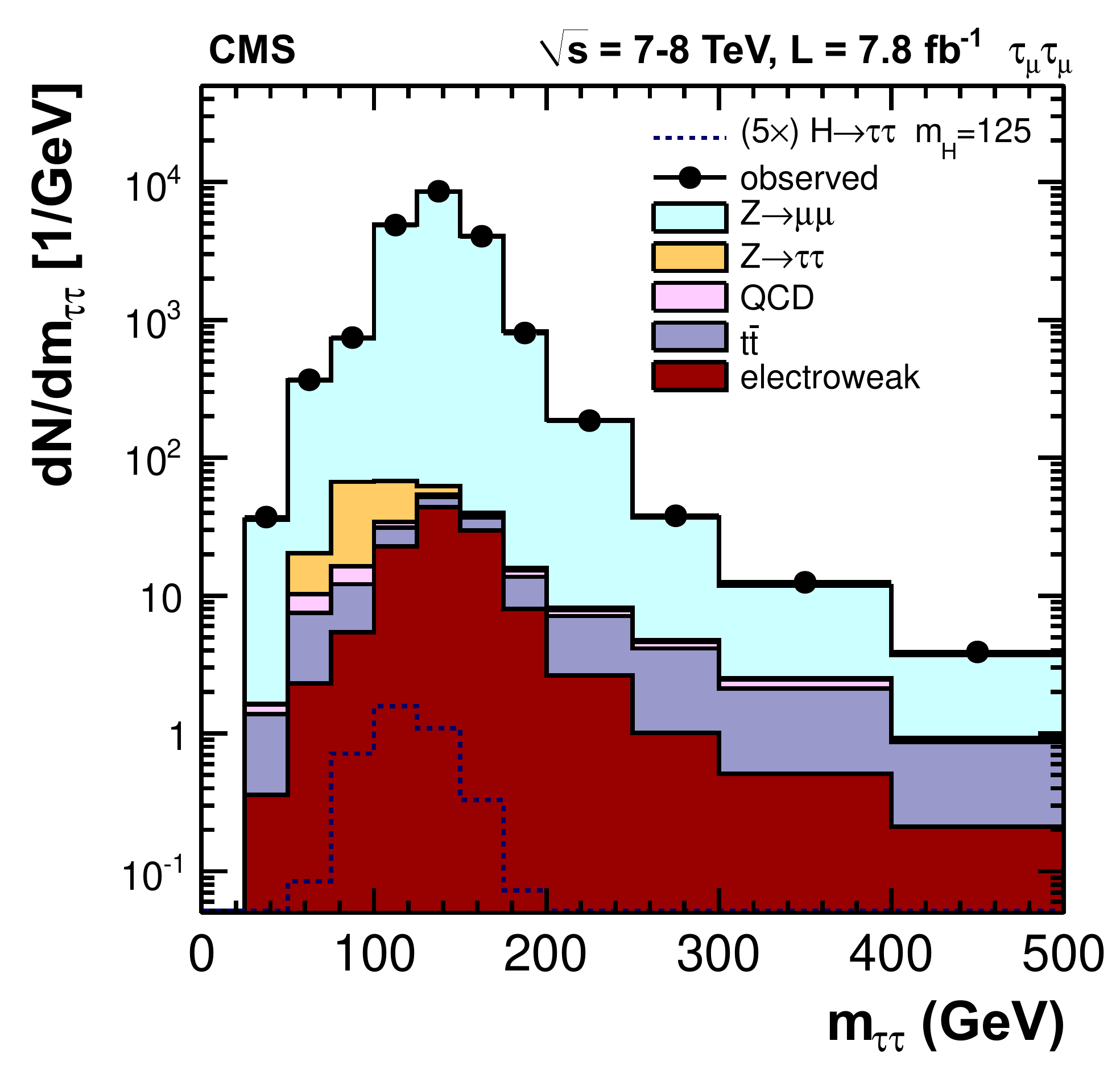} \\
\includegraphics[width=0.42\textwidth]{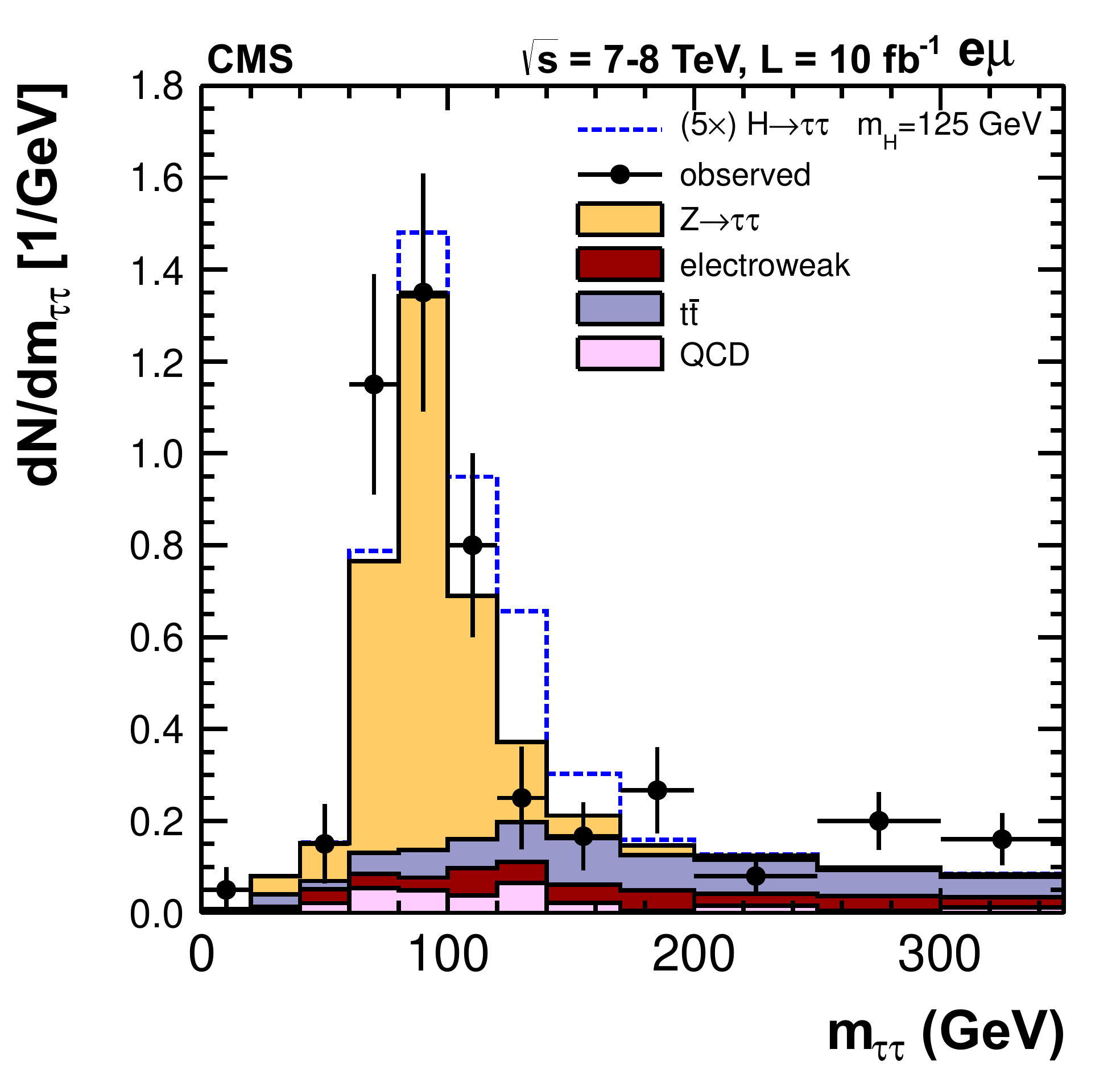}
\includegraphics[width=0.42\textwidth]{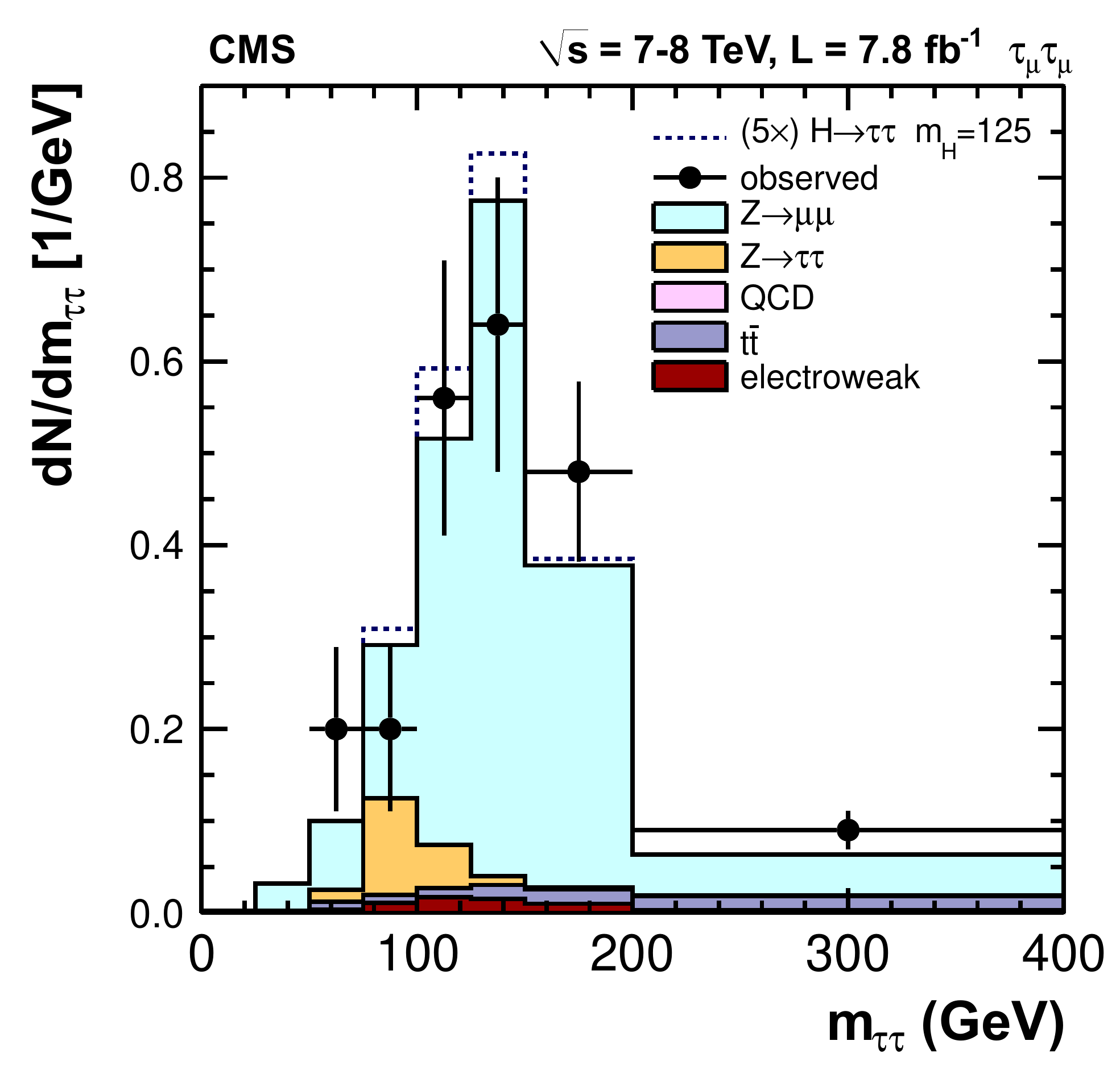}
\end{center}
\caption{Observed (points with error bars) and expected (histograms) $m_{\tau\tau}$ distributions for the $\Pe\mu$ (left) and $\mu\mu$ (right) channels, and,
from top to bottom, the 0-jet, 1-jet, and VBF categories for the combined 7  and 8\TeV data sets.
In the 0- and 1-jet categories, the low- and high-$\pt$ subcategories have been summed.
The electroweak background combines the contributions from $\PW$+jets, $\cPZ$+jets, and diboson processes.
In the case of $\mu\mu$, the $\cPZ\to\mu\mu$ background is shown separately.
The dotted histogram shows the expected distribution for a SM Higgs boson with $\mH=125$\GeV (multiplied by a factor of 5 for clarity).
}
\label{fig:htt_mtt_leplep}
\end{figure}

Figures~\ref{fig:htt_mtt_leptau} and~\ref{fig:htt_mtt_leplep} show the observed  $m_{\tau\tau}$ distributions in the $\Pe\Pgt_h$, $\Pgm\Pgt_h$, $\Pe\Pgm$, and $\Pgm\Pgm$ channels, for each event category, compared with the background predictions.
The  7 and 8\TeV data sets are merged, as well as the low- and high-\PT subcategories of the 0- and 1-jet categories.
The binning given in the figures corresponds to the binning used in the likelihood.
The background mass distributions are the result of the global maximum-likelihood fit under the background-only hypothesis.
This fit finds the best set of values for the nuisance parameters to match the data,
assuming no signal is present.
The variation of the nuisance parameters is limited by the systematic uncertainties estimated for each of the background contributions and used as input to the fit.
For example, in the VBF category of the $\Pe\Pgt_h$ channel,
the most important nuisance parameters related to background normalization are the ones affecting
the $Z\to \tau \tau$ yield ($\Pgt_h$ selection efficiency),
the $\cPZ\to \Pe\Pe$ yield ($\Pe \to \Pgt_h$ misidentification rate),
the $\PW$+jets yield (extrapolation from the high $\MT$ to the low $\MT$ region),
and the QCD yield (ratio between the yields in the opposite-charge and same-charge regions).
The fit makes use of the high-$m_{\tau\tau}$ region of the VBF category to constrain the nuisance parameters affecting the $\PW$+jets yield.
The nuisance parameter related to the $\Pgt_h$ identification efficiency is mostly constrained by the 0- and 1-jet categories,
where the number of events in the $Z\to \tau\tau$ peak is much larger.
It is also the case for the nuisance parameter related to the $\Pgt_h$ energy scale, which affects the shape of the $Z\to \tau \tau$ distribution.

The interpretation of the results in terms of upper limits on the Higgs boson production cross section is given in Section 10.

\section{\texorpdfstring{$\PH\to\cPqb\cPqb$}{H to bb}\label{sec:hbb}}

The decay $\PH\to\cPqb\cPqb$ has the
largest branching fraction of the five search modes for $\mH\leq135$\GeV, but the
signal is overwhelmed by the QCD multijet production of \cPqb\ quarks.  The analysis
is therefore designed to search for a dijet resonance in events
where a Higgs boson is produced at high $\pt$, in association
with a $\PW$ or $\cPZ$ boson that decays leptonically, which largely suppresses the QCD
multijet background.
The following final states are
included in the search:
$\PW(\mu\nu)\PH$, $\PW(\Pe\nu)\PH$, $\cPZ(\mu\mu)\PH$,
$\cPZ(\Pe\Pe)\PH$, and $\cPZ(\nu\nu)\PH$, all with the Higgs
boson decaying to \cPqb\cPqb. Backgrounds arise from the production of vector bosons in association with
jets (from all quark flavours), singly- and pair-produced top quarks, dibosons, and
QCD multijet processes.
Simulated samples of signal and background events are used to
optimize the analysis. Control regions in data
are selected to adjust the predicted event yields from simulation for the main
background processes  and to estimate their contribution in the signal
region.

Several different high-level triggers are used to collect events consistent with
the signal hypothesis in all five channels. For the WH channels, the trigger paths consist of several single-lepton
triggers with tight lepton identification. Leptons are also required
to be isolated from other tracks and calorimeter energy depositions to maintain an acceptable trigger
rate. For the \WmnH\ channel, in the 7\TeV data set, the trigger
thresholds for the muon transverse
 momentum, \PT, vary from  17 to 40\GeV. The higher
thresholds are implemented  for periods of higher instantaneous
luminosity. For the 8\TeV  data set, the muon \PT threshold is 24\GeV
for the isolated-muon trigger, and 40\GeV for muons without any isolation requirements.
The combined single-muon trigger efficiency is $\approx$$90\%$ for
signal events that pass all offline requirements, described
in Section~\ref{sssec:hbb_Event_Selection}.
For the \WenH\ channel, in the 7\TeV data set,  the electron \PT\
threshold ranges from 17 to 30\GeV. In addition, two jets and a minimum
value on the missing transverse energy are required.
These additional requirements
help to maintain acceptable trigger
rates during the periods of high instantaneous luminosity. For the
8\TeV data set, a single-isolated-electron trigger is used with
a 27\GeV \PT\ threshold.
The combined efficiency for these triggers for signal events that
pass the final offline selection criteria is larger than 95\%.

The \ZmmH\ channel uses the same single-muon triggers as the \WmnH\ channel.
For the \ZeeH\ channel, dielectron triggers with lower-\PT\
thresholds of $17$ and $8$\GeV  and tight isolation requirements are used.
These triggers are $\approx 99\%$ efficient for  \ZH\  signal events that pass the
final offline selection criteria.  For the \ZnnH\ channel, combinations of several triggers are
used, all with the requirement that the missing transverse energy be
above a certain threshold. Additional jet requirements are made to
keep the trigger rates acceptable as the luminosity increases and to reduce
the $\ETmiss$ thresholds, in order to increase the signal
acceptance.
A trigger with $\ETmiss$ $>150$\GeV requirement  is
implemented for both the 7 and 8\TeV
data sets. For the 7\TeV data, triggers that require the
presence of two jets with $|\eta|<2.6$,  \PT $>20$\GeV, and
\MET thresholds
of 80 and 100\GeV, depending on the instantaneous luminosity, are also used.
For the 8\TeV data set, a trigger that requires two jets,
each with $|\eta| < 2.6$ and \PT $> 30$\GeV, and $\MET > 80$\GeV is
also implemented.  As the instantaneous luminosity
increased further, this trigger was replaced by one
requiring $\MET > 100$\GeV, two jets with $|\eta| < 2.6$, one with \PT $>60$\GeV
and the other with \PT $> 25$\GeV, the dijet \PT $> 100$\GeV,
and no jet with \PT $> 40$\GeV within 0.5 radians in azimuthal angle of the \ETmiss vector.
For \ZnnH\ signal events with
missing transverse energy $>160$\GeV, the overall
trigger efficiency is $\approx 98\%$ with
respect to the offline event reconstruction and selection described
below. The corresponding efficiency for $120< \ETmiss < 160\GeV$ is about 66\%.

\subsection{Event selection}\label{sssec:hbb_Event_Selection}

The final-state objects used in the $\PH \to \,\cPqb \cPqb$  event reconstruction
are described in Section~\ref{sec:reconstruction}. Electron candidates are
considered in the pseudorapidity range $\left | \eta \right | < 2.5$,
excluding the  $1.44 <\left | \eta \right | < 1.57$ transition
region between the ECAL barrel and endcaps. Tight muon candidates are
considered in the  $\left | \eta \right | < 2.4$ range.
An isolation requirement on $R_\text{Iso}^{\ell}$ of approximately 10\%, as calculated in
Eq.~(\ref{eq:reconstruction_isolation}), that is consistent
with the expectation for leptons originating from W and Z boson decays,
is applied to electron and muon candidates. The exact requirement depends on the lepton $\eta$, $\pt$,
and flavour. To identify \cPqb\ jets, different values for the CSV output discriminant, which can
range between 0 and 1, are used, with
corresponding different efficiencies and misidentification rates.  For example, with
a CSV $>0.90$ requirement, the efficiencies to tag
b quarks, c quarks, and light quarks, are 50\%,
6\%, and 0.15\%, respectively~\cite{CMS-PAS-BTV-12-001}. The corresponding efficiencies for
CSV $>0.50$ are 72\%, 23\%, and 3\%. All events from data and simulation are required to
pass the same trigger and event reconstruction algorithms. Scale
factors that account for differences in the performance of these
algorithms between data and simulation are computed and used in
the analysis.

The background processes to VH production are V+jets, \ttbar,
single-top-quark, diboson (VV),
and QCD multijet production. These overwhelm the signal by several orders of
magnitude. The event selection is based on the kinematic
reconstruction of the vector boson and the Higgs boson decay
into two \cPqb-tagged jets. Backgrounds are then substantially reduced by
requiring a significant boost of the \pt of the vector boson and the
Higgs boson~\cite{PhysRevLett.100.242001}, which tend
to recoil from each other with a large azimuthal opening angle,  \dphiVH,
between them.  For each channel, two ranges
  of \ptV\ are considered. These are
        referred to as ``low'' and ``high''.
        Owing to different signal and background
        compositions, each \ptV\ range  has a different sensitivity, and
        the analysis is performed separately for each range. The
        results from all the ranges are then combined for each channel. The ranges
        for the WH channels are $120<\ptV<170$\GeV and $\ptV>170$\GeV,
        for the \ZnnH\ channel $120<\ptV<160$\GeV and
        $\ptV>160$\GeV, and for the \ZllH\ channel $50<\ptV<100$\GeV
        and $\ptV>100$\GeV.

Candidate \WtoLN\ decays are identified by requiring
the presence of a single isolated lepton and missing
transverse energy. Muons (electrons) are required to have a \pt\ above
20 (30)\GeV.  
For the
\WenH\ channel only, to reduce contamination from QCD multijet
processes,
\MET is required to be greater than 35\GeV.
Candidate \ZtoLL\ decays are reconstructed by combining
isolated, oppositely charged pairs of electrons or muons with  \pt\ $> 20$\GeV
and a dilepton invariant
mass satisfying $75<m_{\ell\ell}<105\GeV$.  The identification of \ZtoNN\ decays
requires the \MET in the event to be within the \ptV\ ranges
described above.
Two requirements suppress events from QCD multijet processes
with an $\ETmiss$ arising from mismeasured jets.
First, the $\ETmiss$ vector must be isolated from jet activity, using
the requirement that the azimuthal angle difference $\dphiMJ$ between
the  $\ETmiss$ direction and any jet with $|\eta|<2.5$ and $\pt>$20
(30)\GeV  be greater that 0.5 radians for the 7 (8)\TeV data sample.
Second, the azimuthal angle between the
$\ETmiss$ vector
calculated using only charged particles with \pt$>0.5$\GeV and $\left |
  \eta \right |<2.5$ and the direction of the standard $\ETmiss$ vector (calculated using all particles, charged
and neutral) must be greater than 0.5 radians.
Subject to these two requirements, background from QCD multijet processes
is reduced to a negligible level in the \ZnnH\ channel.
To reduce the \ttbar and \WZ\ background in the \WH\ and \ZnnH\ channels,
events where the number of additional isolated leptons with \pt$>20$\GeV 
is greater than 0  are rejected.

Reconstruction of the \HBB\ decay is done by requiring two
jets above the minimum \pt thresholds listed in Table~\ref{tab:BDTsel}, having $|\eta|<2.5$, and tagged by the
CSV algorithm.
If more than two such jets are
found in the event, the pair with the
highest total dijet transverse momentum, \ptjj, is selected. The background from V+jets and
dibosons is reduced significantly
through \cPqb\ tagging, and subprocesses where the two jets originate from
genuine \cPqb\ quarks dominate the final selected data sample. After all
the event selection criteria are applied, the invariant-mass
resolution for the Higgs boson decay to \cPqb\cPqb\ is
approximately 10\%, as found in a previous CMS analysis~\cite{VHbb_PLB}. The mass resolution
is improved here by applying regression techniques similar to those used by
the CDF experiment~\cite{1107.3026}. Through this procedure, a further correction, beyond the
standard jet energy corrections, is computed for
individual \cPqb\ jets in order to better measure the true parton
energy. A BDT algorithm is trained on
simulated  \HBB\ signal events, with
inputs that include detailed information about each jet
that helps to differentiate \cPqb-quark jets from light-flavour jets.
The resulting improvement in the \cPqb\cPqb\ invariant-mass resolution is approximately
15\%, resulting in an increase in the analysis sensitivity of
10--20\%, depending on the specific channel. The BDT regression is
implemented in the TMVA framework~\cite{Hocker:2007ht}.
The complete set of input variables is
(though not all variables are used for every channel):
\begin{itemize}
 \item transverse momentum of the jet before  and after energy corrections;
\item  transverse energy and mass of the jet after energy
  correction;
 \item  uncertainty in the jet energy correction;
 \item transverse momentum of the highest-\PT constituent in the jet;
\item pseudorapidity of the jet;
\item total number of jet constituents;
 \item length and uncertainty  of the displacement of the jet's secondary vertex;
\item mass and transverse momentum of the jet's secondary vertex;
\item number and fraction of jet constituents that are charged;
 \item event energy density, $\rho$, calculated using constituents with $\left | \eta \right | < 2.5$;
\item missing transverse energy in the event;
 \item  azimuthal angle between the missing transverse
   energy vector and the direction of the nearest jet in pseudorapidity.
\end{itemize}

To better discriminate the signal from background for
different Higgs boson mass hypotheses, an event classification BDT algorithm is trained separately for
each mass value using simulated samples of signal and background events
that pass the  selection criteria described above, together with the requirements listed in Table~\ref{tab:BDTsel}.
The set of input
variables used in training this BDT is chosen by iterative optimization from a larger number of
potentially discriminating variables. Table~\ref{tab:BDTvars}
lists these variables. The number $N_{aj}$ of additional jets in an
event counts jets that satisfy
$\pt>20\GeV$ and $\left | \eta \right | < 4.5$
for \WlnH,   $\pt>20\GeV$ and $\left | \eta \right | < 2.5$
for \ZllH, or  $\pt>30\GeV$ and $\left | \eta \right | < 4.5$
for \ZnnH.
The output distribution of this BDT algorithm is fitted to search for
events from Higgs boson production. Fitting this distribution,
rather than simply counting events in a range of the distribution with
a good signal-to-background ratio, as in Ref.~\cite{VHbb_PLB}, improves the sensitivity
of the analysis by approximately 20\%.

\begin{table}[tbp]
\topcaption{Selection criteria for the simulated event samples used in
training of the signal and background
  BDT algorithm.
Variables marked ``--'' are not used in the given
channel. Entries in parentheses indicate the selection for the
  high-\ptV\ range.
The second and third rows refer to the \pt\ threshold for the highest-
  and second-highest-\PT jet, respectively, for the pair with the
highest total dijet transverse momentum, \ptjj.  The parameter \Nal\ is the number of additional isolated leptons in the
  event. Kinematic variables are
  given in \GeVns and angles in radians. }
\label{tab:BDTsel}
\begin{center}
\begin{tabular}{lccc} \hline
 Variable      & \WlnH           & \ZllH                 & \ZnnH    \\ \hline \hline
$m_{\ell\ell}$ &  --             & $\in[75-105]$ & --       \\
$\pt(\rm{j}_1)$     & $>30$           & $>20$                 & $>80$    \\
$\pt(\rm{j}_2)$     & $>30$           & $>20$                 & $>30$    \\
\ptjj          & $>120$          & --                    & $\in[120-160]\, (>160)$   \\
\Mjj           & $<250$          & $\in[80-150]$ (--)   & $<250$   \\
\ptV           & $\in[120-170]\, (>170)$ & $\in[50-100]\, (>100)$    & --       \\
CSV$_{\mathrm{max}}$         & $>0.40$         & $0.50\, (0.244)$            & $>0.679$  \\
CSV$_{\mathrm{min}}$            & $>0.40$         & $0.244$                  & $>0.244$  \\
\Nal           & $=0$	         & --	                 & $=0$     \\
\MET          & $>35 (\Pe)$   & --	                 & $\in[120-160]\, (>160)$ \\
\dphiMJ        & --	         & --	                 & $>0.5$   \\
\dphiVH    & --	         & --	                 & $>2.0$   \\
\hline
\end{tabular}
\end{center}
\end{table}

\begin{table}[tbp]
\topcaption{Variables used for training the signal and background
  BDT algorithm.}
\label{tab:BDTvars}
\begin{center}
\begin{tabular}{ll} \hline
Variable& definition \\\hline\hline
$p_{\mathrm{T_\mathrm{j}}}$& transverse momentum of each \cPqb\ jet from the Higgs boson decay      \\
\Mjj& dijet invariant mass                                \\
\ptjj& dijet transverse momentum                          \\
\ptV& vector boson transverse momentum         \\
CSV$_{\text{max}}$& value of CSV for the \cPqb-tagged jet with the largest
CSV value                  \\
CSV$_{\text{min}}$& value of CSV for the \cPqb-tagged jet with the second
largest CSV value           \\
\dphiVH& azimuthal angle between the vector boson  (or \MET vector)
and the dijet direction   \\
\dEtaJJ& difference in $\eta$ between \cPqb\ jets from Higgs boson decay     \\
\dRJJ& distance in $\eta$--$\phi$ between \cPqb\ jets from Higgs boson decay (not for
\ZllH ) \\
\Naj& number of additional jets \\
\dphiMJ& azimuthal angle between \MET and the closest jet (only for \ZnnH )                                 \\
\hline
\end{tabular}
\end{center}
\end{table}

\subsection{Background control regions}\label{sssec:hbb_Background_Control_Regions}

Control regions are identified in the data and used to correct
the estimated yields from the MC simulation
for two of the important background processes:  \ttbar\ production and
V+jets, originating from light-flavour partons (u, d, s, or c
quarks and gluons) or from heavy-flavour (b quarks).
Simultaneous fits are then performed to the distributions of the discriminating
variables in the control regions to obtain scale factors by which the simulation
yields are adjusted. This procedure is performed separately for
each channel. For the \ZllH\ and \WH\ modes the scale factors
derived for the electron and muon decay channels are combined.
These scale factors  account not only for possible simulation cross-section discrepancies
with the data, but also for potential
differences in the selection efficiencies for the various physics object.
Therefore, separate scale factors are used for each background
process in the different channels. The uncertainties in the scale factor determination include
a statistical uncertainty from the fits (owing to the finite size of the samples)
 and an associated systematic uncertainty. The latter is estimated by refitting the
 distributions in the control regions after applying estimates
for sources of potential systematic shifts
such as \cPqb-jet-tagging efficiency, jet energy scale, and jet energy resolution.

Tables~\ref{tab:ZllControl}--\ref{tab:WlnControl} list the selection criteria
used for the control regions in the \ZllH\ , \ZnnH\, and \WH\ channels,
respectively. Table~\ref{tab:SFs} summarizes the fit results
for all channels separately for the 7\TeV and 8\TeV data sets.  The fit results are
found to be robust and the fitted scale factors are consistent with the values
from the previous analysis~\cite{VHbb_PLB}.

\begin{table}[tbp]
\topcaption{Definitions of the control regions for the simulated sample
of Z+jets and \ttbar backgrounds in the  \ZllH\ channel. The
  same selection is used for  the low- and high-\ptV\ ranges.
The values of  kinematical variables are in \GeVns{}.}
\label{tab:ZllControl}
\begin{center}
\begin{tabular}{lcc} \hline
 Variable       & Z+jets              & \ttbar                  \\ \hline\hline
$m_{\ell\ell}$  & $\in[75-105]$       & $\notin [75-105]$             \\
$\pt(\rm{j}_1)$     & $>20$              & $>20$                   \\
$\pt(\rm{j}_2)$     & $>20$             & $>20$                   \\
\ptV            & $\in[50-100]$       & $\in[50-100]$	      \\
CSV$_{\mathrm{max}}$            &  $>0.244$    & $>0.244$              	      \\
CSV$_{\mathrm{min}}$           &  $>0.244$      & $>0.244$                     \\
\Mjj       & $\notin [80-150]$, $<250$        & $\notin [80-150]$, $<250$ \\
\hline
\end{tabular}
\end{center}
\end{table}

\begin{table}[tbp]
\topcaption{Definitions of the control regions for the simulated samples of
V+jets and \ttbar background processes  in the \ZnnH\ channel for the
  low- and high-\ptV\ regions. The values in parentheses are for
  the high-\ptV\ region. The labels LF and HF refer to light- and heavy-flavour
  jets. The parameter \Nal\ is the number of additional isolated leptons in the
  event. The values for kinematical variables are in \GeVns{}.}
\label{tab:ZnnControl}
\begin{center}
\scalebox{0.75}{
\begin{tabular}{lccccc} \hline
 Variable      & Z+jets (LF) & Z+jets (HF) & \ttbar & W+jets (LF) &
 W+jets (HF) \\ \hline\hline
$\pt(\rm{j}_1)$     & $>60 (>80)$        & $>60 (>80)$  	   &
$>60 (>80)$        & $>60 (>80)$         & $>60 (>80)$ 	\\
$\pt(\rm{j}_2)$     & $>30$        & $>30$ 	   &$>30$        &$>30$         &$>30$ 	\\
\ptjj          &$>120 (>160)$        &$>120 (>160)$  	   &$>120 (>160)$
&$>120 (>160)$         &$>120 (>160)$	\\
CSV$_{\mathrm{max}}$          &-         & $>$0.898  	   & $>$0.898        &-         & $>$0.898	\\
\Naj           &-        &-  	           &1        &0         &0 	\\
\Nal           &0         &0  	           &1        &1         &1 	\\
\MET     & $\in$[120--160] ($>160$)  &       $\in$[120--160] ($>160$) & $\in$[120--160] ($>160$)
         &$\in$[120--160] ($>160$)         &$\in$[120--160] ($>160$)	\\
$\Mjj$         &-         & $\notin$[90--150]  	   &  $\notin$[90--150]
&-         & $\notin$[90--150] 	\\
\hline
\end{tabular}
}
\end{center}
\end{table}

\begin{table}[tbp]
\topcaption{Definitions of the control regions for the simulated samples of
 three background processes in the  \WlnH\ channel for the
  low- and high-\ptV\ regions. The values in parentheses are used for
  the high-\ptV\ region. The labels LF and HF refer to light- and heavy-flavour
  jets. The parameter \Nal\ is the number of additional isolated leptons in the
  event,  and  METsig is the ratio of the $\ETmiss$ value to its
  uncertainty~\cite{PFMEtSignAlgo}. The values
for kinematical variables are in \GeVns{}. The symbols $\Pe$ and $\mu$  mean that the
selection is used only for the \WenH\ mode or \WmnH\ mode, respectively.}
\label{tab:WlnControl}
\begin{center}
\begin{tabular}{lccc} \hline
 Variable      & W+jets (LF)          & \ttbar             & W+jets
 (HF)                \\ \hline \hline
$\pt(\rm{j}_1)$     &$>$30  	      & $>$30		    & $>$30		   \\
$\pt(\rm{j}_2)$     & $>$30  	      & $>$30		    & $>$30		   \\
\ptjj	       & $>$120 	       & $>$120 		    & $>$120		   \\
\ptV	       & $\in[120-170]$ ($>$170)    & $\in[120-170]$ ($>$170)  & $\in[120-170]$ ($>$170)   \\
CSV$_{\mathrm{max}}$ 	       & --		      & $>$0.898		    & $>$0.898	   \\
\Naj	       & $<$2		      & $>$1		    & $=$0		   \\
\Nal	       & $=$0		      & $=$0		    & $=$0		   \\
\MET	       & $>$35 ($\Pe$)        & $>$35 ($\Pe$)      & $>$35 ($\Pe$)	    \\
METsig       & $>$2.0($\mu$), $>$3.0($\Pe$) & --		     & --		    \\
$\Mjj$         & $<$250 	       & $<$250 		    &
$\notin[90-150]$ \\
\hline
\end{tabular}
\end{center}
\end{table}

\begin{table}[tbp]
\topcaption{Data/MC scale factors for the control region in each Higgs
boson production process with the 7\TeV and  8\TeV
  data sets in the low- and high-\ptV\ ranges.  The uncertainties shown are
 statistical and systematic, respectively.  The labels LF and HF refer to light- and heavy-flavour
  jets.
}
\label{tab:SFs}
\begin{center}
\scalebox{0.68}{
\begin{tabular}{lcccccccc} \hline
 Process     & WH   &WH               & \ZllH\  	&\ZllH 	 & \ZnnH    & \ZnnH                  \\ \hline\hline
Low-\ptV  & 7\TeV & 8\TeV   & 7\TeV & 8\TeV & 7\TeV & 8\TeV \\ \hline
W+jets (LF)      & $0.88\pm 0.01\pm 0.03$  & $0.97\pm 0.01 \pm 0.03$ & --
& --  & $ 0.89 \pm  0.01 \pm  0.03 $  & $0.91  \pm 0.03  \pm 0.03  $
\\
W+jets (HF)        & $1.91\pm 0.14\pm 0.31$  & $2.05\pm 0.21\pm 0.33$   & --
& --  & $ 1.36 \pm  0.10  \pm 0.15 $ & $1.63  \pm 0.29  \pm 0.14 $\\
Z+jets (LF)      & --    & --       & $1.11 \pm 0.03\pm 0.11$   & $1.41\pm
0.03\pm 0.16$ & $ 0.87 \pm  0.01 \pm  0.03 $  & $1.01  \pm 0.05  \pm
0.03  $ \\

Z+jets (HF)         & --    & --      & $0.98 \pm 0.05\pm 0.12$   & $1.04\pm
0.05\pm 0.20$ & $ 0.96 \pm  0.02 \pm  0.03 $  & $ 1.00 \pm 0.10  \pm
0.04 $ \\
\ttbar       & $0.93\pm 0.02\pm 0.05$  & $1.12\pm 0.01\pm 0.06$ &
$1.03 \pm 0.04\pm 0.11$   & $1.06\pm 0.03\pm 0.11$  & $ 0.97 \pm  0.02 \pm 0.04 $  & $ 1.02 \pm 0.03 \pm 0.03  $ \\
\hline\hline
High-\ptV  & 7\TeV & 8\TeV   & 7\TeV & 8\TeV & 7\TeV & 8\TeV \\ \hline
W+jets (LF)      & $0.79\pm 0.01\pm 0.02$   & $0.88\pm 0.01\pm 0.02$ & --
& --  & $ 0.78 \pm  0.02 \pm  0.03 $    & $ 0.86  \pm 0.03 \pm  0.03 $\\
W+jets (HF)        & $1.49\pm 0.14\pm 0.19$  & $1.30\pm 0.20\pm 0.17$  & --
& -- & $ 1.48 \pm  0.15  \pm 0.20 $ & $ 1.43  \pm 0.28 \pm  0.18 $ \\
Z+jets (LF)      & --     & -- & $1.11 \pm 0.03\pm 0.11$  & $1.41\pm
0.03\pm 0.16$ & $ 0.97 \pm  0.02 \pm  0.04 $ & $ 1.01 \pm 0.04 \pm  0.04 $ \\
Z+jets (HF)         & --    & -- & $0.98 \pm 0.05\pm 0.12$  & $1.03\pm
0.05\pm 0.20$ & $ 1.08 \pm  0.09 \pm  0.06 $ & $ 1.06  \pm 0.06 \pm  0.07 $  \\
\ttbar       & $0.84\pm 0.02\pm 0.03$  & $0.97\pm 0.02\pm 0.03$  &
$1.03 \pm 0.04\pm 0.11$  & $1.06\pm 0.03\pm 0.11$ & $ 0.97 \pm  0.02 \pm 0.04 $ & $ 1.03  \pm 0.04 \pm  0.04 $ \\
\hline
\end{tabular}
}
\end{center}
\end{table}

\subsection{Systematic uncertainties}\label{sssec:hbb_Uncertainties}

Sources of systematic uncertainty
in the expected signal and background yields and distribution shapes
are listed in Table~\ref{tab:syst}.
The uncertainty in the integrated  luminosity measurement is $2.2\%$
for the 7\TeV data~\cite{CMS-PAS-EWK-11-001} and $4.4\%$ for the 8\TeV data~\cite{CMS:2012jza}.
Muon and electron trigger,
       reconstruction, and identification efficiencies are
       determined in data from samples of
       leptonic Z boson decays. The uncertainty in the yields due to the trigger efficiency is
       2\% per charged lepton and the uncertainty in the
       identification efficiency is also 2\% per lepton. The parameters describing the
       \ZnnH\ trigger efficiency turn-on curve are
       varied within their statistical uncertainties and for different
       assumptions on the methodology.
       A 2\% systematic uncertainty in the yield is estimated.

The jet energy scale is
       varied by $\pm$1  standard deviation as a function of the jet $\pt$ and
       $\eta$, and the efficiency of the analysis selection is
       recomputed. A 2--3\%
       yield variation is found, depending on the particular decay channel
       and production process. The effect of the uncertainty
       in the jet energy resolution is evaluated by
       smearing the jet energies by the measured
       uncertainty, giving a 3--6\% variation in yields.
      The uncertainties in the jet energy scale and
       resolution also affect the shape of the BDT
       output distribution. The impact of the jet energy scale uncertainty is determined by recomputing
the BDT distribution after shifting the energy scale up and down by its
uncertainty.   Similarly, the impact of the jet energy resolution is
determined by recomputing the BDT distribution after increasing or
       reducing the jet energy
 resolution.

Data-to-simulation \cPqb-tagging-efficiency scale factors, measured in
\ttbar events and multijet events,
   are applied  to the jets in signal and background
       events. The estimated systematic uncertainties in the \cPqb-tagging scale factors are: $6\%$
       per \cPqb\ tag, $12\%$ per c tag, and $15\%$ per
       mistagged jet (originating from gluons and light quarks)~\cite{CMS-PAS-BTV-12-001}. These translate into yield uncertainties in the 3--15\%
       range, depending on the channel and the production process. The
       shape of the BDT output distribution is also affected by the shape of
       the CSV distribution, and therefore recomputed according to the
      range of variations of the CSV distributions.

 The theoretical VH signal cross section is
        calculated to NNLO,
       and the systematic uncertainty is $4\%$~\cite{LHCHiggsCrossSectionWorkingGroup:2011ti}, including
       the effects of scale and PDF
       variations~\cite{Botje:2011sn,Alekhin:2011sk,Lai:2010vv,Martin:2009iq,Ball:2011mu}.
       The analysis described in this paper is performed
       in the regime where the V and H have a significant boost in $\pt$, and thus, potential differences in
       the \pt\ spectrum of the V and H between the  data and the MC
       simulation generators could introduce systematic effects in the
       estimates of the signal acceptance and efficiency.  Theoretical
       calculations are available that estimate the
       NLO electroweak (EW)~\cite{HAWK1,HAWK2,Denner:2011id} and NNLO
       QCD~\cite{Ferrera:2011bk}
       corrections to VH production in the boosted regime.  The estimated
       effect from electroweak corrections for a boost of $\approx$$150\GeV$
       are $5\%$ for ZH and $10\%$ for WH.  For the QCD correction, a $10\%$ uncertainty
       is estimated for both ZH and WH, which includes effects due
       to additional jet activity from initial- and final-state radiation. The finite size of the
       signal MC  simulation samples, after all selection criteria are
       applied, contributes an uncertainty  of 1--5\% in the various channels.

The total uncertainty in the prediction of the background yields from
estimates using data is approximately 10\%. For the V+jets background, the
differences in the BDT output distribution for events from
the \MADGRAPH and \HERWIG{++} MC simulation generators are considered.
For the single-top-quark and diboson yield predictions, which are obtained solely
from simulation, a $30\%$ systematics uncertainty in the cross
sections is used.

\begin{table}[tbp]
\topcaption{Systematic uncertainties in the predicted signal and
  background yields from  the sources listed. The ranges
  give the variations over the 7 and 8\TeV data sets, different search
  channels, specific processes, and Higgs
  boson mass hypotheses. The acronym EWK stands for electroweak.}
\label{tab:syst}
\begin{center}
\scalebox{0.90}{
\begin{tabular}{lc} \hline

Source                                                  &     Range (\%)  \\ \hline\hline
Integrated luminosity                                            &      2.2--4.4  \\
Lepton identification and trigger efficiency (per lepton) &      3     \\
\ZnnH\ triggers                                       &      2    \\
Jet energy scale                                      &     2--3 \\
Jet energy resolution                               &     3--6  \\
Missing transverse energy                               &      3   \\
\cPqb-tagging efficiency                                           &     3--15   \\
Signal cross section (scale and PDF)         &  4      \\
Signal cross section (\pt boost, EWK/QCD)    &        5--10/10 \\
Statistical precision of signal simulation                     &       1--5 \\
Backgrounds estimated from data                        &     10   \\
Backgrounds estimated from simulation              &  30  \\

\hline
\end{tabular}
}
\end{center}
\end{table}

\subsection{Results}\label{sssec:Results}

Maximum-likelihood fits are performed to the output distributions of the BDT
algorithms, trained separately for each channel and each Higgs
boson mass value hypothesis in the 110--135\GeV range.
In the fit, the BDT shapes and normalizations, for signal and each background component,
are allowed to vary within the systematic and statistical uncertainties described in
Section~\ref{sssec:hbb_Uncertainties}. These uncertainties  are treated
as nuisance parameters,
with appropriate correlations taken into account.

\begin{figure}[tbp]
  \begin{center}
    \includegraphics[width=0.49\textwidth]{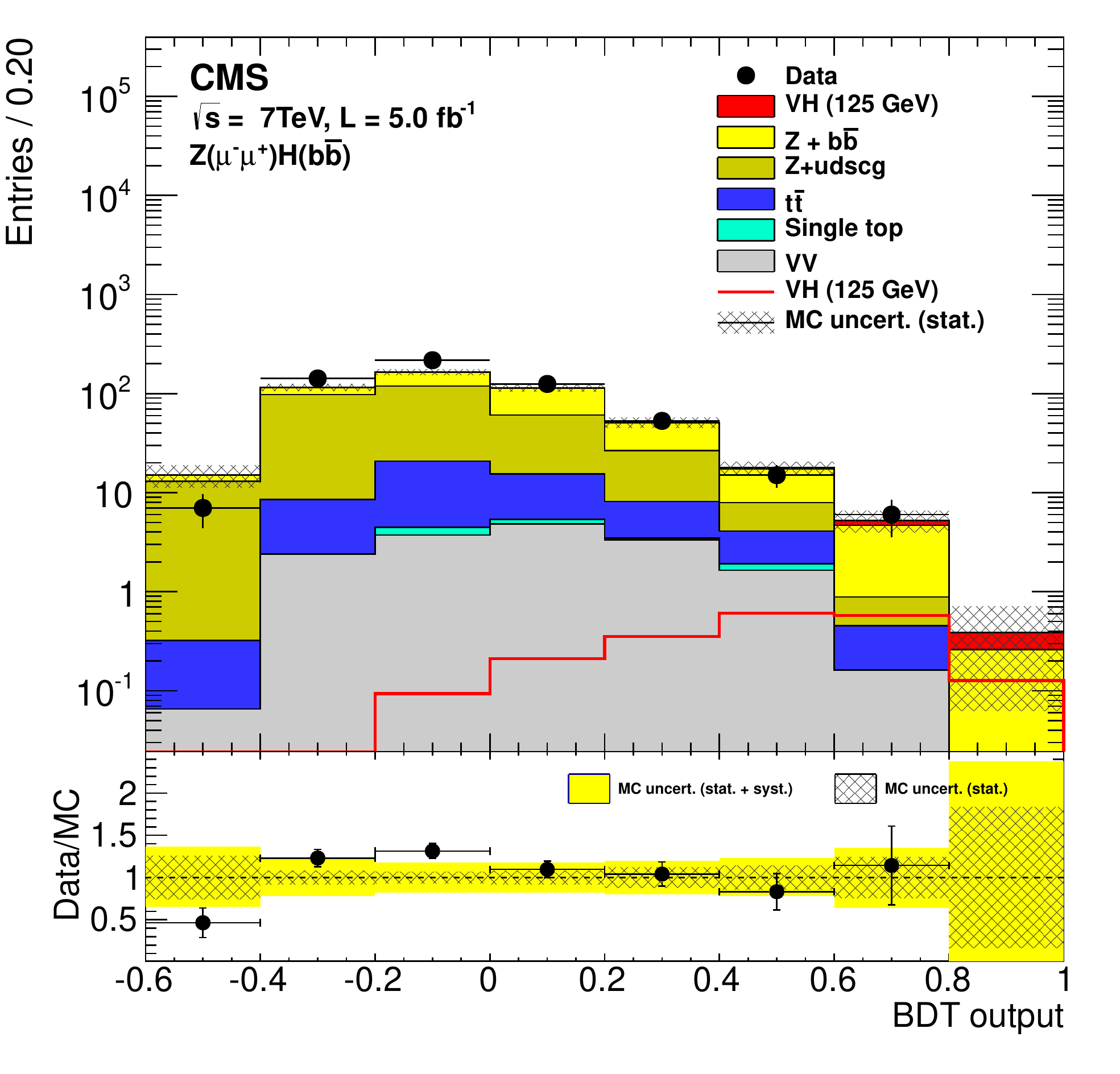}
    \includegraphics[width=0.49\textwidth]{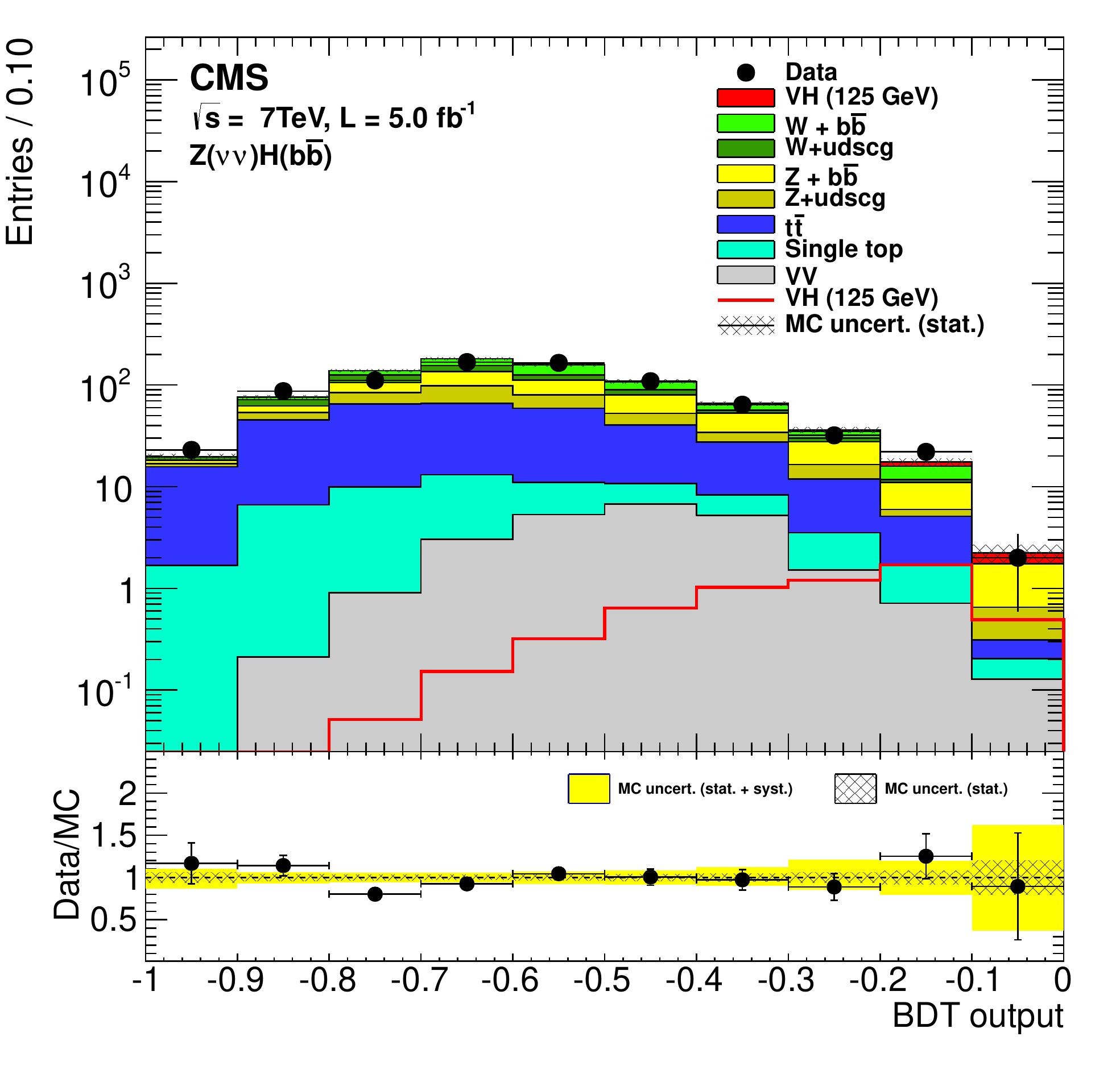}
    \includegraphics[width=0.49\textwidth]{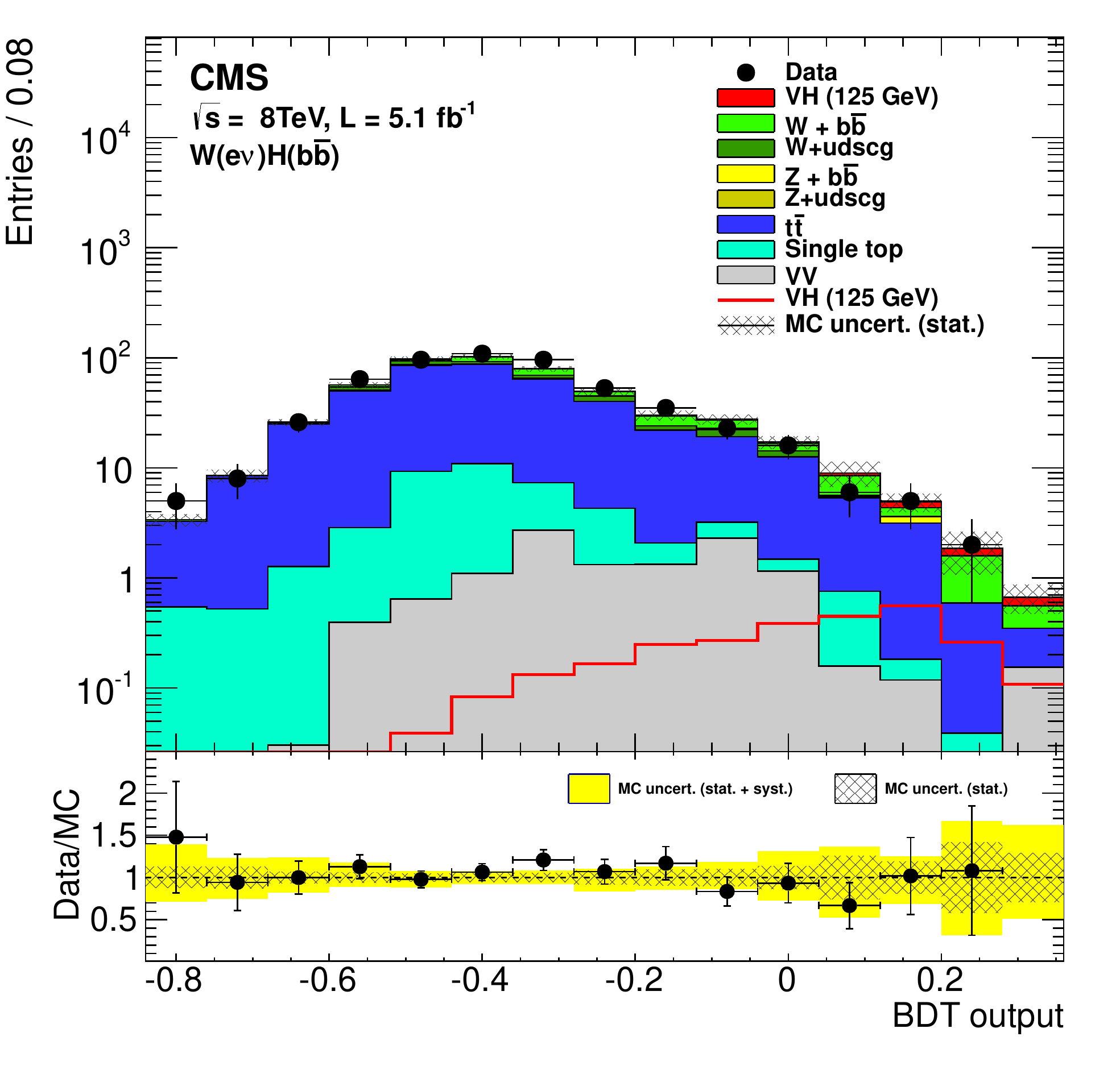}
    \includegraphics[width=0.49\textwidth]{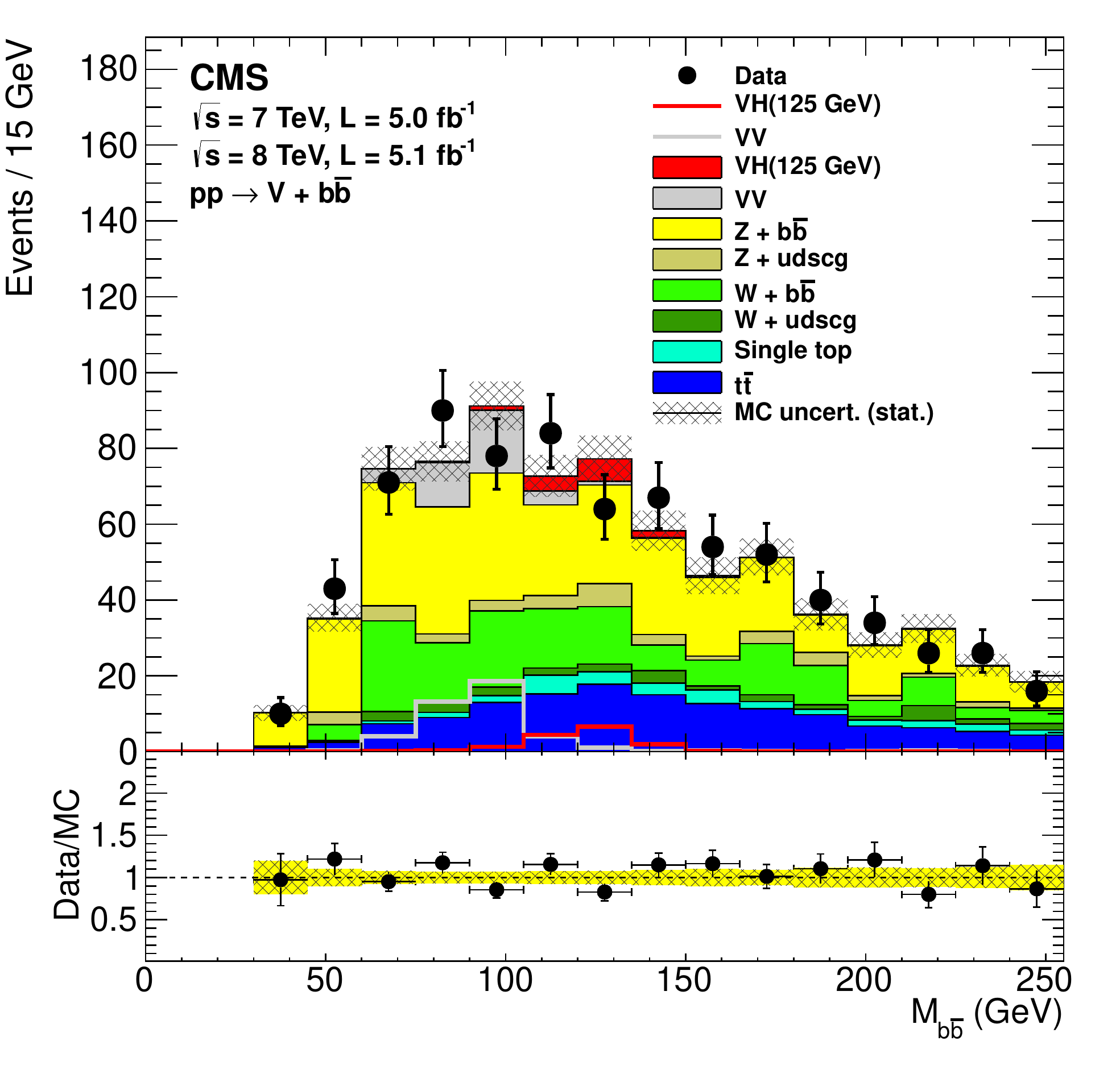}
    \caption{Example of BDT output distributions in the
    high \ptV\ bin, after all the selection criteria have been applied, for \ZmmH\ (top left),
       \ZnnH\ (top right), and \WenH\ (bottom left). Bottom right: the
       \cPqb-tagged dijet invariant-mass distribution from the combination of all
  VH channels for the combined  7 and 8\TeV
  data sets. Only events that pass
    a more restrictive selection are included (see text). For all figures the solid
    histograms show the signal and the various
      backgrounds, with the hatched region denoting the statistical
    uncertainties in the MC simulation. The data are represented by points
      with error bars. The VH signal is represented by a red line
      histogram. The ratio of the data to the sum of the
    expected background distributions is shown at the bottom of each
    figure.}
    \label{fig:Hbb_figs}
  \end{center}
\end{figure}

Tables~\ref{tab:LoPtBDTYields}--\ref{tab:HiPtBDTYields8TeV} summarize the expected
signal and background yields for both \ptV\ bins in each channel from the 7\TeV and 8\TeV data.
All the data/MC scale factors determined in
Section~\ref{sssec:hbb_Background_Control_Regions} have been applied to the corresponding
background yields. Examples of output BDT distributions, for the $\mH=125$\GeV training and
for the high \ptV\ bin, are shown in Figure~\ref{fig:Hbb_figs}. The
signal and background shapes and normalizations are those returned by
the fits. Figure~\ref{fig:Hbb_figs} also shows the
dijet invariant-mass distribution for the combination of all five channels in the
combined $7$ and $8\TeV$ data sets, using an event selection that is more
restrictive than the one used in the BDT analysis and that is more
suitable for a counting experiment in just this observable. The events considered
are those in the high \ptV\ bin with tighter b-tagging requirements on
both jets, and with requirements that there be no additional jets in
the events and that the azimuthal opening angle between the
dijet system and the reconstructed vector boson be large. The
$\PH\to\cPqb\cPqb$ search
with such a selection is significantly less sensitive
than the search using the BDT discriminant and it is therefore not
elaborated on further in this article.

The interpretation of the results from the BDT discriminant analysis, in terms of upper limits on the
Higgs boson production cross section, is given in Section 10.

\begin{table}[tbp]
\topcaption{Predicted signal and background yields (statistical
  uncertainty only) in the BDT output distribution for the low-\ptV\
  range  with the 7\TeV data for each of the five channels. The labels LF and HF refer to light- and heavy-flavour
  jets. The numbers in parentheses refer to the Higgs boson mass
  hypothesis in \GeVns{}.}
\label{tab:LoPtBDTYields}
\begin{center}
\scalebox{0.80}{
\begin{tabular}{lcccccc} \hline

Process & $\ZmmH$ & $\ZeeH$ & $\ZnnH$ & $\WmnH$ & $\WenH$ \\\hline\hline
Z+jets (LF) & $176 \pm 14$ & $255 \pm 18$ & $ 158.3 \pm 6.1 $ & $11.0 \pm 1.5$ & $1.87 \pm 0.56$   \\
Z+jets (HF) & $235 \pm 16$ & $225 \pm 16$ & $ 254.9 \pm 5.5 $ & $23.2 \pm 2.1$ & $2.71 \pm 0.68$   \\
W+jets (LF) & -- & -- & $ 133.1 \pm 8.1 $ & $124.6 \pm 4.6$ & $58.5 \pm 3.1$  \\
W+jets (HF) & -- & -- & $ 171.85 \pm 7.1 $ & $248.3 \pm 9.5$ & $135.3 \pm 7.0$  \\
\ttbar & $74.2 \pm 1.9$ & $64.3 \pm 1.7$ & $ 898.5 \pm 5.2 $ & $894.6 \pm 4.1$ & $575.5 \pm 3.3$ \\
Single Top & $3.73 \pm 0.72$ & $2.67 \pm 0.64$ & $ 98.5 \pm 5.9 $ & $123.1 \pm 3.0$ & $67.7 \pm 2.2$ \\
VV & $10.77 \pm 0.53$ & $10.07 \pm 0.55$ & $ 33.5 \pm 1.5 $ & $15.10 \pm 0.72$ & $7.89 \pm 0.54$ \\\hline
ZH(110) & $2.72 \pm 0.03$ & $2.19 \pm 0.03$ & $6.19\pm 0.05$ & $0.28\pm 0.02$ & $0.08\pm 0.01$ \\
WH(110) & -- & -- & $3.19\pm 0.04$ & $4.98\pm 0.08$ & $2.96\pm 0.06$    \\
ZH(115) & $2.34 \pm 0.03$ & $1.88 \pm 0.03 $ & $4.52\pm 0.05$ & $0.21\pm 0.01$ & $0.07\pm 0.01$ \\
WH(115) & -- & -- & $2.36\pm 0.03$ & $4.57\pm 0.07$ & $2.58\pm 0.05$    \\
ZH(120) & $1.93 \pm 0.02$ & $1.56 \pm 0.02$ & $4.10\pm 0.04$ & $0.19\pm 0.01$ & $0.07\pm 0.01$ \\
WH(120) & -- & -- & $2.15\pm 0.04$ & $3.90\pm 0.05$ & $2.17\pm 0.04$    \\
ZH(125) & $1.52 \pm 0.02$ & $1.23 \pm 0.02$ & $3.67\pm 0.04$ & $0.18\pm 0.01$ & $0.06\pm 0.01$ \\
WH(125) & -- & -- & $1.94\pm 0.04$ & $3.19\pm 0.04$ & $1.90\pm 0.03$    \\
ZH(130) & $1.19 \pm 0.01$ & $0.95 \pm 0.01$ & $2.81\pm 0.04$ & $0.15\pm 0.01$ & $0.05\pm 0.01$ \\
WH(130) & -- & -- & $1.25\pm 0.03$ & $2.56\pm 0.04$ & $1.50\pm 0.03$    \\
ZH(135) & $0.83 \pm 0.01$ & $0.67 \pm 0.01$ & $2.10\pm 0.02$ & $0.11\pm 0.01$ & $0.03\pm 0.01$ \\
WH(135) & -- & -- & $0.87\pm 0.02$ & $1.92\pm 0.02$ & $1.13\pm 0.02$    \\
\hline
Sum  &$500 \pm 22$ &$558 \pm 24$ & $1749 \pm 16$ & $1440\pm 12$ & $850\pm 9$ \\\hline
Data &    $ 493 $ & $512$ & $1793$ &  $1411$ & $925$  \\\hline

\end{tabular}
}
\end{center}
\end{table}

\begin{table}[tbp]
\topcaption{Predicted signal and background yields (statistical
  uncertainty only) in the BDT output distribution for the high-\ptV\
  range with the 7\TeV data for each of the five channels. The labels LF and HF refer to light- and heavy-flavour
  jets. The numbers in parentheses refer to the Higgs boson mass
  hypothesis in \GeVns{}.}
\label{tab:HiPtBDTYields}
\begin{center}
\scalebox{0.80}{
\begin{tabular}{lcccccc} \hline
Process & $\ZmmH$ &  $\ZeeH$ & $\ZnnH$ & $\WmnH$ & $\WenH$ \\\hline\hline
Z+jets (LF) & $291 \pm 15$ & $275 \pm 15$ & $ 107.7 \pm 3.1 $ & $3.47 \pm  0.79$ & $1.63 \pm 0.52$ \\
Z+jets (HF) & $180 \pm 11$ & $160 \pm 10$ & $ 117.0 \pm 4.6 $ & $6.7 \pm  1.2$ & $2.13 \pm 0.59$ \\
W+jets (LF) & -- & -- & $  81.4 \pm 3.8$ & $61.9 \pm 3.0$ & $41.4 \pm 2.5$ \\
W+jets (HF) & -- & -- & $ 171.7 \pm 5.9 $ & $129.5 \pm  6.1$ & $67.8 \pm 4.4$ \\
\ttbar & $41.7 \pm 1.4$ & $39.4 \pm 1.3$ & $ 275.7 \pm 3.0 $ & $302.4 \pm  2.3$ & $225.0 \pm 2.0$ \\
Single Top & $1.49 \pm 0.45$ & $3.44 \pm 0.71 $ & $ 37.9 \pm 3.4 $ & $60.8 \pm  2.1$ & $41.6 \pm 1.7$ \\
VV & $14.02 \pm 0.67$ & $11.68 \pm 0.60$ & $ 24.6 \pm 2.8 $ & $9.71 \pm  0.58$ & $6.28 \pm 0.47$  \\\hline
ZH(110) & $3.19 \pm 0.04 $ & $2.69 \pm 0.03$ & $5.75\pm 0.04$ & $0.14\pm 0.01$ & $0.07\pm 0.01$    \\
WH(110) & -- & -- & $1.88\pm 0.06$ & $4.39\pm 0.07$ & $3.18\pm 0.06$   \\
ZH(115) & $2.78 \pm 0.03$ & $2.37 \pm 0.027$ & $5.87\pm 0.05$ & $0.08\pm 0.01$ & $0.04\pm 0.01$   \\
WH(115) & -- & -- & $1.71\pm 0.05$ & $3.93\pm 0.06$ & $2.82\pm 0.05$   \\
ZH(120) & $2.41 \pm 0.02$ & $2.09 \pm 0.023$ & $5.15\pm 0.04$ & $0.10\pm 0.01$ & $0.06\pm 0.01$   \\
WH(120) & -- & -- & $1.42\pm 0.04$ & $3.57\pm 0.05$ & $2.51\pm 0.04$   \\
ZH(125) & $1.99 \pm 0.02$ & $1.67 \pm 0.02$ & $4.46\pm 0.04$ & $0.08\pm 0.01$ & $0.04\pm 0.01$ \\
WH(125) & -- & -- & $1.15\pm 0.03$ & $3.04\pm 0.04$ & $2.14\pm 0.04$   \\
ZH(130) & $1.58 \pm 0.02$ & $1.37 \pm 0.01$ & $3.54\pm 0.03$ & $0.06\pm 0.01$ & $0.04\pm 0.01$ \\
WH(130) & -- & -- & $0.70\pm 0.02$ & $2.51\pm 0.04$ & $1.83\pm 0.03$   \\
ZH(135) & $1.24 \pm 0.01$ & $1.03 \pm 0.01$ & $2.76\pm 0.02$ & $0.05\pm 0.01$ & $0.03\pm 0.01$ \\
WH(135) & -- & -- & $0.77\pm 0.02$ & $1.94\pm 0.03$ & $1.39\pm 0.02$   \\
\hline
Sum & $529 \pm 19$ & $490 \pm 18$ & $816 \pm 10$ & $575\pm 6$ & $386\pm 6$ \\\hline
Data &    $565$ &   $491$ & $783$ & $551$& $383$ \\\hline
\end{tabular}
}
\end{center}
\end{table}

\begin{table}[tbp]
\topcaption{Predicted signal and background yields (statistical
  uncertainty only) in the BDT output distribution for the low-\ptV\
  range with the 8\TeV data for each of the five channels. The labels LF and HF refer to light- and heavy-flavour
  jets. The numbers in parentheses refer to the Higgs boson mass
  hypothesis in \GeVns{}.}
\begin{center}
\scalebox{0.80}{
\begin{tabular}{lcccccc} \hline
Process & $\ZmmH$ & $\ZeeH$ & $\ZnnH$ & $\WmnH$ & $\WenH$   \\\hline\hline
Z+jets (LF) & $296 \pm 20$ & $254 \pm 23$ & $156.3 \pm 2.6$ & $13.7 \pm 2.5$ & $6.7 \pm 1.9$ \\
Z+jets (HV) & $250 \pm 15$ & $228 \pm 17$ & $355.1 \pm 4.7$ &  $21.7 \pm 2.9$ &   $8.5 \pm 2.0$ \\
W+jets (LF) & -- & -- & $202.6\pm 3.1$ &  $92.8 \pm 6.9$ &  $58.2\pm 5.7$ \\
W+jets (HV) & -- & -- & $384.6 \pm 5.0$ &   $177.7 \pm 14.1$ &    $102.4 \pm 10.7$ \\
\ttbar & $86.3 \pm 3.7$ & $75.7 \pm 3.6$ &  $1573 \pm 29$ &   $1308 \pm 15$ &  $970 \pm 13$ \\
Single Top & $5.4 \pm 1.9$ & $2.45 \pm 0.82$ & $102.2 \pm 2.2$ &   $64.3 \pm 5.2$ &  $49.6 \pm 4.8$ \\
VV & $13.7 \pm 1.1$ & $12.4 \pm 1.0$ & $48.3 \pm 2.5$ &   $19.0 \pm 1.8$ &  $13.0 \pm 1.8$ \\\hline
ZH(110) & $2.83 \pm 0.06$ & $2.21 \pm 0.05$ & $7.78\pm 0.02$ &  $0.31 \pm 0.02$ & $0.14 \pm 0.01$ \\
WH(110) & -- & --& $1.14\pm 0.02$& $4.87 \pm 0.17$ & $3.39\pm 0.14$ \\
ZH(115) & $2.37 \pm 0.05$ & $1.89 \pm 0.04$ & $6.64\pm 0.02$ &  $0.25 \pm 0.01$ &  $0.11 \pm 0.01$ \\
WH(115) & -- & -- & $1.11\pm 0.02$ & $4.73 \pm 0.15$ & $3.28 \pm 0.13$ \\
ZH(120) & $1.92 \pm 0.04$ & $1.54 \pm 0.03$ & $5.78\pm 0.04$ &  $0.23 \pm 0.01$ &  $0.10 \pm 0.01$ \\
WH(120) & -- & -- & $1.07\pm 0.02$ & $3.79 \pm 0.12$ & $2.59 \pm 0.10$ \\
ZH(125) & $1.52 \pm 0.03$ & $1.24 \pm 0.03$ & $4.39\pm 0.02$ &  $0.18 \pm 0.01$ &  $0.08 \pm 0.01$ \\
WH(125) & -- & -- & $0.95\pm 0.03$ & $3.19 \pm 0.10$ & $2.41 \pm 0.09$ \\
ZH(130) & $1.15 \pm 0.02$ & $0.92 \pm 0.02$ & $3.37\pm 0.04$ &  $0.15 \pm 0.01$ &  $0.05 \pm 0.01$ \\
WH(130) & -- & -- & $0.79\pm 0.03$ & $2.61 \pm 0.09$ & $1.85 \pm 0.08$ \\
ZH(135) & $0.83 \pm 0.02$ &   $0.65 \pm 0.02$ & $2.31\pm 0.03$ &  $0.11 \pm 0.01$ &  $0.04 \pm 0.01$ \\
WH(135) & -- & -- & $0.61\pm 0.02$ & $1.85 \pm 0.06$ & $1.40 \pm 0.05$ \\
\hline
Sum &$651 \pm 26$& $572 \pm 29$ & $2822 \pm 30$ & $1697 \pm 22$ & $1208 \pm 19$ \\\hline
Data &  $707$ &  $547$ & $2804$ & $1727$ &$1289$  \\ \hline
\end{tabular}
}
\end{center}
\end{table}

\begin{table}[tbp]
\topcaption{Predicted signal and background yields (statistical
  uncertainty only) in the BDT output distribution for the high-\ptV\
  range with the 8\TeV data for each of the five channels. The labels LF and HF refer to light- and heavy-flavour
  jets. The numbers in parentheses refer to the Higgs boson mass
  hypothesis in \GeVns{}.}
\label{tab:HiPtBDTYields8TeV}
\begin{center}
\scalebox{0.80}{
\begin{tabular}{lcccccc} \hline
Process & $\ZmmH$ & $\ZeeH$ & $\ZnnH$ & $\WmnH$ & $\WenH$ \\\hline\hline
Z+jets (LF) & $426 \pm 17$ & $353 \pm 16$ & $ 109.6 \pm 3.0 $ & $4.1 \pm 1.2$ & $1.33 \pm 0.41$\\
Z+jets (HF) & $238 \pm 11$ & $199 \pm 10$ & $ 182.0 \pm 3.6 $ & $6.3 \pm 1.4$& $3.17 \pm 0.99$  \\
W+jets (LF) & -- & -- & $ 79.0 \pm 2.8$ & $42.8 \pm 4.8$ & $32.2 \pm 4.4$ \\
W+jets (HV) & -- & -- & $ 97.4 \pm 4.9 $ & $64.4 \pm 7.1$ & $45.7 \pm 5.9$ \\
\ttbar & $55.0 \pm 3.0$ & $48.0 \pm 2.8$ & $ 488 \pm 16 $ & $458.8 \pm 8.1$ & $361.8 \pm 7.4$ \\
Single Top & $4.5 \pm 1.5$ & $5.9 \pm 2.2$ & $ 43.2 \pm 2.4 $ & $35.6 \pm 4.4$ & $28.6 \pm 4.1$   \\
VV & $16.5 \pm 1.3$ & $13.4 \pm 1.2$ & $ 34.8 \pm 1.8 $ & $16.1 \pm 1.7$ & $9.0 \pm 1.2$  \\\hline
ZH(110) & $3.66 \pm 0.06$ & $2.95 \pm 0.06$ & $8.05\pm 0.07$ &  $0.14 \pm 0.01$ &  $0.08 \pm 0.01$ \\
WH(110) & -- & -- & $2.63\pm 0.06$ & $4.49 \pm 0.16$ & $3.92 \pm 0.16$ \\
ZH(115) &  $3.17 \pm 0.05$ & $2.64 \pm 0.05$ & $6.81\pm 0.05$ &  $0.12 \pm 0.01$ &  $0.06 \pm 0.01$ \\
WH(115) &  -- & -- & $1.52\pm 0.05$ & $4.30 \pm 0.14$ & $3.52 \pm 0.13$ \\
ZH(120) &  $2.77 \pm 0.04$ & $2.26 \pm 0.04$ & $5.81\pm 0.04$ &  $0.10 \pm 0.01$ &  $0.06 \pm 0.01$ \\
WH(120) &  -- & -- & $1.00\pm 0.04$ & $3.86 \pm 0.12$ & $3.09 \pm 0.11$ \\
ZH(125) &  $2.31 \pm 0.04$ & $1.84 \pm 0.03$ & $5.40\pm 0.04$ &  $0.09 \pm 0.01$ &  $0.06 \pm 0.01$ \\
WH(125) &  -- & -- & $0.74\pm 0.03$ & $3.29 \pm 0.10$ & $2.67 \pm 0.09$ \\
ZH(130) &  $1.84 \pm 0.03$ & $1.53 \pm 0.03$ & $3.99\pm 0.03$ &  $0.07 \pm 0.01$ &  $0.04 \pm 0.01$ \\
WH(130) &  -- & -- & $0.70\pm 0.02$ & $2.56 \pm 0.09$ & $2.07 \pm 0.08$ \\
ZH(135) &  $1.39 \pm 0.02$ &  $1.16 \pm 0.02$ & $2.80\pm 0.02$ &  $0.06 \pm 0.01$ &  $0.03 \pm 0.01$ \\
WH(135) &  -- & -- & $0.67\pm 0.02$ & $2.00 \pm 0.06$ & $1.76 \pm 0.06$ \\
\hline
Sum &$740 \pm 20$&$620 \pm 19$ & $1034 \pm 18$ & $628\pm 13$ &$482\pm 11$\\\hline
Data &    $776$ &$635$ & $1045$ &  $689$ & $544$\\
\hline
\end{tabular}
}
\end{center}
\end{table}

\section{Combined results}\label{sec:results}

In this section, we present the results obtained by combining the measurements  from all
five search channels described above.
We begin with a short summary of the statistical method used to combine
the analyses.

\subsection{Combination methodology}
\label{sec:method}
Combining the Higgs boson search results
requires a simultaneous analysis of the data selected by the individual decay modes,
accounting for their correlations and for all the statistical and systematic uncertainties.
The statistical methodology used in this combination was developed
by the ATLAS and CMS Collaborations in the context of the LHC Higgs Combination Group.
A description of the general methodology can be found in Refs.~\cite{LHC-HCG-Report, Chatrchyan:2012tx}.
Results presented in this paper are obtained using asymptotic formulae from Ref.~\cite{Cowan:2010st}
and recent updates available in the \textsc{RooStats} package~\cite{RooStats}.
The Higgs boson mass is tested in steps accordant with the expected Higgs boson width 
and the experimental mass resolution~\cite{LHC-HCG-Report}.

\subsubsection{Characterizing the absence of a signal: limits}

For the calculation of exclusion limits, we adopt the modified frequentist
criterion $\CLs$~\cite{Junk:1999kv,Read1}. The chosen test statistic $q$, used to determine
how signal- or background-like the data are, is based on a  profile likelihood ratio.
Systematic uncertainties are incorporated via nuisance parameters and
are treated according to the frequentist paradigm, as described in Ref~\cite{LHC-HCG-Report}.
The profile likelihood ratio is defined as

\begin{equation}
q_{\mu} = - 2 \, \ln \frac {\mathcal{L}(\mathrm{obs} \, | \, \mu \cdot s + b, \, \hat \theta_{\mu} ) }
                           {\mathcal{L}(\mathrm{obs} \, | \, \hat \mu \cdot s + b, \, \hat \theta ) } ,
\end{equation}

where ``obs'' stands for the observed data;
$s$ stands for the number and distribution of signal events expected under the SM Higgs boson hypothesis;
$\mu$ is a signal-strength modifier,
introduced to accommodate deviations from the SM Higgs boson predictions;
$b$ is the number and distribution of background events;
$\mu \cdot s + b$ is the signal-plus-background hypothesis,
with the expected SM signal event yields $s$
multiplied by the signal-strength modifier $\mu$;
$\theta$ are nuisance parameters describing the systematic uncertainties.
The value $\hat \theta_{\mu}$ maximizes the likelihood in the numerator for a given $\mu$,
while $\hat \mu$ and $\hat \theta$ define the point at which the likelihood reaches its global maximum.

The ratio of the probabilities to observe a value of the test statistic
at least as large as the one observed in data, $q_{\mu}^{\mathrm{obs}}$,
under the signal+background ($\mu \cdot s + b$) and background-only ($b$) hypotheses,

\begin{equation}
\CLs (\mu) = \frac {\mathrm{P}(q_{\mu} \geq q_{\mu}^{\mathrm{obs}} \, | \, \mu \cdot s+b)}
             {\mathrm{P}(q_{\mu} \geq q_{\mu}^{\mathrm{obs}} \, | \, b )} \, \leq \alpha ,
\end{equation}

is used as the criterion for excluding the presence of a signal at the $1 - \alpha$ confidence level.

A signal with a cross section $\sigma = \mu \cdot
\sigma_{\mathrm{SM}}$ is defined to be excluded at 95\% CL
if $\CLs (\mu)\, \leq \, 0.05$. Here, $\sigma_{\mathrm{SM}}$ stands for the SM Higgs boson cross section.

\subsubsection{Characterizing an excess of events: p-values and significance}

To quantify the presence of an excess of events beyond what is expected for the background,
we use a test statistic:

\begin{equation}
\label{eq:method_q0}
 q_{0} = - 2 \, \ln \frac {\mathcal{L}(\mathrm{obs} \, | \, b, \, \hat \theta_{0} ) }
                       {\mathcal{L}(\mathrm{obs} \, | \, \hat \mu \cdot s + b, \, \hat \theta ) } ,
\end{equation}

where the likelihood in the numerator is for the background-only hypothesis.
The local statistical significance $Z_\text{local}$ of a signal-like excess is computed from the probability $p_0$

\begin{equation}
p_0 = \mathrm{P}(q_0 \geq q_0^\text{obs} \, | \, b),
\end{equation}

henceforth referred to as the local $p$-value, using the one-sided Gaussian-tail convention:

\begin{equation}
\label{eq:Z}
p_0 = \int_{Z_{\text{local}}}^{+\infty} \frac{1}{\sqrt{2\pi}} \exp(-x^2/2) \, \rd x.
\end{equation}

In the Higgs boson search, we scan over the Higgs boson mass hypotheses and
find the value giving the minimum local $p$-value $p_{\text{local}}^{\text{min}}$,
which describes the probability of a background fluctuation for that particular Higgs boson mass
hypothesis. The probability to find a fluctuation with a local $p$-value lower or equal to the observed
$p_{\text{local}}^{\text{min}}$ anywhere in the explored mass range
is referred to as the global $p$-value, $p_{\text{global}}$:

\begin{equation}
p_{\mathrm{global}}= \mathrm{P}(p_0 \leq p_{\text{local}}^{\text{min}} \, | \, b).
\end{equation}

The fact that the global $p$-value can be significantly larger than $p_{\text{local}}^{\text{min}}$
is often referred to as the ``look-elsewhere effect'' (LEE).
The global significance (and global $p$-value) of an observed excess can be evaluated
following the method described in Ref.~\cite{LEE}, using:

\begin{equation}
\label{eq:LEE1}
p_{\text{global}}= p_{\text{local}}^{\text{min}} \, + \, C \cdot \re^{ - Z^2_{\text{local}} / 2 }.
\end{equation}

The constant $C$ is found by generating a set of pseudo-experiments and using it to
evaluate the global $p$-value corresponding to the $p_{\mathrm{local}}^{\mathrm{min}}$
value observed in the data. Pseudo-experiments are a simulated outcome of an experiment obtained by randomly varying the
average expected event yields and their distributions according to a specified model of statistical and
systematic uncertainties.
For example, a Poisson distribution is used to model statistical variations,
while a Gaussian distribution is used to describe the systematic uncertainties.

\subsubsection{Extracting signal-model parameters}

The values of a set of signal-model parameters $a$ (the signal-strength modifier $\mu$ is one of them)
are evaluated from a scan of the profile likelihood ratio $q(a)$:

\begin{equation}
q(a) = - 2 \, \ln \frac {\mathcal{L}(\text{obs} \, | \, s(a) + b,      \, \hat \theta_{a} ) }
                      {\mathcal{L}(\text{obs} \, | \, s(\hat a) + b, \, \hat \theta ) } .
\end{equation}

The values of the parameters $\hat a$ and $\hat \theta$
that maximize the likelihood
$\mathcal{L}(\text{obs} \, | \, s(\hat a) + b, \, \hat \theta )$,
are called the best-fit set.
The 68\%~(95\%)~CL interval for a given signal-model parameter $a_i$ is evaluated from $q(a_i)=1$~(3.84),
with all other unconstrained model parameters treated as nuisance parameters.
The two-dimensional (2D) 68\%~(95\%)~CL contours for pairs of signal-model parameters $a_i,\, a_j$ are derived
from $q(a_i, a_j) = 2.3$~(6.0).
Note that the boundaries of the 2D confidence-level region projected onto
either parameter axis are not identical to the one-dimensional (1D) confidence intervals for this parameter.

\subsection{Exclusion limits on the SM Higgs boson}

\subsubsection{Results of searches in the five decay modes}
\label{sec:SubchannelLimits}

\begin{figure}[htbp]
    \begin{center}
      \includegraphics[width=0.49\linewidth]{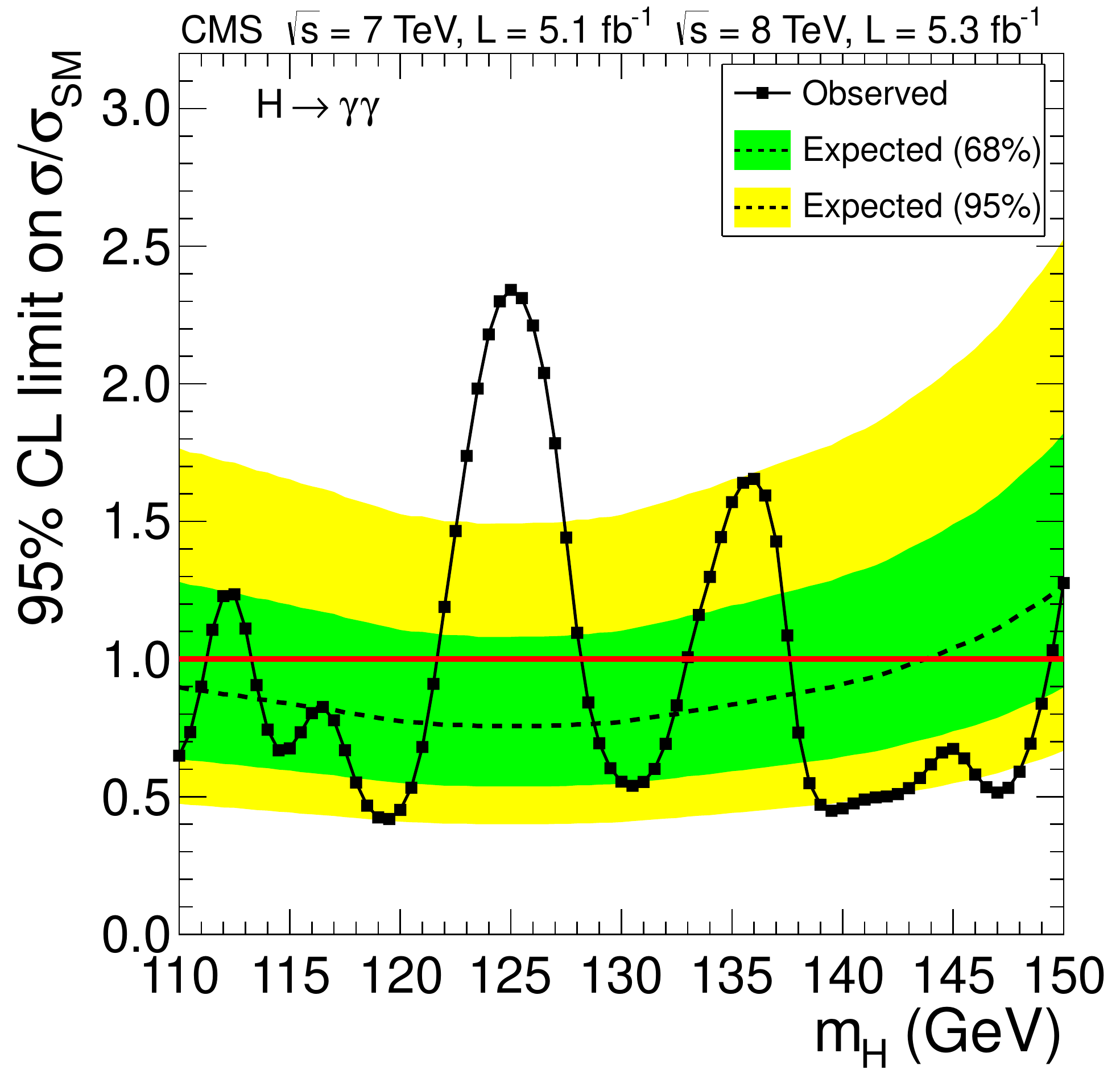}  \\
     \includegraphics[width=0.49\linewidth]{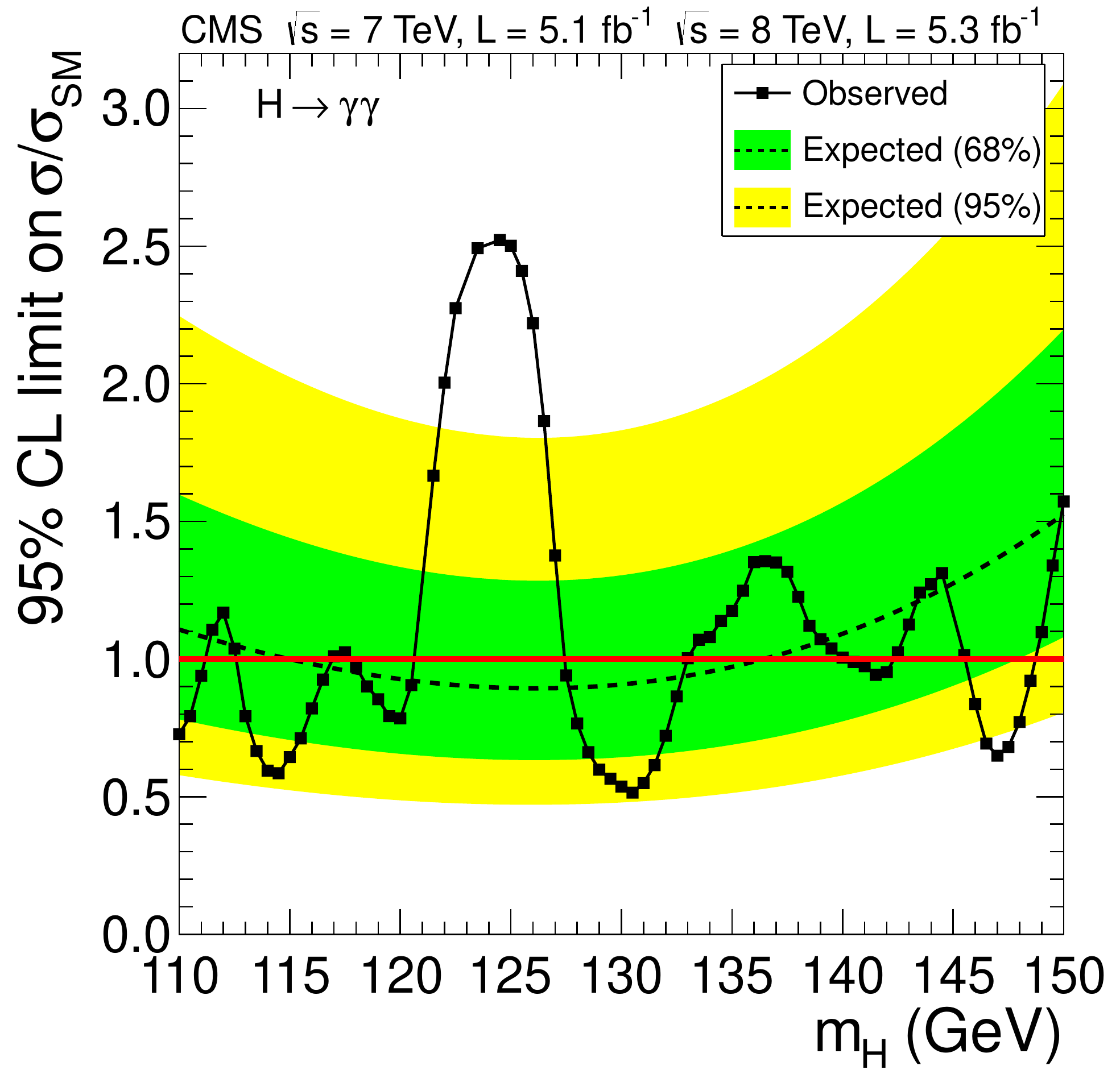}
      \includegraphics[width=0.49\linewidth]{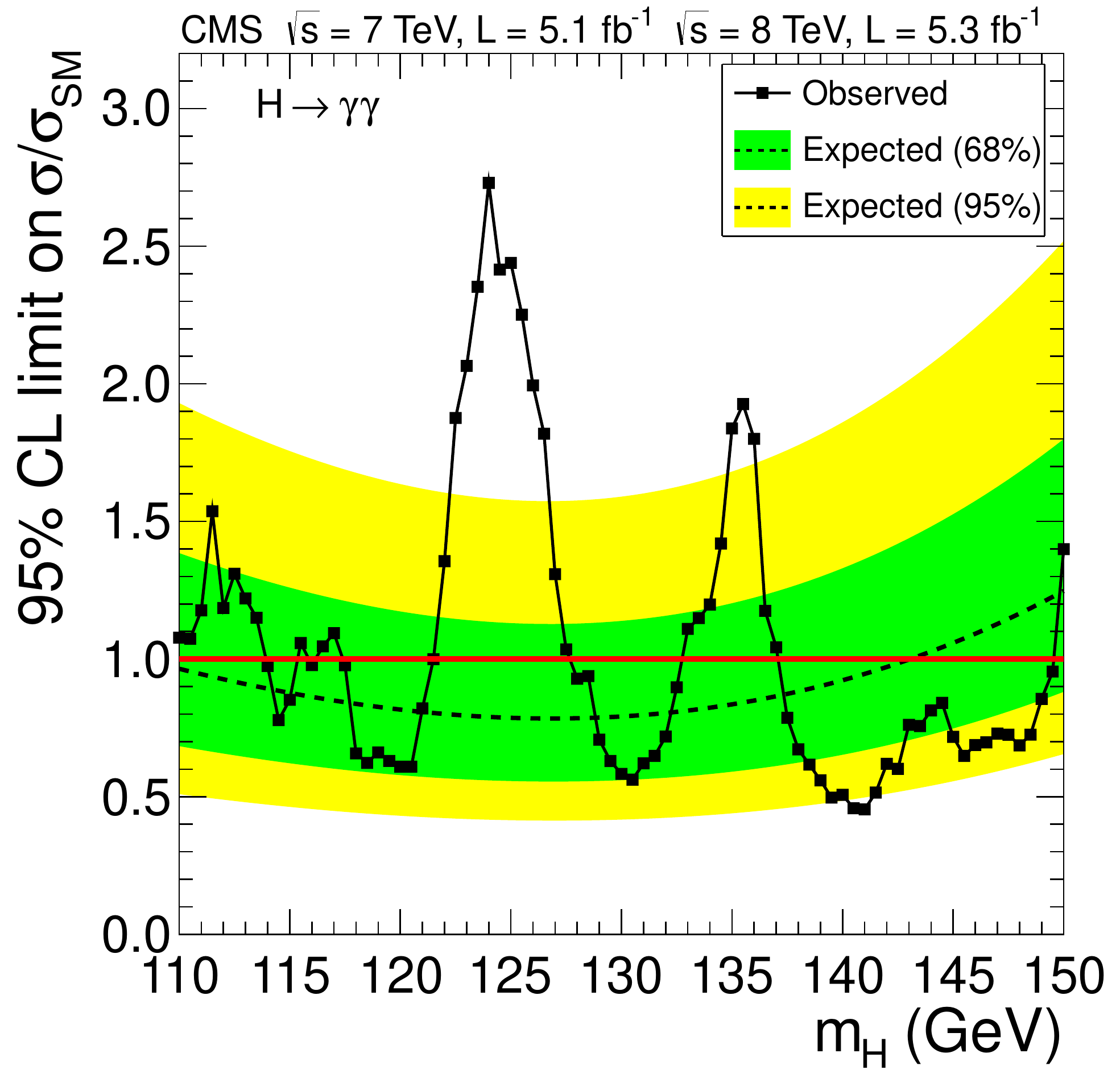}
     \caption{The 95\% CL upper limits
on the production cross section of a Higgs boson
expressed in units of the SM Higgs boson production cross section,
$\sigma / \sigma_\text{SM}$,
     as obtained in the $\PH \to \Pgg\Pgg$ search channel for
     (top) the baseline analysis,
     (lower left) the cut-based analysis, and
     (lower right) the sideband analysis.
     The solid lines represent the observed limits;
     the background-only hypotheses are represented by their median (dashed
     lines) and by their 68\% (dark) and 95\% (light) CL bands.
       }
      \label{fig:LimitGaGa}
    \end{center}
\end{figure}

\begin{figure}[htbp]
    \begin{center}
 \includegraphics[width=0.49\linewidth]{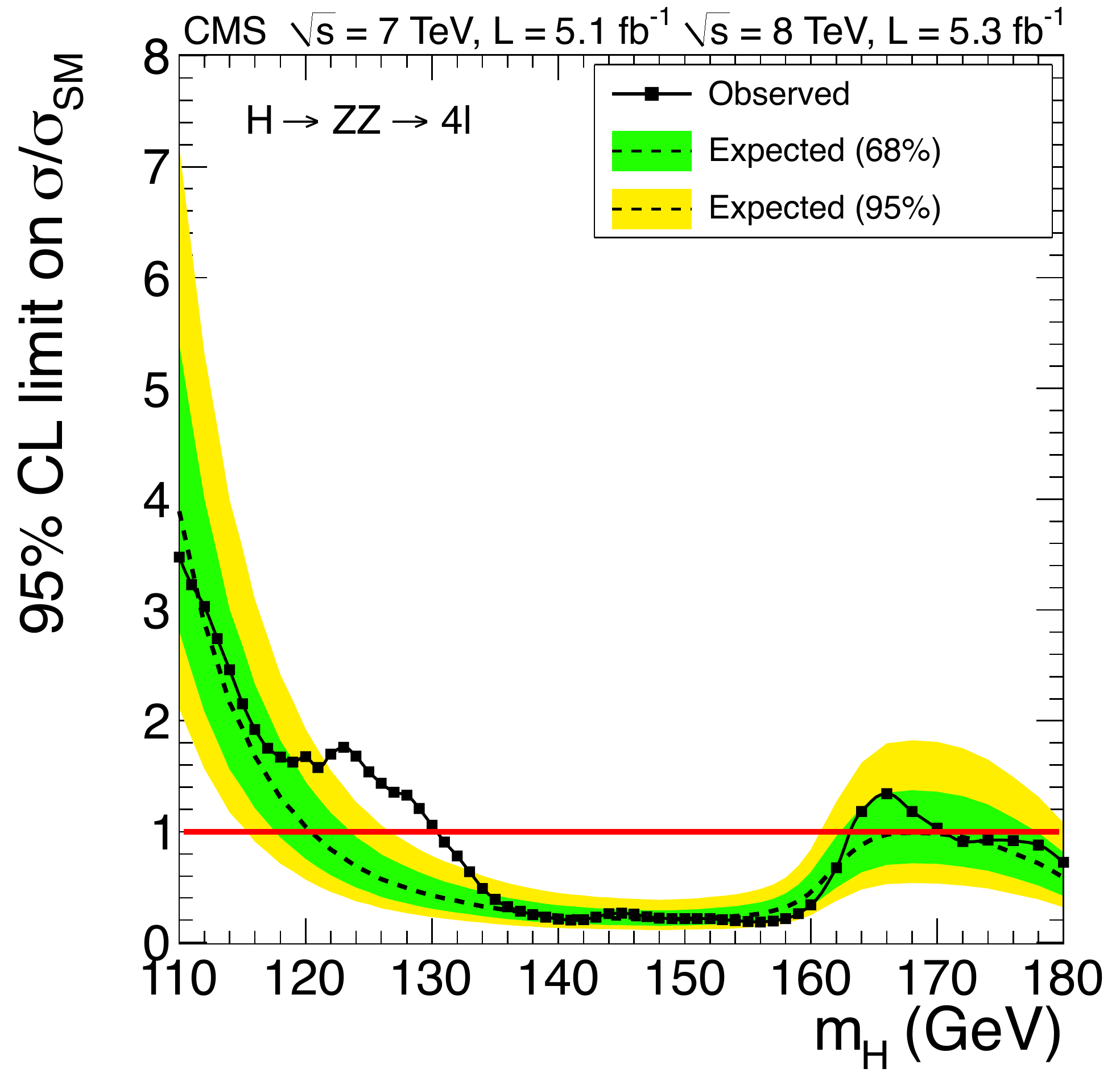} \includegraphics[width=0.49\textwidth]{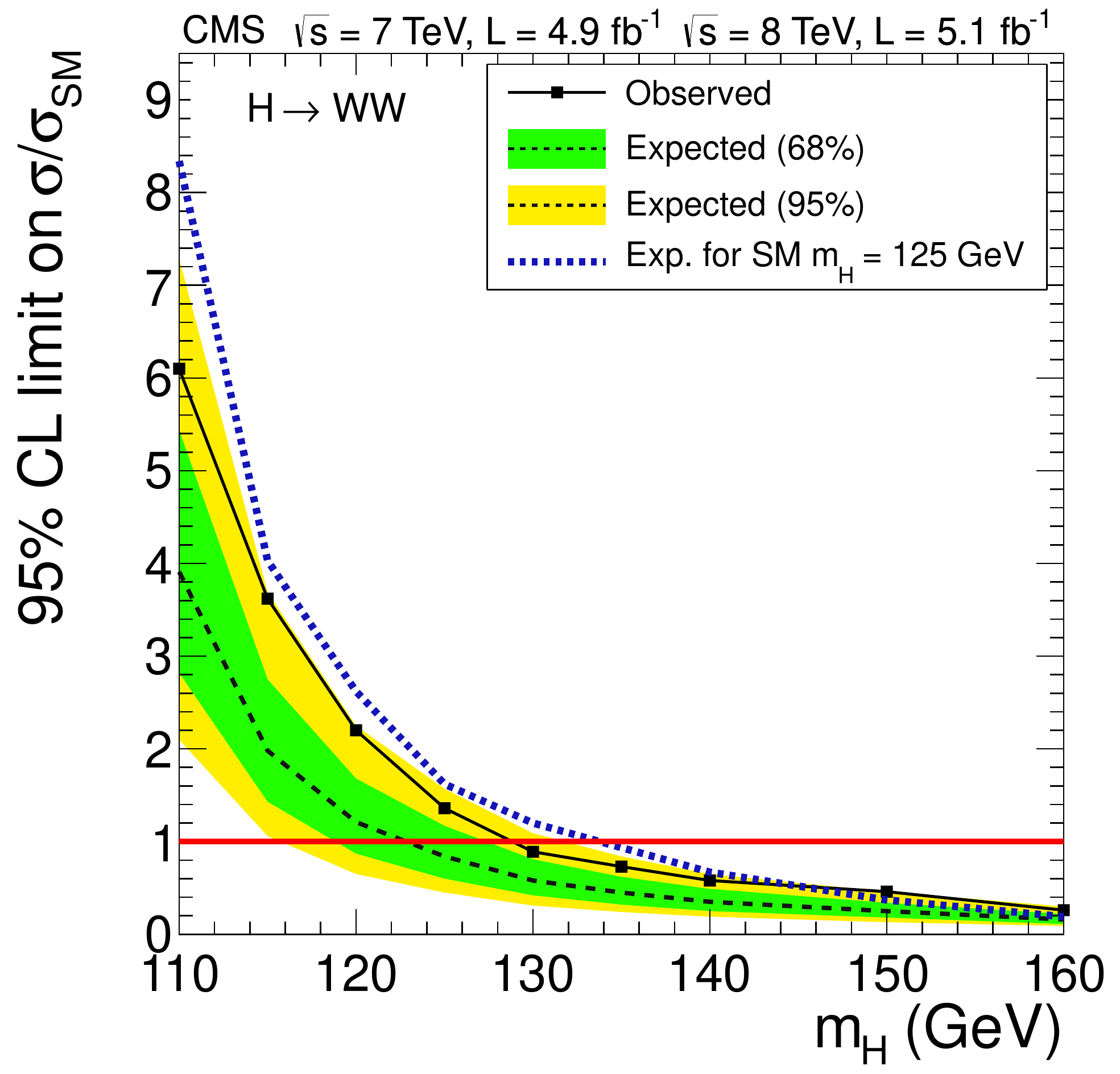} \\
 \includegraphics[width=0.49\textwidth]{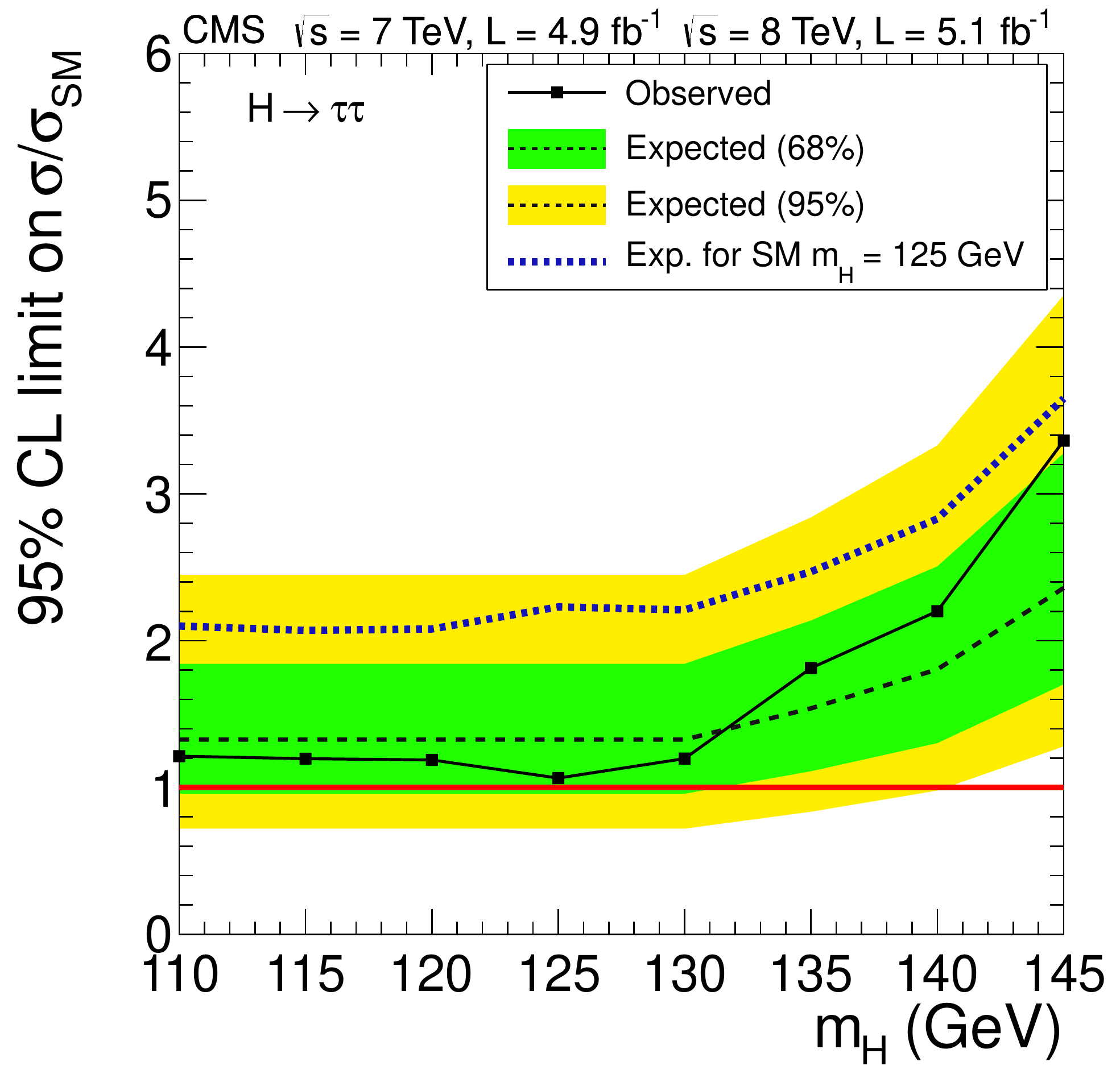}
 \includegraphics[width=0.49\textwidth]{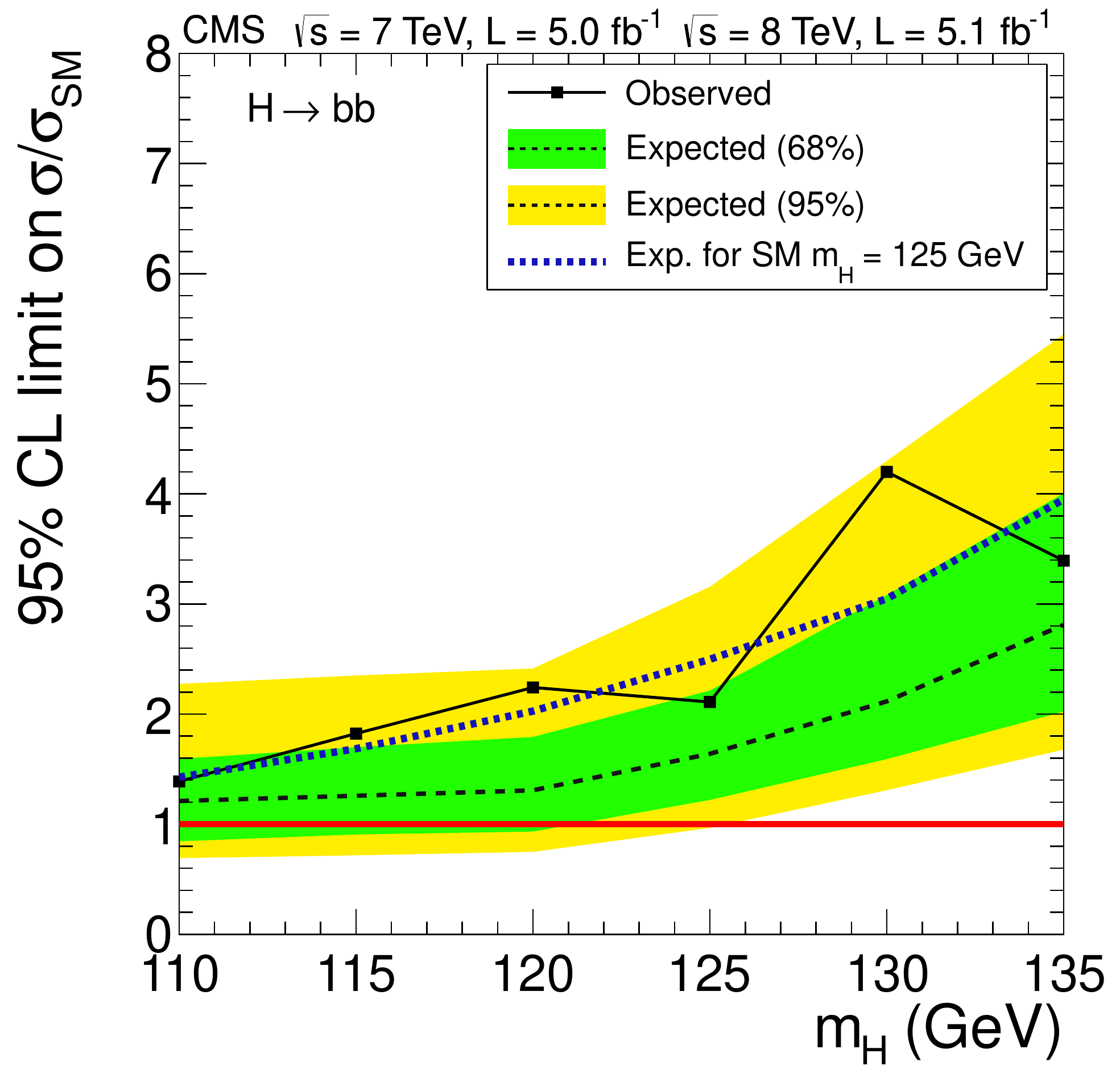}
     \caption{The 95\% CL upper limits
on the production cross section of a Higgs boson
expressed in units of the SM Higgs boson production cross section,
$\sigma / \sigma_\text{SM}$,
for the following search modes:
     (upper left) $\PH \to \cPZ\cPZ \to 4\ell$,
     (upper right) $\PH \to \PW\PW$,
     (lower left) $\PH \to \Pgt\Pgt$, and
     (lower right) $\PH \to \cPqb\cPqb$.
     The solid lines represent the observed limits;
     the background-only hypotheses are represented by their median
 (dashed lines) and by the 68\% and 95\% CL bands.
     The signal-plus-background expectation (dotted lines) from a Higgs boson
     with mass $\mH = 125$\GeV is also shown for the final states
     with a poor mass resolution, $\PW\PW$, $\tau\tau$, and $\cPqb\cPqb$.
     }
      \label{fig:Limit}
    \end{center}
\end{figure}

Figures~\ref{fig:LimitGaGa} and \ref{fig:Limit}
show the 95\% CL upper limits on the signal-strength modifier,
$\mu = \sigma / \sigma_{\mathrm{SM}}$,
as a function of $\mH$ for the five decay modes:
$\Pgg\Pgg$, $\cPZ\cPZ$, $\PW\PW$, $\Pgt\Pgt$, and $\cPqb\cPqb$.
The observed values are shown by the solid lines.
The SM Higgs boson mass regions where the line is below $\sigma / \sigma_{\mathrm{SM}}= 1$ are excluded at 95\% CL.
The dashed lines indicate the median of the expected results for
the background-only hypothesis. The dark  and light bands
indicate the ranges in which
the observed results are expected to reside in
68\% and 95\% of the experiments, should multiple experiments be performed under the background-only hypothesis.
The probabilities for an observation to lie above and below the 68\% (95\%) bands are each 16\% (2.5\%).

In the $\PH \to \Pgg\Pgg$ analysis,
the SM Higgs boson signal is searched for in a simultaneous statistical analysis of the diphoton
invariant-mass distributions for the eleven exclusive event classes:
five classes (four untagged and one VBF-tagged) for the 7\TeV data and
six classes (four untagged and two VBF-tagged) for the 8\TeV data, as described in Section~\ref{sec:hgg}.
Figure~\ref{fig:LimitGaGa} shows the 95\% CL upper limits on
the Higgs boson production cross section obtained in
(a) the \emph{baseline} analysis and the two alternatives analyses:
(b) the \emph{cut-based} analysis and
(c) the \emph{sideband} analysis.
The observed limits in the sideband analysis [Fig.~\ref{fig:LimitGaGa} (lower right)]
are not smooth because, when changing the mass hypothesis, the event class boundaries move as well.
This is true for the ${\pm}2\%$ signal window and each sideband window.
This leads to events moving in and out of the classes in a discrete manner.
Figure~\ref{fig:LimitGaGa} (top) shows that the $\PH \to \Pgg\Pgg$ search has reached
the sensitivity for excluding the SM Higgs boson at 95\% CL
in the mass range 110--144\GeV,
while the observed data exclude it in the following three mass ranges:
113--122\GeV, 128--133\GeV, and 138--149\GeV.
All three diphoton analyses give observed exclusion limits near $\mH=125\GeV$
that are much weaker than the expected for the background-only hypothesis,
which implies a significant excess of events with diphoton masses around 125\GeV.
The consistency of the results obtained with the three alternative approaches
confirms
the robustness of the measurement.

In the $\PH \rightarrow  \cPZ\cPZ   \rightarrow 4\ell$ analysis,
the SM Higgs boson signal is searched for in a simultaneous statistical analysis of six 2D distributions
of the four-lepton invariant mass $m_{4\ell}$ and the matrix-element-based kinematic discriminant $K_D$,
as described in Section~\ref{sec:hzz4l}.
The six distributions correspond to the three lepton final states ($4\Pe$, $4\mu$, $2\Pe2\mu$) and
the 7 and 8\TeV data sets.
Figure~\ref{fig:Limit} (upper left) shows the 95\% CL upper limits
on the Higgs boson production cross section. The $\PH \to \cPZ\cPZ  \to 4\ell$ search has reached
the sensitivity for excluding the SM Higgs boson at 95\% CL
in the mass range 120--180\GeV,
while the observed data exclude it in the following two mass ranges:
130--164\GeV and 170--180\GeV.
The observed exclusion limits for $\mH=120$--$130\GeV$
are much weaker than the expected limits for the background-only hypothesis,
suggesting a significant excess of four-lepton events in this mass range.
As a cross-check, the statistical analysis using only the  $m_{4\ell}$ distributions has been performed.
The results are found to be consistent with the 2D analysis, although with less sensitivity.

In the $\PH \to \PW\PW \to \ell\nu\ell\nu$ analysis,
the SM Higgs boson signal is searched for in a simultaneous statistical analysis
of eleven exclusive final states:
same-flavour ($\Pep\Pem$ and $\mu^+\mu^-$) dilepton events with 0 and 1 jet for the 7 and~8\TeV data sets,
different-flavour $\Pe^{\pm}\mu^{\mp}$ dilepton events with 0 and 1 jet for the 7 and~8\TeV data sets,
dilepton events in the VBF-tag category for the 7\TeV data set,
and same-flavour and different-flavour dilepton events in the VBF-tag category for the 8\TeV data set.
All analysis details can be found in Section~\ref{sec:hww2l2nu}.
Figure~\ref{fig:Limit} (upper right) shows the 95\% CL upper limits
on the Higgs boson production cross section. The $\PH \to \PW\PW \to \ell\nu\ell\nu$ search has reached
a sensitivity for excluding the SM Higgs boson at 95\% CL
in the mass range 122--160\GeV (the higher-mass range is not discussed in this paper),
while the observed data exclude it in the mass range 129--160\GeV.
The observed exclusion limits are weaker than the expected ones
for the background-only hypothesis in the entire mass range,
suggesting an excess of events in data.
However, given the mass resolution of about 20\% in this channel, owing to the presence of the two undetectable neutrinos,
a broad excess is observed across the mass range from 110 to about 130\GeV.
The dotted line in Fig.~\ref{fig:Limit} (b) indicates the median expected exclusion limits
in the presence of a SM Higgs boson with a mass near 125\GeV. The observed limits in this channel
are consistent with the expectation for a SM Higgs boson of 125\GeV.

In the $\PH \to \Pgt\Pgt$ channel, the 0-, 1-jet, and VBF
categories are used to set 95\% CL upper limits
on the Higgs boson production. The ditau system is reconstructed in four final states:
$\Pe\tau_{\mathrm{h}}$, $\mu\tau_{\mathrm{h}}$, $\Pe\mu$, $\mu\mu$, where the leptons come
from $\tau \to \Pe\nu\nu$ or $\tau \to \mu\nu\nu$
decays.
The 0- and 1-jet categories are further split into two categories of low or
high ditau transverse momentum.
The 7 and 8\TeV data are treated independently giving a total of
40 ditau mass distributions.
All analysis details can be found in Section~\ref{sec:htt}.
Figure~\ref{fig:Limit} (lower left) shows the 95\% CL upper limits
on the Higgs boson production cross section in this channel.
The $\PH \to \Pgt\Pgt$ search
has not yet reached the SM Higgs boson exclusion sensitivity;
the expected limits on the signal event rates are
1.3--2.4 times larger than the event rates expected for the SM Higgs boson in this channel.

In the $\PH \to \cPqb\cPqb$ analysis, five final states are considered:
two $\cPqb$-tagged jets with \met\ ($\cPZ \to \nu\nu$), $\Pep\Pem$,
$\mu^+\mu^-$ ($\cPZ \to \ell^+\ell^-$), $\Pe + \met$, and $\mu + \met$($\PW \to \ell\nu$).
Each of these categories is further split into two categories of low or high $\cPqb\cPqb$ transverse momentum.
The 7 and~8\TeV data are treated independently giving a total of
20 BDT-output distributions. All analysis details can be found in Section~\ref{sec:hbb}.
Figure~\ref{fig:Limit} (lower right) shows the 95\% upper CL limits
on the Higgs boson production cross section in this channel.
The $\PH \to \cPqb\cPqb$ search
has not yet reached the SM Higgs boson exclusion sensitivity;
the expected limits on the signal event rates are
1.2--2.8 times larger than the event rates expected for the SM Higgs boson in this channel.

\subsubsection{Combined results}

\begin{figure*} [b]
\centering
\includegraphics[width=0.49\textwidth]{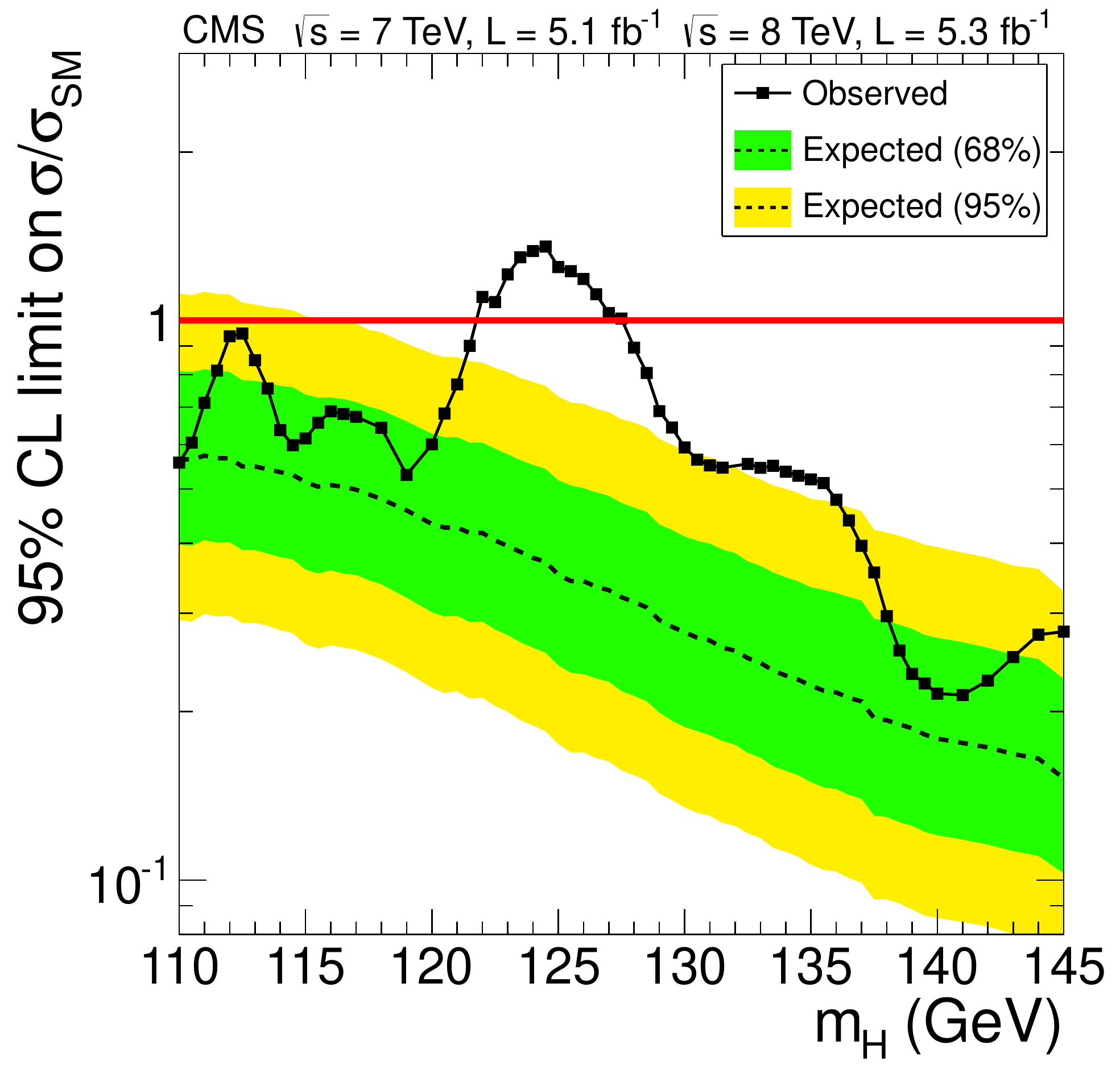} \hfill
\includegraphics[width=0.49\textwidth]{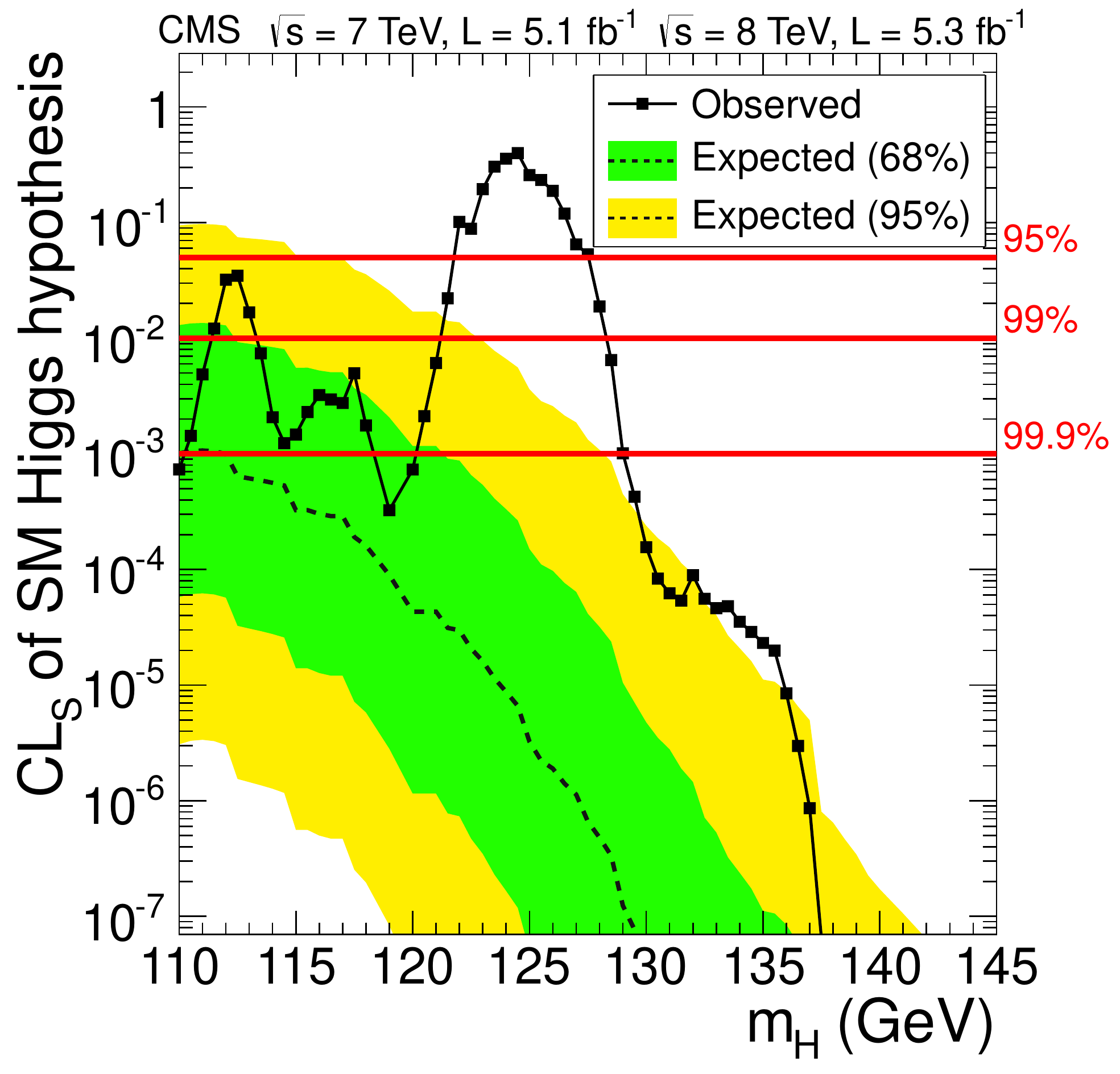}
\caption{ The 95\% CL upper limits on
the production cross section of a Higgs boson
expressed in units of the SM Higgs boson production cross section,
$\sigma / \sigma_\text{SM}$, (left)
and the $\CLs$ values (right) for the SM Higgs boson hypothesis,
as a function of the Higgs boson mass for the five decay modes and the 7 and 8\TeV data sample combined.
The solid lines represent the observed limits;
the background-only hypotheses are represented by their median
(dashed lines) and by the 68\% and 95\% CL bands.
The three horizontal lines on the right plot show the $\CLs$
values 0.05, 0.01, and 0.001,
corresponding to 95\%, 99\%, and 99.9\% confidence levels, defined as $(1-\CLs)$.
    }
\label{fig:CLsMu95}
\end{figure*}

The five individual search channels described above are combined
into a single search for the SM Higgs boson.
Figure~\ref{fig:CLsMu95} (left) shows the 95\% CL upper limits on the signal-strength modifier,
$\mu = \sigma / \sigma_{\mathrm{SM}}$,
as a function of $\mH$.
We exclude a SM Higgs boson at 95\% CL in two mass ranges:
110--\ObsNFL\GeV and \ObsNFH--\MHmax\GeV.

The $\CLs$ value for the SM Higgs boson hypothesis as a function of its mass
is shown in Fig.~\ref{fig:CLsMu95} (right).
The horizontal lines indicate $\CLs$ values of 0.05, 0.01, and 0.001.
The mass regions where the observed $\CLs$ values are below these lines are excluded
with the corresponding ($1-\CLs$) confidence levels of 95\%, 99\%, and 99.9\%, respectively.
The 95\% CL exclusion range for the SM Higgs boson is identical to that shown
in Fig.~\ref{fig:CLsMu95} (left), as both results
are simply different representations of the same underlying information.
At 99\% CL, we exclude the SM Higgs boson in three mass ranges:
\ObsOneNNL --\ObsOneNNH\GeV, \ObsTwoNNL --\ObsTwoNNH\GeV, and \ObsThreeNNL --\ObsThreeNNH\GeV.

Figure~\ref{fig:CLsMu95} (right) shows that,
in the absence of a signal, we would expect to exclude
the entire $m_{\PH}$ range of 110--145\GeV at the 99.9\% CL or higher.
In most of the Higgs boson mass range,
the differences between the observed and expected limits are consistent
since the observed limits are generally within the
68\% or 95\% bands of the expected limit values.
However, in the range $\ObsNFL < m_{\PH}< \ObsNFH$\GeV, we observe an excess of events,
making the observed limits considerably weaker than expected in the absence of the SM Higgs boson
and, hence, not allowing the exclusion of the SM Higgs boson in this range.

\subsection{Significance of the observed excess}

\subsubsection{Results of searches in the $\PH \to \Pgg\Pgg$ and $\PH \to \cPZ\cPZ \to 4\ell$ decay  modes}
\label{sec:SubchannelSignificance}

As presented in Section~\ref{sec:SubchannelLimits},
the searches for the SM Higgs boson in the $\Pgg\Pgg$ and $ \cPZ\cPZ \to 4\ell$ modes
reveal a substantial excess of events with diphoton and four-lepton invariant masses near 125\GeV.

\begin{figure}[htbp]
    \begin{center}
      \includegraphics[width=0.49\linewidth]{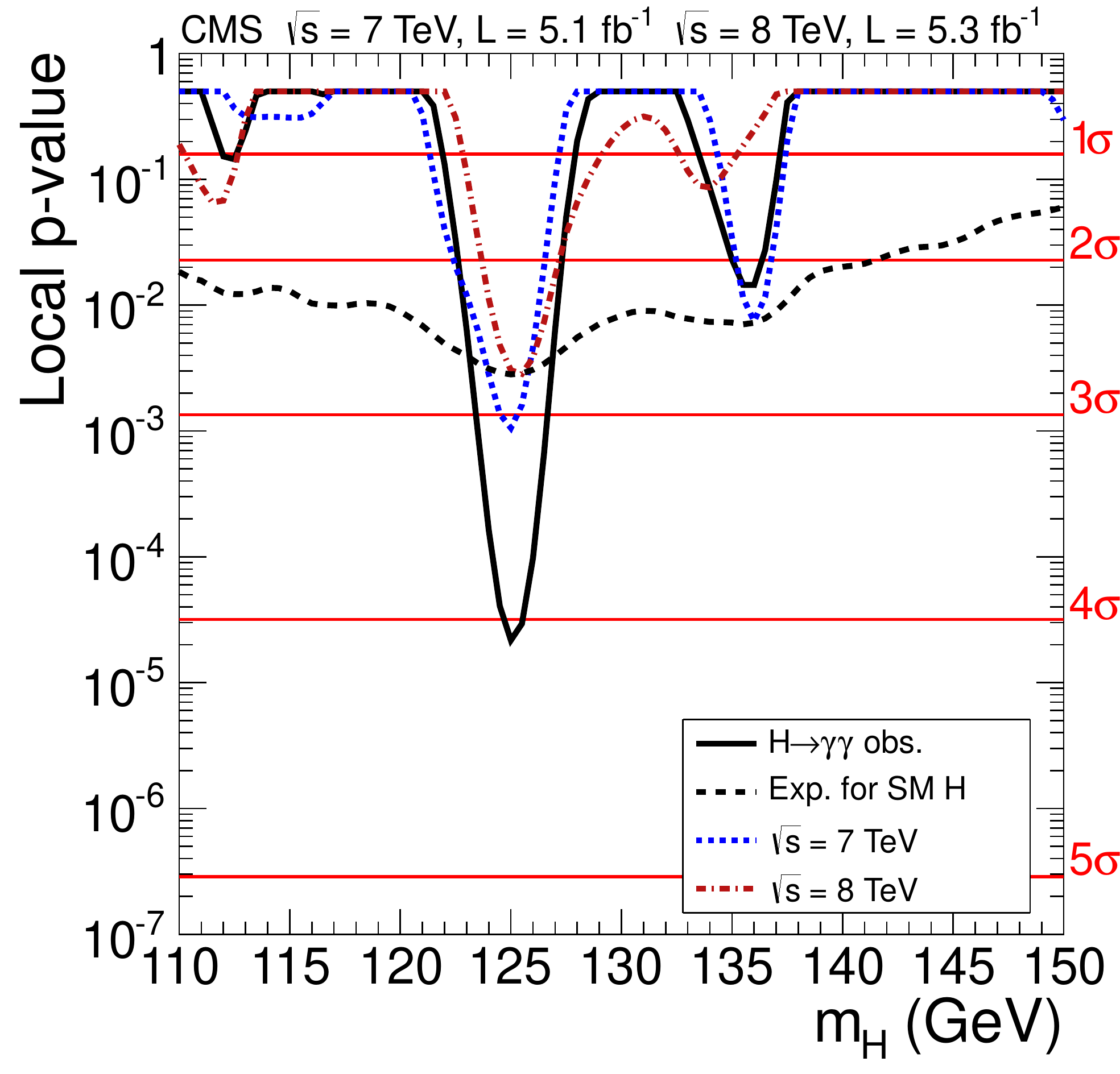}  \\
      \includegraphics[width=0.49\linewidth]{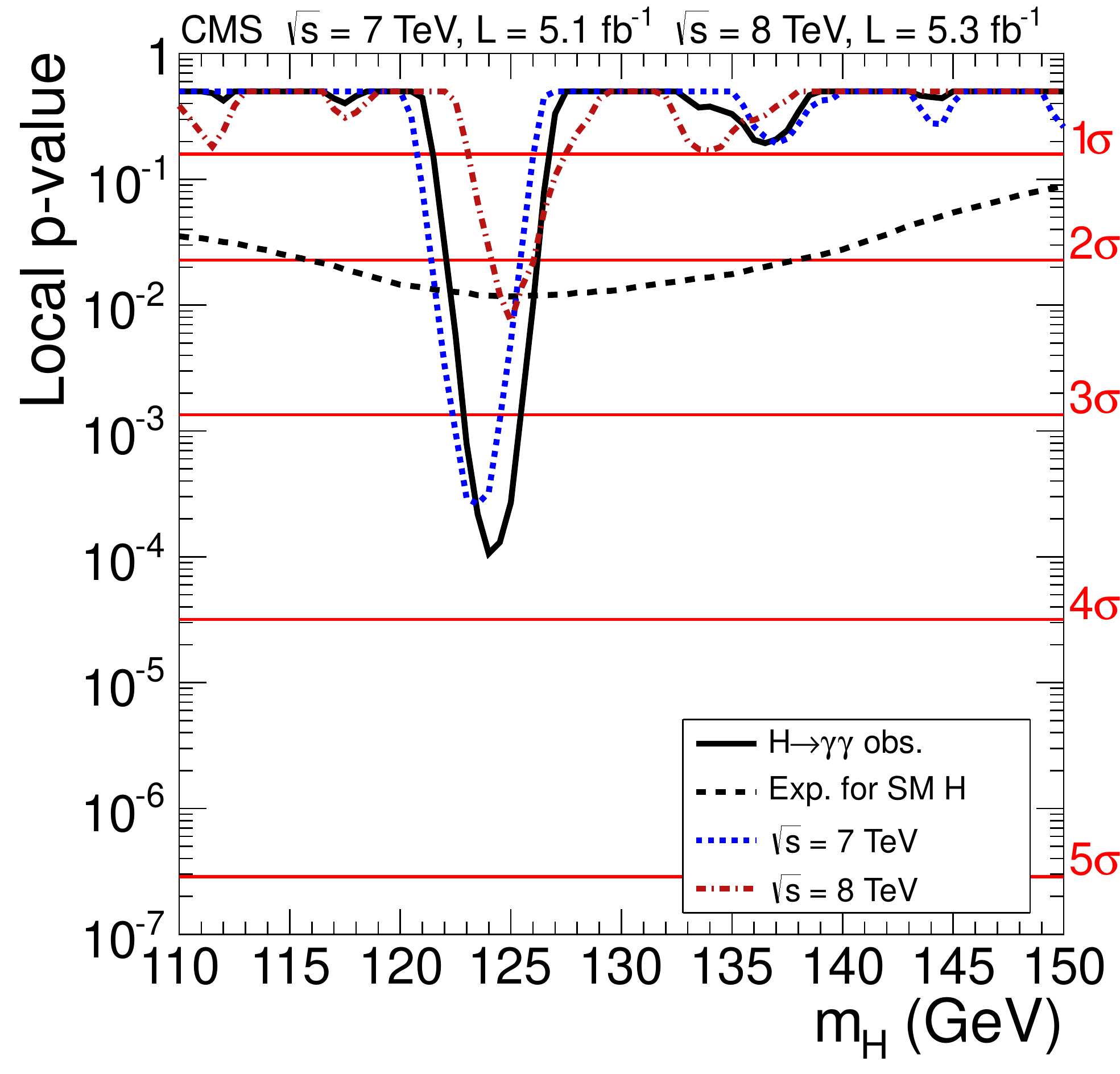}
      \includegraphics[width=0.49\linewidth]{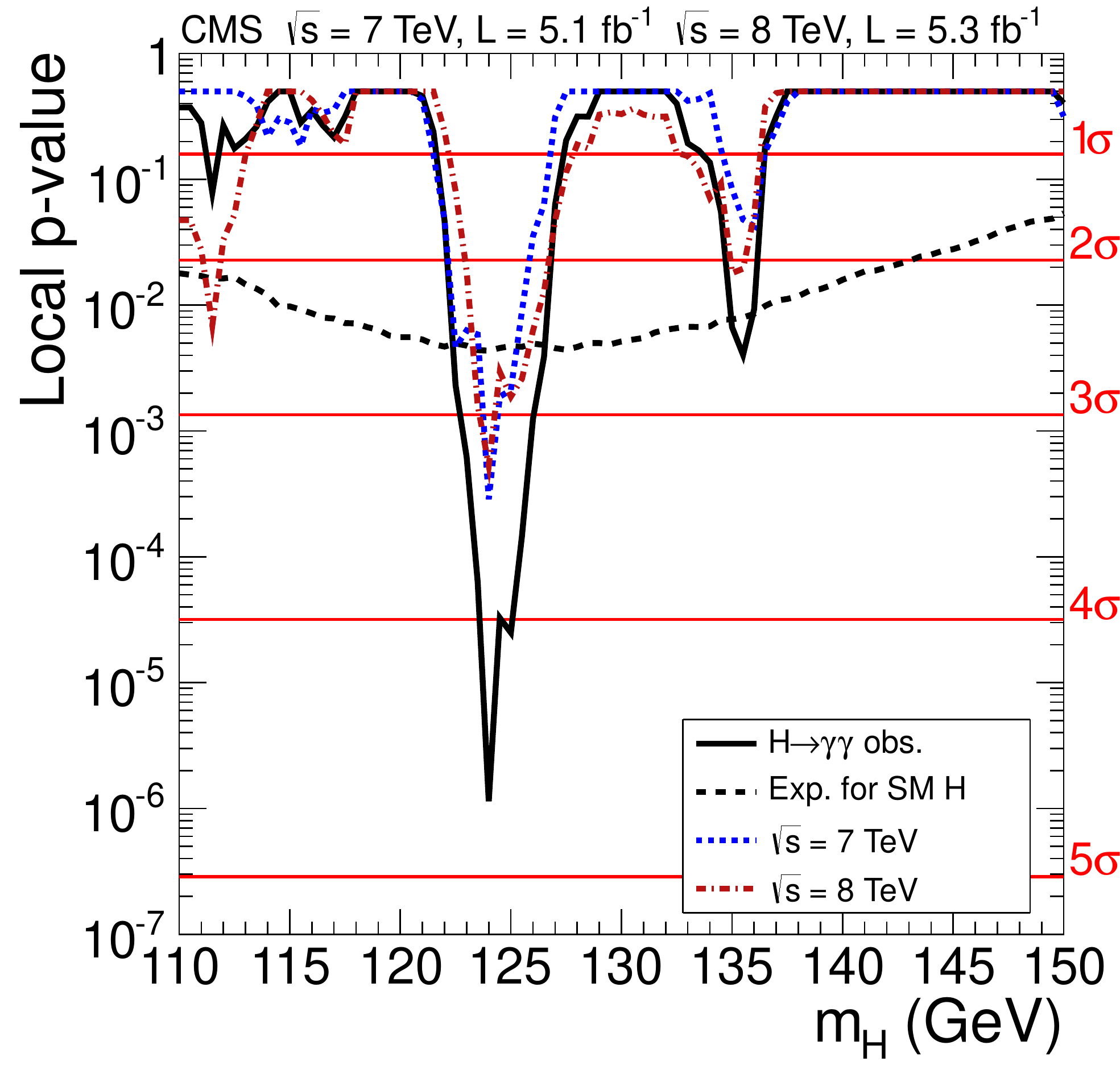}
     \caption{The local $p$-value as a function of $\mH$ for the 7 and 8\TeV data sets and their combination
      for the  $\Pgg\Pgg$ mode from
      (top) the primary analysis,
     (lower left) the cut-based analysis, and (lower right) the side-band analysis.
     The observed $p$-values for the combined 7 and 8\TeV data sets  are shown by the solid lines;
     the median expected $p$-values for a SM Higgs boson with mass $\mH$, are shown by the dashed lines.
     The horizontal lines show the relationship between the $p$-value
     (left $y$ axis) and
     the significance in standard deviations (right $y$ axis).
       }
      \label{fig:PValueGaGa}
    \end{center}
\end{figure}

Figure~\ref{fig:PValueGaGa} shows the local $p$-value as a function of the SM Higgs boson mass
in the $\Pgg\Pgg$ channel.
The results are presented for the three analyses:
(a) \emph{baseline} analysis,  and in the two alternative analyses:
(b) \emph{cut-based} analysis,
and (c) \emph{sideband} analysis.
Figure~\ref{fig:PValueGaGa} (top) shows about a $3\sigma$ excess near 125\GeV
in both the 7 and 8\TeV data.
The minimum local $p$-value $p_0 = 1.8\ten{-5}$, corresponding to a local
maximum significance of 4.1$\sigma$, occurs at a mass of 125.0\GeV for the combined 7 and 8\TeV data sets.
The median expected significance for a SM Higgs boson of this mass
is 2.7$\sigma$. In the asymptotic approximation, 68\% (95\%) of repeated experiments would
give results within ${\pm}1 \sigma$ (${\pm}2 \sigma$) around the median expected
significance. Therefore, the excess seen in data, even being larger than the expected  median for a Higgs boson signal,
is consistent with a SM Higgs boson with a probability
of about 16\%.
The consistency of the results from the three analyses is a good check on the robustness
of the measurement.

\begin{figure}[htbp]
  \begin{center}
    \includegraphics[width=0.49\linewidth]{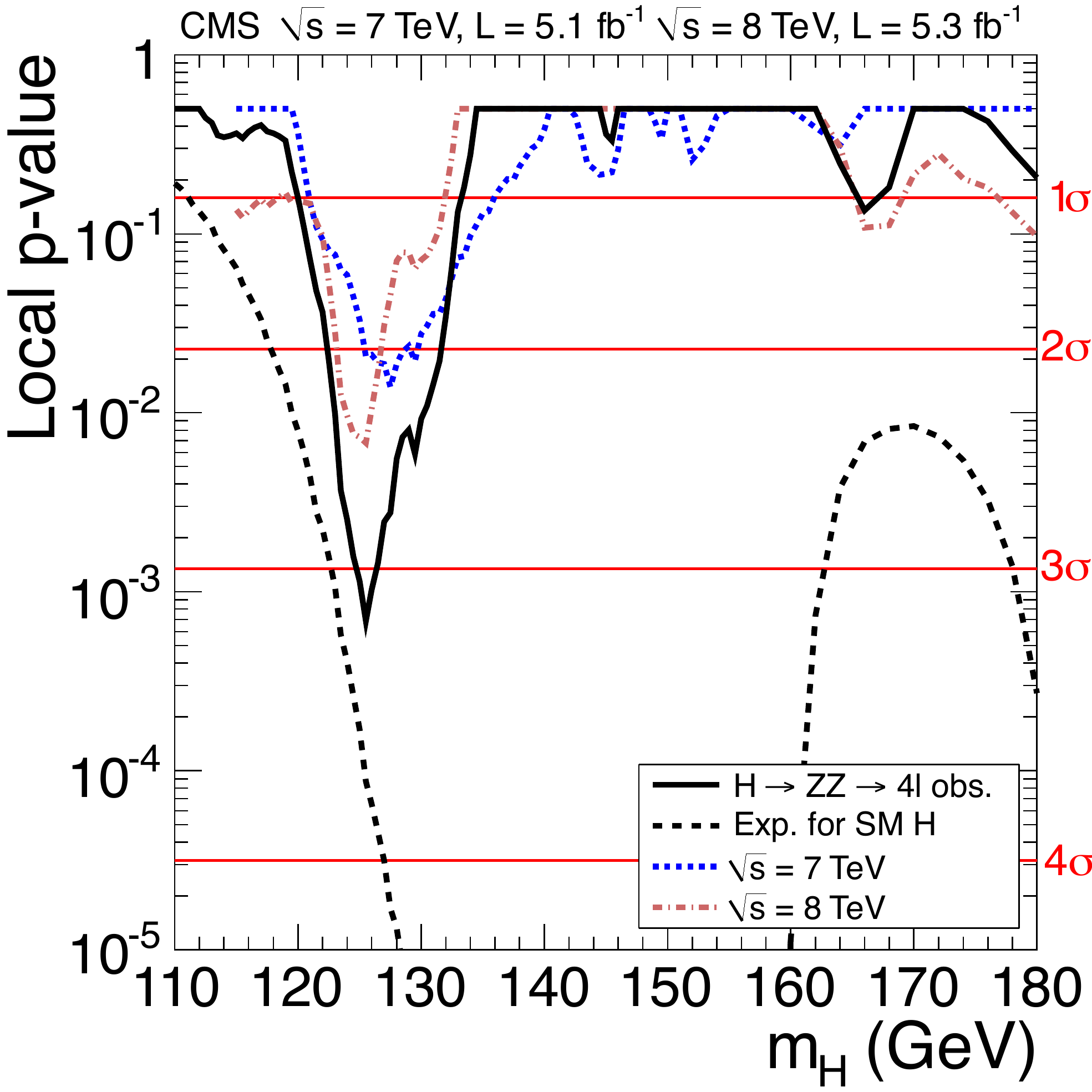}
    \caption{The local $p$-value as a function of $\mH$ for the 7 and 8\TeV data sets and their combination
     for the $\cPZ\cPZ \to 4\ell$ channel.
     The observed $p$-values  for the combined 7 and 8\TeV data sets are shown by the solid line;
     the median expected $p$-values for a SM Higgs boson with mass $\mH$ are shown by the dashed line.
     The observed $p$-values for the 7 and 8\TeV data sets are shown by the dotted lines.
     The horizontal lines show the relationship between the $p$-value (left $y$ axis) and
     the significance in standard deviations (right $y$ axis).
     }
   \label{fig:PValue}
  \end{center}
\end{figure}

The local $p$-value as a function of the Higgs boson mass $\mH$
for the $\cPZ\cPZ   \rightarrow 4\ell$ channel
is shown in Fig.~\ref{fig:PValue}.
The minimum of the local $p$-value is at $\mH=125.5\GeV$ and
corresponds to a local significance of $3.2\sigma$.
A local significance of $2.2\sigma$ is found for a  1D fit of the invariant mass without using the $K_{D}$ discriminant.
The median expected significance for a SM Higgs boson of this mass
is $3.8\sigma$ and 3.2$\sigma$ for the 2D and 1D fits, respectively.

\subsubsection{Combined results}

\begin{figure*} 
\centering
\includegraphics[width=0.49\textwidth]{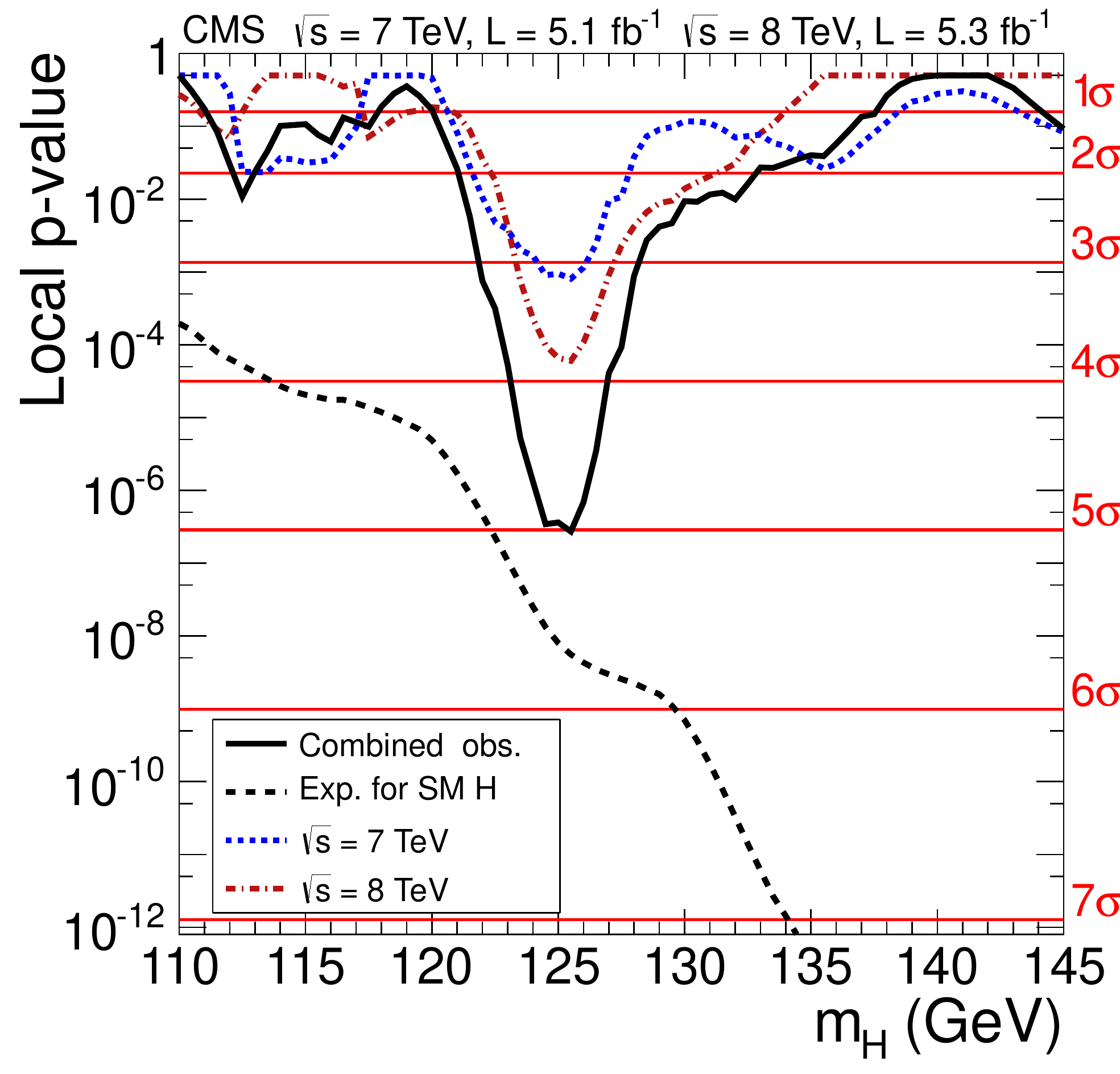} \hfill
\includegraphics[width=0.49\textwidth]{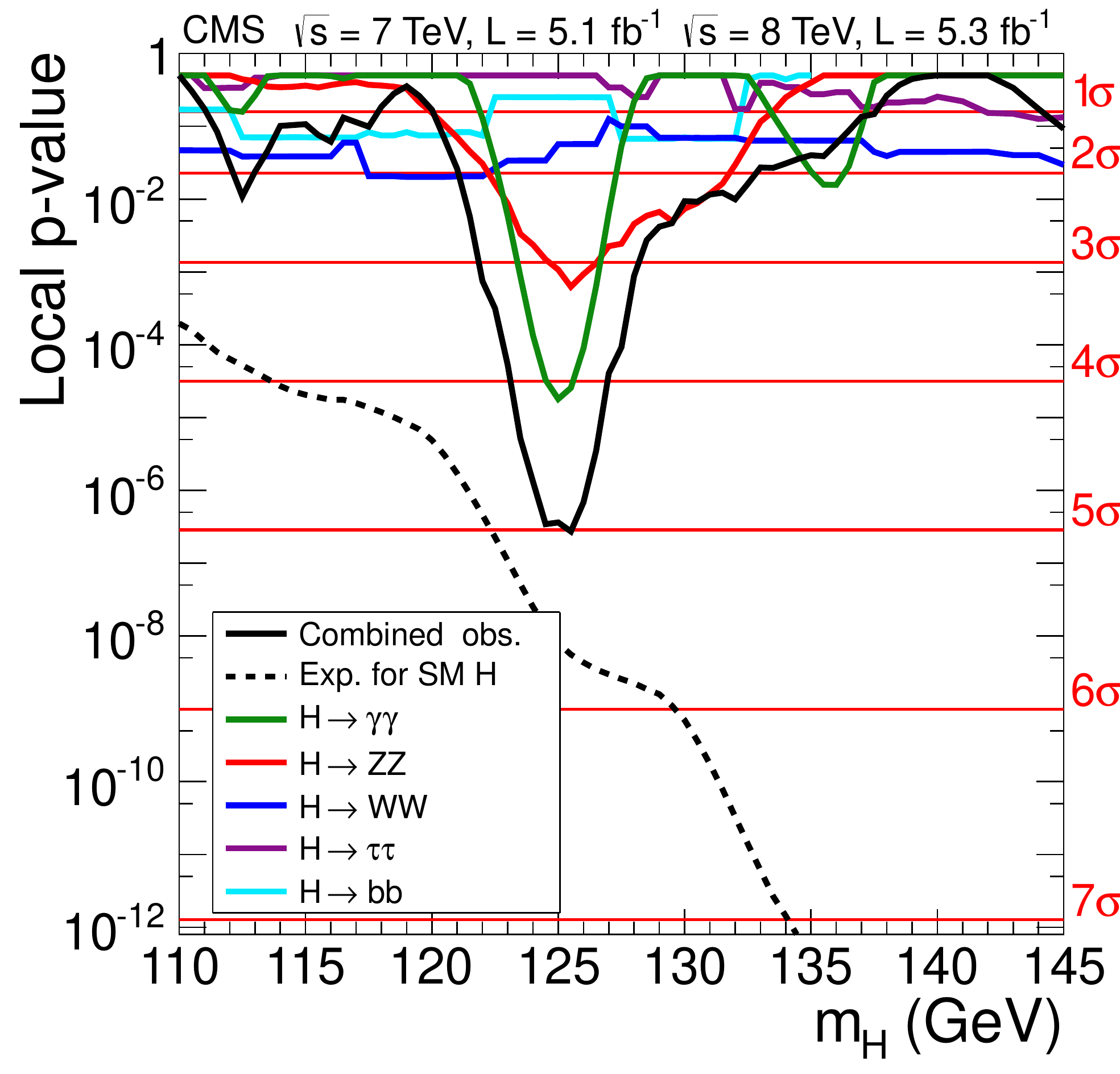}
\caption{
(Left) The observed local $p$-value for the combination of all five decay modes with the 7 and 8\TeV data sets,
and their combination
as a function of the Higgs boson mass.
(Right) The observed local $p$-value for each separate  decay mode
and their combination, as a function of the Higgs boson mass.
The dashed lines show the mean expected local $p$-values for
a SM Higgs boson with  mass $\mH$.
}
\label{fig:pvalue}
\end{figure*}

To quantify the inconsistency of the observed excesses with the background-only hypothesis,
we show in Fig.~\ref{fig:pvalue} (left) the
local $p$-value $p_0$ for the five decay modes combined for the 7 and 8\TeV data sets.
The 7 and 8\TeV data sets
exhibit excesses of $\MaxLocalZseven \sigma$ and $\MaxLocalZeight  \sigma$, respectively,
for a SM Higgs boson with a mass near {\color{black}125}\GeV.
In the combination, the minimum local $p$-value of $p_{\min}$ = \MinLocalP ,
corresponding to a local significance of $\MaxLocalZ  \sigma$, occurs at $\mH = 125.5$\GeV.

\begin{figure*} 
\centering
\includegraphics[width=0.49\textwidth]{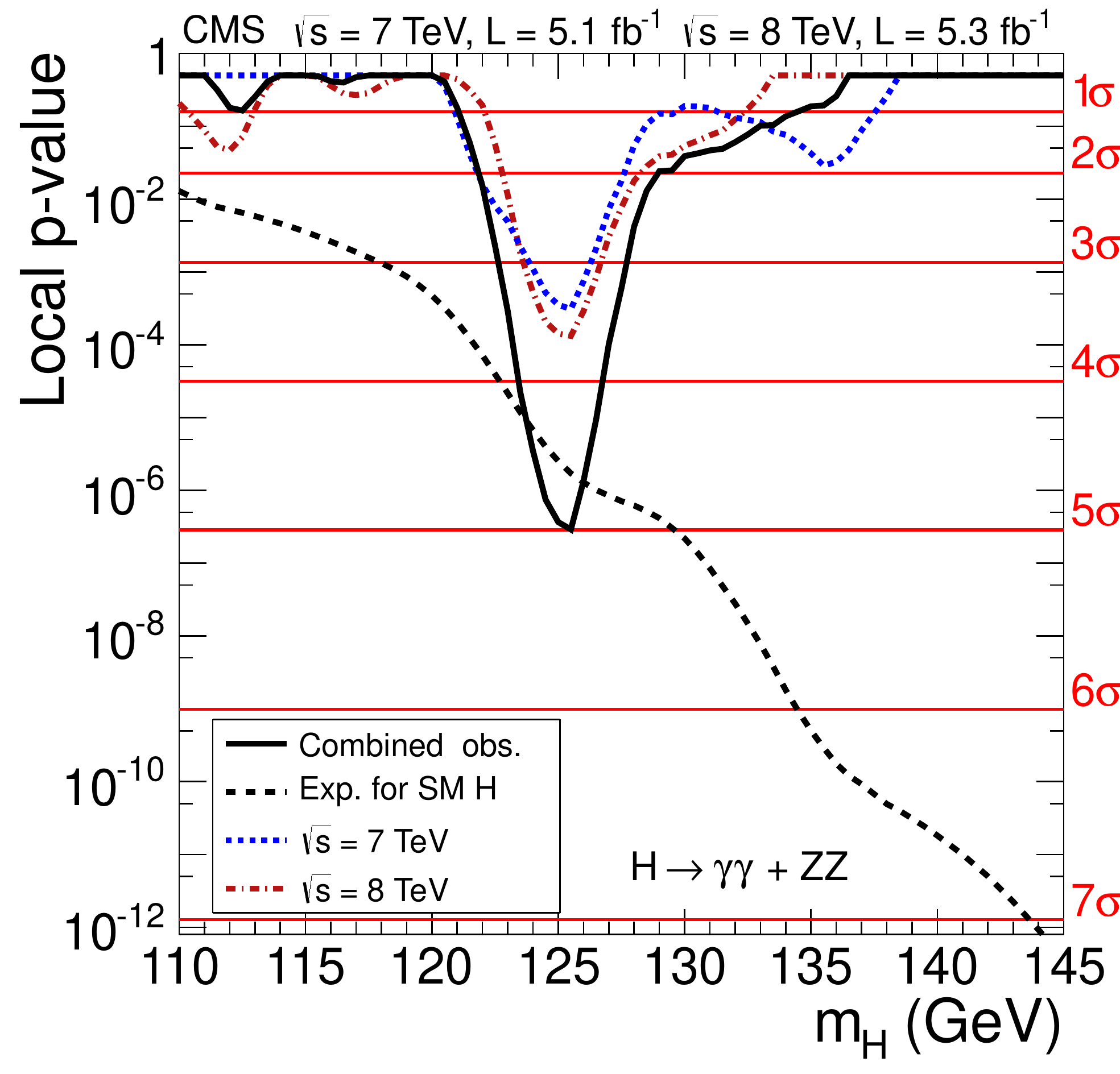} \hfill
\includegraphics[width=0.49\textwidth]{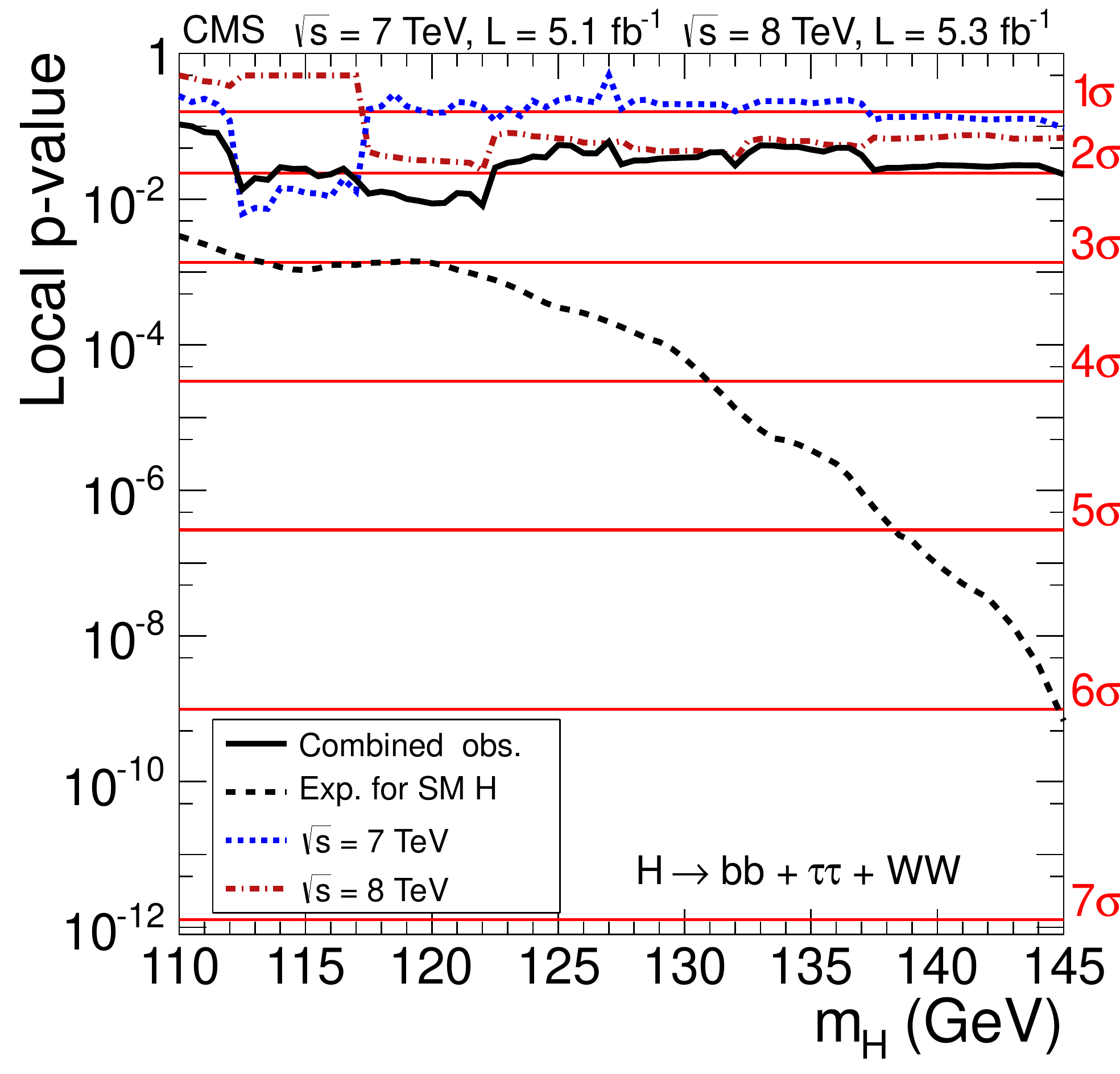}
\caption{
The observed local $p$-value for the $\Pgg\Pgg$ and $\cPZ\cPZ \to 4\ell$ decay channels with good  mass resolution (left)
and the  $\PW\PW$, $\cPqb\cPqb$, and $\Pgt\Pgt$ modes with poorer mass resolution (right), as a function of the
Higgs boson mass for the 7 and 8\TeV data sets and their combination.
The dashed lines show the expected local $p$-values for
a SM Higgs boson with  mass $\mH$.
    }
\label{fig:pvalue_subcomb}
\end{figure*}

Figure~\ref{fig:pvalue} (right) gives the $p$-value distribution for each of the
decay channels. The largest contributions to the overall excess
are from the  $\Pgg\Pgg$ and $\cPZ\cPZ \to 4\ell$ channels. Both channels have good
mass resolution and allow a precise measurement of the mass of
the resonance corresponding to the excess. Their combined
significance is \ZhighRes$\sigma$, as displayed in Fig.~\ref{fig:pvalue_subcomb} (left).
Figure~\ref{fig:pvalue_subcomb} (right) shows the combined $p$-value distribution for the channels
with poorer mass resolution: $\PW\PW$, $\Pgt\Pgt$, and $\cPqb\cPqb$.

Table~\ref{tab:Signif} summarizes the median expected and observed local significance
for a SM Higgs boson mass hypothesis of 125.5\GeV
from the individual decay modes and their combinations.
In the $\Pgt\Pgt$ channel, we do not observe an excess of events at this mass.
The expected significance is evaluated assuming
the expected background and signal rates.
The observed significance is expected to be within $\pm 1\sigma$  of
the expected significance with a 68\% probability.

\begin{table}[htbp]
\begin{center}
\topcaption{
The median expected and observed significances of the excesses
in the individual decay modes and their various combinations
for a SM Higgs boson mass hypothesis of 125.5\GeV.
There is no observed excess in the $\Pgt\Pgt$ channel.
}
\label{tab:Signif}
\begin{tabular}{l|c|c}
\hline
Decay mode or combination & Expected ($\sigma$) & Observed ($\sigma$) \\
\hline\hline
$\cPZ\cPZ$    &  3.8  & 3.2  \\ %
$\Pgg\Pgg$    &  2.8  & 4.1  \\ %
$\PW\PW$      &  2.5 &  1.6  \\ %
$\cPqb\cPqb$  &  1.9 &  0.7  \\ %
$\Pgt\Pgt$    &  1.4 &  --   \\ %
\hline
$\Pgg\Pgg$ + $\cPZ\cPZ$                 & 4.7  & 5.0 \\
$\PW\PW$ + $\Pgt\Pgt$ + $\cPqb\cPqb$    & 3.4  & 1.6 \\
\hline
$\Pgg\Pgg$ + $\cPZ\cPZ$ + $\PW\PW$ + $\Pgt\Pgt$ + $\cPqb\cPqb$ & 5.8 & 5.0 \\%
\hline
\end{tabular}
\end{center}
\end{table}

The LEE-corrected significance
is evaluated by generating $10\,000$ pseudo-experiments.  After fitting for
the constant $C$ in Eq.~(\ref{eq:LEE1}),
we find that the global significance of the signal at $\mH=125.5$\GeV is
$\GlobalZsmall \sigma$ ($\GlobalZmedium \sigma$) for the mass search
range 115--130\GeV (110--145\GeV).

The low probability for an excess at least as large as the observed one
to arise from a statistical fluctuation of the background leads  to the conclusion that
we observe a new particle with a mass near 125\GeV.
The  $\Pgg\Pgg$ and $\cPZ\cPZ \to 4\ell$ decay modes indicate that the new particle is a boson, and
the diphoton decay implies that its spin is different from 1~\cite{Landau,Yang}.

\subsection{Mass of the observed state}

\begin{figure*} [thbp]
\centering
\includegraphics[width=0.49\textwidth]{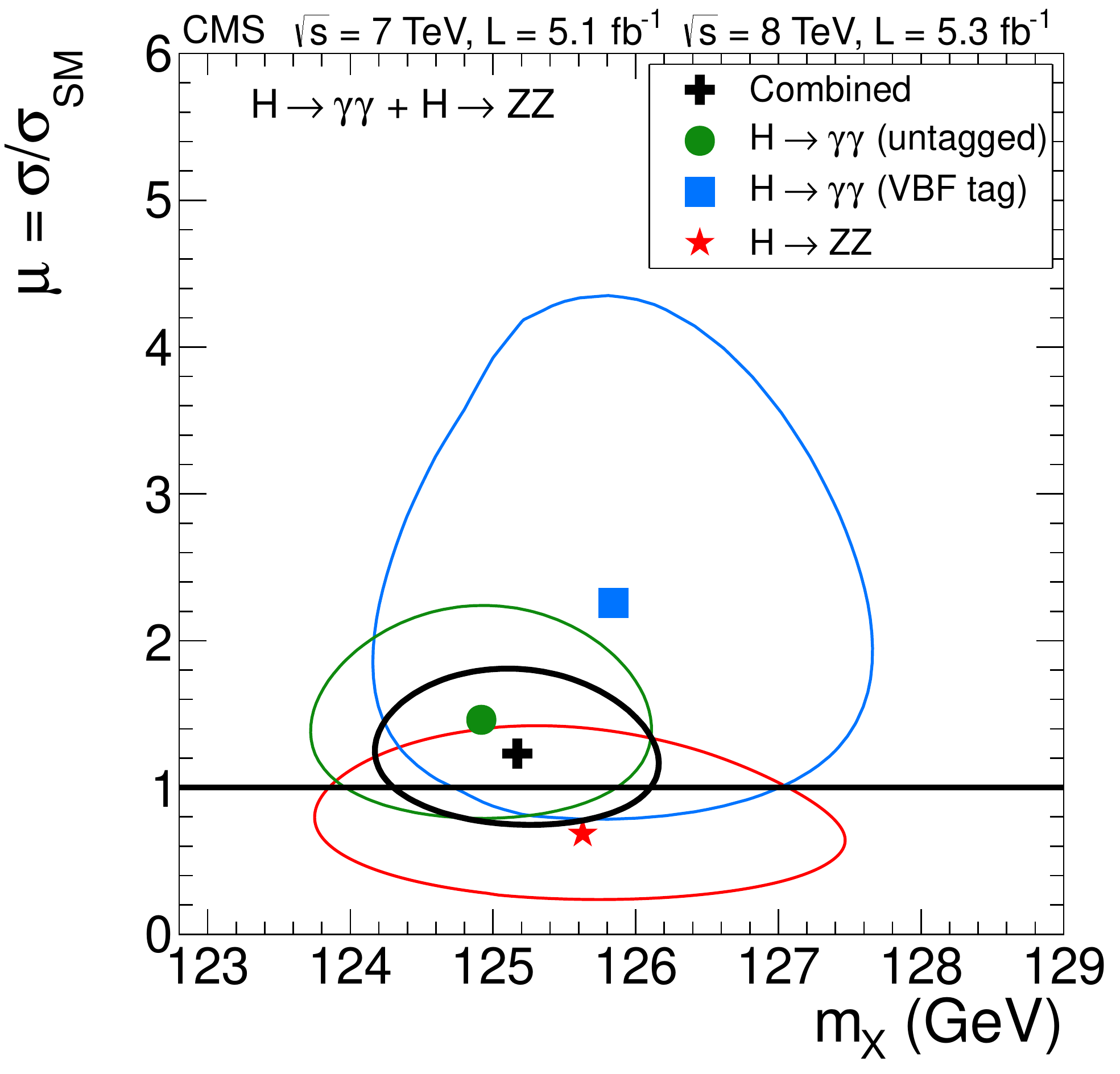} \hfill
\includegraphics[width=0.49\textwidth]{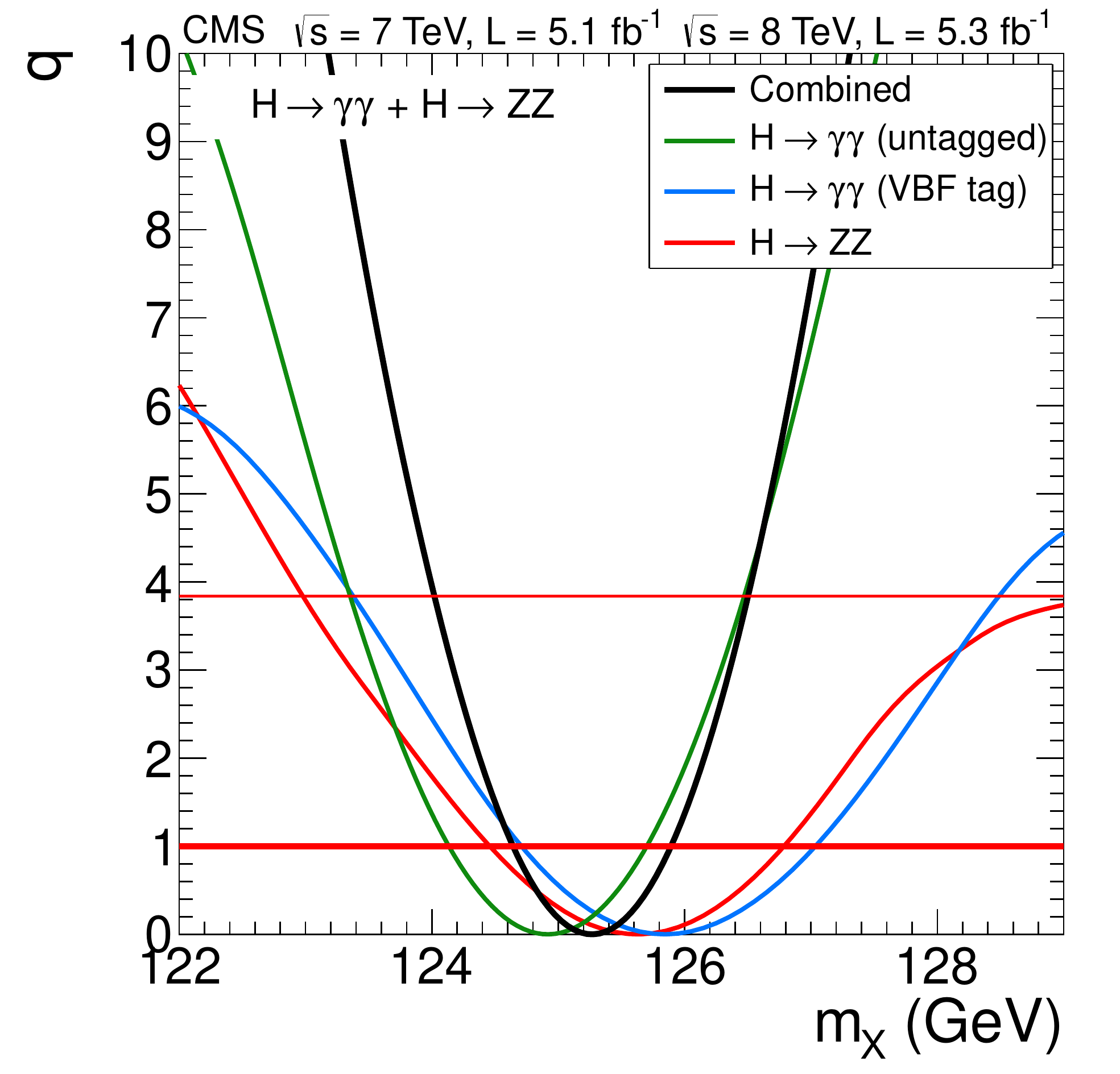}
\caption{
(Left) The 2D 68\% CL contours for a hypothesized boson mass $m_{X}$
   versus $\mu= \sigma / \sigma_{\mathrm{SM}}$  for the untagged $\gamma \gamma$,
   VBF-tagged $\gamma \gamma$, and $\cPZ\cPZ \to 4\ell$ decay channels, and their combination from the combined
    7 and 8\TeV data.
   In the combination, the relative signal strengths for the three final states
   are fixed to those for the SM Higgs boson.
(Right) The maximum-likelihood test statistic $q$ versus $m_{\mathrm{X}}$
   for the untagged $\gamma \gamma$, VBF-tagged $\gamma \gamma$, and  $\cPZ\cPZ \to 4\ell$ final states,
   and their combination from the combined 7 and 8\TeV data.
   Neither the absolute nor the relative signal strengths for the three final states
   are constrained to the SM Higgs boson expectations.
The crossings with the thick (thin) horizontal line $q=1\,(3.8)$ define the 68\% (95\%) CL interval
for the measured mass, shown by the vertical lines.}

\label{fig:fit_mass}
\end{figure*}

To measure the mass of the observed state,
we use the $\gamma\gamma$ and $\cPZ\cPZ \to 4\ell$ decay modes.
Figure~\ref{fig:fit_mass} (left) shows
the 2D~68\%~CL regions for
the signal cross section (normalized to the SM Higgs boson cross section) versus the new boson's mass $m_{\mathrm{X}}$, separately for
untagged $\gamma\gamma$, VBF-tagged $\gamma\gamma$, and $\cPZ\cPZ \to 4\ell$ events, and their combination.
The combined 68\%~CL contour shown with a solid line in Fig.~\ref{fig:fit_mass} (left)
assumes that the relative event yields between the three channels are fixed to the SM expectations,
while the overall signal strength is a free parameter.

The energy scale uncertainties for photons, electrons, and muons are treated as independent.
The $\cPZ \to \Pe\Pe$ peak is used for correcting both photon and electron energy scales. However,
we find that they have a very weak correlation, since photons in $\PH \to \gamma\gamma$
decays and electrons in $\PH \to \cPZ\cPZ \to 4\ell$ decays have substantially different energy scales.
Moreover, the photons have an additional systematic uncertainty associated with the extrapolation
of the energy scale corrections derived for the electrons to the energy scale corrections
to be used for the photons.

To measure the value of $m_{\mathrm{X}}$ in a model-independent way,
the untagged $\gamma\gamma$, VBF-tagged $\gamma\gamma$, and $\cPZ\cPZ \to 4\ell$
channels are assumed to have independent signal cross sections.
This is achieved by scaling the expected SM Higgs boson event yields
in these channels by independent factors $\mu_i$,
where $i$ denotes the individual channel.
The signal is assumed to be a particle with a unique mass $m_{\mathrm{X}}$.
The mass and its uncertainty are extracted from a scan of the combined
test statistic $q$, frequently referred to as $-2 \, \Delta \ln \mathcal{L}$, versus $m_{\mathrm{X}}$.
The signal-strengths $\mu_i$ in such a scan are treated in the same way as the other nuisance parameters.
Figure~\ref{fig:fit_mass} (right) shows the test statistic
as a function of $m_{\mathrm{X}}$ for the three final states separately and their combination.
The crossing of the $q(m_{\mathrm{X}})$ curves with the horizontal thick (thin) lines at $q=$1 (3.8) defines
the 68\% (95\%) CL interval for the mass of the observed particle. These intervals include
both the statistical and systematic uncertainties.
The resulting mass measurement and 68\% CL interval in such a combination is $m_{\mathrm{X}} = 125.3 \pm 0.6$\GeV.

\begin{figure*}[thbp]
\centering
\includegraphics[width=0.49\textwidth]{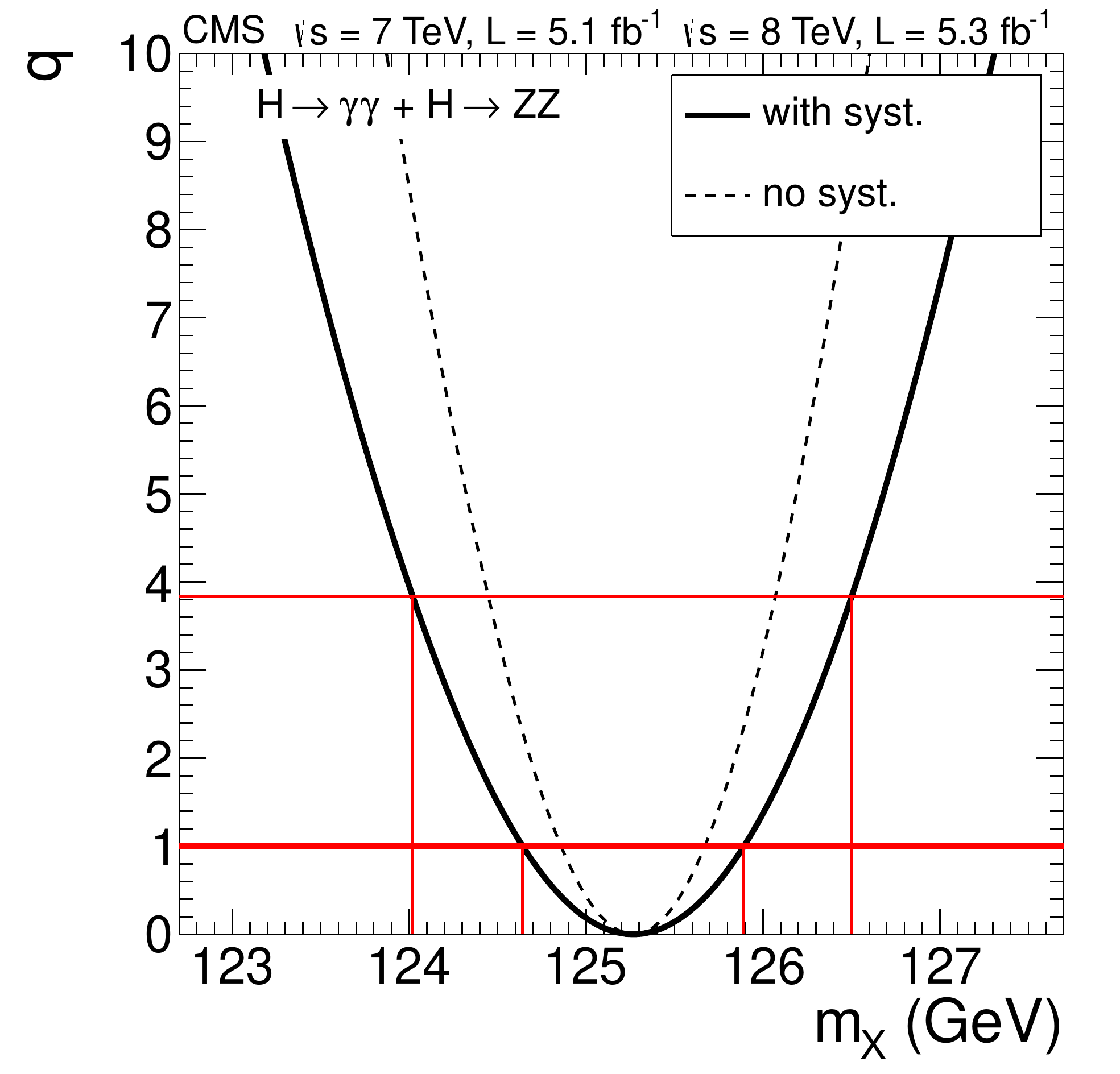}
\caption{
The maximum-likelihood test statistic $q$ versus the  hypothesized boson mass $m_{\mathrm{X}}$
for the combination of the $\gamma \gamma$ and  $\cPZ\cPZ \to 4\ell$
modes  from the combined 7 and 8\TeV data.
The solid line is obtained including all the nuisance parameters
and, hence, includes both the statistical and systematic uncertainties.
The dashed line is found with all nuisance parameters fixed to their best-fit values
and, hence, represents the statistical uncertainties only.
The crossings with the thick (thin) horizontal line $q=1$ (3.8) define the 68\% (95\%) CL interval
for the measured mass, shown by the vertical lines.
}
\label{fig:fit_mass_statsyst}
\end{figure*}

To determine the statistical component in the overall uncertainty, we evaluate
the test statistic $q(m_{\mathrm{X}})$ with all the nuisance parameters fixed to their best-fit values.
The result is shown by the dashed line in Fig.~\ref{fig:fit_mass_statsyst}.
The crossing of the dashed line with the thick horizontal line
$q=$1 gives the statistical uncertainty
(68\% CL interval) in the
mass measurements: $\pm 0.4$\GeV. The quadrature difference
between the overall and statistical-only uncertainties determines
the systematic uncertainty component in the mass measurements: ${\pm}0.5$\GeV.
Therefore, the final result for the mass measurement is $m_{\mathrm{X}}= \MASS\GeV$.

\subsection{Consistency of the observed state with the SM Higgs boson hypothesis}

The $p$-value characterizes the probability of the background
producing the observed excess of events or greater, but
it does not give information about the consistency of the observed excess
with the expected signal. The current
data sample allows for only a limited number of such consistency tests, which we present in this section.
These consistency tests do not constitute measurements of any physics parameters per se, but rather
show the consistency of the various observations with the expectations for the SM Higgs boson.
Unless stated otherwise, all consistency tests presented in this section are for the hypothesis of the
SM Higgs boson with mass 125.5\GeV and all quoted uncertainties include both the statistical and systematic ones.

\subsubsection{Measurement of the signal strength}

\begin{figure*} [htbp]
\centering
\includegraphics[width=0.49\textwidth]{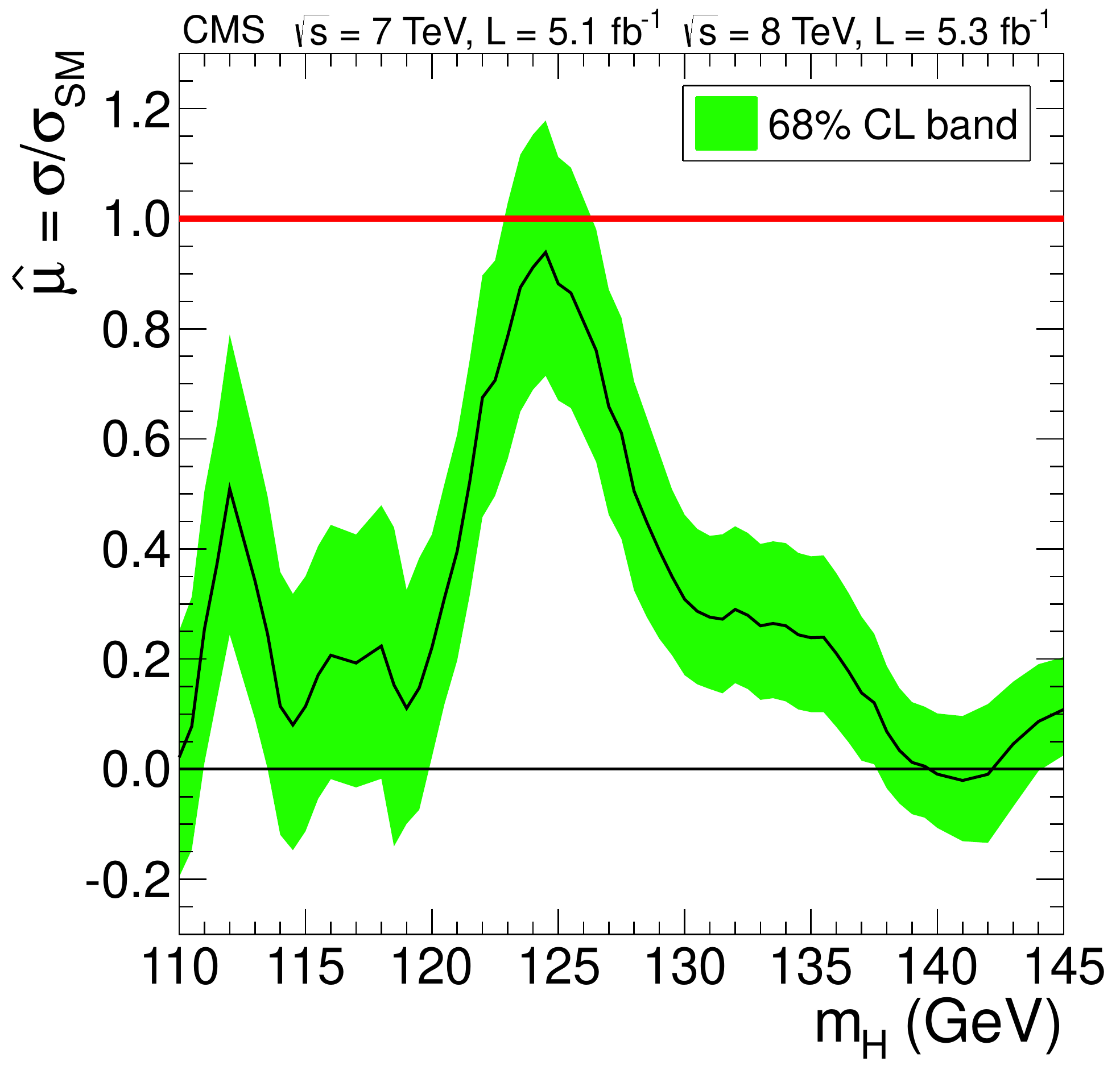}
\caption{
	The signal-strength $\hat \mu = \sigma / \sigma_\mathrm{SM}$
	as a function of the hypothesized
        SM Higgs boson mass $\mH$ using all the decay modes and the combined 7 and 8\TeV data sets.
	The bands correspond to ${\pm}1$ standard deviation including both statistical and systematic uncertainties.
    }
\label{fig:muhat}
\end{figure*}

The value for the signal-strength modifier $\hat \mu = \sigma / \sigma_{\mathrm{SM}}$,
obtained by combining  all the search channels,
provides the first consistency test.
Note that $\hat \mu$ becomes negative if the observed number of events is
smaller than the expected rate for the background-only hypothesis.
Figure~\ref{fig:muhat} shows
the $\hat \mu$ value versus the hypothesized
Higgs boson mass \mH. The band corresponds to the 68\% CL region when including the statistical and
systematic  uncertainties.
The value of $\mu$ is found in 0.5\GeV steps of \mH. The measured $\hat \mu$ value
for a Higgs boson mass of 125.5\GeV is  \MUHAT,
consistent with the value $\mu=1$ expected for the SM Higgs boson.

\begin{figure*} [htbp]
\centering
\includegraphics[width=0.49\textwidth]{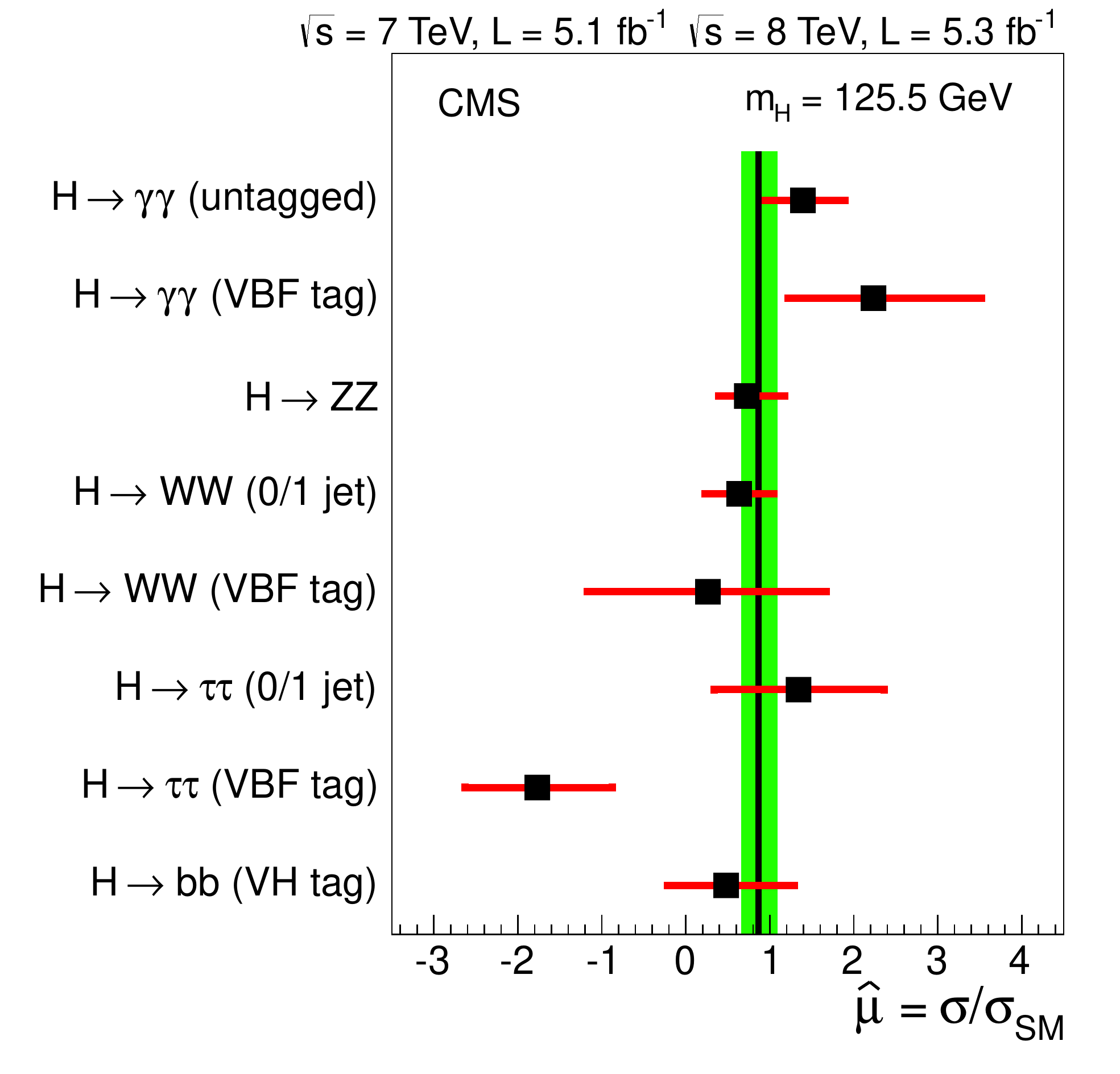} \\
\includegraphics[width=0.49\textwidth]{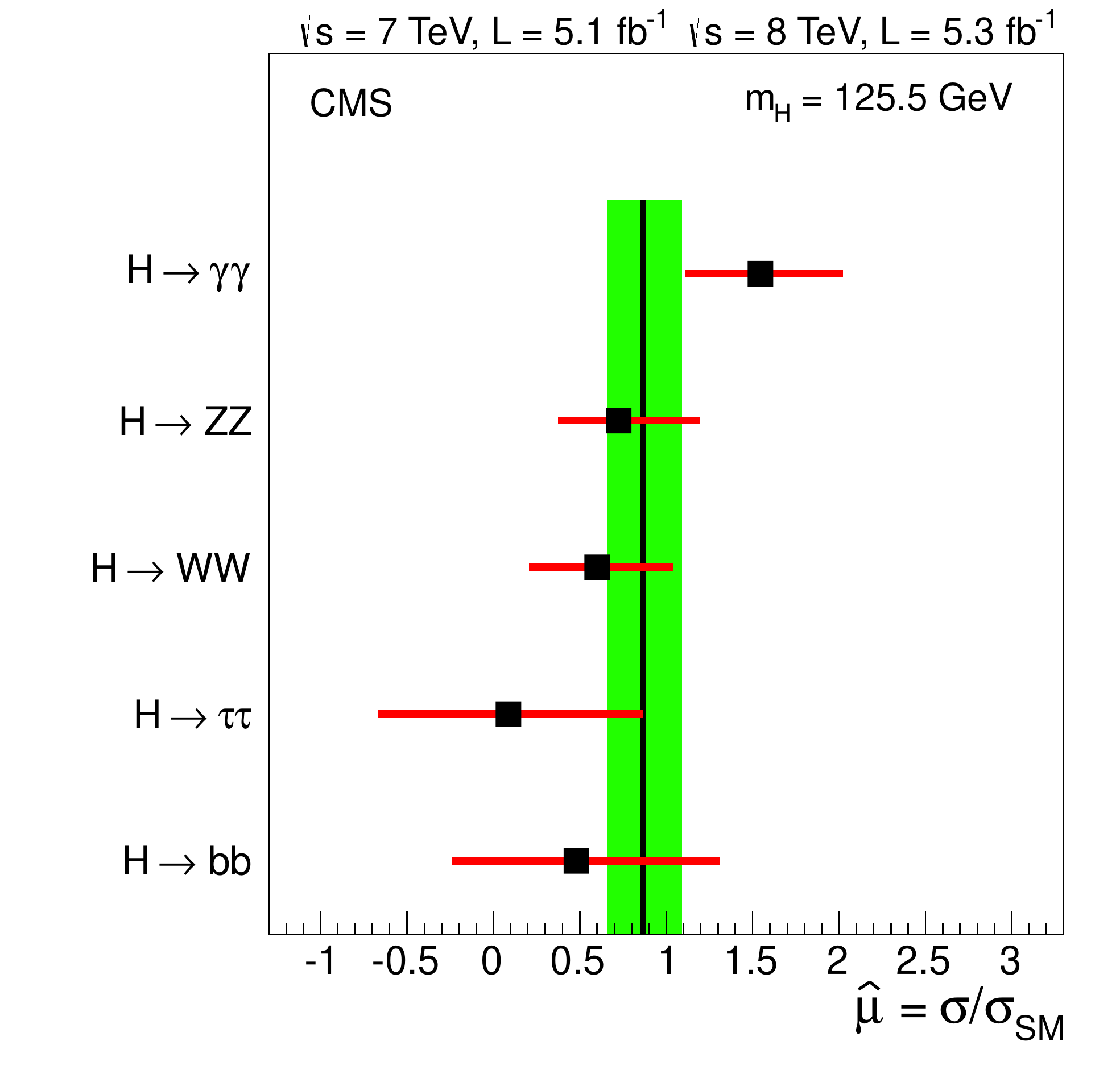}\hfill
\includegraphics[width=0.49\textwidth]{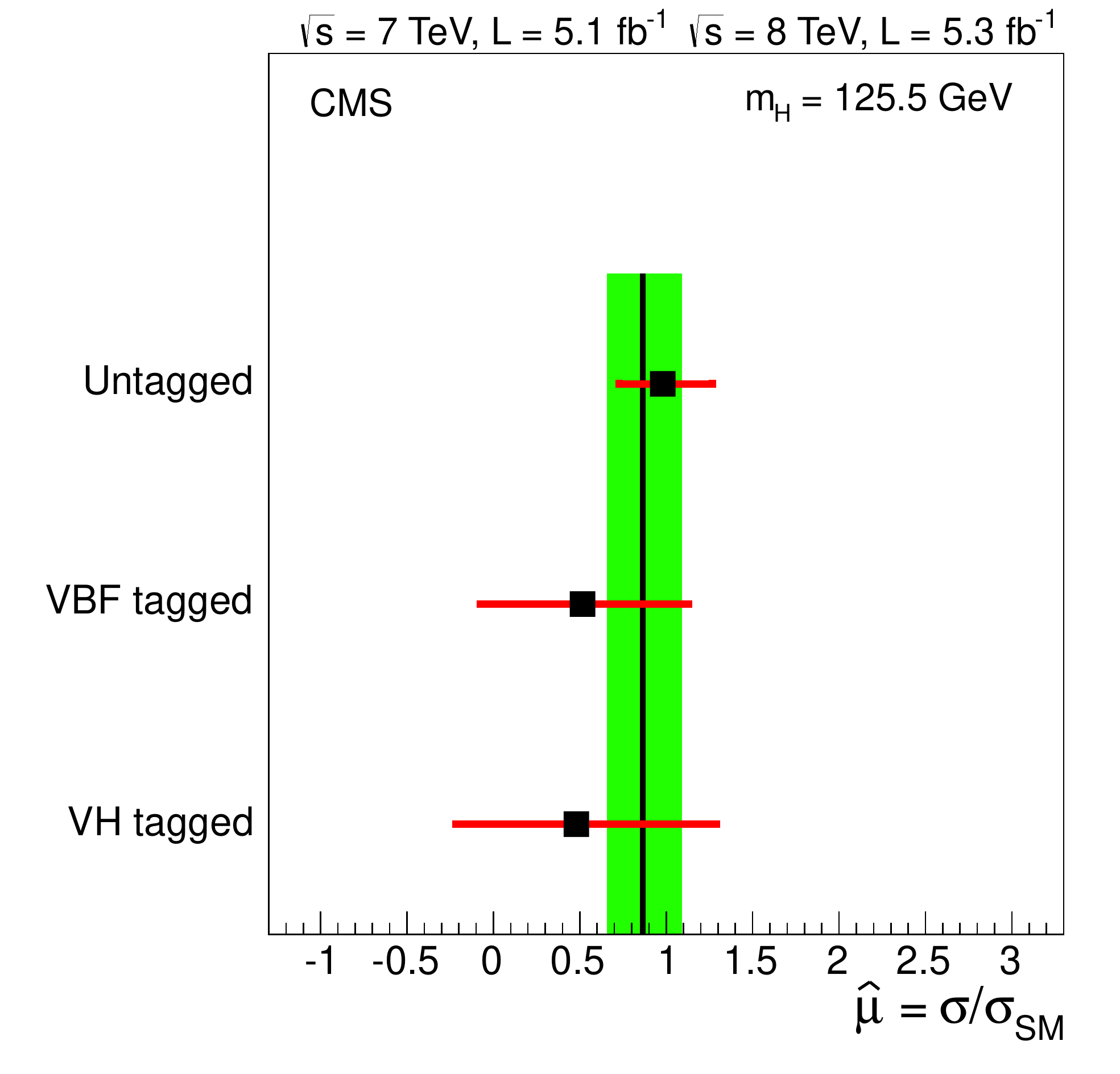}
\caption{
	Signal-strength values $\hat \mu = \sigma / \sigma_\mathrm{SM}$
	for various combinations of the search channels with $\mH=125.5\GeV$.
	The horizontal bars indicate the $\pm 1 \sigma$ statistical-plus-systematic uncertainties.
	The  vertical line with the band shows the combined $\hat \mu$ value with its uncertainty.
(Top) Combinations by decay mode and additional requirements
      that select events with an enriched purity of a particular production mechanism.
(Bottom-left) Combinations by decay mode.
(Bottom-right) Combinations by selecting events with additional requirements that select events with an enriched purity of a
 particular production mechanism.
    }
\label{fig:muhat_compatibility}
\end{figure*}

Figure~\ref{fig:muhat_compatibility} shows a consistency test of
the $\hat \mu$ values obtained in different combinations of search channels.
The combinations are organized by decay mode and additional features that
allow the selection of events with an enriched purity of a particular production mechanism.
The expected purities of different combinations
are discussed in the sections describing the individual analyses.
For example, assuming the SM Higgs boson cross sections,
the channels with the VBF dijet requirements have a substantial fraction (20--50\%)
of gluon-gluon fusion events. There is consistency
among all the channels contributing to the overall measurement and their various combinations.

\begin{figure*}[bhtp]
\centering
\includegraphics[width=0.49\textwidth]{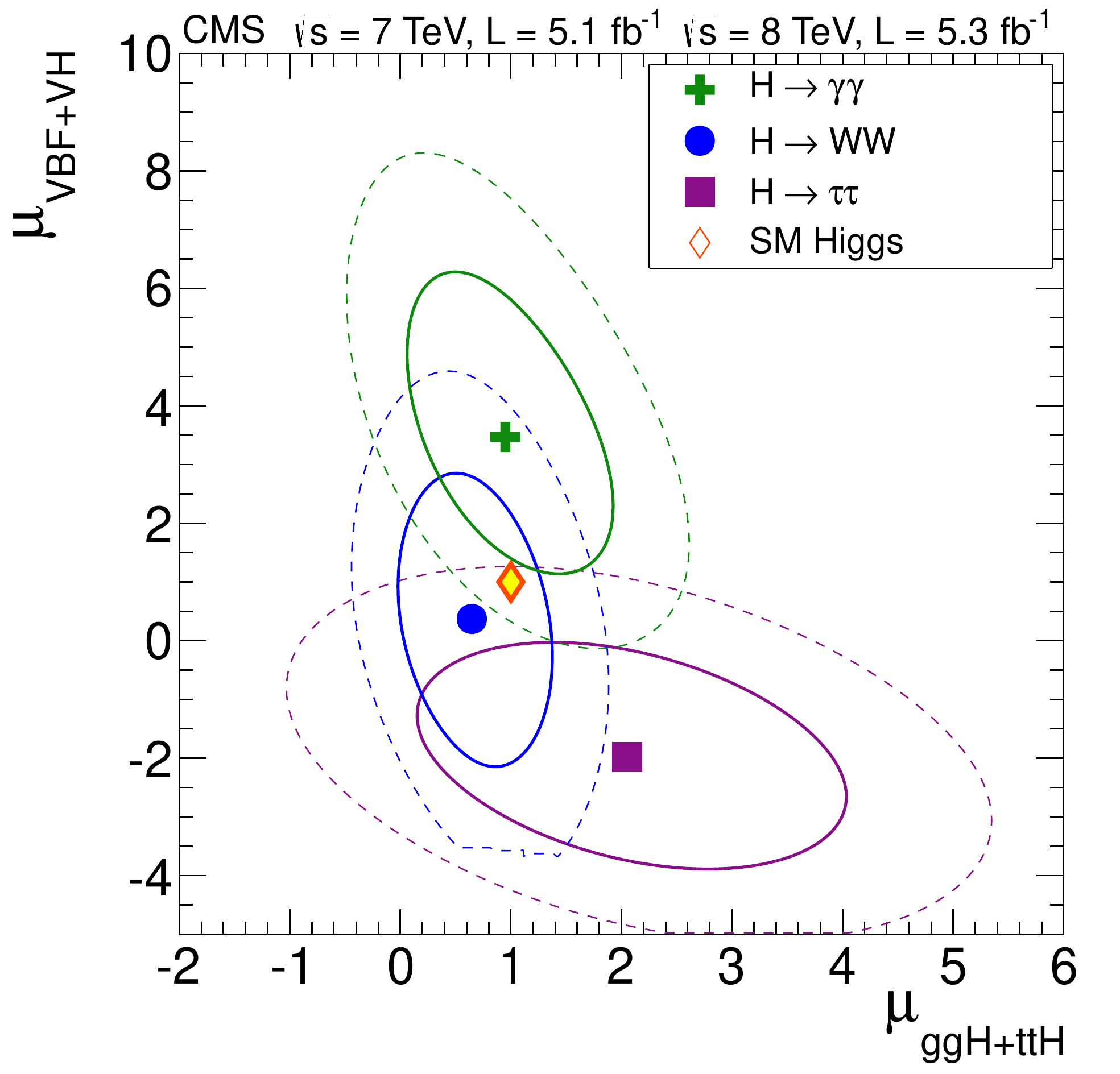}
\caption{
The 68\% (solid lines) and 95\% (dashed lines) CL contours for the
signal strength of the gluon-gluon fusion plus $\cPqt\cPqt\PH$  production mechanisms ($\mu_{\cPg\cPg\PH+\cPqt\cPqt\PH}$), versus
VBF plus VH ($\mu_{\mathrm{VBF+VH}}$).
 The three different lines show the results for the decay modes: $\gamma\gamma$, $\PW\PW$,
and $\tau\tau$. The markers indicate the best-fit values for each mode. The diamond
at (1,1) indicates the expected values for the SM Higgs boson.
}
\label{fig:rvrf}
\end{figure*}

The four main Higgs boson production mechanisms can be associated with either top-quark
couplings (gluon-gluon fusion and \cPqt\cPqt\PH) or vector-boson couplings (VBF and VH).
Therefore, combinations of channels associated with a particular decay mode
and explicitly targeting different production mechanisms
can be used to test the relative strengths of the couplings of the new state to the vector bosons and top quark.
Figure~\ref{fig:rvrf} shows the 68\% and 95\% CL contours for the signal-strength modifiers $\mu_{\cPg\cPg\PH+\cPqt\cPqt\PH}$  of
the gluon-gluon fusion plus \cPqt\cPqt\PH, and $\mu_{\mathrm{VBF+VH}}$ of the VBF plus VH production mechanisms.
The three sets of contours correspond to
the channels associated with the $\Pgg\Pgg$, $\tau\tau$, and $\PW\PW$
decay modes; searches in these decay modes have subchannels with VBF dijet tags.
The SM Higgs boson point shown by the diamond at $\mu_{\mathrm{ggH}+\mathrm{ttH}},\mu_{\mathrm{VBF+VH}}=$ (1, 1)
is within the 95\% CL intervals for each of the three decay modes.

\subsubsection{Consistency of the data with the SM Higgs boson couplings}

The event yield $N$ of Higgs bosons produced in collisions of partons $x$ ($xx \to \PH$)
and decaying to particles $y$ ($\PH \to yy$), is proportional to the
partial and total Higgs boson decay widths as follows:

\begin{equation}
N \propto \sigma(xx \to \PH) \cdot \mathcal{B}(\PH \to yy)
  \propto \frac {\Gamma_{xx} \, \Gamma_{yy} }
                { \Gamma_{\mathrm{tot}} },
\end{equation}
where $\sigma(xx \to \PH)$ is the Higgs boson production cross section, $\mathcal{B}(\PH \to yy)$
is the branching fraction for the decay mode, $\Gamma_{xx}$ and
$\Gamma_{yy}$ are the partial widths
associated with the $\PH \to xx$ and $\PH \to yy$ processes, and $\Gamma_{\mathrm{tot}}$ is the total width.

Seven partial widths
($\Gamma_{\PW\PW}$,
$\Gamma_{\cPZ\cPZ}$,
$\Gamma_{\cPqt\cPqt}$,
$\Gamma_{\cPqb\cPqb}$,
$\Gamma_{\tau\tau}$,
$\Gamma_{\cPg\cPg}$,
$\Gamma_{\gamma\gamma}$)
and the total width $\Gamma_{\text{tot}}$ are relevant for the current analysis,
where $\Gamma_{\cPg\cPg}$ is the partial width for the Higgs boson decay to two gluons.
The partial widths $\Gamma_{\cPg\cPg}$ and $\Gamma_{\gamma\gamma}$ are generated by loop diagrams and thus
are directly sensitive to the presence of new physics.
The possibility of Higgs boson decays to beyond-the-standard-model (BSM) particles,
with a partial width $\Gamma_{\mathrm{BSM}}$, is
accommodated by making $\Gamma_{\text{tot}}$ equal to
the sum of all partial widths of allowed decays to the SM particles plus $\Gamma_{\mathrm{BSM}}$.

The partial widths are proportional to the square of the effective Higgs boson
couplings to the corresponding particles.
To test for possible deviations of the measurements from the rates expected in different channels
for the SM Higgs boson, we introduce different sets of coupling scale factors $\kappa$
and fit the data to these new parameters. One can introduce up to eight independent parameters
relevant for the current analysis.
Significant deviations of the scale factors from unity
would imply new physics beyond the SM Higgs boson hypothesis.

The current data set is insufficient to measure all
eight independent parameters.
Therefore, we measure different subsets, with the remaining
unmeasured parameters either constrained to equal the SM Higgs boson expectations
or included in the likelihood fit as unconstrained nuisance parameters.

\textit{A. Test of custodial symmetry}

In the SM, the Higgs boson sector possesses a global $\mathrm{SU(2)_L \times SU(2)_R}$
symmetry, which is broken by the Higgs boson vacuum expectation value down to
the diagonal subgroup $\mathrm{SU(2)_{L+R}}$. As a result, the tree-level relations
between the ratios of the \PW\ and \cPZ\ boson  masses, $m_{\PW} / m_{\cPZ}$,
and their couplings to the Higgs boson, $g_{\PW} / g_{\cPZ}$,
are protected against large radiative corrections,
a phenomenon known as ``custodial symmetry''~\cite{Veltman:1977kh,Sikivie:1980hm}.
However, large violations of custodial symmetry are possible in BSM theories.
To test custodial symmetry, we introduce two scaling factors
$\kappa_{\PW}$ and $\kappa_{\cPZ}$ that modify
the SM Higgs boson couplings to $\PW$ and $\cPZ$ bosons,
and perform two different procedures to determine the consistency of
the ratio $\lambda_{\PW\cPZ} = \kappa_{\PW} / \kappa_{\cPZ}$ with unity.

\begin{figure*}[bhtp]
\centering
\includegraphics[width=0.49\textwidth]{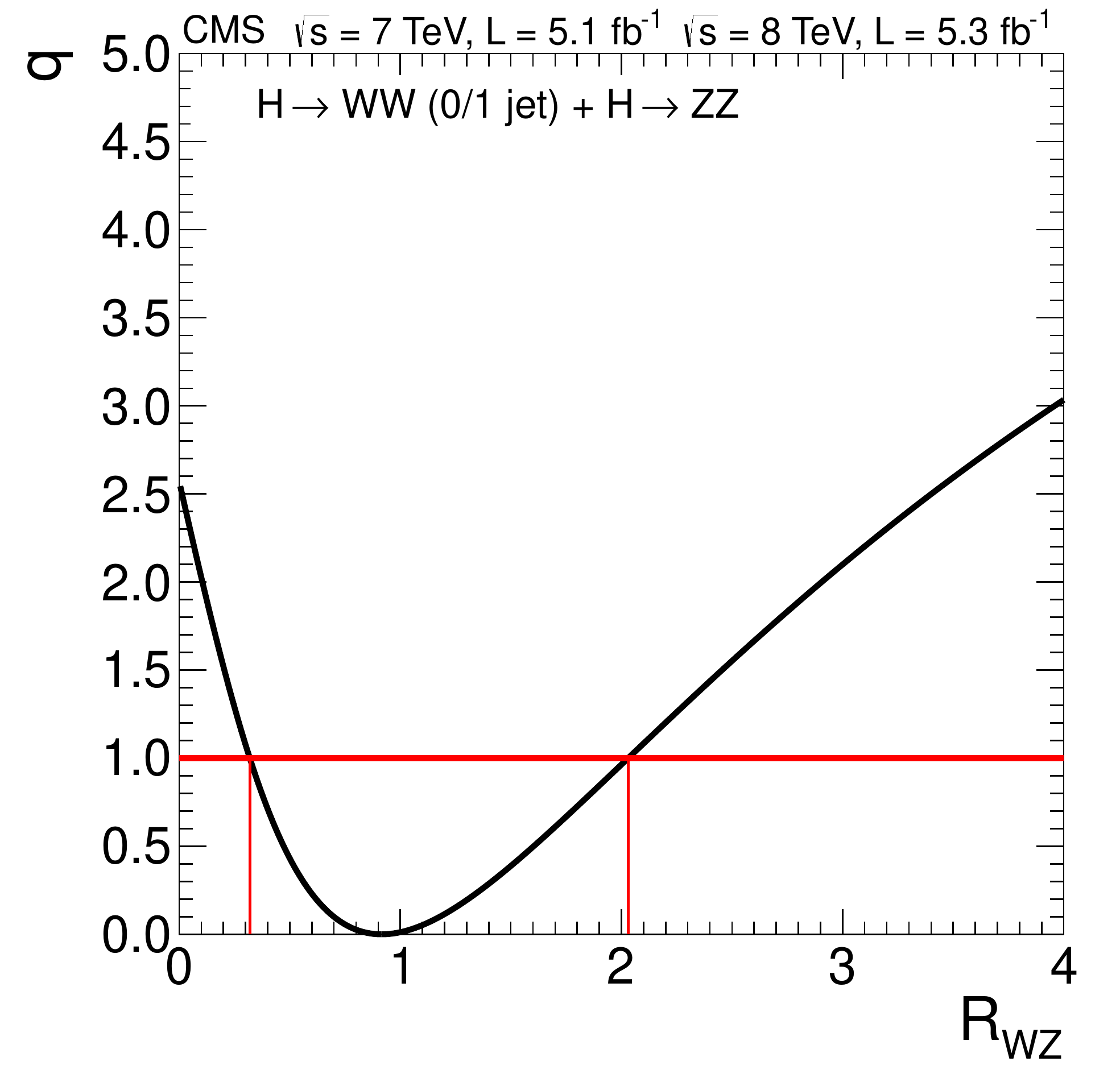} \hfill
\includegraphics[width=0.49\textwidth]{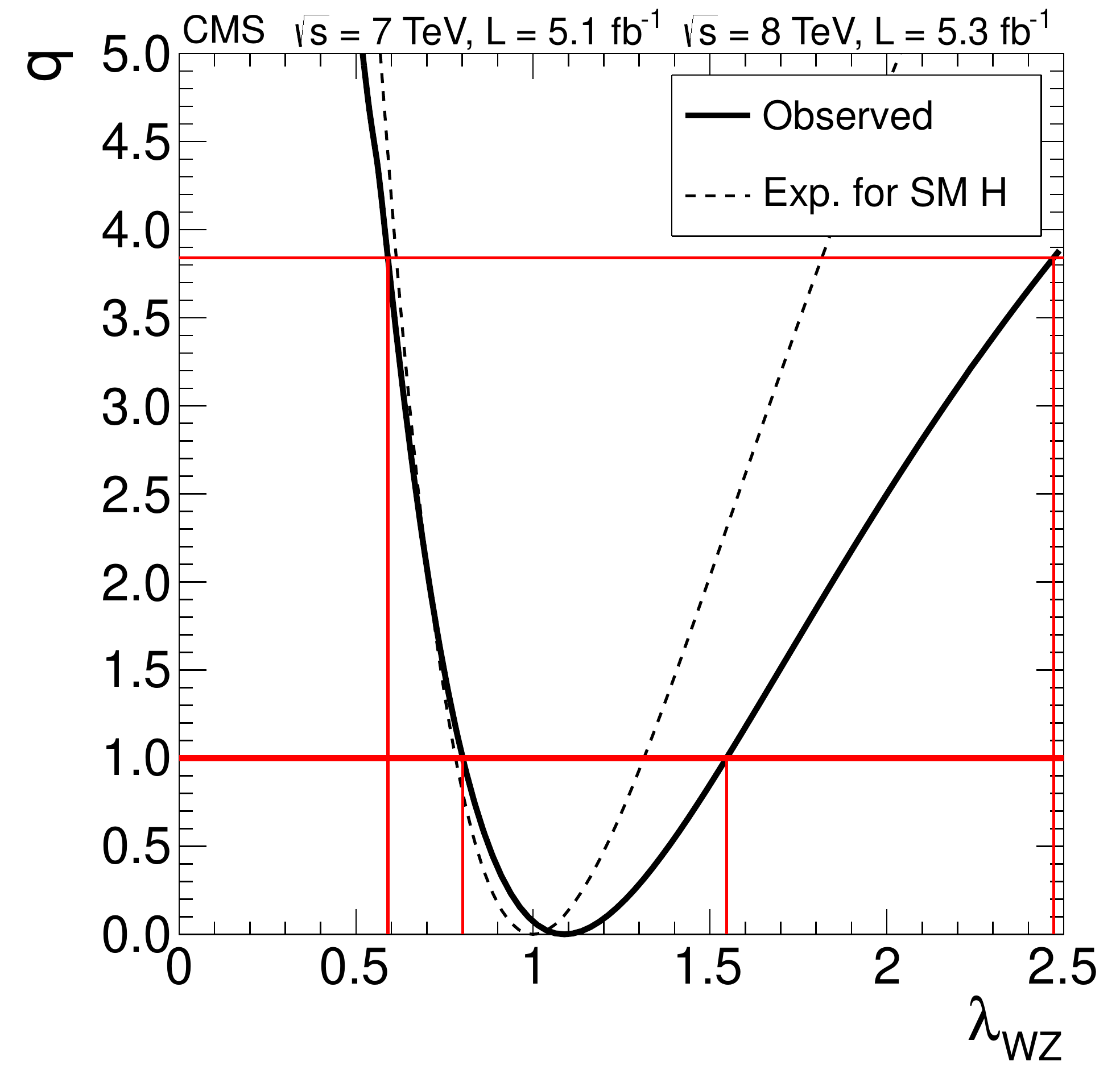}
\caption{
(Left) The likelihood test statistic $q(R_{\PW\cPZ})$ as a function of the event-rate modifier $R_{\PW\cPZ}$
from the combined untagged $\PH \to \PW\PW \to \ell\cPgn\ell\cPgn$ and
inclusive $\PH \to \cPZ\cPZ \to  4\ell$ searches.
(Right) The test statistic  $q(\lambda_{\PW\cPZ})$  as a function of the ratio of the couplings to $\PW$ and $\cPZ$
 bosons, $\lambda_{\PW\cPZ}$,
   from the combination of all channels.
The intersection  of the curves with the horizontal lines $q= 1$ and 3.8 give the 68\% and 95\% CL intervals, respectively.
}
\label{fig:fit_rwz_scan}
\end{figure*}

The dominant Higgs boson production mechanism for the inclusive $\PH \to \cPZ\cPZ$
and untagged $\PH \to \PW\PW$ channels is $gg \to \PH$. Therefore,
the ratio of the event yields for these channels provides a test of custodial symmetry.
To quantify the test, we introduce two event-rate modifiers $\mu_{\cPZ\cPZ}$ and $R_{\PW\cPZ}$.
The expected $H \to \cPZ\cPZ \to 4\ell$ event yield is scaled by $\mu_{\cPZ\cPZ}$,
while the expected untagged $H \to \PW\PW \to \ell\nu\ell\nu$ event yield is
scaled by $R_{\PW\cPZ} \cdot \mu_{\cPZ\cPZ}$.
The mass of the observed state is fixed to {\color{black} 125.5}\GeV.
The test statistic $q(R_{\PW\cPZ})$ as a function of $R_{\PW\cPZ}$, with $\mu_{\cPZ\cPZ}$ included
with the other nuisance parameters,
is shown in Fig.~\ref{fig:fit_rwz_scan} (left) and yields
$R_{\PW\cPZ} =$ \Rwz, where the uncertainty is the combined
statistical and systematic.
The contributions from VBF and VH production to the fit
give a small bias of 0.02  when relating the observed event-yield ratio
$R_{\PW\cPZ}$ to the square of the ratio of the couplings $\lambda^2_{\PW\cPZ}$.
Hence, the current measurements
are consistent, within the uncertainties, with the expectation from custodial symmetry.

In the second method, we extract $\lambda_{\PW\cPZ}$ directly from the combination of all search
channels.  In this approach, we use  three parameters: $\lambda_{\PW\cPZ}$,
$\kappa_{\cPZ}$, and $\kappa_F$. The latter variable is a single event-rate modifier for
all Higgs boson couplings to fermions. The BSM Higgs boson width $\Gamma_{\mathrm{BSM}}$ is set to zero.
The partial width $\Gamma_{\cPg\cPg}$, induced by quark loops, scales as $\kappa_F^2$.
The partial width $\Gamma_{\Pgg\Pgg}$ is also induced via
loop diagrams, with the \PW\ boson and top quark being the dominant contributors;
hence, it scales as $| \alpha \, \kappa_{\PW} + \beta \, \kappa_F |^2$,
where $\kappa_{\PW} = \lambda_{\PW\cPZ} \cdot \kappa_{\cPZ}$ and
the ratio of the factors $\alpha$ and $\beta$,  $\beta / \alpha \approx -0.22$, is taken from the prediction
for the SM Higgs boson with $\mH=125.5\GeV$~\cite{Spira:1995rr}.
In the evaluation of $q (\lambda_{\PW\cPZ})$,
both $\kappa_{\cPZ}$ and $\kappa_F$ are included with the other nuisance
parameters. Assuming a common scaling factor for all fermions makes this measurement
model dependent, but using all the channels gives it
greater
sensitivity.
The results are shown in Fig.~\ref{fig:fit_rwz_scan} (right)
by the solid line. The dashed line indicates the median expected result for the SM Higgs boson,
given the integrated luminosity.
The measured value is $\lambda_{\PW\cPZ} = 1.1^{+0.5}_{-0.3}$, where the uncertainty is
the combined statistical and systematic. The result is
consistent with the expectation of $\lambda_{\PW\cPZ} =1$ from
custodial symmetry.  In all further combinations
presented below, we assume $\lambda_{\PW\cPZ}=1$ and use a common factor $\kappa_V$
to modify the Higgs boson couplings to $\PW$ and $\cPZ$ bosons.

\textit{B. Test of the couplings to vector bosons and fermions}

We further test the consistency of the measurements with the SM
Higgs boson hypothesis by fitting for the two free parameters $\kappa_V$ and $\kappa_F$ introduced above.
We assume  $\Gamma_{\mathrm{BSM}}=0$, \ie no BSM Higgs boson decay modes.
At lowest order, all partial widths,
except for $\Gamma_{\Pgg\Pgg}$, scale either as
$\kappa^2_V$ or $\kappa^2_F$.
As discussed above, the partial width $\Gamma_{\Pgg\Pgg}$
scales as $| \alpha \, \kappa_V + \beta \, \kappa_F |^2$.
Hence, $\Pgg\Pgg$ is the only channel sensitive to the relative sign of
$\kappa_V$ and $\kappa_F$.

\begin{figure*}
\centering
\includegraphics[width=0.49\textwidth]{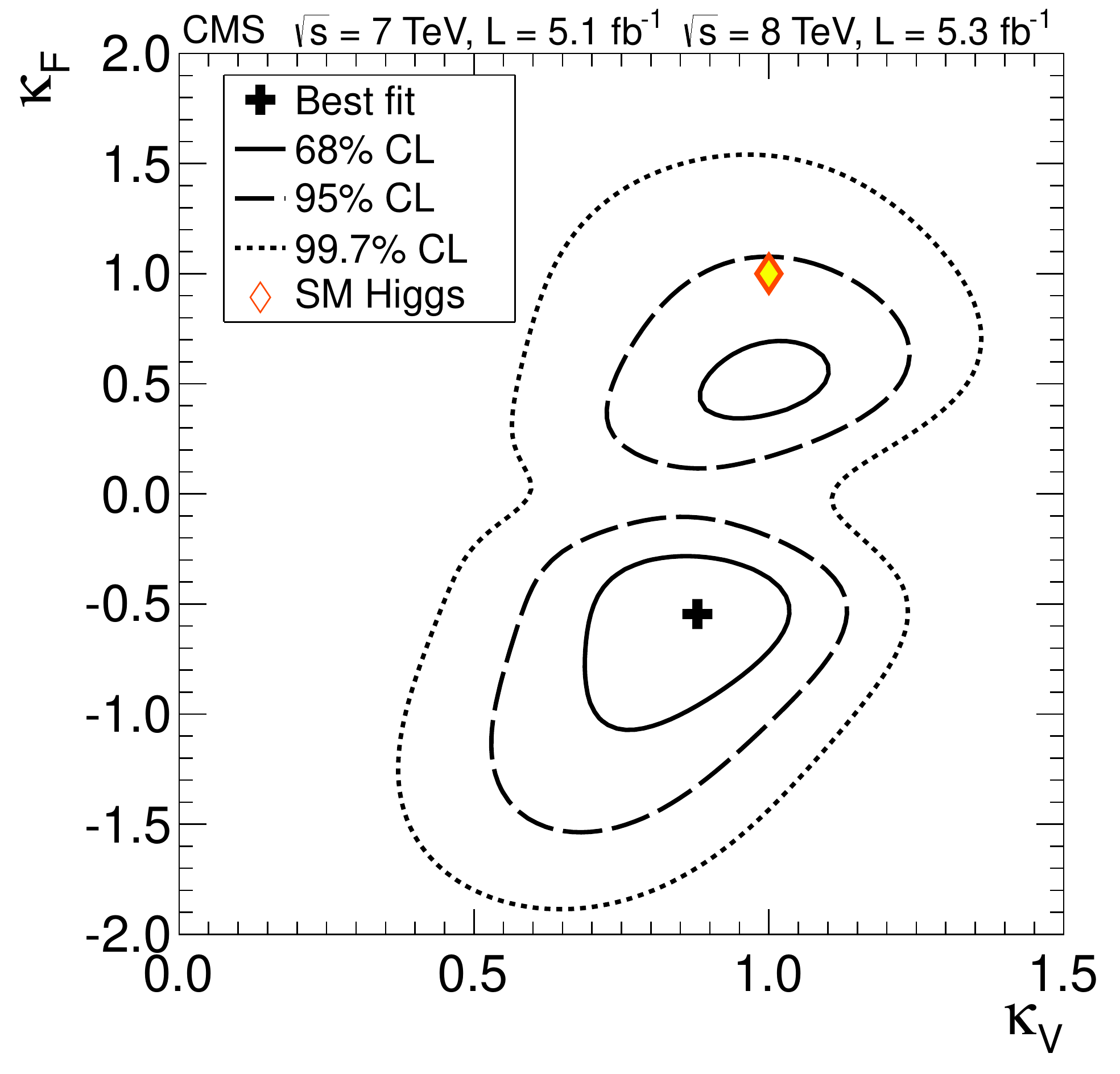} \hfill
\includegraphics[width=0.49\textwidth]{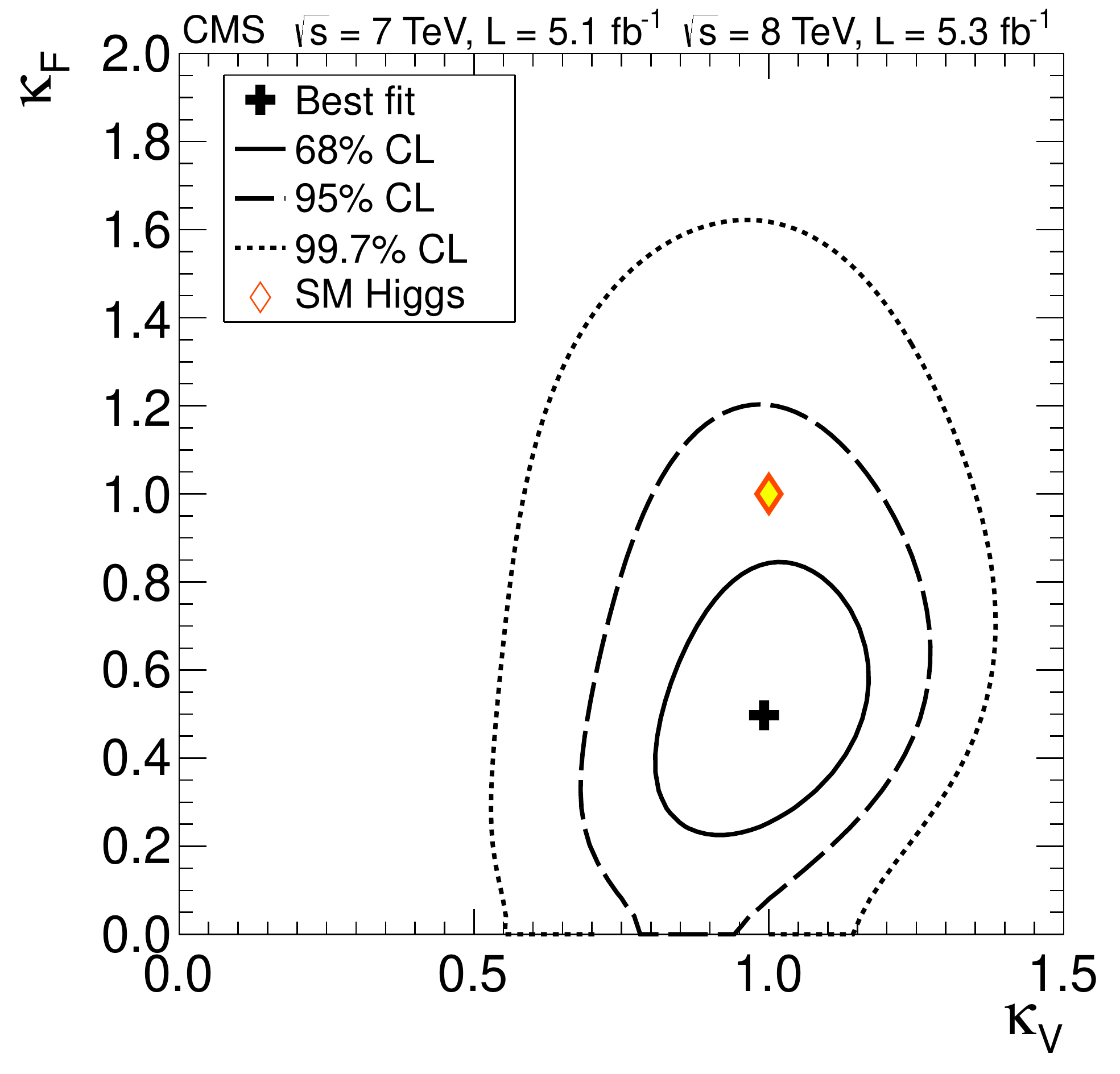}
\caption{
The likelihood test statistic in the  $\kappa_V$ versus $\kappa_F$ plane.
The cross indicates the best-fit values. The solid, dashed, and dotted lines  show the
68\%, 95\%,  and 99.7\% CL contours, respectively.
The diamond shows the SM point
$(\kappa_V, \kappa_F)$ = (1, 1). The left plot allows for different signs of $\kappa_V$ and $\kappa_F$, while the
right plot constrains them both to be positive.
}
\label{fig:cVcF_2D}
\end{figure*}

Figure~\ref{fig:cVcF_2D} shows the 2D likelihood test statistic
over the $(\kappa_V,\,\kappa_F)$ plane.
The left plot allows for different signs of $\kappa_V$ and $\kappa_F$, while the
right plot constrains both of them to be positive.
The 68\%,  95\%, and 99.7\% CL contours
are shown by the solid, dashed, and dotted lines, respectively.
The global minimum in the left plot
occurs in the $(+,-)$ quadrant, which is due to the observed excess in the $\Pgg\Pgg$ channel.
If the relative sign between
$\kappa_V$ and $\kappa_F$ is negative, the interference term between
the $\PW$ and top-quark loops responsible for the $\PH \to \Pgg\Pgg$ decays becomes positive
and helps boost the $\Pgg\Pgg$ branching fraction.
However, the difference between the global minimum in the $(+,-)$ quadrant and
the local minimum in the $(+,+)$ quadrant is not statistically significant
since the 95\% CL contours encompass both of them.
The data are consistent with the expectation for the
SM Higgs boson: the point at ($\kappa_V,\kappa_F$) = (1, 1), shown by the diamond, is within the 95\% CL contour.
Any significant deviation from ($\kappa_V,\kappa_F$) = (1, 1) would imply BSM physics, with
the magnitude and sign of the $\kappa_V$ and $\kappa_F$ measurements
providing a clue to the most plausible BSM scenarios.

\begin{figure*}
\centering
\resizebox{!}{0.44\textwidth}{\includegraphics{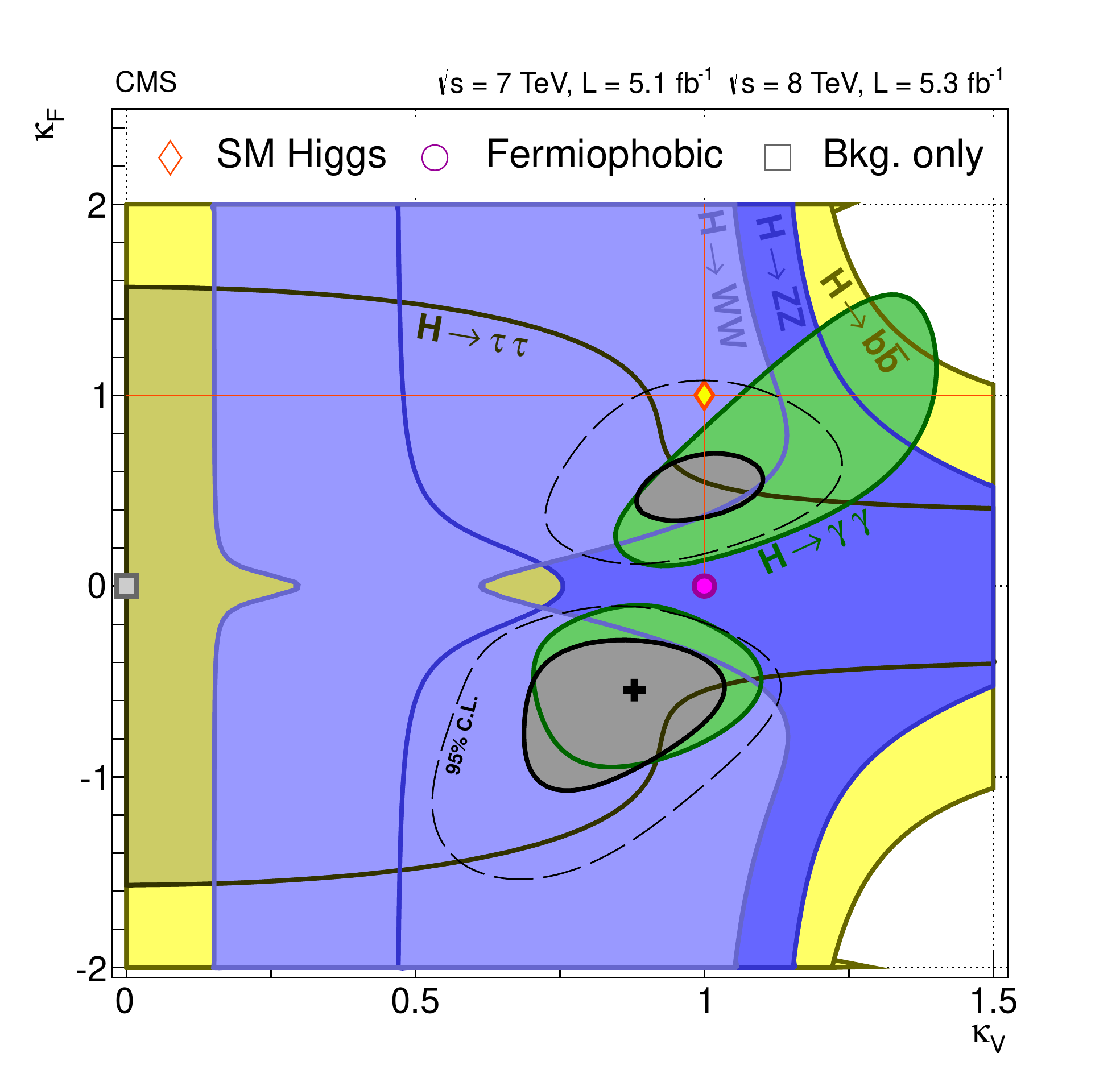}} \hfill
\resizebox{!}{0.44\textwidth}{\includegraphics{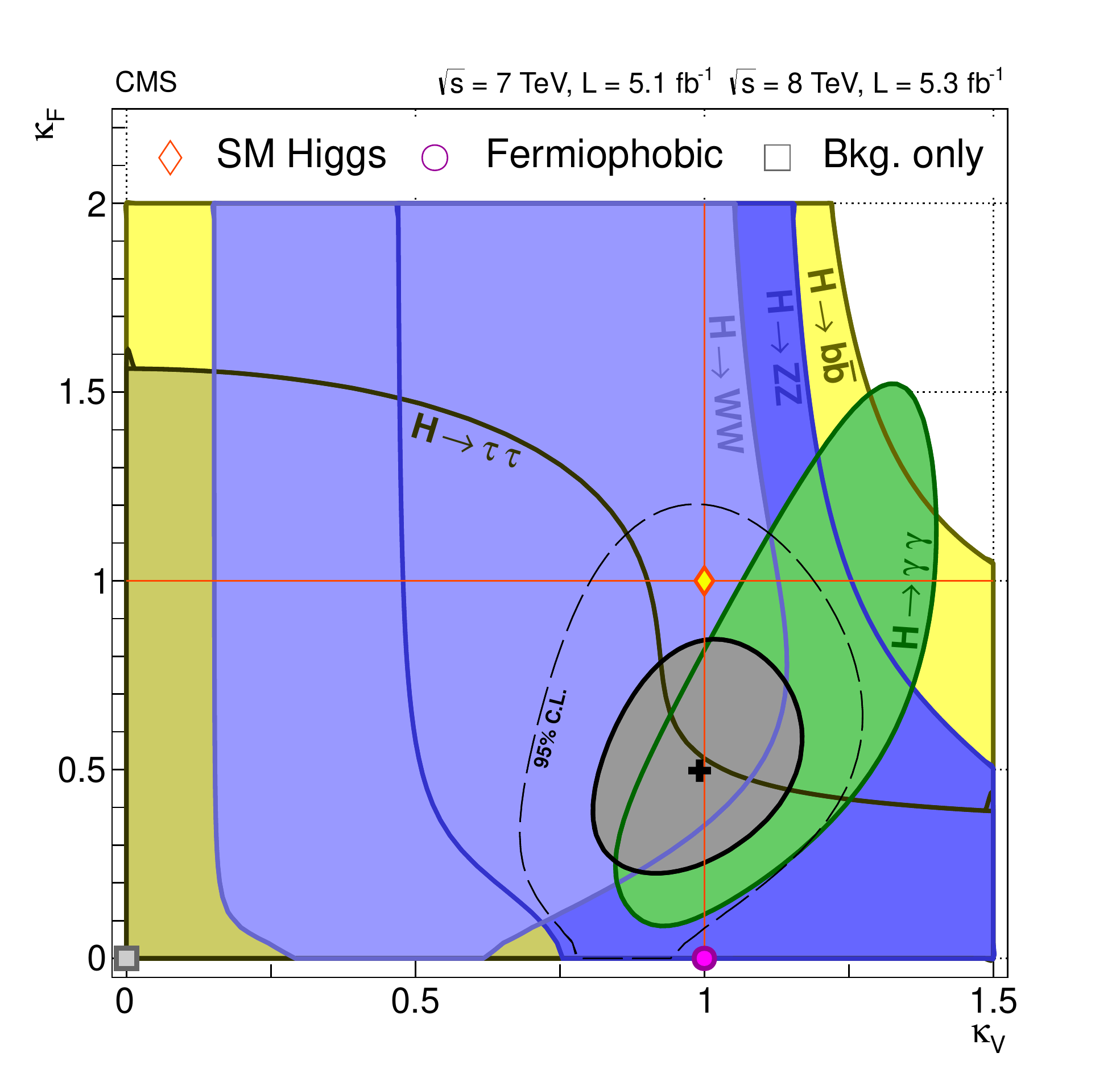}}
\caption{
The 68\% CL contours for the test statistic in the $(\kappa_V$ versus
$\kappa_F)$ plane for individual channels (coloured regions) and the
overall combination (solid thick lines).
The thin dashed lines show the 95\% CL range for the overall combination.
The black cross indicates the global best-fit values. The diamond shows the SM Higgs boson
point $(\kappa_V, \kappa_F)$ = (1, 1).
The point $(\kappa_V, \kappa_F)$ = (1, 0), indicated by the circle, corresponds to the fermiophobic Higgs boson scenario.
The left plot allows for different signs of $\kappa_V$ and $\kappa_F$, while the
right plot constrains them both to be positive.
}
\label{fig:cVcF_subchannels}
\end{figure*}

Figure~\ref{fig:cVcF_subchannels} displays the corresponding 68\% and 95\% contours of $\kappa_V$ versus $\kappa_F$
from each of the individual decay modes,
restricting the parameters to the $(+,+)$ and $(+,-)$ quadrants (left), and the $(+,+)$ quadrant (right).
The hypothesis of a ``fermiophobic'' Higgs boson that couples only to bosons
is represented by the point at (1, 0). The point is just outside the 95\% CL contour,
which implies that  a fermiophobic Higgs boson with \mH\ = 125.5\GeV is excluded at 95\% CL.

\begin{figure*}
\centering
\includegraphics[width=0.49\textwidth]{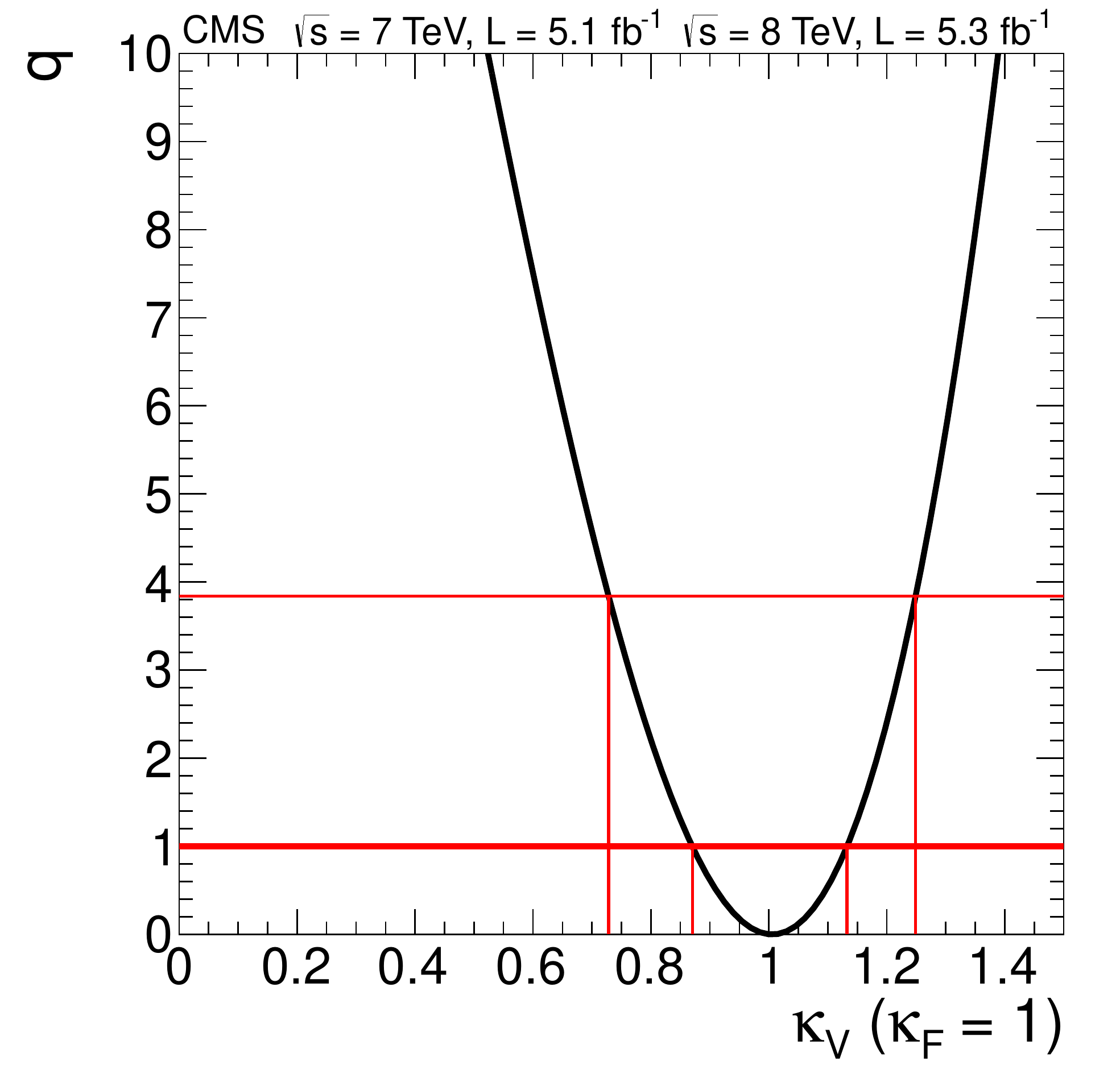} \hfill
\includegraphics[width=0.49\textwidth]{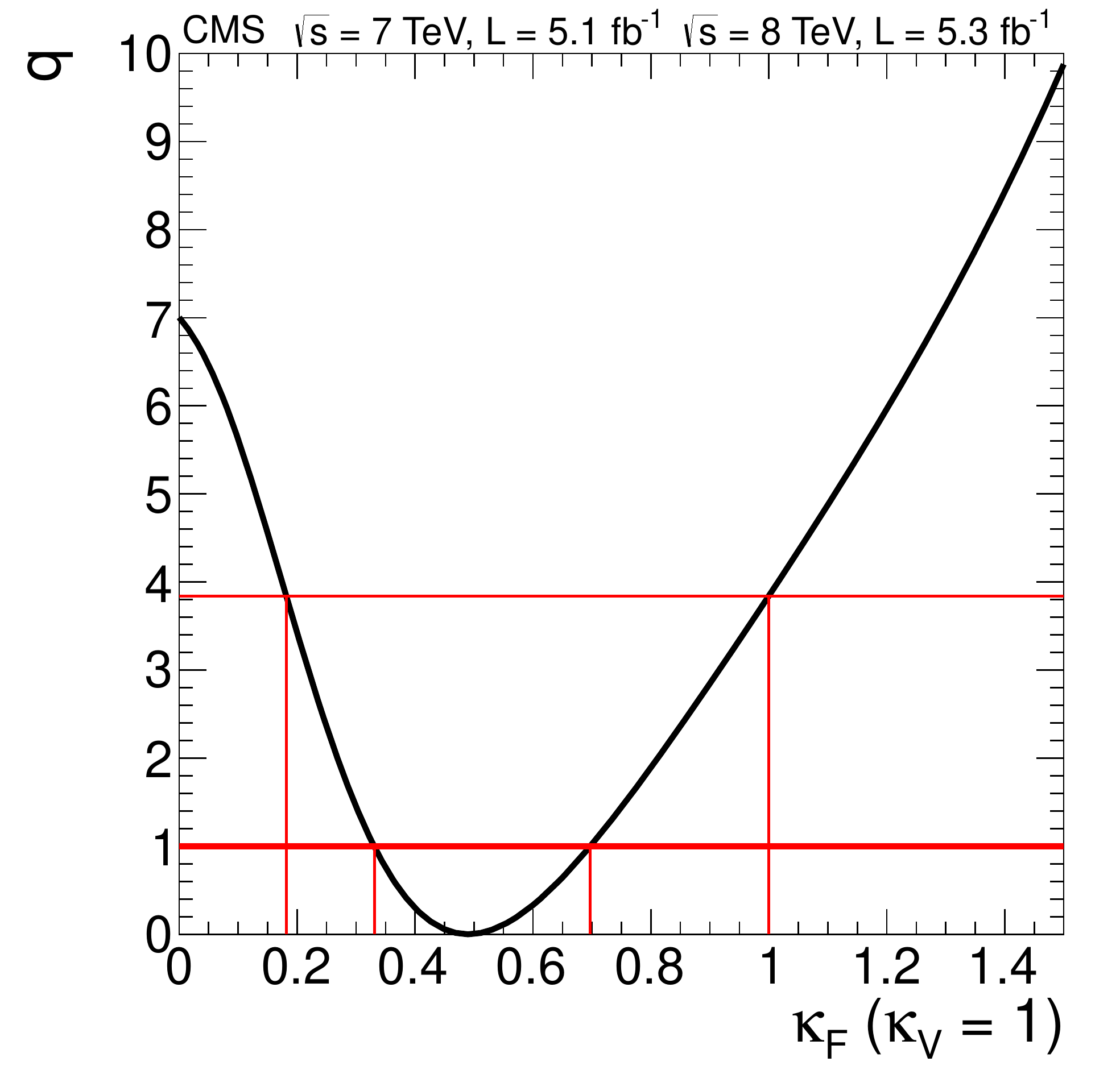}
\caption{
The likelihood test statistic $q(\kappa_V;\kappa_F=1)$ (left) and $q(\kappa_F;\kappa_V=1)$ (right).
The intersections with the horizontal lines $q=1$ and $q=3.84$ mark the 68\% and 95\% CL intervals, respectively,
as shown by the vertical lines.
}
\label{fig:cVcF_1D}
\end{figure*}

The 1D likelihood scans versus $\kappa_V$ and $\kappa_F$,
setting one parameter at a time to the SM value of 1,
are given in the left and right plots of Fig.~\ref{fig:cVcF_1D}, respectively.
The resulting fit values are:
$\kappa_V =1.00 \pm 0.13$ and
$\kappa_F= 0.5 \pm 0.2$,
where the uncertainties are combined statistical and
systematic, with corresponding
95\% CL intervals of \CVNF\ and \CFNF, respectively.

\textit{C. Test for the presence of BSM particles}

The presence of BSM  particles can considerably modify the Higgs boson phenomenology,
even if the underlying Higgs boson sector in the model remains unaltered.
Processes induced by loop diagrams ($\PH \to \Pgg\Pgg$ and $\cPg\cPg \to \PH$)
can be particularly sensitive to the presence of new particles.
Therefore, we combine and fit the data to the scale factors $\kappa_{\Pgg}$ and $\kappa_\cPg$
for these two processes. The partial widths associated with the tree-level production processes
and decay modes are assumed to be unaltered.

\begin{figure*}[bhtp]
\centering
\includegraphics[width=0.49\textwidth]{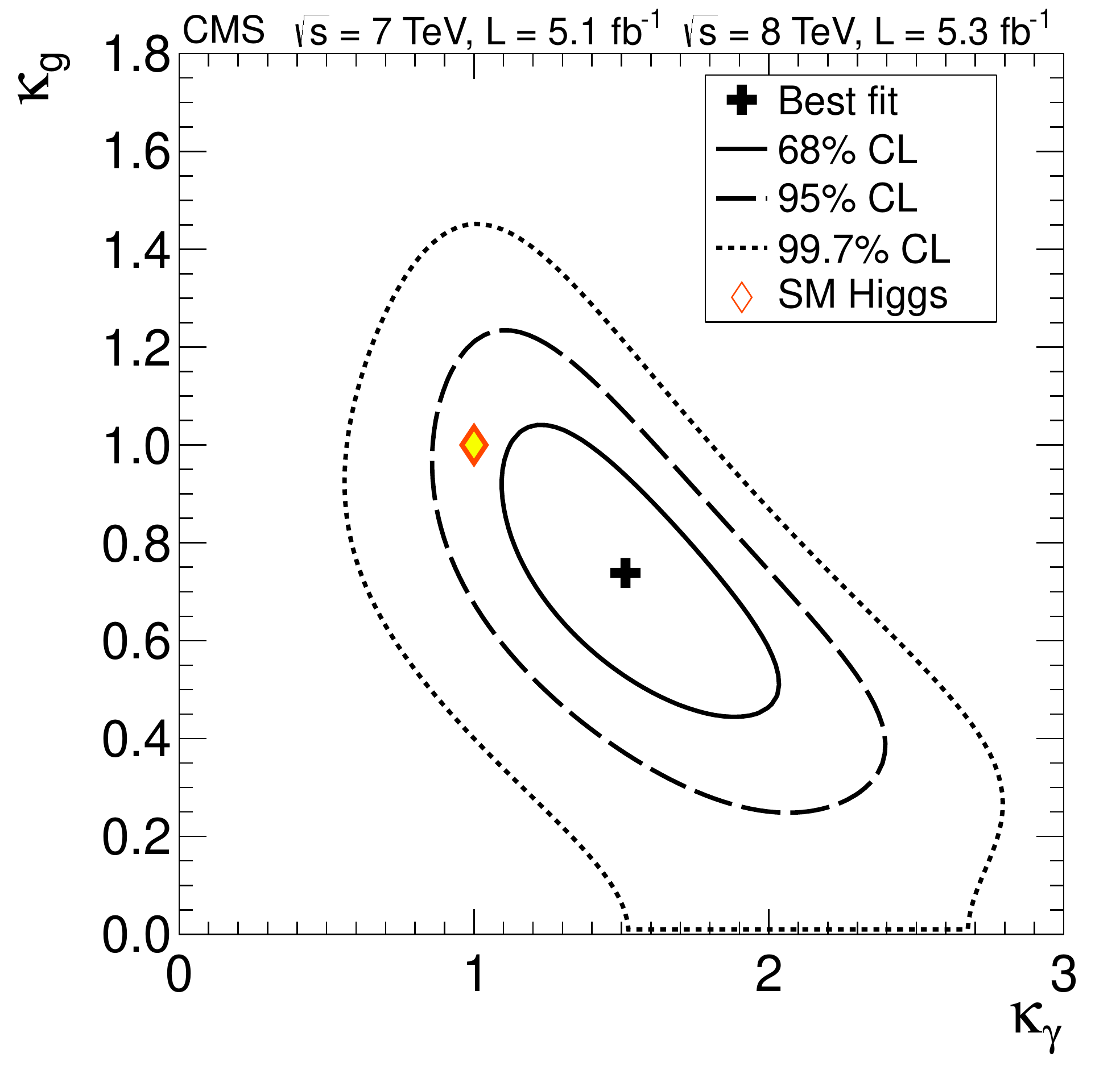}
\caption{
The likelihood test statistic  $q(\kappa_{\Pgg}, \kappa_\cPg)$ assuming $\Gamma_{\mathrm{BSM}}=0$.
        The cross indicates the best-fit values. The solid, dashed, and dotted contours show the
        68\%, 95\%,  and 99.7\% CL contours, respectively. The diamond shows the SM point
        $(\kappa_{\Pgg}, \kappa_\cPg)$ = (1, 1).
The partial widths associated with the tree-level production processes
and decay modes are assumed to be unaltered ($\kappa = 1$).
}
\label{fig:BSM1}
\end{figure*}

Figure~\ref{fig:BSM1} displays  the likelihood test statistic in the $\kappa_\cPg$ versus $\kappa_{\Pgg}$ plane, under
the assumption that $\Gamma_{\mathrm{BSM}}=0$.
The results are consistent with the expectation for the SM Higgs boson
of $(\kappa_{\Pgg}, \kappa_\cPg)=(1, \,1)$.
The best-fit value is $(\kappa_{\Pgg}, \kappa_\cPg)=(1.5,\,0.75)$.

\begin{figure*}[bhtp]
\centering
\includegraphics[width=0.49\textwidth]{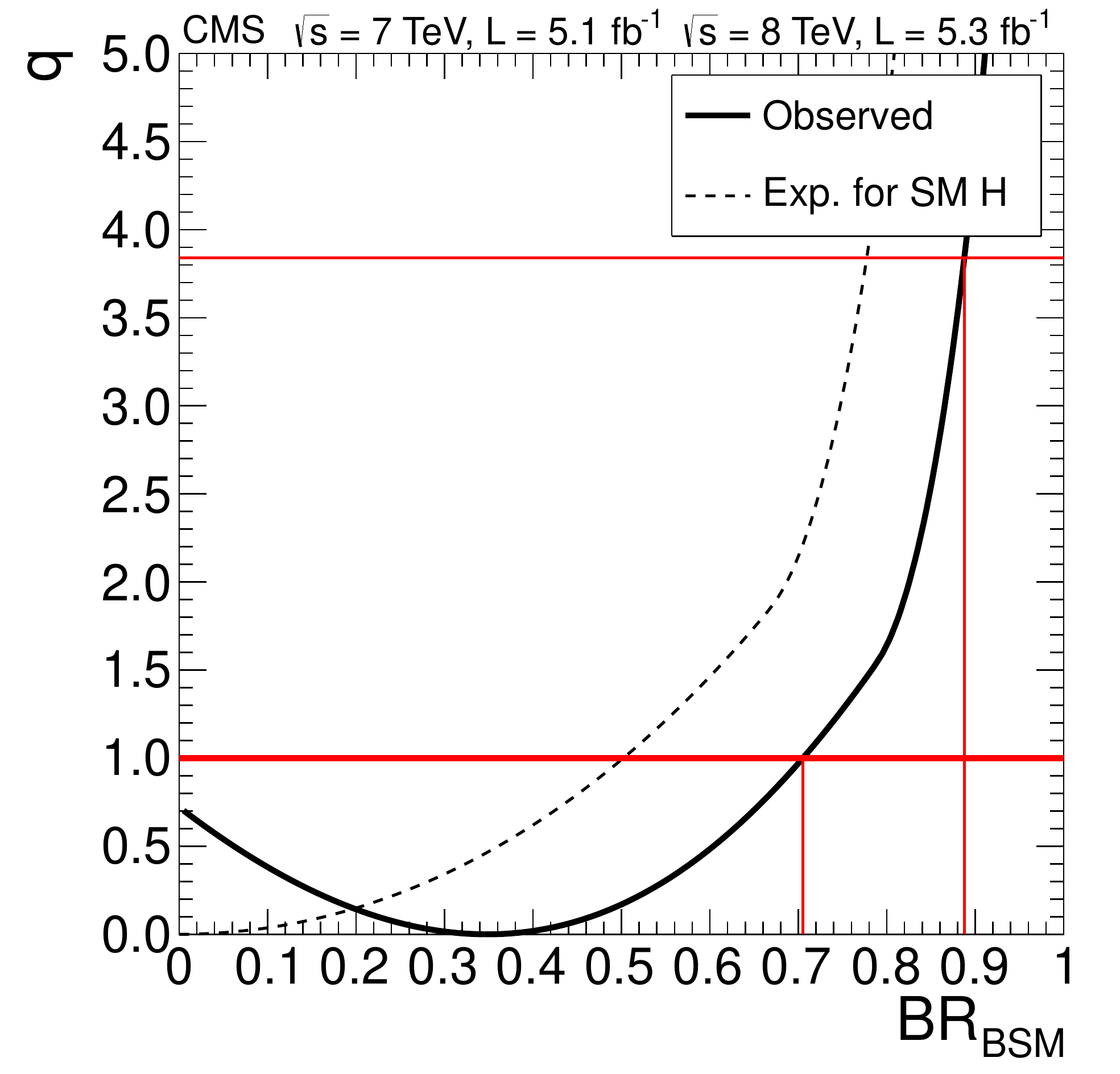}
\caption{
The likelihood test statistic $q$ versus $\mathrm{BR}_{\mathrm{BSM}}=\Gamma_{\mathrm{BSM}}/\Gamma_{\mathrm{tot}}$,
        with the parameters $\kappa_\cPg$ and $\kappa_{\Pgg}$ included as  nuisance parameters.
        The solid curve is the data; the dashed curve indicates the expected median results in the presence of the SM Higgs boson.
The intersections with the horizontal lines $q=$ 1 and 3.8 give the 68\% and 95\% CL intervals, respectively.
The partial widths associated with the tree-level production processes
and decay modes are assumed to be unaltered ($\kappa = 1$).
}
\label{fig:BSM2}
\end{figure*}

Figure~\ref{fig:BSM2} gives the likelihood test statistic versus
$\mathrm{BR}_{\mathrm{BSM}}=\Gamma_{\mathrm{BSM}}/\Gamma_{\text{tot}}$,
with $\kappa_\cPg$ and $\kappa_{\Pgg}$ included as unconstrained nuisance parameters.
The resulting 95\% CL upper limit is  $\mathrm{BR}_{\mathrm{BSM}} < 0.89$.

\textit{D. Test for differences  in the couplings to fermions}

In  two-Higgs-boson doublet models (2HDM) \cite{Branco:2011iw}, the couplings of the neutral Higgs bosons to fermions
can be substantially modified with respect to the Yukawa couplings of the SM Higgs boson. For example,
in the minimal supersymmetric model (MSSM), the couplings of the neutral Higgs bosons to up-type and down-type fermions are modified,
with the modification being the same for all three generations and for quarks and leptons. In more
general 2HDMs, leptons can be nearly decoupled from the Higgs boson that otherwise would behave
like a SM Higgs boson with respect to the $\PW$ and $\cPZ$ bosons and the quarks.
To test for such modifications to the fermion couplings,
we evaluate two different combinations of the corresponding parameters: one in which we allow
different ratios of couplings to the up- and down-type fermions
($\lambda_{\mathrm{du}} = \kappa_{\mathrm{d}} / \kappa_{\mathrm{u}}$),
and the other where we allow different ratios of the couplings to the leptons and quarks
($\lambda_{\ell\mathrm{q}} = \kappa_{\ell} / \kappa_{\mathrm{q}}$).
We assume that $\Gamma_{\mathrm{BSM}}=0$.

\begin{figure*}[bhtp]
\centering
\includegraphics[width=0.49\textwidth]{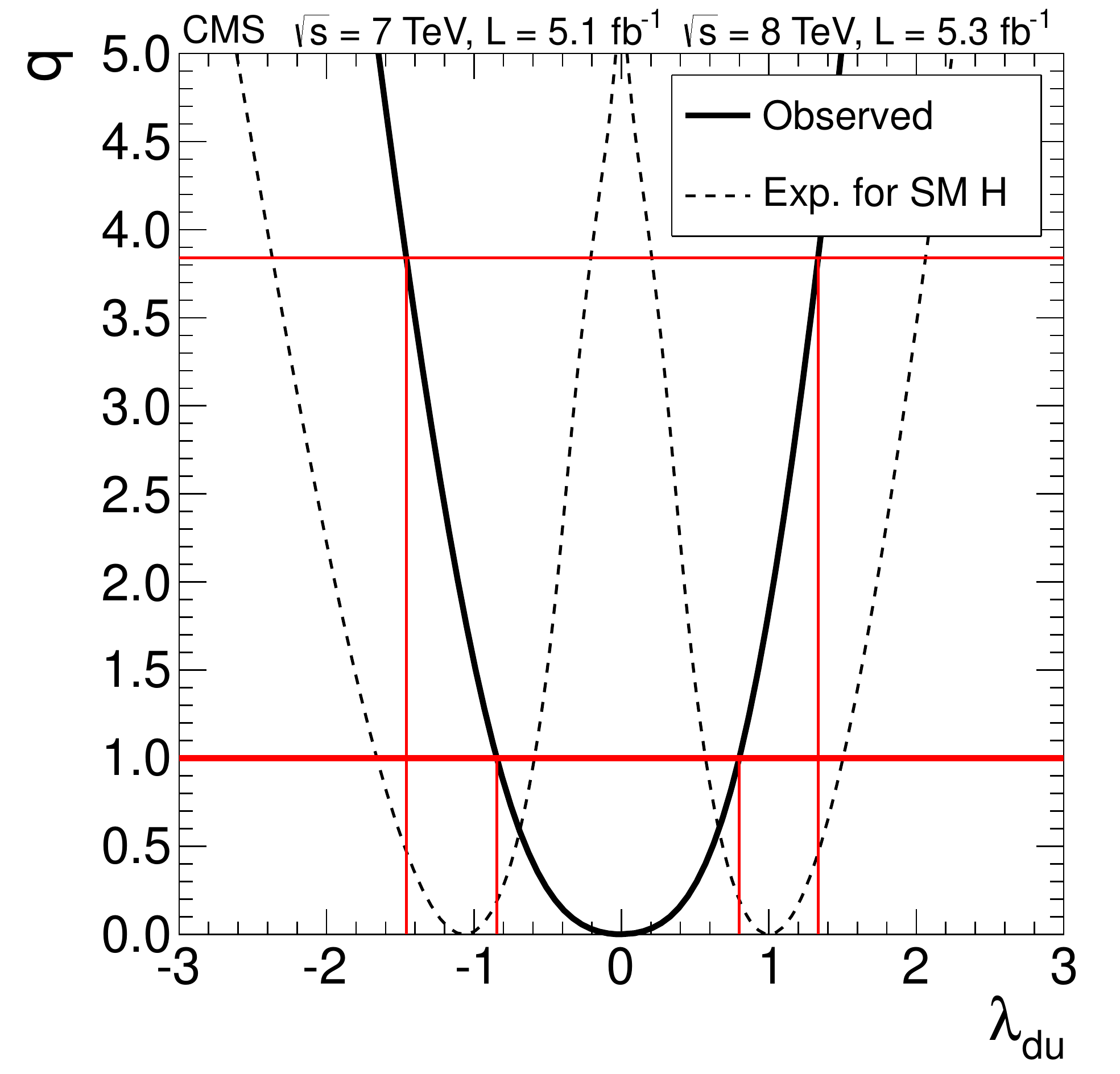} \hfill
\includegraphics[width=0.49\textwidth]{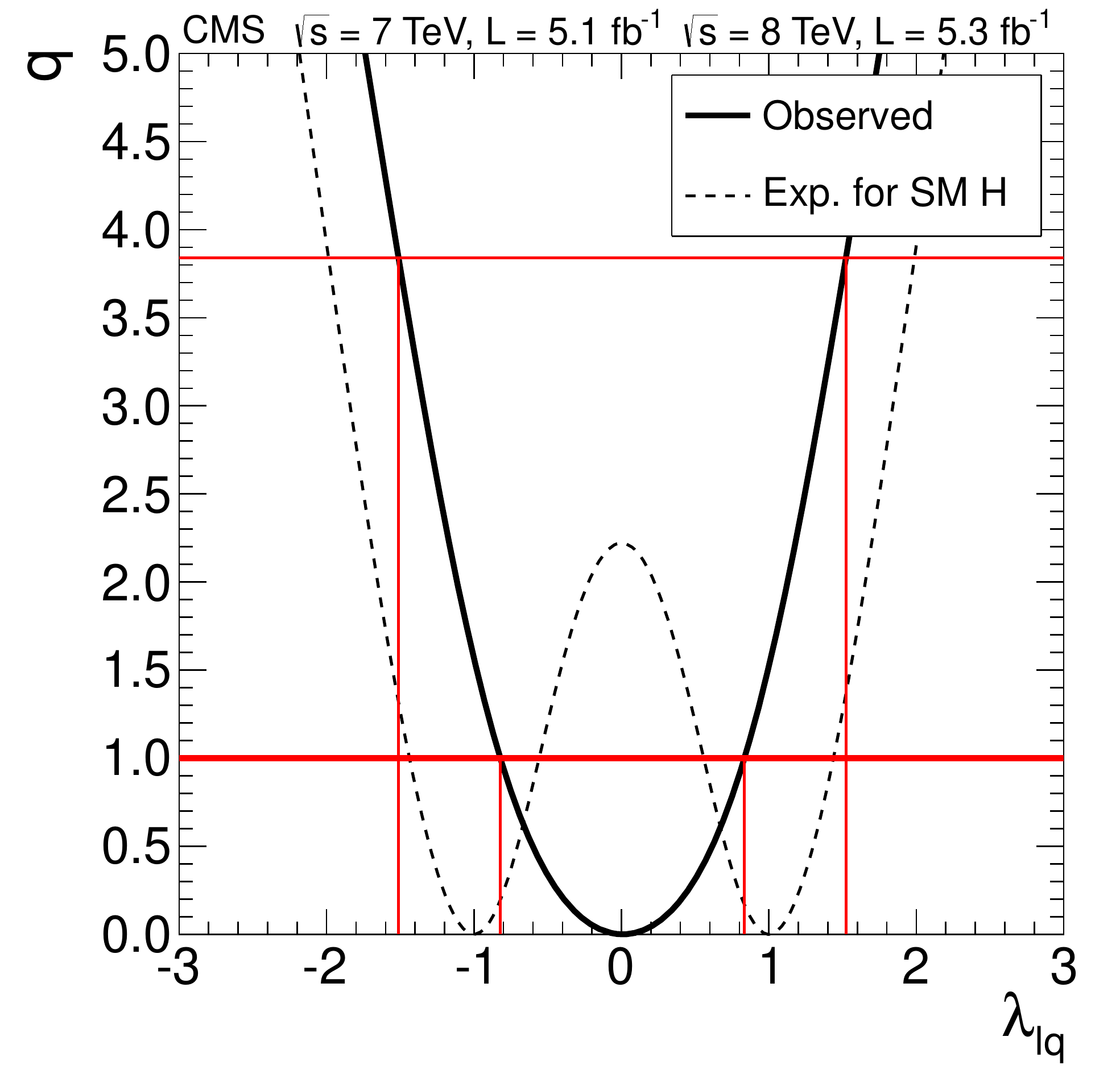}
\caption{
(Left)  Likelihood test statistic $q$ as a function of
the ratio $\lambda_{\mathrm{du}}$ of the coupling to the up- and down-type fermions
with the coupling modifiers $\kappa_V$ and $\kappa_{\mathrm{u}}$
treated as nuisance parameters.
(Right) The likelihood test statistic as a function of
the ratio $\lambda_{\ell \mathrm{q}}$ of the couplings to leptons and quarks
with the coupling modifiers $\kappa_V$ and $\kappa_{\mathrm{q}}$
treated as nuisance parameters. The solid curves are the results from the
data. The dashed curves show the expected distributions for the SM Higgs boson. The intersection
of the curves with the horizontal lines $q=$1 and 3.8 give the 68\% and 95\% CL intervals, respectively.
}
\label{fig:fit_ldu_llq_scan}
\end{figure*}

Figure~\ref{fig:fit_ldu_llq_scan} (left) shows the resulting test statistic versus
$\lambda_{\mathrm{du}}$, with the other free coupling modifiers,
$\kappa_V$ and $\kappa_{\mathrm{u}}$, included as
unconstrained nuisance parameters. The relative sign between the couplings to up- and down-type fermions
is nearly degenerate, which manifests itself in the left-right symmetry
observed in the plot. The symmetry is not perfect since there is some sensitivity
to the sign of $\lambda_{\mathrm{du}}$ because of the nonvanishing role
of the $\cPqb$ quark (in comparison to the top quark) in generating the Higgs boson coupling to gluons.
Figure~\ref{fig:fit_ldu_llq_scan} (right) displays the corresponding results versus
$\lambda_{\ell \mathrm{q}}$, with the two coupling modifiers,
$\kappa_V$ and $\kappa_{\mathrm{q}}$,
treated as unconstrained nuisance parameters. There are no loop-induced processes
measurably sensitive to the relative sign of the couplings to leptons and quarks; hence,
the plot exhibits a perfect left-right symmetry.
Both $| \lambda_{\mathrm{du}} |$ and $| \lambda_{\ell \mathrm{q}} |$ are
consistent with 0 and 1, with a 95\% CL upper limit of 1.5 for both.
The main reason for both parameters  having their best-fit values close to 0
is the lack of any event excess in the $H \to \Pgt\Pgt$
channel. However, neither the $H \to \Pgt\Pgt$
nor the $H \to \cPqb\cPqb$ channels have reached sufficient sensitivity to place strong
constraints on the parameters associated with the corresponding Higgs boson couplings.

\section{Summary}\label{sec:Conclusion}

In this paper, the analyses that were the basis for the discovery of a new
boson at a mass of approximately 125\GeV have been described in detail.
The data were collected by the CMS experiment
at the LHC in proton-proton collisions at $\sqrt{s}=7$~and~8\TeV, corresponding
to integrated luminosities of up to 5.1\fbinv and 5.3\fbinv, respectively.

The particle is observed in the search for the SM Higgs boson
using five decay modes $\Pgg\Pgg$, $\cPZ\cPZ$, $\PW\PW$, $\Pgt\Pgt$, and $\cPqb\cPqb$.
An excess of events is found above the expected background, with a local significance of
5.0$\sigma$, signaling the production of a new particle.  The expected
significance for a SM Higgs boson of that mass is 5.8$\sigma$.

The excess is most significant in the two decay modes with the best mass resolution, $\Pgg\Pgg$ and
$\cPZ\cPZ \to 4\ell$, and a fit to these invariant-mass peaks gives a mass of $125.3\pm 0.4\stat\pm
0.5\syst\GeV$.  The decay to two photons indicates that the new particle is a boson
with spin different from one.
Within the SM hypothesis, the couplings of the new particle to vector bosons, fermions,
gluons, and photons have been measured. All the results are consistent, within their uncertainties, with
expectations for a SM Higgs boson.
More data are needed to ascertain whether the properties of this new state imply physics beyond the SM.

\section*{Acknowledgements}
\hyphenation{Bundes-ministerium Forschungs-gemeinschaft Forschungs-zentren} We congratulate our colleagues in the CERN accelerator departments for the excellent performance of the LHC and thank the technical and administrative staffs at CERN and at other CMS institutes for their contributions to the success of the CMS effort. In addition, we gratefully acknowledge the computing centres and personnel of the Worldwide LHC Computing Grid for delivering so effectively the computing infrastructure essential to our analyses. Finally, we acknowledge the enduring support for the construction and operation of the LHC and the CMS detector provided by the following funding agencies: the Austrian Federal Ministry of Science and Research and the Austrian Science Fund; the Belgian Fonds de la Recherche Scientifique, and Fonds voor Wetenschappelijk Onderzoek; the Brazilian Funding Agencies (CNPq, CAPES, FAPERJ, and FAPESP); the Bulgarian Ministry of Education, Youth and Science; CERN; the Chinese Academy of Sciences, Ministry of Science and Technology, and National Natural Science Foundation of China; the Colombian Funding Agency (COLCIENCIAS); the Croatian Ministry of Science, Education and Sport; the Research Promotion Foundation, Cyprus; the Ministry of Education and Research, Recurrent financing contract SF0690030s09 and European Regional Development Fund, Estonia; the Academy of Finland, Finnish Ministry of Education and Culture, and Helsinki Institute of Physics; the Institut National de Physique Nucl\'eaire et de Physique des Particules~/~CNRS, and Commissariat \`a l'\'Energie Atomique et aux \'Energies Alternatives~/~CEA, France; the Bundesministerium f\"ur Bildung und Forschung, Deutsche Forschungsgemeinschaft, and Helmholtz-Gemeinschaft Deutscher Forschungszentren, Germany; the General Secretariat for Research and Technology, Greece; the National Scientific Research Foundation, and National Office for Research and Technology, Hungary; the Department of Atomic Energy and the Department of Science and Technology, India; the Institute for Studies in Theoretical Physics and Mathematics, Iran; the Science Foundation, Ireland; the Istituto Nazionale di Fisica Nucleare, Italy; the Korean Ministry of Education, Science and Technology and the World Class University program of NRF, Republic of Korea; the Lithuanian Academy of Sciences; the Mexican Funding Agencies (CINVESTAV, CONACYT, SEP, and UASLP-FAI); the Ministry of Science and Innovation, New Zealand; the Pakistan Atomic Energy Commission; the Ministry of Science and Higher Education and the National Science Centre, Poland; the Funda\c{c}\~ao para a Ci\^encia e a Tecnologia, Portugal; JINR (Armenia, Belarus, Georgia, Ukraine, Uzbekistan); the Ministry of Education and Science of the Russian Federation, the Federal Agency of Atomic Energy of the Russian Federation, Russian Academy of Sciences, and the Russian Foundation for Basic Research; the Ministry of Science and Technological Development of Serbia; the Secretar\'{\i}a de Estado de Investigaci\'on, Desarrollo e Innovaci\'on and Programa Consolider-Ingenio 2010, Spain; the Swiss Funding Agencies (ETH Board, ETH Zurich, PSI, SNF, UniZH, Canton Zurich, and SER); the National Science Council, Taipei; the Thailand Center of Excellence in Physics, the Institute for the Promotion of Teaching Science and Technology of Thailand and the National Science and Technology Development Agency of Thailand; the Scientific and Technical Research Council of Turkey, and Turkish Atomic Energy Authority; the Science and Technology Facilities Council, UK; the US Department of Energy, and the US National Science Foundation.

Individuals have received support from the Marie-Curie programme and the European Research Council and EPLANET (European Union); the Leventis Foundation; the A. P. Sloan Foundation; the Alexander von Humboldt Foundation; the Belgian Federal Science Policy Office; the Fonds pour la Formation \`a la Recherche dans l'Industrie et dans l'Agriculture (FRIA-Belgium); the Agentschap voor Innovatie door Wetenschap en Technologie (IWT-Belgium); the Ministry of Education, Youth and Sports (MEYS) of Czech Republic; the Council of Science and Industrial Research, India; the Compagnia di San Paolo (Torino); and the HOMING PLUS programme of Foundation for Polish Science, cofinanced from European Union, Regional Development Fund.

\bibliography{auto_generated}
\cleardoublepage \appendix\section{The CMS Collaboration \label{app:collab}}\begin{sloppypar}\hyphenpenalty=5000\widowpenalty=500\clubpenalty=5000\textbf{Yerevan Physics Institute,  Yerevan,  Armenia}\\*[0pt]
S.~Chatrchyan, V.~Khachatryan, A.M.~Sirunyan, A.~Tumasyan
\vskip\cmsinstskip
\textbf{Institut f\"{u}r Hochenergiephysik der OeAW,  Wien,  Austria}\\*[0pt]
W.~Adam, T.~Bergauer, M.~Dragicevic, J.~Er\"{o}, C.~Fabjan\cmsAuthorMark{1}, M.~Friedl, R.~Fr\"{u}hwirth\cmsAuthorMark{1}, V.M.~Ghete, N.~H\"{o}rmann, J.~Hrubec, M.~Jeitler\cmsAuthorMark{1}, W.~Kiesenhofer, V.~Kn\"{u}nz, M.~Krammer\cmsAuthorMark{1}, I.~Kr\"{a}tschmer, D.~Liko, I.~Mikulec, D.~Rabady\cmsAuthorMark{2}, B.~Rahbaran, C.~Rohringer, H.~Rohringer, R.~Sch\"{o}fbeck, J.~Strauss, A.~Taurok, W.~Treberer-treberspurg, W.~Waltenberger, C.-E.~Wulz\cmsAuthorMark{1}
% ...
\vskip\cmsinstskip
% ...
1:~~Also at Vienna University of Technology, Vienna, Austria\\
2:~~Also at CERN, European Organization for Nuclear Research, Geneva, Switzerland\\
% ...

\end{sloppypar}
\end{document}